\documentclass[twocolumn]{aa}  

\usepackage{txfonts}
\usepackage[colorlinks,linkcolor=blue,anchorcolor=blue,citecolor=blue]{hyperref}
\usepackage{graphicx}
\usepackage{textgreek} 
\usepackage{subcaption}
\usepackage{rotating}
\usepackage{longtable}
\usepackage{threeparttablex} 
\usepackage{booktabs}        
\usepackage{lscape}
\usepackage{multirow}
\usepackage{pifont} 
\usepackage{placeins}
\usepackage{nccmath}
\usepackage{amssymb}
\usepackage{supertabular}
\usepackage{tikz}
\usepackage{transparent}
\usepackage{wasysym} 
\usepackage{float}
\usepackage{graphicx}
\usepackage{subcaption}
\usepackage{textcomp}
\usepackage{textgreek}
\usepackage{enumitem} 
\usepackage{float}

\usepackage{tikz}

\newcommand{\cmark}{\ding{51}}  
\newcommand{\xmark}{\ding{55}}  

\def\ion#1#2{#1$\;${\footnotesize\rm{#2}}\relax}

\usepackage{xcolor, soul}
\sethlcolor{green}

\usepackage{bm}
\begin{document}

   \title{eROSITA-RU Tidal Disruption Events with Keck-I/LRIS: Sample Selection, Optical Properties, and Host Galaxy Demographics}
   \titlerunning{eROSITA TDEs with Keck-I/LRIS}

    \author{Zirui Zhang\inst{1,2}
    \and Yuhan Yao\inst{3,1,4}
    \and Marat Gilfanov\inst{5,6}
    \and Sergey Sazonov\inst{5}
    \and Pavel Medvedev\inst{5}
    \and Georgii Khorunzhev\inst{5}
    \and Rashid  Sunyaev\inst{5,6}
    \and Vikram Ravi\inst{7}
    \and S. R. Kulkarni\inst{7}
    \and Jean Somalwar\inst{1,4}
    \and Ryan Chornock\inst{1,4}
    \and Ilfan Bikmaev\inst{8}
    \and Mark A. Gorbachev\inst{8}
    }
    
    \institute{
    Department of Astronomy, University of California, Berkeley, CA 94720-3411, USA
    \and
    College of Physics, Sichuan University, Chengdu, 610065, China
    \and
    Miller Institute for Basic Research in Science, 206B Stanley Hall, Berkeley, CA 94720, USA\\
    \email{yuhanyao@berkeley.edu}
    \and
    Berkeley Center for Multi-messenger Research on Astrophysical Transients and Outreach (Multi-RAPTOR), University of California, Berkeley, CA 94720, USA
    \and 
    Space Research Institute (IKI), Russian Academy of Sciences, Profsoyuznaya 84/32, Moscow 117997, Russia
    \and
    Max-Planck-Institut f\"{u}r Astrophysik, Karl-Schwarzschild-Str. 1, D-85741 Garching, Germany
    \and
    Cahill Center for Astrophysics, California Institute of Technology, MC 249-17, 1200 E California Boulevard, Pasadena, CA 91125, USA
    \and 
    Kazan Federal University,  Kremlevskaya Str., 18, Kazan, 420008, Russia
    }

   \date{}

  \abstract
   {We select seventy tidal disruption event (TDE) candidates among X-ray transients discovered during the eROSITA all-sky surveys in the Eastern Galactic hemisphere ($0^\circ < l < 180^\circ$) between December 2020 and February 2022 (eRASS1--5). }
   {We perform a systematic analysis of this sample to characterize the properties of their optical counterparts and host galaxies.}
  {We cross-match each X-ray source to a host galaxy in archival optical surveys using Bayesian likelihood-ratio techniques and obtain Keck/LRIS spectroscopy for all 70 host galaxies. Host properties are inferred through SED fitting with \texttt{Prospector} and emission line analysis with \texttt{pPXF}. We develop a robust classification scheme using X-ray and broad line luminosities, narrow-line ionization diagnostics, and optical variability to identify high-confidence TDEs, for which we analyze optical spectral features, light curve properties, and host galaxy demographics.}
  {Our final sample contains 52 TDEs with redshifts of $0.018 \leq z\leq0.714$, comprising 41 gold (high-confidence) and 11 silver (lower-confidence) events. The vast majority (93\%) of gold TDEs are intrinsically brighter in the X-ray band, with $L_{\rm X,peak} > L_{\rm opt,peak}$. Among 23 events with detected optical flares, delayed X-ray peak is commonly observed. We identify transient spectral features in eight events, including six with prominent broad \ion{He}{II} $\lambda$4686 and/or H$\alpha$ emission and two coronal-line TDEs. Host galaxy demographics reveal modest over-representation in green valley ($\times1.8$) and quiescent Balmer-strong ($\times5.3$) galaxies, significantly weaker than previous TDE samples, demonstrating greater diversity in star formation histories than previously recognized. Most TDE hosts exhibit suppressed star formation relative to the main sequence, consistent with X-ray selection biases against dusty star-forming galaxies.}
   {}

   \maketitle
%

\section{Introduction}

Tidal disruption events (TDEs) of stars by massive black holes (MBHs) produce a dazzling cornucopia of astrophysical phenomena. Accretion of stellar material gives rise to soft X-ray and extreme-ultraviolet (EUV) thermal flares \citep{Komossa2015}. Reprocessing of accretion emission and stream-disk shocks powers flares in the optical band \citep{Andalman2022, Ryu2023, Price2024, Steinberg2024}. Relativistic jets are launched in $\lesssim 1$\% of events, resulting in spectacular radio and hard X-ray transients \citep{Alexander2020, DeColle2020}. TDEs represent unique laboratories for studying MBH demographics \citep{Stone2020, Yao2023, Hannah2025}, as well as the accretion and jet physics of MBHs \citep{Guolo2024, Yao2024_22cmc}.

The first observational evidence for TDEs came from the detection of soft X-ray flares from the centers of quiescent galaxies during the ROSAT (0.1--2.4 keV) all-sky survey (RASS) in 1990--1991 \citep{Bade1996, Komossa1999}. Their X-ray luminosity was observed to increase over $\sim\!1$\,month and decay over years, with the decline rate generally following the theoretically expected mass fall-back rate of $\propto t^{-5/3}$ \citep{Rees1988, Phinney1989}. Subsequently, using RASS as a reference, the search for soft X-ray nuclear transients in the Chandra and XMM-Newton archives yielded $\sim\!20$ TDEs and TDE candidates (see \citealt{Saxton2020} for a review). 

Since $\sim\!2010$, optical sky surveys have dominated the TDE discovery rate (see, e.g.,  Fig. 1.5 of \citealt{Yao2023_thesis}). The continuum emission of these events is characterized by blackbody radiation with effective temperatures of $\sim\!20,000$\,K. Depending on the presence or absence of broad emission lines around \ion{He}{II}, \ion{He}{I}, and the Balmer series, they are broadly classified into TDE-H, TDE-He, TDE-H+He, and TDE-featureless \citep{vanVelzen2021, Hammerstein2023}. Within individual events, the evolution of relative line strengths is common, and transitions between spectral subtypes are observed \citep{Charalampopoulos_2022}. 
This spectroscopic diversity might be attributed to the stratified structure of line emitting region, where the effective photospheres of different lines arise from distinct layers. Lines can be broadened with electron scattering \citep{Roth_2016}, bulk kinetic motion in optically thick outflow \citep{Roth_2018, Nicholl_2020}, or Keplerian rotation of a disk \citep{Short2020}.

Population studies of host galaxy properties of early optical TDE samples show that the host galaxies generally show low levels of current star formation (weak H$\alpha$ emission) and substantial star formation in the last $\sim{\rm Gyr}$ (strong H$\delta$ absorption) \citep{Arcavi_2014, French2016}. These so-called quiescent Balmer-strong (QBS) or E+A galaxies appeared to dominate early TDE samples. More recent studies using the ZTF sample confirm this E+A/QBS over-representation, though the enhancement is less extreme than initially suggested \citep{Hammerstein_2021}. 
In the galaxy color-mass diagram, optical TDEs occur preferentially in green valley galaxies, with rates exceeding those in both red sequence and blue cloud populations \citep{vanVelzen2021, Hammerstein2023, Yao2023}.

A small subset of TDEs (probably those in gas-rich environments) exhibit prominent coronal emission lines and are referred to as CrL-TDE \citep{Onori2022, Newsome2024, Callow2024, Clark2025}. They form a subset of the population of extreme coronal line emitters (ECLEs; \citealt{Komossa2008, Wang2011, Wang2012}). These CrL-TDEs typically show accompanying X-ray or UV flares that ionize gas located well beyond the immediate disruption region, producing both the high-ionization coronal lines and reprocessed infrared (IR) emission \citep{Jiang2021_tde_echo, Short2023, Newsome2024, Clark2025}. As such, they are considered as light echos of TDEs. The star formation rates (SFRs) of the host galaxies of CrL-TDEs are statistically higher than that of other optically selected TDEs \citep{Hinkle2024_CrLTDE}.

The X-ray discovery of TDEs experienced a renaissance from December 12 2019 to February 26 2022, when the eROSITA (0.2--8\,keV; \citealt{Predehl2021}) telescope onboard the Spektrum-Roentgen-Gamma (SRG) mission \citep{Sunyaev2021} was conducting all-sky surveys. Each eROSITA all-sky survey (eRASS) takes six months, with eRASS1--eRASS4 fully completed and nearly 40\% of eRASS5 conducted before the telescope was put in safe mode. The 6-month survey cadence and the large grasp (product of field of view and effective area) of eROSITA allow for effective TDE search \citep{Khabibullin2014}. 
The selection of eROSITA TDE is independently performed by two consortia: eROSITA-RU for the eastern Galactic hemisphere ($0^{\circ}<l<180^{\circ}$) and eROSITA-DE for the western hemisphere ($180^{\circ} < l < 360^{\circ}$). The first reports of TDEs candidates found by eROSITA began to appear soon after the start of the all-sky survey \citep{2020ATel13494....1K,2020ATel13499....1K}. Single-object studies have revealed various phenomena, from repeating nuclear transients consistent with partial disruptions \citep{Liu2023_pTDE} to TDEs with novel optical or X-ray properties \citep{Yao_2022, Malyali2023, Malyali2024}.

While individual discoveries have revealed the diversity of eROSITA TDEs, understanding the population as a whole requires uniform selection and statistical analysis. Such selection is challenging because, unlike most optically selected TDEs with characteristic broad emission lines, X-ray–selected (and optically faint) events must be distinguished from active galactic nuclei (AGN) that can produce similar soft X-ray flares (e.g., \citealt{Brandt1995, Grupe1995}), as witnessed by eROSITA \citep{2022AstL...48..735M,2024AstL...50..744K,2025A&A...693A..62G}. Despite these challenges, both eROSITA consortia have released TDE samples that probe population-level properties. From the eROSITA-RU side, \citet{Sazonov2021} presented 13 TDEs that were undetected in eRASS1 but appeared in eRASS2 at flux levels exceeding more than ten times the upper limits of eRASS1, with follow-up optical spectroscopy confirming their non-AGN nature. 
\citet{Khorunzhev2022} presented an additional sample of 5 TDEs detected by eROSITA in eRASS1--eRASS4 and confirmed by optical spectroscopy. From the eROSITA-DE side, \citet{Grotova2025} reported 31 TDEs selected for X-ray variability amplitudes greater than four between eRASS1 and eRASS2, among which 11 sources have only photometric redshifts. Together, these studies established some key results: (1) the volumetric rate of X-ray TDEs is comparable to that of optically selected events, (2) X-ray TDE hosts are, like optical TDEs, preferentially found in green-valley galaxies, and (3) the majority of X-ray selected TDEs are intrinsically faint in the optical. 

A much larger eROSITA TDE sample, with detailed characterization of both transient properties and host galaxies, is essential to address several outstanding questions. First, volumetric rates remain poorly constrained. Integrating the best-fit X-ray luminosity functions from \citet{Sazonov2021} and \citet{Grotova2025} above the peak luminosities of $L_{\rm X}>10^{42.5}\,{\rm erg\,s^{-1}}$ yields TDE rates that differ by a factor of four: $\approx\!204_{-104}^{+113}\,{\rm Gpc^{-3}\,yr^{-1}}$ from \citet{Sazonov2021} versus $\approx\!52\pm15\,{\rm Gpc^{-3}\,yr^{-1}}$ from \citet{Grotova2025}. As noted by \citet{Mondal2025}, the eROSITA-DE sample is probably contaminated by AGN due to its lower detection threshold. Second, existing eROSITA studies have not yet examined how TDE rates vary with host galaxy stellar mass or black hole mass. Such constraints are essential for using TDE samples to probe black hole mass functions and loss cone physics. Finally, potential correlations between UV/optical broad-line properties and X-ray loudness offer powerful diagnostics to distinguish between competing models of the TDE radiative emission mechanisms \citep{Roth2020}, but remain largely unexplored.

In this paper, we present optical spectroscopic observations using the Low Resolution Imaging Spectrometer (LRIS; \citealt{Oke1995}) on the Keck I telescope of 70 TDE candidates identified by the eROSITA-RU consortium. For most candidates, the Keck/LRIS spectra show only host galaxy light, with transient features detected in a small subset. We note that the 13 TDEs published by \citet{Sazonov2021} are all observed by LRIS, and are therefore included here for both comparison and completeness. These observations provide a comprehensive spectroscopic dataset for studying the host galaxies of TDEs. 

The remainder of this paper is organized as follows. In Section~\ref{sec:selection}, we outline the sample selection and host galaxy association procedures. Section~\ref{sec:data} describes the archival photometry of the host galaxies and our new Keck/LRIS spectroscopic observations. In Section~\ref{sec:analysis}, we combine spectroscopy and photometry to derive host galaxy properties, address AGN contamination, and define a gold sample of the most robust TDEs. Finally, we discuss our findings in Section~\ref{sec:discuss} and summarize in Section~\ref{sec:conclusion}.

All coordinates are given in J2000. 
Optical observations are corrected for Galactic extinction using the \citet{1998ApJ...500..525S} dust maps 
and the \citet{Cardelli1989} extinction curves with $R_V=3.1$. 
We adopt a standard $\Lambda$CDM cosmology with matter density $\Omega_{\rm M} = 0.3$, dark energy density $\Omega_{\Lambda}=0.7$, and the Hubble constant $H_0=70\,{\rm km\,s^{-1}\,Mpc^{-1}}$.
Unless otherwise noted, the uncertainties quoted represent 68\% confidence intervals. 

\section{Initial Sample Selection}
\label{sec:selection}

\begin{table*}
\caption{List of eROSITA TDE candidates with unique reliable host galaxy association. \label{tab:sample}}
\small
\vspace{-15pt}
\begin{center}
\begin{tabular}{cccccccccc}
    \toprule
    ID &
    Name &
    $\alpha_{\rm X}$ &
    $\delta_{\rm X}$ &
    $r_{98}$ (\arcsec) &
    $\alpha_{\rm opt}$ &
    $\delta_{\rm opt}$ &
    $\Delta_{\rm pos}$ (\arcsec) &
    $f_{\rm Xpeak}$ &
    $t_{\rm Xpeak}$ \\
  \midrule
  1 & SRGe J003524.8-263550 & 8.853195 & -26.597256 & 6.22 & 8.85251 & -26.59819 & 4.02 & 31.32 & 2021-12-14.17 \\
  2 & SRGe J004123.2-153705 & 10.346576 & -15.618171 & 7.67 & 10.346609 & -15.618729 & 2.01 & 26.12 & 2021-06-16.71 \\
  3 & SRGe J010301.0-130120 & 15.754132 & -13.022233 & 5.0 & 15.75312 & -13.021384 & 4.68 & 69.01 & 2021-12-26.09 \\
  4 & SRGe J010445.8+044319 & 16.190908 & 4.721981 & 7.01 & 16.189389 & 4.722495 & 5.76 & 20.3 & 2022-01-02.60 \\
  5 & SRGe J011603.1+072258 & 19.013054 & 7.382774 & 5.85 & 19.012677 & 7.382249 & 2.32 & 32.55 & 2021-07-04.87 \\
  6 & SRGe J011943.5-024142 & 19.931051 & -2.695102 & 6.55 & 19.930647 & -2.696627 & 5.68 & 16.26 & 2021-07-01.62 \\
  7 & SRGe J013204.4+122235 & 23.018508 & 12.376507 & 5.77 & 23.018675 & 12.376562 & 0.62 & 22.2 & 2020-07-08.99 \\
  8 & SRGe J015353.9+372945 & 28.474627 & 37.495725 & 6.17 & 28.475665 & 37.496032 & 3.16 & 10.36 & 2021-07-30.11 \\
  9 & SRGe J015444.7-070012 & 28.686206 & -7.0033 & 10.57 & 28.68661 & -7.003443 & 1.53 & 14.87 & 2021-07-11.67 \\
  10 & SRGe J015754.6-154214 & 29.477387 & -15.704019 & 6.21 & 29.477263 & -15.70359 & 1.6 & 26.15 & 2021-01-02.88 \\
  11 & SRGe J015907.1-150323 & 29.779444 & -15.05652 & 5.52 & 29.779383 & -15.056262 & 0.95 & 24.16 & 2021-01-03.46 \\
  12 & SRGe J021213.6+310535 & 33.056503 & 31.093126 & 7.08 & 33.056536 & 31.092021 & 3.98 & 14.85 & 2020-08-02.07 \\
  13 & SRGe J021939.7+361819 & 34.915471 & 36.305141 & 5.0 & 34.916264 & 36.305054 & 2.32 & 24.7 & 2020-08-08.36 \\
  14 & SRGe J023017.3+283606 & 37.572164 & 28.601784 & 5.0 & 37.571364 & 28.601323 & 3.02 & 84.78 & 2022-02-02.19 \\
  15 & SRGe J023440.1-021812 & 38.667082 & -2.303241 & 5.0 & 38.667078 & -2.302912 & 1.18 & 43.86 & 2021-07-25.13 \\
  16 & SRGe J025548.1+142800 & 43.950595 & 14.466526 & 5.0 & 43.950275 & 14.466619 & 1.16 & 104.64 & 2022-02-03.68 \\
  17 & SRGe J030747.8+401842 & 46.949092 & 40.311655 & 5.0 & 46.949274 & 40.311363 & 1.16 & 726.75 & 2021-08-17.03 \\
  18 & SRGe J050948.4+695221 & 77.451697 & 69.872474 & 5.97 & 77.451198 & 69.872119 & 1.42 & 19.58 & 2021-09-22.10 \\
  19 & SRGe J060324.7+621112 & 90.853098 & 62.18658 & 5.0 & 90.853499 & 62.186385 & 0.97 & 82.46 & 2021-03-30.14 \\
  20 & SRGe J071310.4+725627 & 108.293384 & 72.940741 & 5.0 & 108.293835 & 72.940751 & 0.48 & 112.16 & 2020-10-11.89 \\
  21 & SRGe J081006.4+681755 & 122.526835 & 68.298718 & 5.42 & 122.525923 & 68.299092 & 1.81 & 23.2 & 2021-04-14.56 \\
  22 & SRGe J083640.9+805410 & 129.170469 & 80.902719 & 5.42 & 129.17618 & 80.903218 & 3.71 & 19.54 & 2021-10-11.09 \\
  23 & SRGe J091747.3+524818 & 139.447148 & 52.805055 & 5.51 & 139.447492 & 52.805635 & 2.22 & 26.39 & 2020-10-29.15 \\
  24 & SRGe J095928.7+643024 & 149.869625 & 64.50679 & 5.0 & 149.869012 & 64.506081 & 2.72 & 44.04 & 2020-10-28.89 \\
  25 & SRGe J113323.2+693635 & 173.346493 & 69.60971 & 6.9 & 173.349376 & 69.609137 & 4.16 & 13.52 & 2021-05-04.90 \\
  26 & SRGe J131014.7+444319 & 197.561406 & 44.72182 & 6.45 & 197.562712 & 44.72188 & 3.35 & 22.67 & 2021-06-07.23 \\
  27 & SRGe J131404.1+515427 & 198.517088 & 51.907552 & 6.24 & 198.517527 & 51.907817 & 1.36 & 24.99 & 2021-12-04.91 \\
  28 & SRGe J132718.1+350437 & 201.825359 & 35.076816 & 5.0 & 201.825306 & 35.076286 & 1.91 & 42.59 & 2021-06-17.72 \\
  29 & SRGe J133053.5+734823 & 202.722784 & 73.806298 & 5.34 & 202.720918 & 73.806739 & 2.46 & 24.75 & 2020-11-03.46 \\
  30 & SRGe J133731.5+601849 & 204.381115 & 60.313748 & 6.07 & 204.380151 & 60.313456 & 2.01 & 13.65 & 2020-11-26.16 \\
  31 & SRGe J135353.7+535949 & 208.473873 & 53.996971 & 5.0 & 208.474197 & 53.997141 & 0.92 & 140.46 & 2021-06-07.15 \\
  32 & SRGe J135515.0+311605 & 208.812591 & 31.267931 & 5.13 & 208.812579 & 31.268121 & 0.68 & 38.35 & 2020-06-23.85 \\
  33 & SRGe J135812.1+195357 & 209.550215 & 19.899209 & 6.87 & 209.550684 & 19.898775 & 2.23 & 20.35 & 2020-12-31.88 \\
  34 & SRGe J144738.3+671818 & 221.909782 & 67.305102 & 5.0 & 221.912771 & 67.305094 & 4.15 & 62.54 & 2021-05-21.74 \\
  35 & SRGe J145226.6+670437 & 223.111019 & 67.076826 & 6.1 & 223.111719 & 67.076509 & 1.51 & 7.51 & 2021-05-22.57 \\
  36 & SRGe J150328.1+495117 & 225.86706 & 49.854703 & 5.34 & 225.867028 & 49.853579 & 4.05 & 11.68 & 2020-12-27.20 \\
  37 & SRGe J152656.3+353317 & 231.734534 & 35.554619 & 7.85 & 231.734127 & 35.553842 & 3.04 & 9.68 & 2021-01-17.12 \\
  38 & SRGe J153134.9+330539 & 232.895476 & 33.094222 & 5.13 & 232.895722 & 33.094928 & 2.65 & 19.92 & 2020-07-28.18 \\
  39 & SRGe J153331.5+390536 & 233.381222 & 39.093288 & 6.47 & 233.381985 & 39.092769 & 2.83 & 12.57 & 2021-01-16.61 \\
  40 & SRGe J153403.5+621851 & 233.514553 & 62.314238 & 5.52 & 233.514195 & 62.314555 & 1.29 & 16.71 & 2021-12-15.06 \\
  41 & SRGe J153503.3+455054 & 233.7639 & 45.848311 & 5.0 & 233.763172 & 45.848598 & 2.1 & 34.21 & 2020-07-14.69 \\
  42 & SRGe J155113.5+515845 & 237.806284 & 51.979265 & 6.62 & 237.805954 & 51.978777 & 1.9 & 6.26 & 2021-01-08.03 \\
  43 & SRGe J155834.0+382528 & 239.641788 & 38.424573 & 5.0 & 239.641623 & 38.424662 & 0.56 & 18.02 & 2021-08-02.65 \\
  44 & SRGe J160943.1+253603 & 242.429451 & 25.600825 & 5.0 & 242.429019 & 25.601028 & 1.58 & 23.05 & 2021-08-26.16 \\
  45 & SRGe J161001.3+330120 & 242.505597 & 33.022304 & 6.45 & 242.50592 & 33.022416 & 1.06 & 14.12 & 2020-08-15.89 \\
  46 & SRGe J161559.1+360156 & 243.996068 & 36.032235 & 5.0 & 243.995909 & 36.031732 & 1.87 & 32.26 & 2021-08-11.06 \\
  47 & SRGe J162159.4+271133 & 245.497537 & 27.192442 & 5.0 & 245.497951 & 27.192419 & 1.33 & 19.95 & 2020-08-24.39 \\
  48 & SRGe J162932.1+280521 & 247.383633 & 28.089035 & 5.0 & 247.38332 & 28.08869 & 1.59 & 53.06 & 2022-02-18.16 \\
  49 & SRGe J163030.3+470125 & 247.626092 & 47.023546 & 5.0 & 247.626052 & 47.02373 & 0.67 & 32.22 & 2020-08-10.55 \\
  50 & SRGe J163831.9+534018 & 249.632855 & 53.671541 & 5.0 & 249.633401 & 53.672931 & 5.14 & 15.22 & 2020-08-04.97 \\
  51 & SRGe J165055.8+301634 & 252.732698 & 30.275976 & 6.4 & 252.732439 & 30.275924 & 0.83 & 15.1 & 2021-02-20.95 \\
  52 & SRGe J170139.0-085911 & 255.412324 & -8.986521 & 6.06 & 255.412083 & -8.98657 & 0.87 & 23.49 & 2021-03-13.86 \\
  53 & SRGe J171337.4+581732 & 258.40603 & 58.292136 & 5.0 & 258.404923 & 58.292426 & 2.34 & 7.7 & 2020-08-18.49 \\
  54 & SRGe J171423.6+085237 & 258.598357 & 8.876814 & 5.0 & 258.598393 & 8.876918 & 0.4 & 116.83 & 2020-09-18.72 \\
  55 & SRGe J174505.5+104700 & 266.273008 & 10.783297 & 8.69 & 266.272965 & 10.782156 & 4.11 & 12.16 & 2021-09-24.64 \\
  56 & SRGe J174513.6+401608 & 266.306768 & 40.269026 & 6.77 & 266.308609 & 40.268842 & 5.1 & 10.18 & 2021-03-22.42 \\
  57 & SRGe J174912.1+595530 & 267.300607 & 59.924895 & 15.13 & 267.302181 & 59.923772 & 4.94 & 11.03 & 2021-03-16.33 \\
  58 & SRGe J175023.7+712857 & 267.598554 & 71.482499 & 5.0 & 267.599017 & 71.48239 & 0.66 & 34.34 & 2021-10-07.50 \\
  59 & SRGe J180757.1+565625 & 271.988035 & 56.940306 & 5.19 & 271.987841 & 56.939758 & 2.01 & 7.61 & 2020-10-10.87 \\
  60 & SRGe J182716.8+044603 & 276.820057 & 4.767595 & 5.61 & 276.819812 & 4.768384 & 2.97 & 24.94 & 2021-04-05.94 \\
  61 & SRGe J192143.8+503853 & 290.432436 & 50.648059 & 6.78 & 290.432057 & 50.647861 & 1.12 & 14.65 & 2021-11-06.65 \\
  62 & SRGe J200953.9+672317 & 302.474514 & 67.388044 & 5.79 & 302.47453 & 67.388565 & 1.88 & 4.76 & 2020-06-28.96 \\
  \bottomrule
  \end{tabular}
  \end{center}
  \textbf{Notes.} $\alpha_{\rm X}$, $\delta_{\rm X}$ are the eROSITA coordinates. $r_{98}$ is the eROSITA 98\%  localization region in the stacked eROSITA data ($r_{98}=2.8\sigma_{\rm pos}$). $\alpha_{\rm opt}$, $\delta_{\rm opt}$ are the coordinates of optical host galaxy associations. $\Delta_{\rm pos}$ (\arcsec) represents the angular separation between the X-ray and optical positions of each source. $f_{\rm X,peak}$, $t_{\rm X,peak}$ are the peak X-ray flux ($10^{-14}$ erg cm$^{-2}$ s$^{-1}$, 0.3--2.3\,keV) from eROSITA multi-epoch observations and the median time of the corresponding epoch with maximum flux.
\end{table*}

\begin{table*}
\caption{Continued table of Tab.~\ref{tab:sample}. \label{tab:sample2}}
\small
\vspace{-15pt}
\begin{center}
\begin{tabular}{cccccccccc}
    \toprule
    ID &
    Name &
    $\alpha_{\rm X}$ &
    $\delta_{\rm X}$ &
    $r_{98}$ (\arcsec) &
    $\alpha_{\rm opt}$ &
    $\delta_{\rm opt}$ &
    $\Delta_{\rm pos}$ (\arcsec) &
    $f_{\rm Xpeak}$ &
    $t_{\rm Xpeak}$ \\
  \midrule
  63 & SRGe J201138.9-210935 & 302.912143 & -21.159756 & 6.26 & 302.912163 & -21.160194 & 1.58 & 36.89 & 2021-10-29.04 \\
  64 & SRGe J204129.5+214409 & 310.372873 & 21.735758 & 5.0 & 310.372816 & 21.735375 & 1.39 & 208.1 & 2021-05-11.52 \\
  65 & SRGe J213214.8-003006 & 323.061768 & -0.501774 & 9.49 & 323.062306 & -0.501395 & 2.37 & 26.09 & 2021-05-14.78 \\
  66 & SRGe J213527.2-181635 & 323.863278 & -18.276355 & 5.0 & 323.86361 & -18.276464 & 1.2 & 136.41 & 2020-11-09.49 \\
  67 & SRGe J223905.0-270551 & 339.77096 & -27.097538 & 5.0 & 339.769487 & -27.097436 & 4.73 & 77.17 & 2021-05-18.46 \\
  68 & SRGe J231834.5-351914 & 349.643947 & -35.320561 & 6.4 & 349.645082 & -35.320365 & 3.41 & 49.11 & 2021-11-22.61 \\
  69 & SRGe J234034.7+293400 & 355.144765 & 29.566581 & 5.0 & 355.144139 & 29.566323 & 2.17 & 25.02 & 2020-06-17.81 \\
  70 & SRGe J235453.0+421711 & 358.720689 & 42.286526 & 5.0 & 358.720023 & 42.286442 & 1.8 & 18.83 & 2022-01-04.87 \\
\bottomrule
  \end{tabular}
  \end{center}
\end{table*}

\subsection{X-ray Selection}

We selected TDE candidates from strongly variable X-ray sources detected by eROSITA-RU sky surveys in the 0.3--2.3\,keV energy band \citep{2022AstL...48..735M}. The source catalog was built and maintained by the eROSITA-RU X-ray source catalog science working group at the Space Research Institute (Moscow, Russia). As data from successive sky scans was collected throughout the mission, the master catalog was being updated and multiple selections were created to support various multiwavelength follow-up programs, including the extensive TDE identification campaign reported here. Selection criteria for TDE candidates evolved throughout the mission. The distribution of TDE candidates among participating optical telescopes also varied throughout the mission. Therefore the sample with Keck spectroscopy presented here is not statistically well-defined and is a not a result of uniform selection across eROSITA sky scans.  This data set should therefore be regarded as a large 
subsample of the full eROSITA TDE sample, which will be defined using strict and uniform criteria and presented in a future publication.

Prior to the end of 2022, we classified an X-ray source as highly variable if the source was not detected in the previous surveys and its  flux in the 0.3--2.3\,keV energy band during some eROSITA scan exceeded its upper limit in the previous scan by a factor of larger than $R_{\rm X, min}$. Our initial TDE candidate samples used $R_{\rm X, min}=10$  \citep{Sazonov2021}, later expanded to $R_{\rm X, min}=7$ to include fainter events \citep{Khorunzhev2022}. At the later stages of the mission we redefined the sample using the formalism described in \citet{2022AstL...48..735M}, which uses theoretical probability distribution for the flux ratio to select sources variable with the factor  larger than $R_{\rm X, min}$ at the specified confidence level (usually corresponding to $3\sigma$). The sample with Keck spectroscopy presented here was largly selected using the former method  with $R_{\rm X, min}=10$ and $7$. 
Note that we do not use X-ray spectral softness in selecting TDE candidates among eROSITA extragalctic transients, as some TDEs are known to exhibit relatively hard spectra (e.g., \citealt{2020MNRAS.497L...1W,Yao_2022, Guolo2024, Ho2025}). 

Our selection process applies several filters to distinguish TDE candidates from other types of Galactiuc and extragalctic transients, following the procedures described in \citet{Sazonov2021} and \citet{Khorunzhev2022}. Specifically, we exclude Galactic sources based on positional coincidence (with the eROSITA 98\% localization radius $r_{98}$) with stars having statistically significant Gaia parallax and/or proper motion  \citep{2023A&A...674A...1G}. We also exclude known and candidate AGN based on mid-infrared colors $W1-W2>0.8$ \citep{2012ApJ...753...30S} in the WISE all-sky survey \citep{2014yCat.2328....0C}, archival X-ray detections, or optical transients with AGN-like stochastic variability. These criteria evolved slightly (in particular, we used a stricter infrared-color condition, $W1-W2<0.5$, in \citealt{Sazonov2021}) over the course of our TDE search; full details will appear in the forthcoming comprehensive catalog paper. 

\subsection{Host Galaxy Association}

We identified the host galaxies of our TDE candidates as optical sources detected in the Panoramic Survey Telescope and Rapid Response System Data Release 1 (PS1; \citealt{Flewelling2020}). For regions outside the PS1 footprint, we used sources detected in the DESI Legacy Imaging Survey (LS; \citealt{Dey2019}) DR10. For PS1, we consider only objects classified as `GALAXY' or `UNSURE' by the machine learning catalog of \citet{Beck2021}, and that have an $i$-band PSF magnitude. 
We assume that all TDE candidates are located at galaxy centers and are not associated with off-nuclear MBHs.

We cross-matched eROSITA sources with optical objects using a search radius of 8$\sigma_{\rm pos}$ (note that $r_{98}=2.8\sigma_{\rm pos}$). For each match, we computed the reliability $R$ of the association using the Bayesian likelihood ratio method \citep{Sutherland1992}. We adopted a flat prior on the magnitude distribution of true associations and modeled the surface density of the background sources as $dN / dm \propto m^{0.6}$, as expected in a simple Euclidean universe with uniform spatial distribution. 

For all optical objects with $R>0.1$, we performed forced photometry using the Zwicky Transient Facility (ZTF; \citealt{Bellm2019, Graham2019, Masci2019, Masci2023}) and the Asteroid Terrestrial-impact Last Alert System (ATLAS; \citealt{Tonry2018, Smith2020, Shingles2021}). If a significant optical flare is detected near the time of the X-ray transient, we set $R=1$. 
Optical spectroscopy (see \S\ref{subsec:opt_spec}) is obtained for X-ray sources with a unique optical counterpart having $R > 0.9$. 
We exclude Galactic sources and obvious broad-line AGN (BLAGN) whose optical spectra exhibit very strong broad emission lines (hydrogen Balmer series, \ion{Mg}{II}, etc.) that significantly exceed the continuum level.\footnote{We later remove BLAGN with weaker broad lines in \S\ref{subsubsec:agn}.} This yields a final sample of 70 TDE candidates with high-confidence optical associations.
This sample, listed in Tables~\ref{tab:sample} and \ref{tab:sample2}, is the focus of our analysis. We show the 25\arcsec~$\times$~25\arcsec\ cutouts of the host galaxy associations in Appendix~\ref{sec:sup_fig}. Of the 70 optical counterparts, 69 lie within the $r_{98}$ region, while 1 (1.4\%) lies outside, consistent with the 2\% of counterparts expected to fall outside $r_{98}$ by construction.

\section{Data} \label{sec:data}

\subsection{Archival Photometry}
\label{subsec:photometry}

We compiled UV-to-IR photometry for all host galaxies, adopting measurements with signal-to-noise ratio (S/N) $>3$ as detections and $3\sigma$ upper limits otherwise. 
UV fluxes were measured from the Galaxy Evolution Explorer (GALEX; \citealp{Martin2005}) using \texttt{gPhoton} \citep{Million2016}, with Kron apertures and optimized background annuli.
For the optical band, we prioritized PS DR2 $grizy$ Kron magnitudes \citep{2016arXiv161205560C} where available, otherwise using the Sloan Digital Sky Survey (SDSS) DR16 \citep{2002AJ....123..485S}. 
Near-infrared $JHK_{\rm s}$ magnitudes were drawn from the Two Micron All-Sky Survey (2MASS) extended source catalog \citep{2006AJ....131.1163S} where available.
For the mid-infrared, we used data provided by the Wide-field Infrared Survey Explorer
(WISE; \citealt{2010AJ....140.1868W}). We prioritized WISE photometry from the LS DR10 Tractor catalog \citep{2020A&C....3300411N}, supplemented by LS DR10 aperture fluxes or entries from the AllWISE \citep{2013wise.rept....1C} and CatWISE \citep{2020ApJS..247...69E} catalogs.

\subsection{Optical Spectroscopy}
\label{subsec:opt_spec}

We obtained optical spectroscopy for the host galaxies of all TDE candidates using LRIS. 
For the blue side, we used the 400/3400 grism with a 1\arcsec\ slit for all but four observations. SRGe J071310.4+725627 (ID 20) was observed with the 600/4000 grism and a 1.5\arcsec\ slit. SRGe J174505.5+104700 (ID 55), SRGe J201138.9-210935 (ID 63), and the +1333-day spectrum of SRGe J175023.7+712857 (ID 58) were taken with the 600/4000 grism and a 1\arcsec\ slit.
This configuration provided spectral coverage of 3200--10250\,\AA. For the majority of objects, only one epoch of LRIS spectroscopy was obtained. Multiple spectra were collected for four objects (IDs 5, 26, 58, 66) where prominent TDE emission lines were detected during the first LRIS epoch (see \S\ref{subsubsec:tde-broad} for details). A log of LRIS observations is presented in Table~\ref{tab:spectral-log}, which also lists the host redshifts measured from the LRIS spectra. 

\tablecaption{Keck-I/LRIS spectroscopic observation log.\label{tab:spectral-log}}
\begin{center}
\tablefirsthead{%
\toprule
 ID & Start Date (UT)& Exp. (s) & Redshift & Phase (d) \\
\midrule
}
\tablehead{%
\toprule
 ID & Start Date (UT) & Exp. (s) & Redshift & Phase (d) \\
\midrule
}
\tabletail{%
\midrule
\multicolumn{5}{r}{Continued in next column}\\
}
\tablelasttail{\bottomrule}
\small
\begin{supertabular}{cccccc}
1 & 2021-10-04.44 & 300 & 0.0727 $\pm$0.0001  \ & -71 \\
2 & 2022-07-21.57 & 600 & 0.0535$\pm$0.0003 \ & 400 \\
3 & 2021-09-07.53 & 600 & 0.1197$\pm$0.0002 \ & -110 \\
4 & 2023-01-16.26 & 600 & 0.1752$\pm$0.0002 \ & 379 \\
\hline
5 & 2021-08-13.43 & 1800 & \multirow{2}{*}{0.1500$\pm$0.0001} \ & 40 \\
5 & 2021-09-07.53 & 900 & \ & 65 \\
\hline
6 & 2022-01-12.18 & 900 & 0.1250$\pm$0.0004 \ & 195 \\
7 & 2021-07-06.55 & 600 & 0.1314$\pm$0.0001 \ & 363 \\
8 & 2021-09-07.57 & 300 & 0.0778$\pm$0.0001 \ & 39 \\
9 & 2022-01-26.27 & 600 & 0.0809$\pm$ 0.0014 \ & 199 \\
10 & 2021-08-13.54 & 1485 & 0.3253$\pm$0.0001 \ & 223 \\
11 & 2021-08-13.54 & 600 & 0.0949$\pm$0.0001 \ & 222 \\
12 & 2022-07-21.59 & 300 & 0.0338$\pm$0.0001 \ & 719 \\
13 & 2021-07-06.61 & 750 & 0.3879$\pm$0.0001 \ & 332 \\
14 & 2022-07-21.60 & 300 & 0.0364$\pm$0.0001 \ & 169 \\
15 & 2021-08-13.62 & 1480 & 0.2682$\pm$0.0005 \ & 19 \\
16 & 2022-07-21.61 & 600 & 0.0735$\pm$0.0006 \ & 168 \\
17 & 2022-07-21.62 & 450 & 0.0180$\pm$0.0005  \ & 339 \\
18 & 2021-10-04.58 & 300 & 0.0845$\pm$0.0001 \ & 12 \\
19 & 2021-04-14.26 & 1200 & 0.2221$\pm$0.0038 \ & 15 \\
20 & 2020-11-20.51 & 900 & 0.1052$\pm$0.0001 \ & 40 \\
21 & 2021-05-13.27 & 774 & 0.1423$\pm$0.0001 \ & 29 \\
22 & 2022-02-06.36 & 1160 & 0.2302$\pm$0.0003 \ & 118 \\
23 & 2021-04-14.30 & 600 & 0.1877$\pm$0.0001 \ & 167 \\
24 & 2021-05-13.32 & 2508 & 0.4553$\pm$0.0004 \ & 196 \\
25 & 2021-05-13.26 & 584 & 0.1931$\pm$0.0001 \ & 8 \\
\hline
26 & 2021-07-06.34 & 2709 & \multirow{3}{*}{0.1982$\pm$0.0004} \ & 29 \\
26 & 2021-08-13.25 & 1800 &  \ & 67 \\
26 & 2022-02-06.46 & 1770 &  \ & 244 \\
\hline
27 & 2022-04-07.46 & 1800 & 0.2782$\pm$0.0001 \ & 124 \\
28 & 2021-06-07.37 & 435 & 0.1491$\pm$0.0003 \ & -10 \\
29 & 2021-04-14.41 & 400 & 0.1503$\pm$0.0001 \ & 162 \\
30 & 2022-08-02.28 & 1200 & 0.3091$\pm$0.0007 \ & 614 \\
31 & 2021-06-07.39 & 285 & 0.0705$\pm$0.0002 \ & 0 \\
32 & 2021-06-07.38 & 435 & 0.1995$\pm$0.0004 \ & 349 \\
33 & 2021-05-13.36 & 1137 & 0.1889$\pm$0.0004 \ & 132 \\
34 & 2021-04-14.42 & 300 & 0.1242$\pm$0.0003 \ & -37 \\
35 & 2021-06-07.41 & 585 & 0.2681$\pm$0.0001 \ & 16 \\
36 & 2021-05-13.42 & 1756 & 0.3521$\pm$0.0006 \ & 137 \\
37 & 2021-05-13.43 & 574 & 0.1540$\pm$ 0.0003\ & 116 \\
38 & 2021-06-07.45 & 285 & 0.0679$\pm$0.0003 \ & 314 \\
39 & 2021-06-07.45 & 285 & 0.1085$\pm$0.0002 \ & 142 \\
40 & 2022-04-07.54 & 1800 & 0.2437$\pm$0.0002 \ & 113 \\
41 & 2021-05-13.44 & 574 & 0.2308$\pm$0.0004 \ & 303 \\
42 & 2021-06-07.47 & 1185 & 0.4385$\pm$0.0001 \ & 150 \\
43 & 2021-07-06.45 & 600 & 0.1684$\pm$0.0003 \ & -27 \\
44 & 2021-05-13.46 & 469 & 0.1468$\pm$0.0003 \ & -105 \\
45 & 2021-04-14.52 & 300 & 0.1304$\pm$0.0002 \ & 242 \\
46 & 2021-06-07.49 & 580 & 0.1534$\pm$0.0003 \ & -65 \\
47 & 2022-05-26.40 & 500 & 0.0965$\pm$0.0002 \ & 640 \\
48 & 2022-05-26.41 & 600 & 0.1415$\pm$0.0002 \ & 97 \\
49 & 2021-04-14.64 & 500 & 0.2949$\pm$0.0009 \ & 247 \\
50 & 2021-04-14.54 & 1250 & 0.5813$\pm$0.0006 \ & 253 \\
51 & 2021-05-13.49 & 254 & 0.2265$\pm$0.0007 \ & 82 \\
52 & 2022-04-07.56 & 900 & 0.4601$\pm$0.0003 \ & 390 \\
53 & 2022-02-06.66 & 600 & 0.2624$\pm$0.0004 \ & 537 \\
54 & 2021-06-07.54 & 280 & 0.0364$\pm$0.0001 \ & 262 \\
55 & 2025-06-25.39 & 625 & 0.0882$\pm$0.0001 \ & 1370 \\
56 & 2021-05-13.55 & 1354 & 0.7140$\pm$0.0002 \ & 52 \\
57 & 2021-05-13.60 & 254 & 0.1089$\pm$0.0003 \ & 58 \\
\hline
58 & 2022-04-07.56 & 900 & \multirow{2}{*}{0.0904$\pm$0.0002} \ & 182 \\
58 & 2025-06-01.55 & 975 &  \ & 1333 \\
\hline
59 & 2022-07-21.44 & 300 & 0.0602$\pm$0.0002 \ & 649 \\
60 & 2022-05-26.46 & 750 & 0.0464$\pm$0.0003 \ & 416 \\
61 & 2022-04-07.60 & 900 & 0.2593$\pm$0.0003 \ & 152 \\
62 & 2022-07-21.47 & 600 & 0.1203$\pm$0.0003 \ & 753 \\
63 & 2025-06-01.61 & 1200 & 0.0804$\pm$0.0001 \ & 1312 \\
64 & 2021-07-06.53 & 600 & 0.1103$\pm$0.0003 \ & 56 \\
65 & 2022-05-26.57 & 600 & 0.1415$\pm$0.0004 \ & 377 \\
\hline
66 & 2020-11-20.22 & 1140 & \multirow{3}{*}{0.0939$\pm$0.0001} \ & 11 \\
66 & 2020-12-12.21 & 600 & & 33 \\
66 & 2021-09-07.43 & 600 &   & 302 \\
\hline
67 & 2021-10-04.43 & 450 & 0.2195$\pm$0.0004  & 139 \\
68 & 2022-07-21.49 & 600 & 0.0549$\pm$0.0001  & 241 \\
69 & 2022-01-12.14 & 600 & 0.1272$\pm$0.0003  & 573 \\
70 & 2022-07-21.05 & 1760 & 0.1990$\pm$0.0001  & 197 \\
\bottomrule
\end{supertabular}
\tablefoot{Phase is observer-frame days since eROSITA X-ray peak (i.e., $t_{\rm X, peak}$ in Tables~\ref{tab:sample} and \ref{tab:sample2}). Redshifts are fitted with \texttt{pPXF} from the Keck/LRIS spectra (see \S\ref{subsec:ppxf}). For sources with multiple spectra (IDs 5, 26, 58, 66), redshifts are determined from the latest, host-dominated spectra. The fitted redshift uncertainties are statistical errors (scaled by $\sqrt{\chi^2}$).}
\end{center}

We also cross-matched our sample with the Dark Energy Spectroscopic Instrument (DESI) survey data release one (DR1; \citealt{DESI2025_dr1}). 
We found DESI spectra for 13 host galaxies in our sample. For 12 of these, the DESI spectra are broadly consistent with the Keck/LRIS data, exhibiting similar continuum shapes and line properties. Only the one remaining DESI spectrum is retained for our analysis which has substantially higher S/N than the Keck spectrum. 
The retained DESI spectrum (3600--9800\,\AA), corresponding to the host galaxy of SRGe\,J234034.7+293400 (ID 69), was obtained with a single 982\,s exposure on 2021-10-16.12 (485\,d after the observed peak X-ray flux) using a 1.5\arcsec\ fiber. 

\section{Analysis} \label{sec:analysis}
\subsection{Host Galaxy SED Fitting} \label{subsec:prospector_fitting}
To infer host galaxy properties, we fit the galaxy data using \texttt{Prospector} \citep{2021ApJS..254...22J} built on \texttt{FSPS} \citep{2010ascl.soft10043C}. We adopt a non-parametric continuity star-formation history (SFH) that captures bursts, quenching, and rejuvenation episodes with flexibility.

Our fits use extinction-corrected host galaxy photometry and optical spectroscopy. 
When jointly fitting photometry and spectroscopy, we employ \texttt{Prospector}'s \texttt{PolySpecModel} with spectroscopic calibration, including a free normalization parameter (\texttt{spec\_norm}) to scale the spectrum to match the photometry.
Only SRGe J163831.9+534018 (ID 50) was fit using photometry alone, as the joint photometry+spectroscopy fit did not converge.
For objects with multiple LRIS spectra (i.e., IDs 5, 26, 58, 66), we use the latest observation. 

The LRIS spectra are typically dominated by host galaxy light, as they were obtained at late epochs when the optical counterparts of most X-ray-selected TDEs had faded. However, in some cases, broad emission lines associated with the transient may still be present. We perform the \texttt{Prospector} fitting twice: once with the \ion{He}{II} $\lambda$4686 and H$\alpha$ regions masked, and once without masking. We adopt the unmasked fits for our final host galaxy parameter estimates, as the transient emission lines are typically weak at these late epochs and their impact on the stellar continuum and broadband photometry is minimal. In \S\ref{subsubsec:tde-broad}, we use the masked fits to identify and characterize any residual broad-line emission from the optical counterparts by subtracting the host model from the observed spectra.

Posterior distributions are sampled with \texttt{Dynesty} \citep{2020MNRAS.493.3132S}, and convergence is monitored via the built-in KL divergence test. We used the nested sampler with 500 live points (both \texttt{nested\_nlive\_init} and \texttt{nested\_nlive\_batch} set to 500), random-walk sampling mode, a convergence criterion of 0.01 for \texttt{nested\_dlogz\_init}, and a maximum of $7.5\times10^{6}$ likelihood calls, keeping the default \texttt{dynesty} bounds. All fits were run in parallel using MPI via \texttt{MPIPool}. We included a multiplicative spectroscopic jitter term (\texttt{spec\_jitter}) as a free parameter that scales the spectroscopic uncertainties by a wavelength-independent factor, and we did not include any additional jitter term for the photometric data.

The model parameters and measured galaxy properties are described as follows.
We adopt a uniform TopHat prior on redshift, constrained to a narrow range of 0.05 around the spectral fitted redshift which is derived from \texttt{pPXF} in \S\ref{subsec:ppxf}.
The galaxy stellar mass ($M_\ast$) is derived from the surviving fraction of stars formed. Stellar metallicity is parameterized as $\log(Z_*/Z_\odot)$. We adopt an empirical mass–metallicity prior from \citet{2005MNRAS.362...41G}.  
The SFH is represented by logarithmically spaced age bins (with 0--30 Myr and 30--100 Myr fixed). We use 6 bins for spectra with ${\rm (S/N) }< 15$, and typically 10 bins for higher S/N spectra, chosen as a compromise between flexibility and constraining power given the data quality.
We adopt a Student's T prior on the logSFR ratios between adjacent time bins, and recent activity is summarized by the average SFR over the past 10 and 100 Myr (i.e., $\mathrm{SFR}_{10}$ and $\mathrm{SFR}_{100}$).  
We report the mass-weighted stellar age, defined as the average lookback time of stars weighted by their formed stellar mass across all SFH bins. 

Dust is modeled with the two-component prescription of \citet{2000ApJ...539..718C}, including diffuse ISM optical depth, a birth-cloud fraction, and a variable slope modifier \citep{2013ApJ...775L..16K}. From these, we derive $A_V$ and $E(B-V)$ for both diffuse and young-star components.  
As the \texttt{FSPS} nebular emission-line grids may not fully capture the diversity of observed line ratios, we model nebular emission with \texttt{FSPS+CLOUDY} and enable line-amplitude marginalization, allowing line strengths to vary independently while keeping the stellar-population parameters primarily constrained by the continuum and photometry \citep{Johnson2021}. These nebular parameters are treated as free parameters and for galaxies with no detectable emission lines, we disable the nebular emission component altogether.  
For dust emission, we adopt the model of \citet{2007ApJ...657..810D}, with $U_{\min}$, $\gamma_{\mathrm{dust}}$, and $Q_{\mathrm{PAH}}$ treated as free parameters when constrained by IR data, and held fixed otherwise.
Possible AGN contamination is modeled with the \texttt{CLUMPY} torus templates \citep{2008ApJ...685..160N}, with $f_{\mathrm{AGN}}$ and $\tau_{\mathrm{AGN}}$ held fixed when unconstrained.

The SED fitting results of our gold and silver TDE sample (to be defined later in \S\ref{subsec:classification}) are shown in Appendix~\ref{sec:sup_fig} (Figure~\ref{fig:sed_fits} and \ref{fig:sed_fits_2}).

\subsection{Emission-line Fitting}
\label{subsec:ppxf}
We fit the stellar continuum and nebular emission lines in our optical spectra using the penalized pixel-fitting code \texttt{pPXF} \citep{Cappellari2023} with the \texttt{EMILES} stellar library \citep{2016MNRAS.463.3409V}. For sources with multiple LRIS spectra (IDs 5, 26, 58, 66), we adopt the most recent, host-dominated spectrum for analysis.
Each spectrum is modeled with three components: stellar continuum, narrow Balmer emission lines, and narrow forbidden lines. For objects exhibiting broad Balmer wings, we include an additional broad component.

For emission lines with ${\rm S/N} \geq 3$, we adopt the best-fit fluxes and uncertainties from \texttt{pPXF}. 
For lines with ${\rm S/N} < 3$, we perform 1000 Monte Carlo realizations by perturbing the spectra with noise, and use the resulting flux distributions to estimate robust fluxes and uncertainties.

We also measure the equivalent width (EW) of emission lines. EW is defined as 
\begin{equation}
    \mathrm{EW} = \int \frac{F_{\lambda,\mathrm{obs}} - F_{\lambda,\mathrm{cont}}}{F_{\lambda,\mathrm{cont}}} \, \mathrm{d}\lambda,
\end{equation}
measured in \AA, where $F_{\lambda, \rm obs}$ is the observed spectrum, and $F_{\lambda, \mathrm{cont}}$ is the continuum level estimated as the median flux in the line-free windows on either side of the emission line.
The EW uncertainty is calculated via standard error propagation.

\subsection{Sample Classification} \label{subsec:classification}
Although the X-ray flares from some AGN can mimic TDEs and contaminate X-ray–selected TDE samples, our ZTF/ATLAS forced photometry and uniform optical spectra for all host galaxies enable a systematic photometric and spectroscopic assessment of AGN interlopers. Below, we examine the available data to remove AGN contaminations, define a gold sample of events that we are most confident are TDEs, and a silver sample of events for which an AGN origin cannot be completely ruled out.

\subsubsection{Identification of TDE Candidates with Optical Flares} \label{subsubsec:optical_flare}

We visually inspected the ATLAS and ZTF forced photometry light curves from 2018 March 1 to 2024 March 1 (Figures~\ref{fig:opt_diff_lc}--\ref{fig:opt_diff_lc_4} in Appendix~\ref{sec:sup_fig}) to identify TDE candidates exhibiting optical flares. We only search for optical flares showing a rise and decline, as opposed to stochastic variability characteristic of AGN.

We detected optical flares in 28 objects (IDs 2, 5, 15, 16, 17, 21, 23, 24, 26, 27, 
30, 31, 38, 39, 40, 41, 43, 47, 48, 49, 55, 58, 63, 64, 65, 66, 67, 70).
Among these, the flares of SRGe J155834.0+382528 (ID 43), SRGe J174505.5+104700 (ID 55) and SRGe J213214.8-003006 (ID 65) exhibit colors ($g-r\gtrsim 0.5$\,mag and $r-i\gtrsim0.3$\,mag) that are much redder than typical TDEs\footnote{In \S\ref{subsubsec:agn}, we show that optical spectra obtained long after the flare faded exhibit typical AGN signatures.}.
The remaining 25 objects display optical flares with colors that are consistent with optically-selected TDEs, with $g-r<0.3$\,mag at peak. 

\subsubsection{Identification of CrL-TDEs} \label{subsubsec:crl-tde}

\begin{figure}[htbp]
    \centering
    \includegraphics[width=\columnwidth]{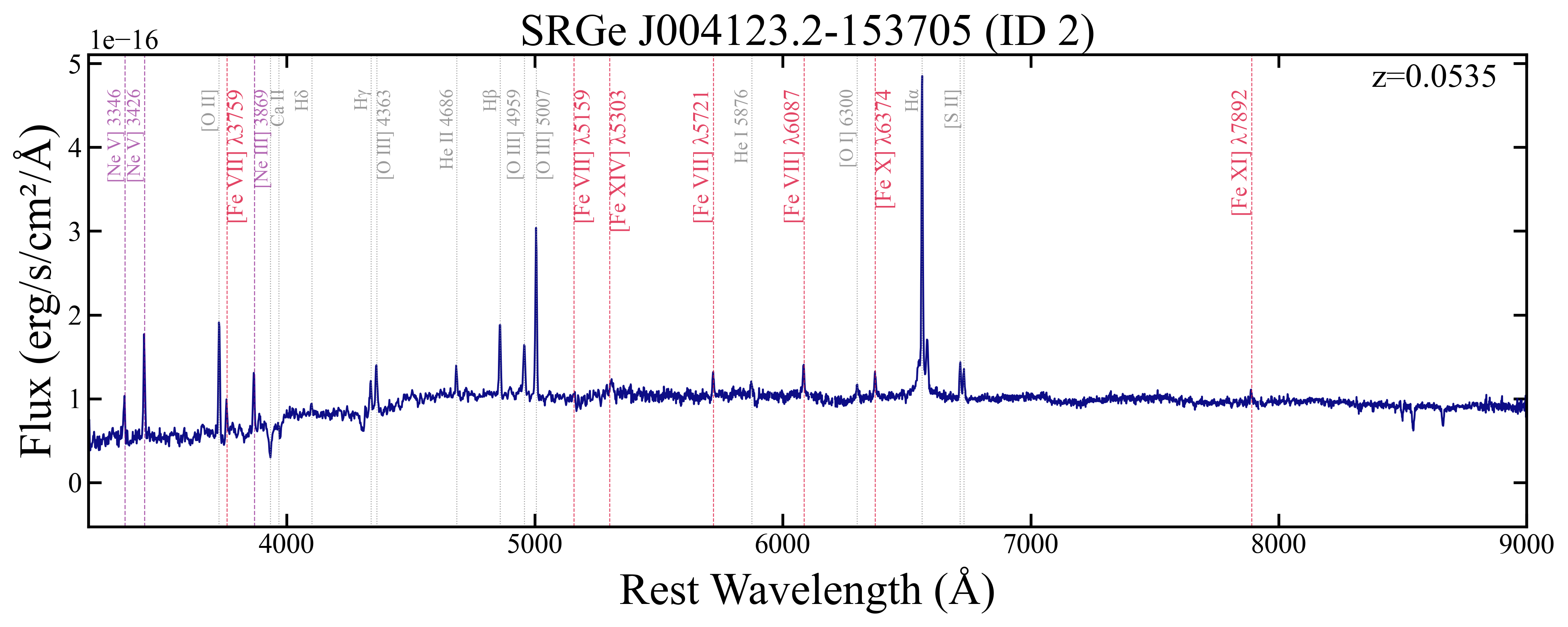}
    \caption{Keck-I/LRIS optical spectrum of SRGe~J004123.2$-$153705 (ID 2), a CrL-TDE. 
}    
\label{fig:SeJ0041_spec}
\end{figure}

After examining all LRIS spectra of our candidates, we identified two objects that exhibit high-ionization coronal lines: \object{SRGe J004123.2-153705} (ID 2) and \object{SRGe J201138.9-210935} (ID 63). The optical counterpart of \object{SRGe J201138.9-210935} (ID 63), TDE\,2021qth/ZTF21abhrchb, has been previously classified as a CrL-TDE by \citet{Yao2023}. The optical counterpart of \object{SRGe J004123.2-153705} (ID 2), AT\,2021swi/ZTF21abkqvdo, exhibited an optical flare with a well-defined rise and decline coincident with the X-ray detection (see Figure~\ref{fig:opt_diff_lc}), along with the characteristic spectroscopic features of a CrL-TDE (Figure~\ref{fig:SeJ0041_spec}). We therefore classify it as a CrL-TDE as well.

\subsubsection{Removal of AGN Interlopers} \label{subsubsec:agn}

\begin{figure}[htbp]
    \centering
        \includegraphics[width=\columnwidth]{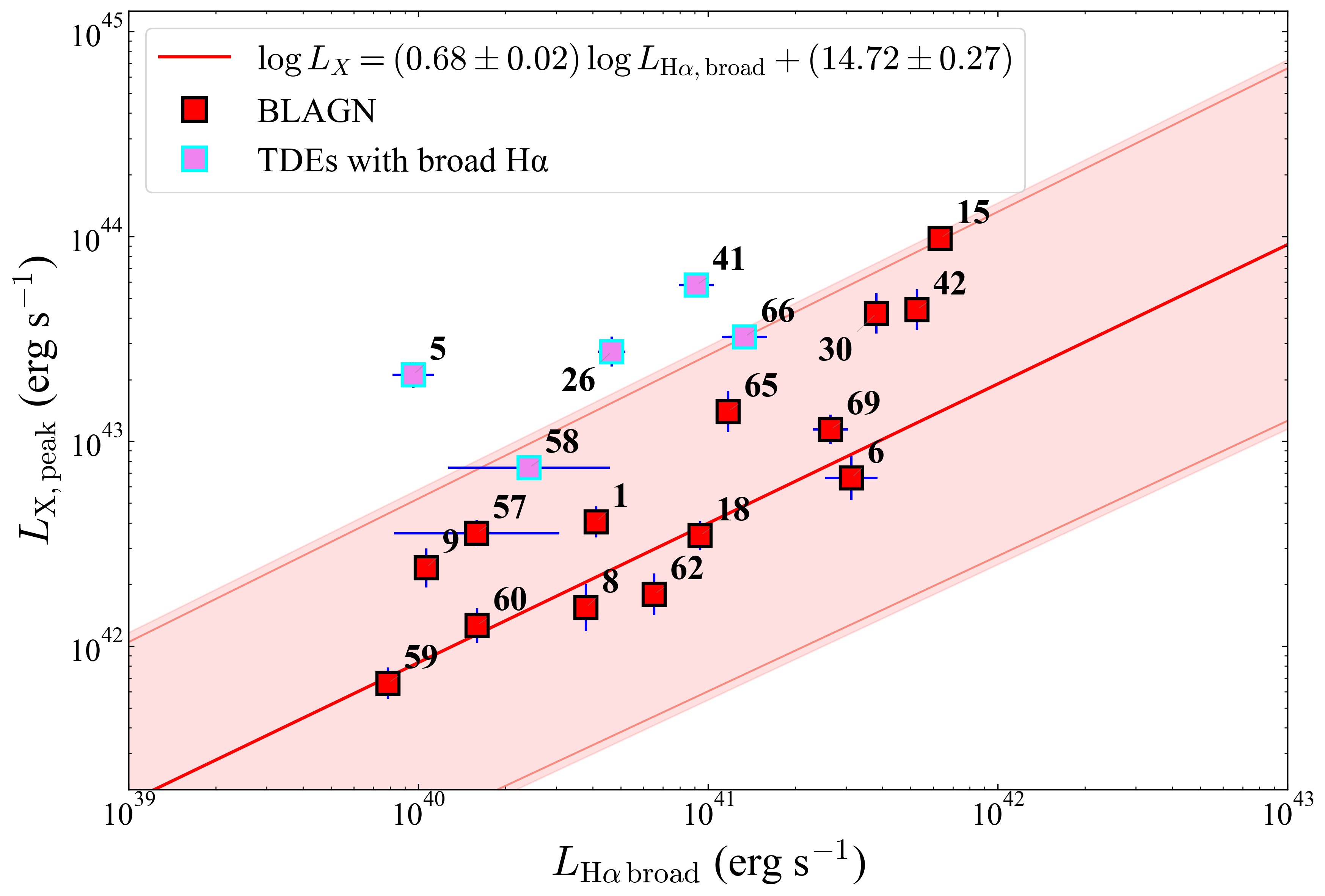}
        \includegraphics[width=\columnwidth]{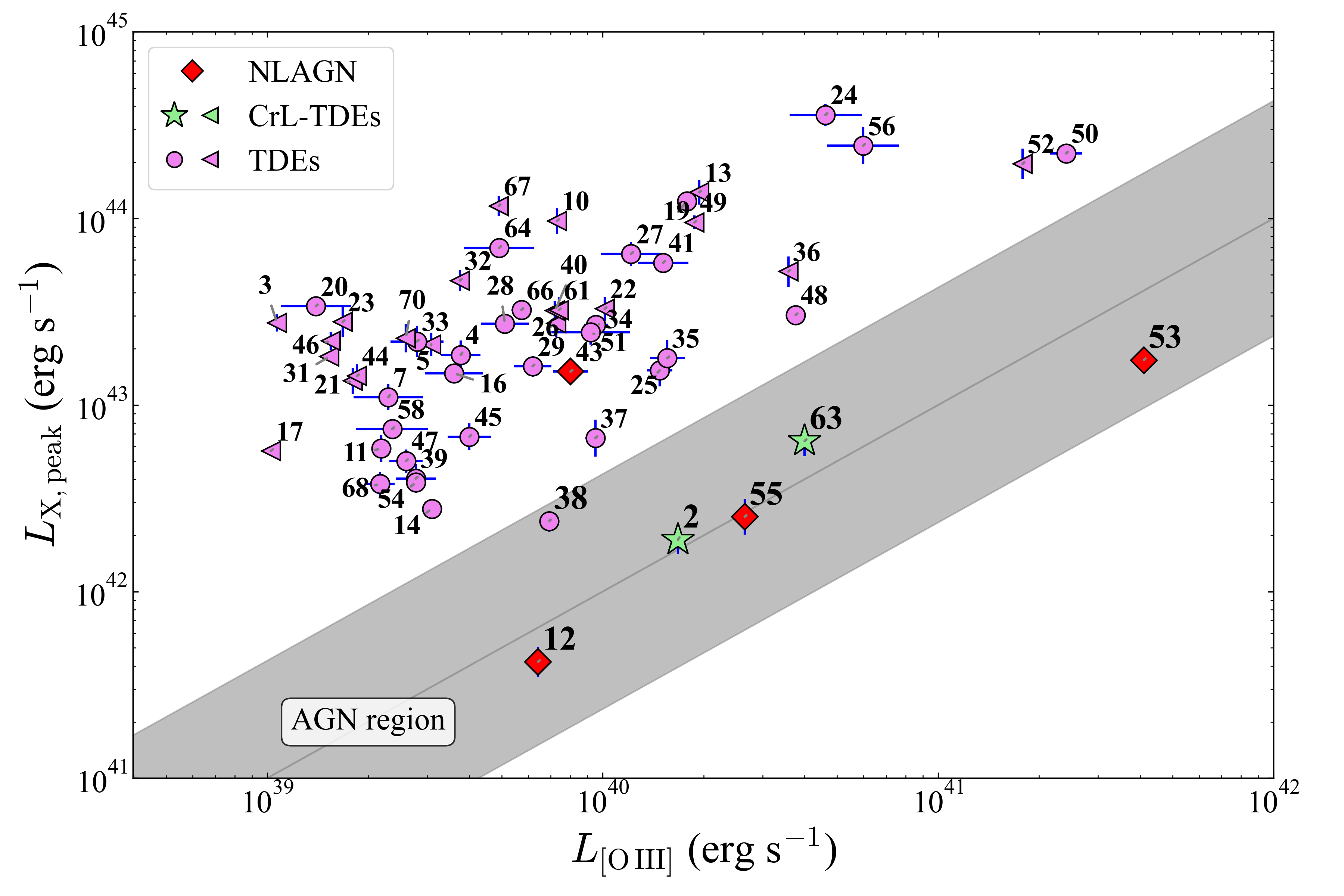}
    \caption{
    Top: Peak 0.3--2.3 keV X-ray luminosity versus broad H$\alpha$ luminosity. 
    The solid line shows the best-fit relation from the corrected calibration of \citet{pulatova2025}, 
    and the shaded region indicates the intrinsic scatter of the correlation.
    Note that the H$\alpha$ luminosity of SRGe J234034.7+293400 (ID 69), is derived from DESI spectrum (see \S\ref{subsec:opt_spec}), while others are all from Keck LRIS spectra. 
    Bottom: $L_{\rm X, peak}$ versus [O\,III]$\lambda5007$ luminosity for the remaining sources. 
    The shaded region shows the AGN locus \citep{Ueda_2015} calibrated in the eROSITA 0.3--2.3\,keV band \citep{Khorunzhev2022}, 
    with the solid line indicating the median relation 
    $\log(L_{[\mathrm{O\,III}]}/L_{\rm X}) = -1.99 \pm 0.63$. 
    Left-pointing triangles denote sources with upper limits on [\ion{O}{III}] line luminosity. 
    }
    \label{fig:Lx_agn_diagnostics}
\end{figure}

\begin{figure*}[htbp]
    \centering
    \includegraphics[width=1\linewidth]{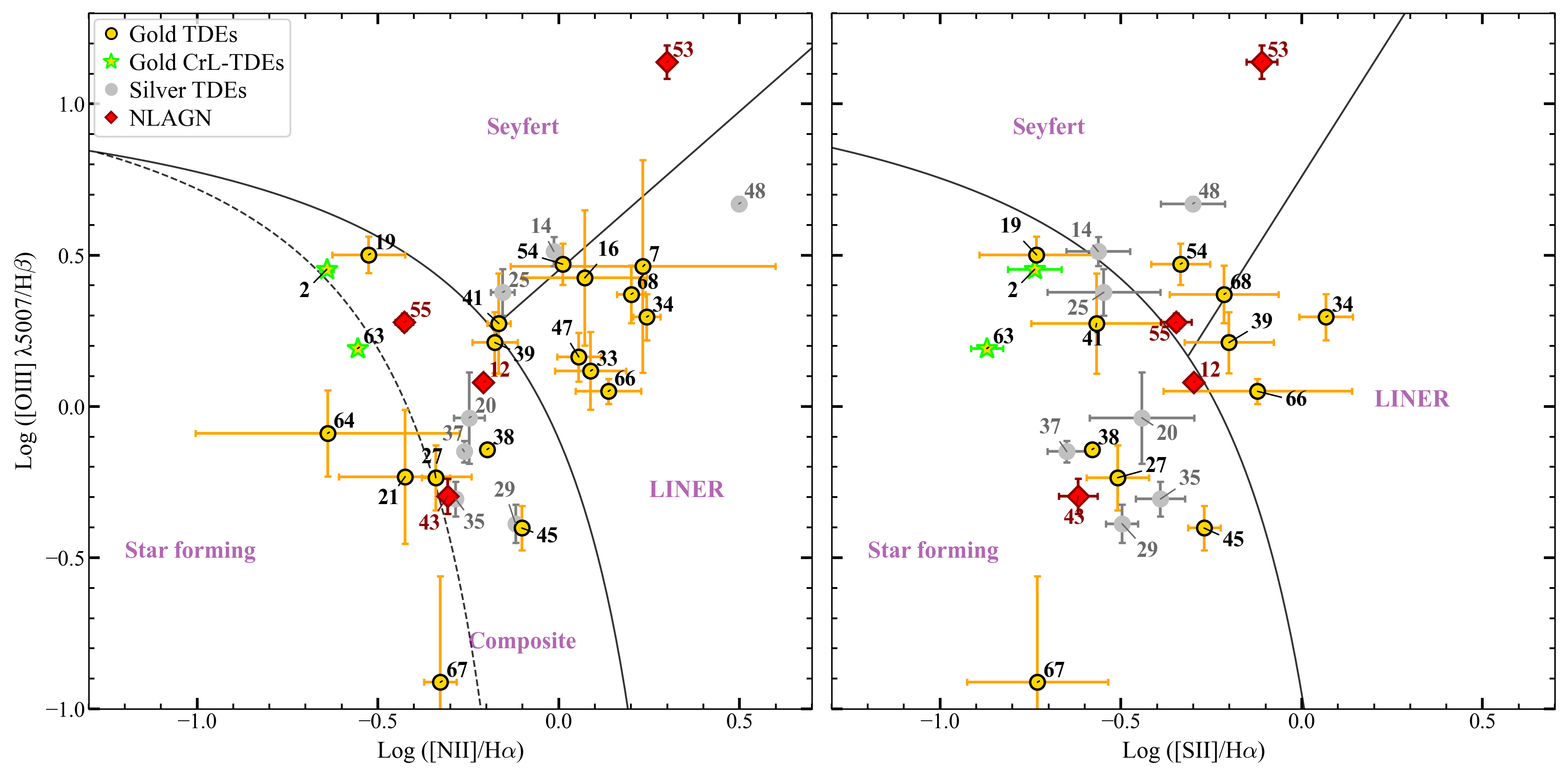}
    \caption{%
    BPT diagrams of host galaxies.
    The solid curves mark the theoretical maximum starburst demarcations of \citet{2006MNRAS.372..961K},
    and the dashed curve in the left panel shows the empirical star-forming/composite boundary from \citet{2003MNRAS.346.1055K}. \label{fig:bpt}}
\end{figure*}

\begin{figure}[htbp]
    \centering
    \includegraphics[width=1\linewidth]{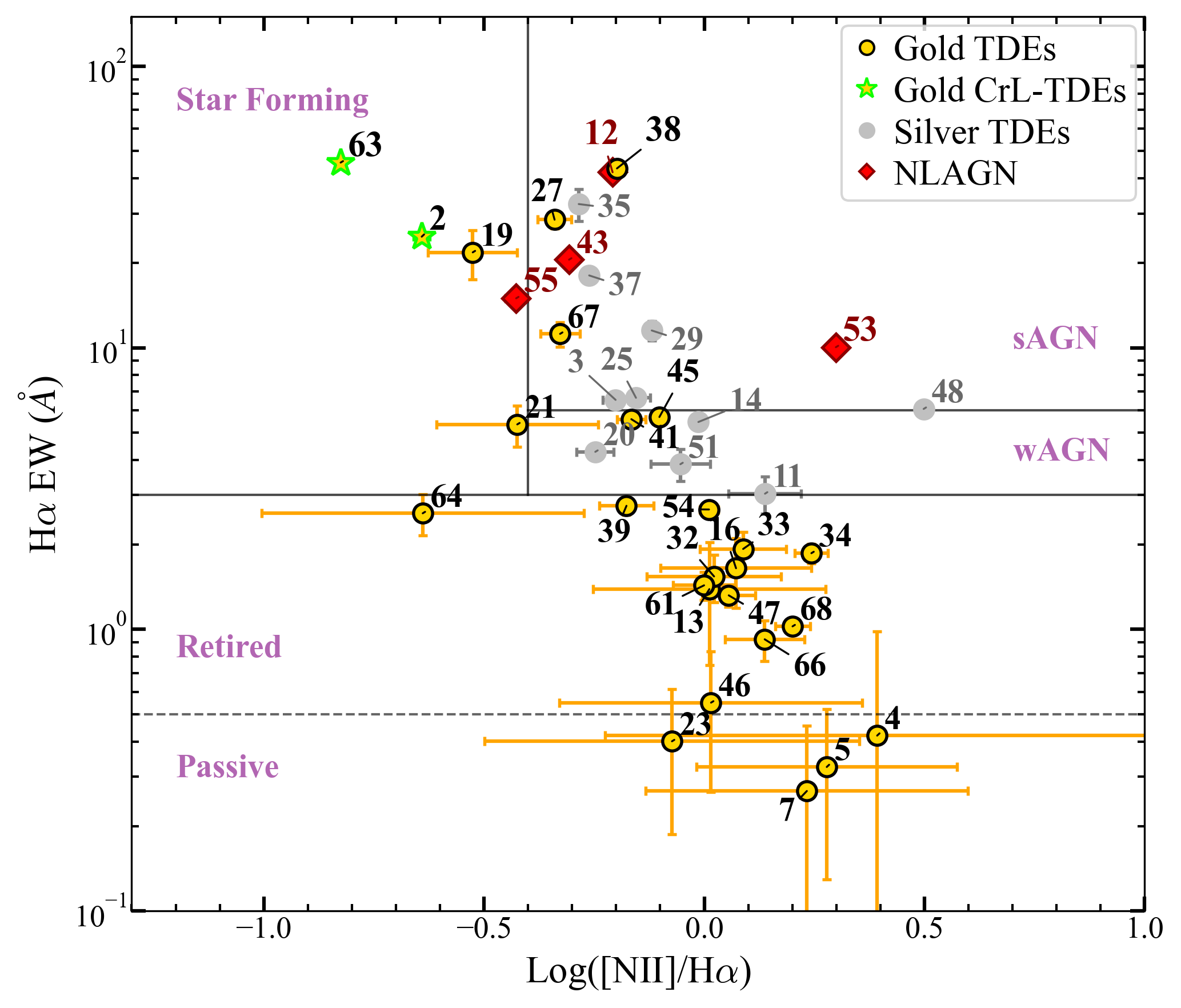}
    \caption{%
    WHAN diagram of host galaxies. 
    The horizontal lines at $\mathrm{EW}(\mathrm{H}\alpha)=6$\,\AA\ and $3$\,\AA\ indicate the sAGN/wAGN and wAGN/retired divisions, respectively, following \citet{2011MNRAS.413.1687C}.
    \label{fig:whan}}
\end{figure}

The AGN population can be broadly divided into BLAGN and narrow-line AGN (NLAGN) based on the full-width at half-maximum (FWHM) of H$\alpha$, with BLAGN exhibiting ${\rm FWHM}>10^3\,{\rm km\,s^{-1}}$ \citep{Stern_2012}. In the AGN unification model \citep{Antonucci1993}, BLAGN are viewed at low inclinations where high-energy X-ray emission from the central engine is directly observable, while NLAGN are viewed at higher inclinations where X-rays are obscured by the dusty torus. We conservatively exclude all BLAGN from our sample, as intrinsic variations in the accretion rate can produce substantial X-ray flux changes that mimic TDE flares. For NLAGN, however, we assess whether the observed X-ray flare luminosity is consistent with typical AGN variability. We retain NLAGN as TDE candidates only if the X-ray flare significantly exceeds the luminosity expected from AGN activity, indicating a distinct transient event rather than normal AGN variability.


We identify BLAGN by requiring a broad H$\alpha$ component with ${\rm FWHM}>1000\,\rm km\,s^{-1}$ in the \texttt{pPXF} fitting results (\S\ref{subsec:ppxf}). \texttt{pPXF} detects 20 sources with broad H$\alpha$ emission\footnote{For sources with multiple spectra, we use the spectrum obtained closest to the X-ray peak to compute H$\alpha$ line luminosities.}. 
The top panel of Figure~\ref{fig:Lx_agn_diagnostics} shows the distribution of these sources in the peak X-ray luminosity versus broad H$\alpha$ line luminosity plane. We use the empirical $L_{\rm X}$--$L_{\mathrm{H}\alpha,\mathrm{broad}}$ relation from \citet{pulatova2025} as the primary diagnostic for identifying BLAGN. A total of 13 objects (IDs 1, 6, 8, 9, 18, 30, 42, 57, 59, 60, 62, 65, 69) fall within the BLAGN locus and are classified accordingly. Two of these (IDs 30 and 65) exhibit optical flares identified in \S\ref{subsubsec:optical_flare}. The optical flare of ID 65 has AGN-like colors, while the colors of ID 30 are poorly constrained. We conservatively classify both as likely BLAGN flares rather than TDEs.

Three sources (IDs 15, 58, 66) fall near the upper edge of the BLAGN locus and require additional scrutiny. SRGe J023440.1-021812 (ID 15) exhibits, in addition to broad Balmer lines, broad H$\beta$ and broad \ion{Mg}{II} $\lambda\lambda2796$, 2803 emission which is rarely detected in known TDEs \citep{Hung2019}, favoring a BLAGN interpretation. In contrast, SRGe J175023.7+712857 (ID 58) and SRGe J213527.2-181635 (ID 66) both display broad helium lines (see \S\ref{subsubsec:tde-broad}) characteristic of TDEs and are therefore classified as TDEs with broad H$\alpha$ rather than BLAGN.

The remaining three sources with detected broad H$\alpha$ (IDs 5, 26, 41)  all exhibit TDE-like optical flares (\S\ref{subsubsec:optical_flare}). For IDs 5 and 26, LRIS spectra obtained during the optical flares show transient broad \ion{He}{II} features (see \S\ref{subsubsec:tde-broad}), indicating the broad lines are transient rather than AGN-driven. For ID 41, LRIS spectra were obtained after the optical flares faded below ZTF sensitivity. Given the absence of other typical BLAGN signatures (broad H$\beta$, stochastic variability, or broad \ion{Mg}{II}), we interpret the detected broad H$\alpha$ as associated with the TDE optical flare.


Some of the X-ray transients are likely powered by NLAGN. To identify this population, we analyze the relation between the [\ion{O}{III}] $\lambda5007$ line luminosity ($L_{[\mathrm{O\,III}]}$) and the peak X-ray luminosity, along with the Baldwin, Phillips \& Terlevich (BPT; \citealt{Baldwin1981}) and WHAN \citep{2011MNRAS.413.1687C} diagnostic diagrams.

The BPT diagram classifies galaxies using flux ratios of emission lines that are close in wavelengths. This diagnostic framework distinguishes four galaxy types: star-forming, composite, Seyfert, and Low-Ionization Nuclear Emission-line Region (LINER; \citealt{Heckman1980}) galaxies. 
Seyfert galaxies are AGN dominated by accretion-powered nuclear emission.
Composite galaxies occupy an intermediate region and show mixed contributions from both star formation and AGN ionization \citep{2006MNRAS.372..961K}.
The ionization source in LINER galaxies remains debated: while many LINERs harbor low-luminosity AGN, alternative mechanisms including shocks, outflows, or photoionization by post-asymptotic giant branch (post-AGB) stellar populations may also contribute \citep{Ho1993, Filippenko1996, Singh2013}.

We employ two BPT diagrams using [\ion{O}{III}]/H$\beta$ versus [N\,\textsc{ii}]/H$\alpha$ and [S\,\textsc{ii}]/H$\alpha$ ratios (Figure~\ref{fig:bpt}). Not all gold and silver sources appear in these diagrams, as BPT classification requires reliable measurements of [\ion{O}{III}], [\ion{N}{II}], [\ion{S}{II}], narrow H$\beta$ and narrow H$\alpha$. We only plot sources for which all required lines are detected with $\mathrm{S/N}\ge 5$. The two BPT panels are generally consistent with each other, with no sources appearing as star-forming on one panel while being classified as Seyfert or LINER on the other. Discrepancies mainly occur among sources classified as Composite, Seyfert, or LINER. For sources that fall into different regions among these three categories between the two BPT panels, we conservatively adopt the classification indicating AGN presence (Seyfert or LINER) in BPT diagnostic to ensure robust AGN exclusion from our gold TDE sample (see \S\ref{subsubsec:final_classification}).

Many systems classified as composite or LINER on the BPT diagram are not genuine AGN. 
The WHAN diagram provides a cleaner separation between true AGN and such contaminants using the EW of H$\alpha$ and the [\ion{N}{II}]/H$\alpha$ line flux ratio \citep{2011MNRAS.413.1687C}. 
This framework classifies galaxies into five categories: star forming, strong AGN (sAGN), weak AGN (wAGN), retired, and passive. As with BPT, WHAN classification requires reliable narrow H$\alpha$ and [\ion{N}{II}] measurements, so sources lacking these measurements are not shown (Figure~\ref{fig:whan}). 
Galaxies classified as LINER in the BPT diagram but falling outside the sAGN or wAGN regions in the WHAN diagram are more consistent with ionization by hot evolved stars (e.g., post-AGB stars) rather than accretion activity. Similarly, BPT–composite galaxies falling outside the WHAN-sAGN region are more consistent with ionization by diffuse gas or evolved stellar populations rather than an active nucleus.

On the $L_{[\mathrm{O III}]}$--$L_{\mathrm{X}}$ diagram (bottom panel of Figure~\ref{fig:Lx_agn_diagnostics}), four sources that are not classified as CrL-TDEs fall within the AGN locus (IDs 12, 38, 53, and 55).
Three (IDs 12, 53, 55) are independently confirmed as AGN by both BPT and WHAN diagnostics (Figures~\ref{fig:bpt} and \ref{fig:whan}) and are classified as NLAGN. The remaining source, \object{SRGe J153134.9+330539} (ID 38; AT\,2020ocn/ZTF20aabqihu), shows evidence for a TDE origin: it was previously identified as a TDE-He by \citet{Hammerstein2023} and exhibits strong early-time X-ray variability \citep{Cao2024_20ocn}.


Finally, we consider SRGe J155834.0+382528 (ID 43), which exhibits a red optical flare (\S\ref{subsubsec:optical_flare}; Figure~\ref{fig:opt_diff_lc_3}). Although it does not fall within the AGN locus in Figure~\ref{fig:Lx_agn_diagnostics}, it is classified as star-forming/composite in the BPT diagram and as sAGN in the WHAN diagram. We therefore conservatively classify it as an AGN flare rather than a TDE.

In summary, we identify 18 objects as AGN interlopers (IDs 1, 6, 8, 9, 12, 15, 18, 30, 42, 43, 53, 55, 57, 59, 60, 62, 65, 69).

\begin{figure*}[htbp!]
    \centering
    \includegraphics[width=\columnwidth]{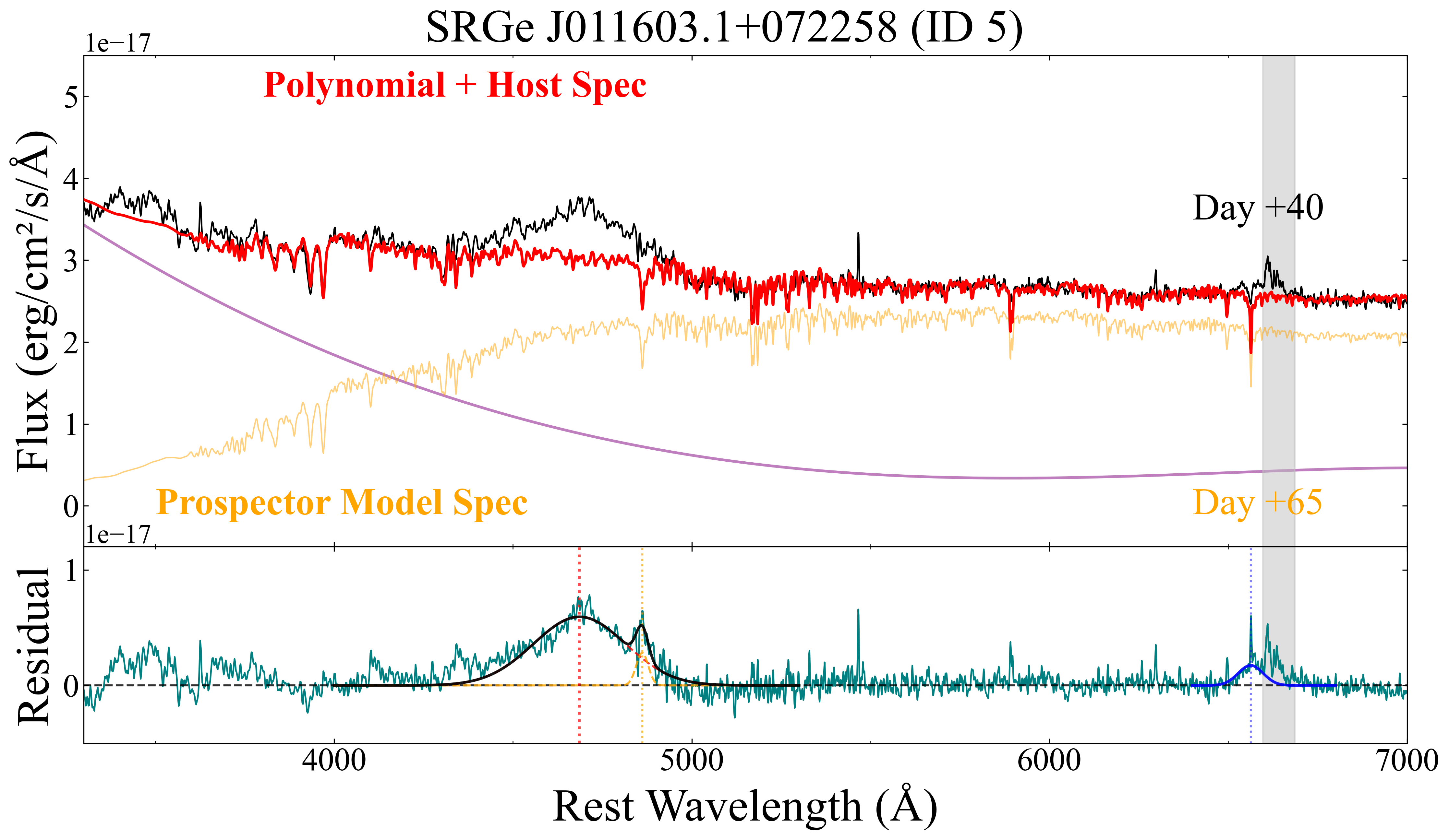}
    \includegraphics[width=\columnwidth]{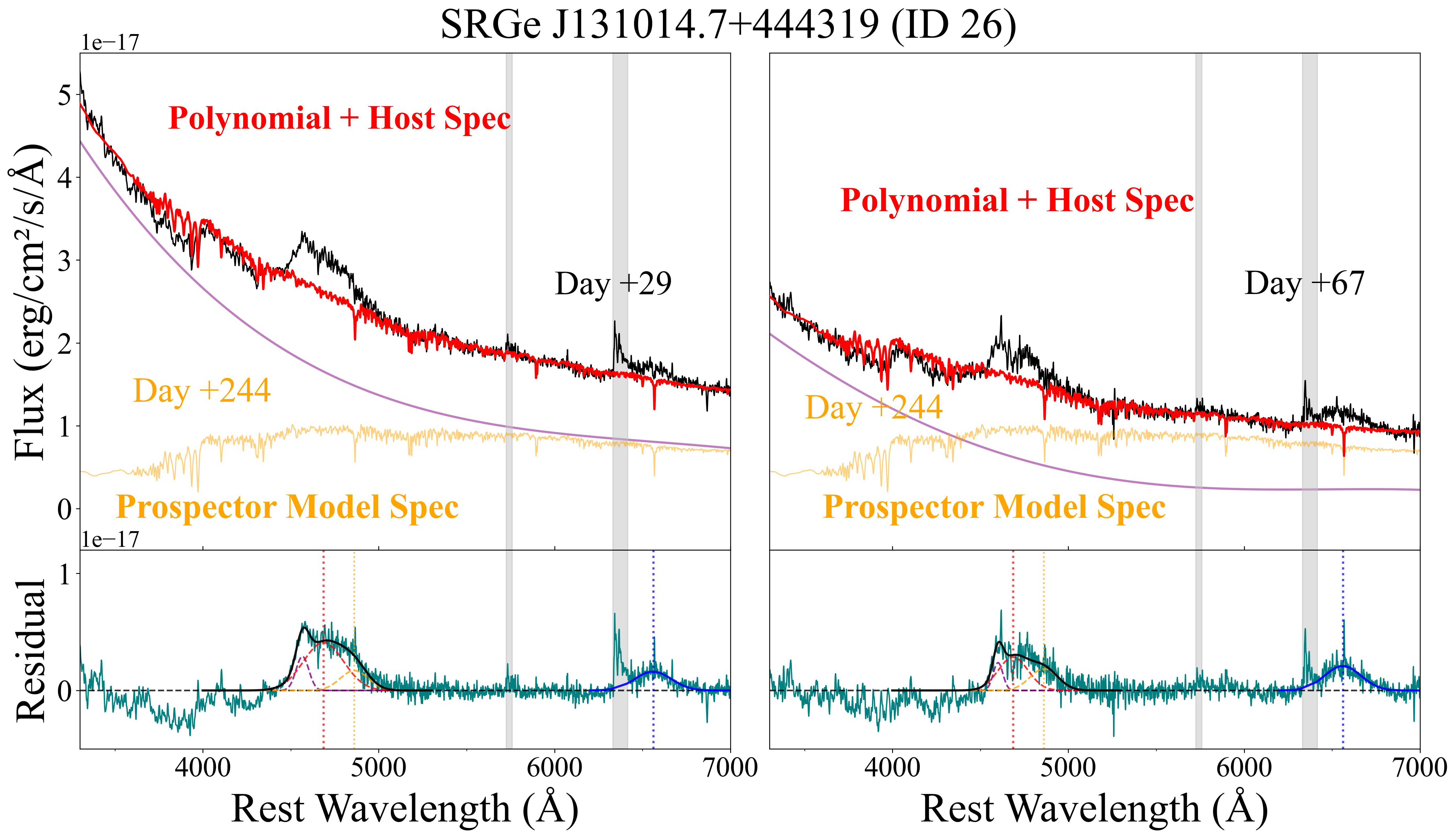}\\
    \includegraphics[width=\columnwidth]{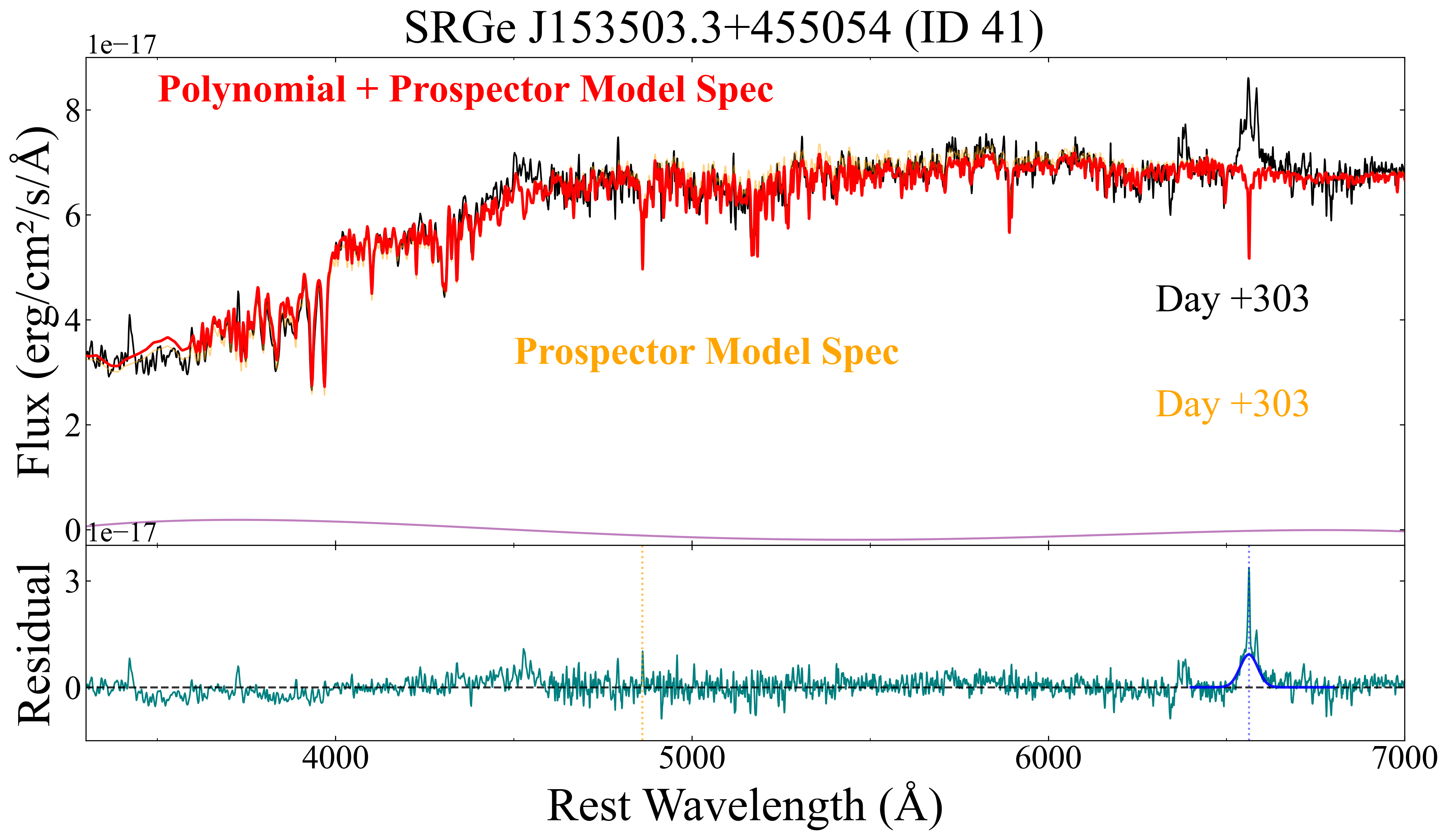}
    \includegraphics[width=\columnwidth]{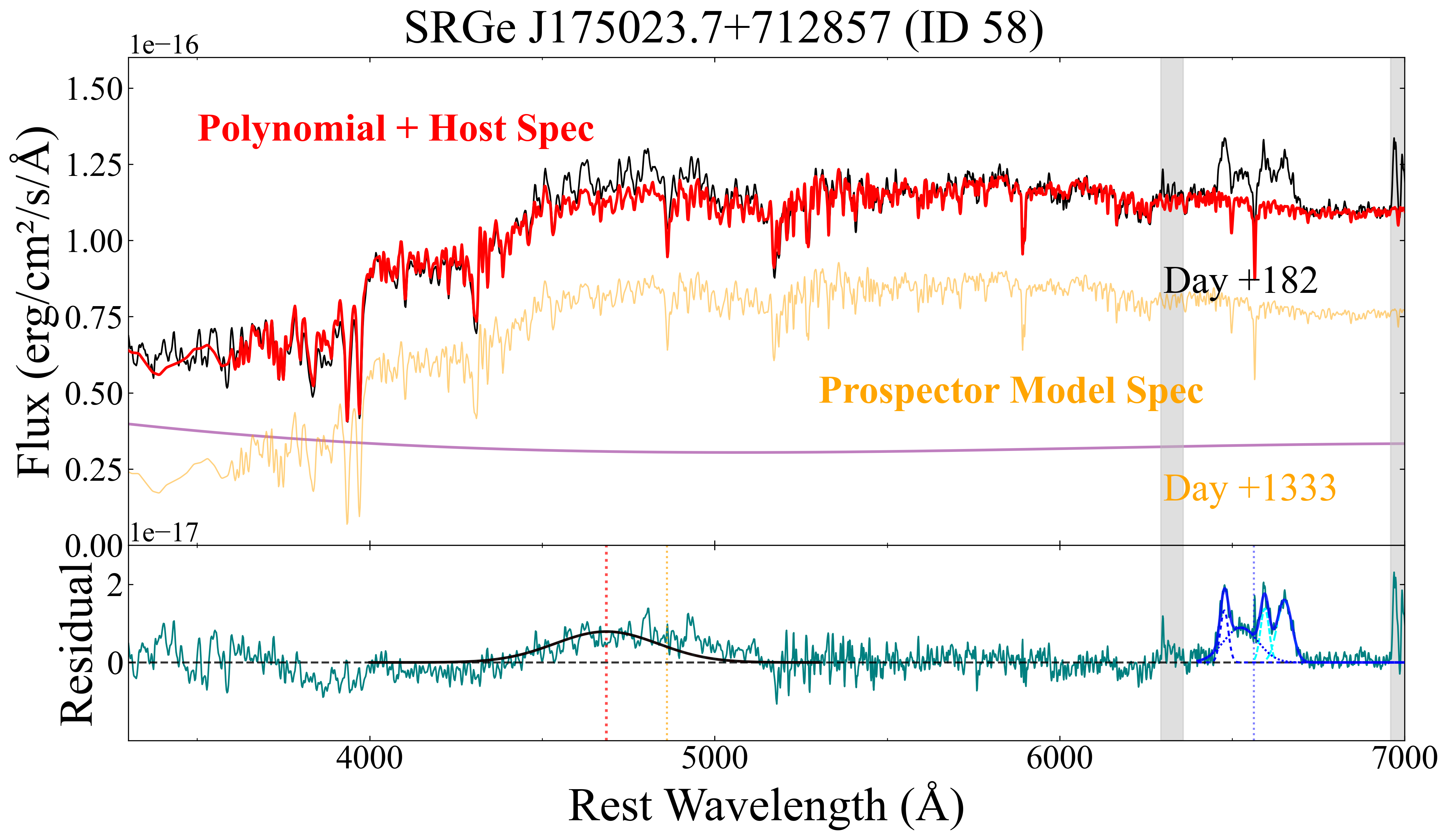}\\
    \includegraphics[width=\columnwidth]{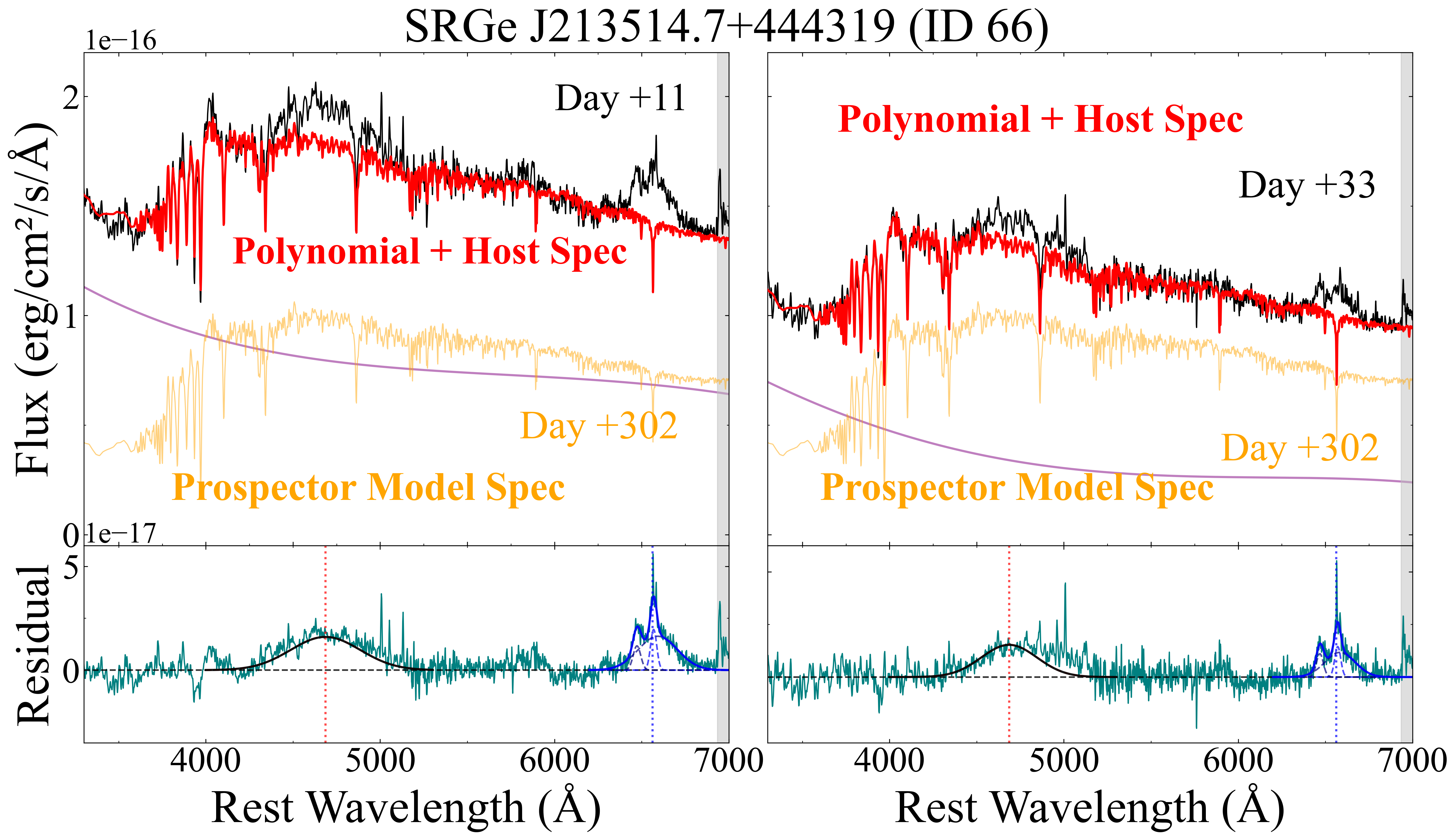}
    \includegraphics[width=\columnwidth]{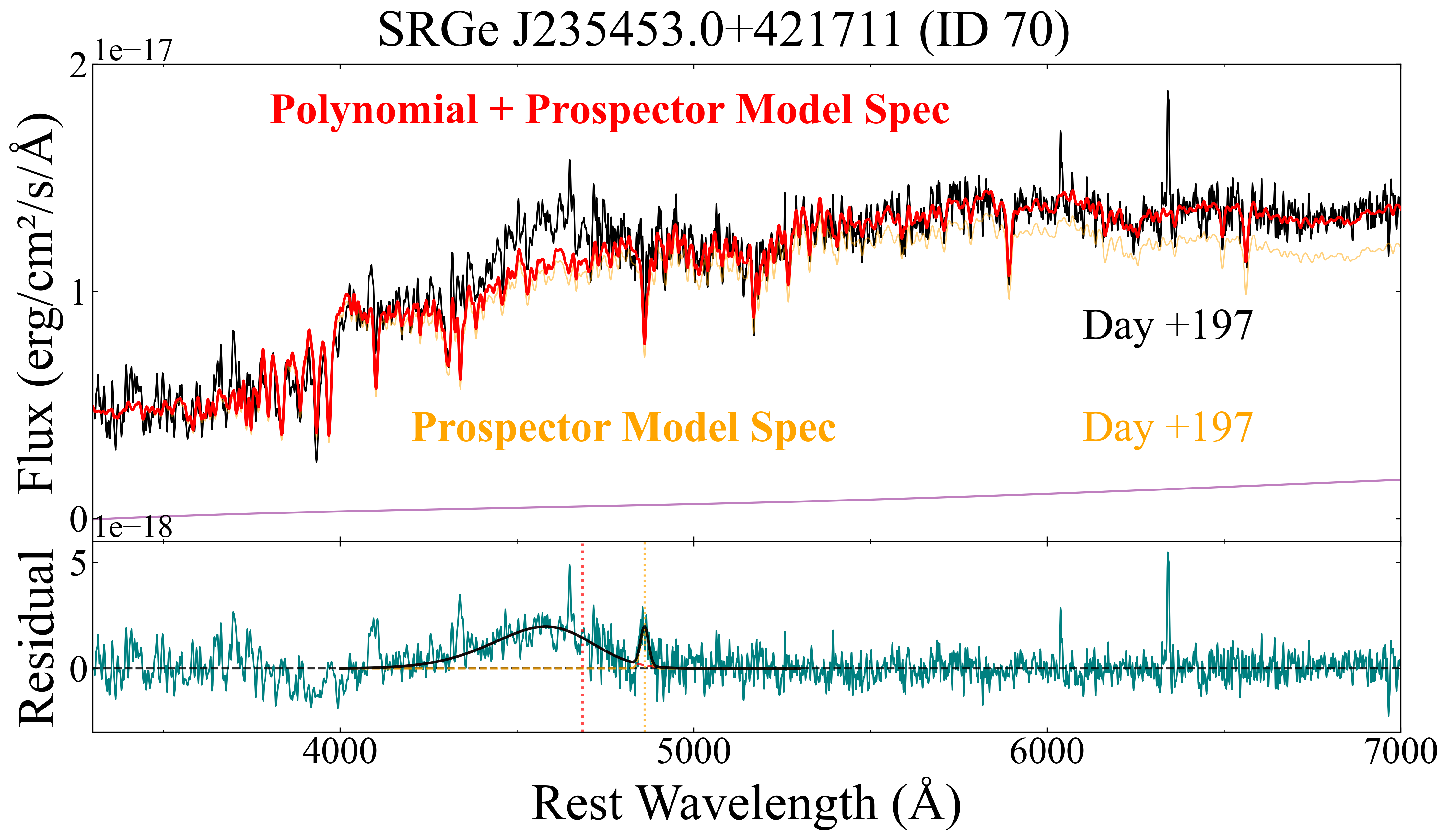}
    \caption{Spectral modeling of six objects with broad lines associated with the TDE. The observed LRIS data is shown in black. The orange line shows the \texttt{Prospector}-derived host model, and the red line shows the final continuum (polynomial fit plus scaled host model). 
    The continuum-subtracted residuals are shown in teal.
    The individual Gaussian fits are shown as dashed curves, with vertical dotted lines marking their rest-frame wavelengths: purple for the Bowen blend, red for He\,\textsc{ii}~$\lambda4686$, orange for H$\beta$, and blue for the broad component of H$\alpha$~$\lambda6563$. The solid black and blue curves show the total models in the He\,\textsc{ii} and H$\alpha$ regions, respectively. 
    Grey bands indicate masked telluric regions.} 
    \label{fig:HeII}
\end{figure*}

\subsubsection{Identification and Modeling of Broad TDE Lines in LRIS Spectra}
\label{subsubsec:tde-broad}
In this section, we examine whether broad emission lines associated with the optical TDEs are detected in the LRIS spectra. We focus on the 50 objects that are not classified as AGN interlopers (§\ref{subsubsec:agn}) or CrL-TDEs (§\ref{subsubsec:crl-tde}). As noted above, four TDEs (IDs 5, 26, 58, 66) have multiple LRIS observations in which prominent broad lines that are well above the continuum and easily identifiable by eye are detected in earlier epochs.

Do we expect to detect broad lines in the remaining 46 candidates? 
In most cases, the LRIS follow-up occurred sufficiently late that the optical flare had faded below the ZTF sensitivity limit. 
According to the forced-photometry light curves in Appendix~\ref{sec:sup_fig}, only three objects (IDs 17, 24, 49) were observed while the optical flare was still detectable by ZTF. ID 17 was previously classified as a featureless TDE based on early-time observations \citep{Yao_2022}, while IDs 24 and 49 have faint continuum fluxes ($f_\nu \lesssim 3\,\mu{\rm Jy}$, $m\gtrsim 22.7$\,mag) such that transient emission might be dominated by host-galaxy light.
Nevertheless, broad lines could remain visible in our spectra even when ZTF forced photometry shows no optical flare, provided their luminosities fall below the ZTF detection threshold.

To systematically search for broad lines around the He\,\textsc{ii} and H$\alpha$ regions in the 46 candidates, we use the two \texttt{Prospector} fits described in §\ref{subsec:prospector_fitting}. 
We first inspect the \texttt{Prospector} fits in which the He\,\textsc{ii} and H$\alpha$ regions are masked. 
We require the significance of EW of the He \,\textsc{ii} (4000--5300~\AA) or H$\alpha$ (6400--6800~\AA) regions to be greater than five. 
This procedure yields four candidates exhibiting broad-line excess (IDs 39, 41, 64, 70). Because template mismatch can produce spurious residuals, we also examine the unmasked \texttt{Prospector} fits. The broad-line excess persists in IDs 39, 41, and 70 but disappears for ID 64.

In total, we therefore identify seven objects with flux excess around common TDE broad line regions in LRIS spectroscopy (IDs 5, 26, 39, 41, 58, 66, 70). Following \citet{Charalampopoulos_2022}, we measure their line fluxes after subtracting two components from the observed LRIS spectra: (1) a scaled \texttt{Prospector} stellar continuum model (with He\,\textsc{ii} and H$\alpha$ masked) representing the host contribution, and (2) a low-order polynomial accounting for the optical flare continuum. In Figure~\ref{fig:HeII}, the black lines in the top panels show the observed spectra ($F_{\lambda,\mathrm{obs}}$), and the red lines show the combined continuum model ($F_{\lambda,\mathrm{cont}}$).

We model the emission lines with Gaussian profiles to determine their fluxes and kinematics. 
In the H$\alpha$ region (6400--6700~\AA), the main components include a narrow and a broad H$\alpha$ component, and the [N\,\textsc{ii}]~$\lambda\lambda$6548,~6583 doublet whose flux ratio is fixed at the theoretical value of 1:3.
We include additional free-centered Gaussian components as needed to capture complex line structure \citep{Nicholl2019,Charalampopoulos_2022,kumar_2024}\footnote{While multiple (3--4) Gaussians do not constitute a physical model, this approach provides a simple mathematical framework for characterizing features with complex morphologies.}.
In the He\,\textsc{ii} region (4000--5300~\AA), we simultaneously fit He\,\textsc{ii}~$\lambda$4686, H$\beta$, and Bowen blend features using both broad and narrow Gaussian components. 
For SRGe J235453.0+421711 (ID 70), which exhibits asymmetric blueshifted He \,\textsc{ii} emission, we employ skewed Gaussian profiles following \citet{kumar_2024}.

Model selection is guided by the Bayesian Information Criterion (BIC). A more complex model is adopted only when the improvement exceeds $\Delta\mathrm{BIC}>10$, a threshold commonly interpreted as very strong evidence in favor of the more complex model \citep{Kass1995_BIC,Lorah2019_BIC}. This criterion ensures that additional components are statistically justified while minimizing the risk of overfitting. The measured line properties (luminosity, FWHM, EW) are presented in Table~\ref{tab:he_h_flux} of Appendix~\ref{sec:sup_tab}. We note that broad line detections in SRGe J153331.5+390536 (ID 39) have rather low significance  ($5.2\sigma$ for \ion{He}{II} and $2.0\sigma$ for H$\alpha$), and we therefore exclude this object from further discussion. The residual spectra and Gaussian fitting results for the other six objects are shown in the bottom panels of Figure~\ref{fig:HeII}.

For spectra where H$\alpha$ or He\,\textsc{ii}~$\lambda4686$ lines are not detected, we computed $3\sigma$ upper limits on line luminosities from the residual spectrum. We integrate the residual flux over $\pm500\,{\rm km\,s^{-1}}$ velocity windows centered on each line. To properly account for measurement uncertainties, we performed 10,000 Monte Carlo realizations by perturbing the spectra with noise. The 99.73rd percentile of the resulting flux distribution defines the $3\sigma$ upper limit. These luminosity upper limits are also listed in Table~\ref{tab:he_h_flux} (Appendix~\ref{sec:sup_tab}). We discuss the results of this analysis in \S\ref{subsec:broad_lines}.

\begin{sidewaystable*}[htbp!]
\begin{center}
\caption{The gold TDE sample. \label{tab:classification_gold}}
\small
\begin{tabular}{ccccccccc}
 \toprule
 ID  &  Name  &  Type in &  Type in &  Optical &  Spectral &  Reference & IAU  & ZTF  \\
    &       &  BPT     &   WHAN    &   flare & subtype    &    &  name & name \\
\midrule
2 & SRGe J004123.2-153705 & SF & SF & \cmark & CrL-TDE &  & AT 2021swi & ZTF21abkqvdo\\
4 & SRGe J010445.8+044319 & --- & Retired/Passive & \xmark & --- &  &  \\
5 & SRGe J011603.1+072258 & --- & Passive & \cmark & TDE-H+He &  &  \\
7 & SRGe J013204.4+122235 & LINER/Seyfert & Passive & \xmark & --- & \citet{Sazonov2021} &  \\
10  &  SRGe J015754.6-154214  &  ---  &  --- & \xmark& & &   \\
13 & SRGe J021939.7+361819 & --- & Retired & \xmark & --- & \citet{Sazonov2021} &  \\
16 & SRGe J025548.1+142800 & LINER/Seyfert & Retired  & \cmark & --- &  &  \\
17 & SRGe J030747.8+401842 & --- & --- & \cmark & TDE-featureless & \citet{Yao_2022} & TDE 2021ehb & ZTF21aanxhjv\\
19 & SRGe J060324.7+621112 & Composite & SF & \xmark & --- &  &  \\
20 & SRGe J071310.4+725627 & Composite & wAGN & \xmark & --- & \citet{Sazonov2021} &  \\
21 & SRGe J081006.4+681755 & SF/Composite & SF/wAGN & \cmark & --- &  &  \\
22  &  SRGe J083640.9+805410  &  ---  &  --- & \xmark & &  &\\

23 & SRGe J091747.3+524818 & --- & Retired/Passive & \cmark & --- & \citet{Sazonov2021} &  \\
24 & SRGe J095928.7+643024 & --- & --- & \cmark & --- & \citet{Sazonov2021} &  \\
26 & SRGe J131014.7+444319 & --- & --- & \cmark & TDE-H+He &  &  \\
27 & SRGe J131404.1+515427 & SF/Composite & sAGN & \cmark & --- &  &  & ZTF21aafkznp\\
28  &  SRGe J132718.1+350437  &  ---  &  --- & \xmark &  &  &\\

31 & SRGe J135353.7+535949 & --- & --- & \cmark & TDE-He & \citet{Hammerstein2023} & AT 2020ocn & ZTF18aakelin\\
32 & SRGe J135515.0+311605 & --- & Retired & \xmark & --- & \citet{Sazonov2021} &  \\
33 & SRGe J135812.1+195357 & LINER & Retired & \xmark & --- &  &  \\
34 & SRGe J144738.3+671818 & LINER & Retired & \xmark & --- & \citet{Sazonov2021} &  \\
38 & SRGe J153134.9+330539 & Composite & sAGN & \cmark & TDE-H+He & \citet{Hammerstein2023} & TDE 2020pj & ZTF20aabqihu\\
39 & SRGe J153331.5+390536 & Comp/Sey/LIN & Retired & \cmark & --- &  &  \\
40 & SRGe J153403.5+621851 & --- & --- & \cmark & &  &  \\
41 & SRGe J153503.3+455054 & Comp/Sey/LIN & wAGN & \cmark & TDE-H? & \citet{Sazonov2021} &  \\
44  &  SRGe J160943.1+253603  &  ---  &  --- & \xmark & --- &  & \\

45 & SRGe J161001.3+330120 & Composite & wAGN & \xmark & --- & \citet{Sazonov2021} &  \\
46 & SRGe J161559.1+360156 & --- & Retired/Passive & \xmark & --- &  &  \\
47 & SRGe J162159.4+271133 & LINER & Retired & \cmark & --- &  &  \\
48 & SRGe J162932.1+280521  & LINER/Seyfert  &  sAGN/wAGN  & \cmark & --- &  &   \\
49 & SRGe J163030.3+470125 & --- & --- & \cmark & --- & \citet{Sazonov2021} &  \\
50 & SRGe J163831.9+534018 & --- & --- & \xmark & --- & \citet{Sazonov2021} &  \\
54 & SRGe J171423.6+085237 & LINER/Seyfert & Retired & \xmark & --- & \citet{Sazonov2021} &  \\
58 & SRGe J175023.7+712857 & --- & --- & \cmark & TDE-H+He &  & & ZTF20achpskf  \\
61 & SRGe J192143.8+503853 & --- & Retired & \xmark & --- &  &  \\
63 & SRGe J201138.9-210935 & SF & SF & \cmark & CrL-TDE & \citet{Yao2023} & TDE 2021qth & ZTF21abhrchb\\
64 & SRGe J204129.5+214409 & SF/Composite & Retired & \cmark & --- & \citet{Khorunzhev2022} & AT 2021imi &ZTF21aatgiaq\\
66 & SRGe J213527.2-181635 & Composite/LINER & Retired & \cmark & TDE-H+He & 
\citet{Wevers2024} & AT 2020ksf & ZTF20abgbdpr\\
67 & SRGe J223905.0-270551 & SF & sAGN & \cmark & --- &  &  \\
68 & SRGe J231834.5-351914 & LINER/Seyfert & Retired & \xmark & --- &  &  \\
70 & SRGe J235453.0+421711 & --- & --- & \cmark & TDE-He &  &  \\
  \bottomrule
  \end{tabular}
\end{center}
\textbf{Notes.} In the ``Spectral subtype'' column, we list subtype classifications only if available from the literature or if we detect coronal lines or broad emission lines associated with the TDE in this work.
\end{sidewaystable*}

\begin{table}[htbp!]
\setlength{\tabcolsep}{2pt}
\begin{center}
\caption{The silver TDE sample. \label{tab:classification_silver}}
\begin{tabular}{cccc}
 \toprule
 ID  &  Name  &  Type in &  Type in  \\
    &       &  BPT     &   WHAN   \\
\midrule
 3  &  SRGe J010301.0-130120  &  ---  &  sAGN/wAGN  \\
 11  &  SRGe J015907.1-150323  &  ---  &  wAGN/Retired  \\
 14  &  SRGe J023017.3+283606  &  Comp/Seyfert  &  wAGN  \\
 25  &  SRGe J113323.2+693635  &  Comp/Sey/LIN  &  sAGN \\
 29 & SRGe J133053.5+734823 & Composite & sAGN \\
 35  &  SRGe J145226.6+670437  &  Composite  &  sAGN  \\
 36  &  SRGe J150328.1+495117  &  ---  &---  \\
 37  &  SRGe J152656.3+353317  &  Composite  &  sAGN  \\
 51  &  SRGe J165055.8+301634  &  ---  &  wAGN  \\
 52  &  SRGe J170139.0-085911  &  ---  &  ---  \\
 56  &  SRGe J174513.6+401608  &  ---  &  ---  \\
\bottomrule
  \end{tabular}
\end{center}
\textbf{Notes.} SRGe J023017.3+283606 (ID 14) has been identified as an X-ray transient with monthly quasi-periodic eruptions \citep{Evans2023, Guolo2024_22d}. SRGe J133053.5+734823 (ID 29) was reported as a possible TDE by \citet{Sazonov2021}, but it does not pass our classification scheme as it is classified as composite in the BPT diagram and sAGN in the WHAN diagram, so we retain it only in the silver sample.
\end{table}

\subsubsection{Final Sample Definition}
\label{subsubsec:final_classification}

Among the 70 eROSITA TDE candidates, we identified 18 AGN interlopers in \S\ref{subsubsec:agn}. To further identify high-confidence TDEs among the remaining 52 objects, we define a gold sample using the following criteria:
\begin{enumerate}
    \item Events with TDE-like optical flares (\S\ref{subsubsec:optical_flare}) that are not classified as AGN interlopers (\S\ref{subsubsec:agn}) are included in the gold sample. A total of 23 events satisfy this criterion (IDs 2, 5, 16, 17, 21, 23, 24, 26, 27, 31, 38, 39, 40, 41, 47, 48, 49, 58, 63, 64, 66, 67, 70).
    \item For the remaining 29 events, those with host galaxies classified as retired or passive on the WHAN diagram are included. A total of 10 events satisfy this criterion (IDs 4, 7, 13, 32, 33, 34, 46, 54, 61, 68).
    \item For the remaining 19 events, those with host galaxies classified as composite on the BPT diagram and non-sAGN on the WHAN diagram are included. A total of 3 events satisfy this criterion (IDs 19, 20, 45).
    \item Among the remaining 16 events, 8 lack detected emission lines and therefore do not appear in the BPT and WHAN diagrams (IDs 10, 22, 28, 36, 44, 50, 52, 56). Five of these (IDs 10, 22, 28, 44, 50) are quiescent galaxies with minimal emission, while three (IDs 36, 52, 56) have low signal-to-noise spectra near the H$\alpha$ and/or H$\beta$ regions. We include the five quiescent galaxy hosts in the gold sample, as they show no AGN signatures.
\end{enumerate}

Our final classification yields a gold sample of 41 objects (Table~\ref{tab:classification_gold}) and a silver sample of 11 objects (Table~\ref{tab:classification_silver}).

\subsection{SDSS Comparison Sample}
\label{subsec:SDSS_comparison_sample}
We construct a volume-limited SDSS comparison sample for TDE host galaxies based on completeness. First, we identify the minimum stellar mass in our TDE sample ($M_{\rm *,min} = 10^{9.3}~M_{\odot}$), then compute the completeness redshift $z_{\rm complete}$ by analyzing the number density $\phi(z)$ of SDSS galaxies in a mass bin ($\pm 0.15$ dex) centered on $M_{\rm *,min}$. The number density is calculated in redshift bins of $\Delta z = 0.01$, accounting for comoving volume effects. $z_{\rm complete}$ is defined as the redshift where the density drops below 70\% of its reference value, computed over $z = 0.02$--$0.06$, yielding $z_{\rm complete} = 0.065$.

We select SDSS galaxies with $z < z_{\rm complete} = 0.065$. The sample is cross-matched with the stellar mass catalog from \citet{Mendel_2014} and the NYU Value-Added Galaxy Catalog \citep{Blanton_2005}, and rest-frame $u-r$ colors are computed from K-corrected photometry. This results in a volume-limited sample of 144,801 SDSS galaxies. For spectral feature analysis, we further require SDSS galaxies to have measurements in the MPA-JHU catalog \citep{Kauffmann2003, Brinchmann_2004, Tremonti2004}.


\section{Results and Discussion} \label{sec:discuss}

\subsection{Optical Light Curves of X-ray selected TDEs}

\begin{figure}[htbp]
    \includegraphics[width=\columnwidth]{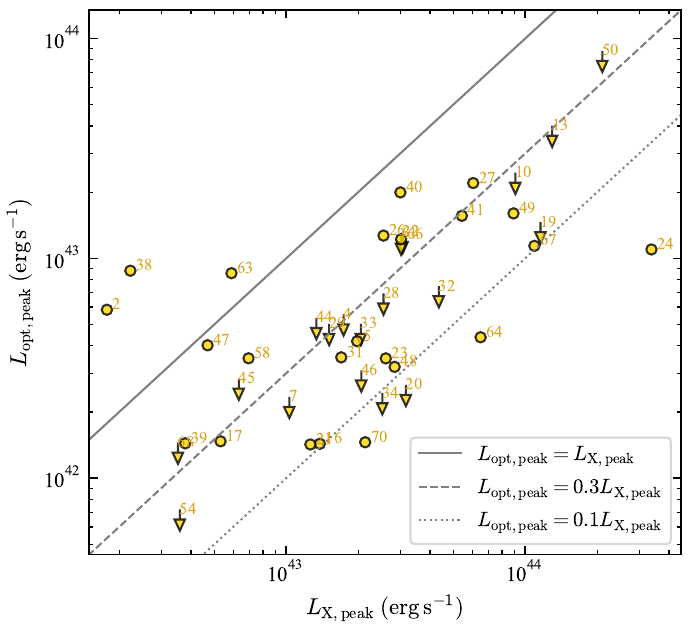}
    \caption{Peak optical luminosity versus peak X-ray luminosity for the 43 TDEs in our gold sample. \label{fig:Lx_Lopt} }
\end{figure}

Here we discuss the optical light curve properties of X-ray-selected TDEs. In our gold TDE sample, 23 of 41 events exhibit optical flares detected in forced photometry. Peak optical luminosities ($L_{\rm opt, peak}$) are computed using the best-fit Gaussian process models shown in Appendix~\ref{sec:sup_fig} (Figures~\ref{fig:opt_diff_lc}--\ref{fig:opt_diff_lc_4}). For events without detected optical flares, we report $3\sigma$ upper limits, where $\sigma$ is the greater of either the median forced photometry flux uncertainty or the standard deviation of the forced photometry flux\footnote{Note that in our previous work \citep{Sazonov2021}, we considered only the median forced photometry flux uncertainty. However, forced photometry uncertainties can be underestimated; the standard deviation provides a more conservative estimate.}. 

Figure~\ref{fig:Lx_Lopt} shows the distribution of our gold sample in the $L_{\rm opt, peak}$ versus $L_{\rm X, peak}$ plane. Note that peak X-ray luminosity $L_{\rm X, peak}$ is estimated from the observed peak flux ($f_{\rm X,peak}$ in Tables~\ref{tab:sample} and \ref{tab:sample2}) captured by the 6-month-cadence eROSITA survey and is therefore likely fainter than the true intrinsic X-ray peak. Nevertheless, 38 of 41 events have $L_{\rm opt, peak}<L_{\rm X, peak}$, demonstrating that the vast majority of X-ray-selected TDEs are intrinsically brighter in X-rays than in the optical.

\begin{figure}
    \includegraphics[width=\columnwidth]{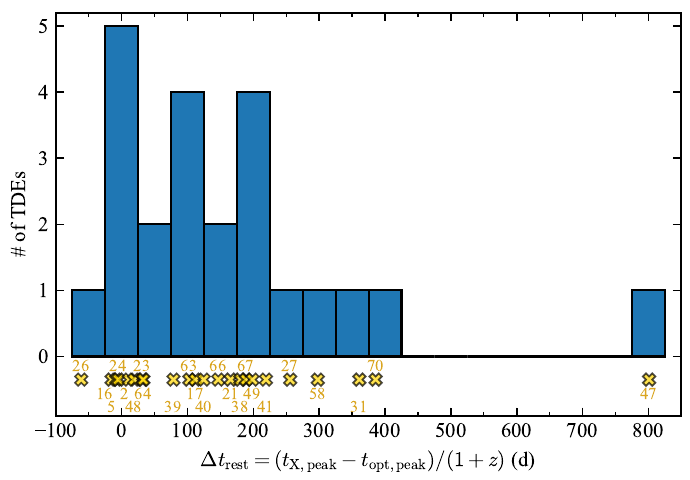}
    \caption{Time difference between the peak X-ray and peak optical emission. The distribution is shown as a histogram, with individual measurements marked along the x-axis. Our sample demonstrates that delayed X-ray peak is common among X-ray selected TDEs. \label{fig:tpeak_diff}}
\end{figure}

Delayed X-ray brightening is commonly observed in optically selected TDEs \citep{Guolo2024}. This raises the question of whether this feature is intrinsic to the TDE population or a result of selection effects. To investigate this, we measure the time difference between the time of the peak X-ray flux detected by eROSITA ($t_{\rm X, peak}$ in Tables~\ref{tab:sample} and \ref{tab:sample2}) and the optical peak from the best-fit Gaussian process models (Appendix~\ref{sec:sup_fig}; Figures~\ref{fig:opt_diff_lc}--\ref{fig:opt_diff_lc_4}). This estimate has large uncertainties for individual events due to eROSITA's 6-month cadence and seasonal gaps in optical light curves. However, with 23 events exhibiting optical flares, the distribution of rest-frame time difference [$\Delta t_{\rm rest} = (t_{\rm X, peak} - t_{\rm opt, peak})/(1+z)$] provides meaningful constraints on the underlying population properties.

Figure~\ref{fig:tpeak_diff} shows the distribution of $\Delta t_{\rm rest}$, which is positively skewed (${\rm skewness} = 1.85$).
The median time difference is 125 days, with 68\% of events falling between 3\,d and 276\,d (16th and 84th percentiles, respectively). The 90\% range spans $-15$\,d to 383\,d. Despite the large uncertainties in the peak time estimates, a number of events exhibit $\Delta t_{\rm rest}$ values that are consistent with zero, indicating nearly simultaneous X-ray and optical peaks. We note that delayed X-ray brightening, while common, exhibits substantial event-to-event variation.

We highlight two events. 
SRGe J131014.7+444319 (ID 26)  has the smallest value of $\Delta t_{\rm rest} = -61$\,d, with the eROSITA detection occurring near the beginning of the optical flare (Figure~\ref{fig:opt_diff_lc_2}). 
This event may be analogous to TDE\,2022dsb, which showed transient fading X-ray emission detected by eROSITA-DE during the rise of the optical light curve \citep{Malyali2024}. 
At the opposite extreme, SRGe J162159.4+271133 (ID 47) exhibits the longest delay, $\Delta t_{\rm rest} = 801$\,d (see Figure~\ref{fig:opt_diff_lc_3}). However, without prompt X-ray follow-up observations, we cannot constrain the physical origin of the X-ray emission in either case.
 
\subsection{Optical Broad-line Properties of X-ray selected TDEs} \label{subsec:broad_lines}

\begin{figure}[htbp]
    \centering
    \includegraphics[width=\columnwidth]{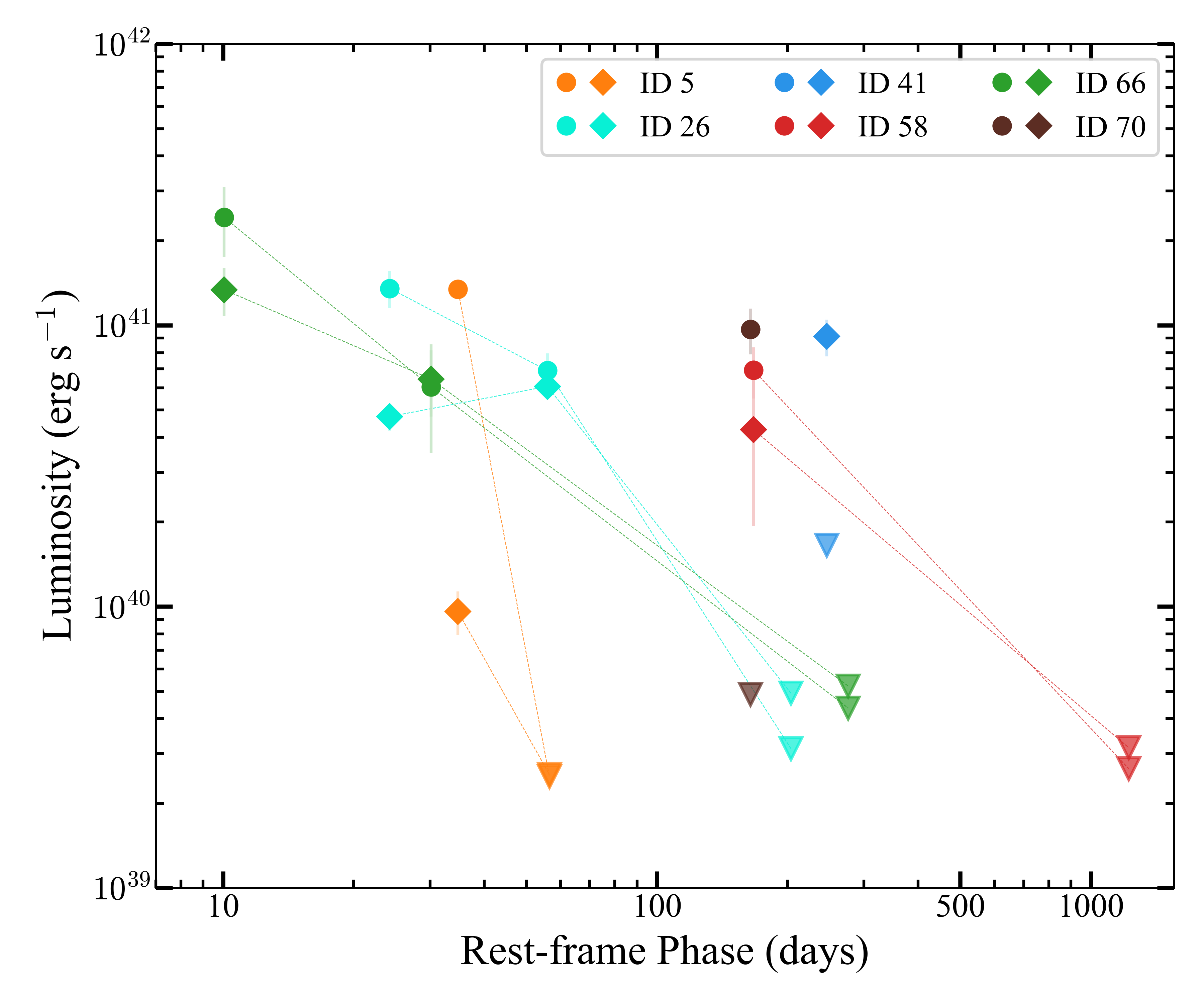}
    \caption{Broad emission line luminosity evolution for six TDEs with significant broad lines detected in our Keck spectra (see \S\ref{subsubsec:tde-broad}). Circles and diamonds represent the fitted luminosity of \ion{He}{II} $\lambda$4686 (broad or shifted components) and H$\alpha$ (broad + shifted components) transient lines, respectively. Downward triangles indicate 3$\sigma$ upper limits. }  
\label{fig:luminosity}
\end{figure}

In \S\ref{subsubsec:tde-broad}, we identified significant broad emission lines around \ion{He}{II} $\lambda$4686 and H$\alpha$  in LRIS spectra of six objects. Figure~\ref{fig:luminosity} displays the evolution of their line luminosities.
Below, we discuss their possible physical origins.

\subsubsection{Broad H$\alpha$ without He (ID 41)} \label{subsubsec:h_no_he}

SRGe J153503.3+455054  (ID 41) exhibits intermediate-width (${\rm FWHM}\sim 2700 \,{\rm km\,s^{-1}}$) luminous ($\sim 10^{41}\,{\rm erg\,s^{-1}}$) H$\alpha$ emission $\sim 300$\,d after the X-ray peak ($\sim 500$\,d after the optical peak). The line width and luminosity are reminiscent of features reported by \citet{Somalwar2025_vlass2}, which may originate from dense clumps of outflowing gas in the circumnuclear medium.

\subsubsection{Broad He without H$\alpha$ (ID 70)}

SRGe J235453.0+421711 (ID 70) exhibits an asymmetric broad feature consistent with luminous ($\sim 10^{41}\,{\rm erg\,s^{-1}}$) He\,\textsc{ii} $\lambda$4686 emission that is blueshifted with pronounced negative skewness\footnote{We cannot completely rule out line blending from Bowen fluorescence features, but since Bowen emission is more commonly observed in the TDE-H+He subtype rather than the TDE-He subtype \citep{Gezari2021}, and given the deep H$\alpha$ line luminosity upper limit, we consider this scenario less likely.}. 
The line exhibits ${\rm FWHM}\sim 2\times 10^4\,{\rm km\,s^{-1}}$ and is blueshifted by $0.04c$ relative to the rest frame, which is comparable to FWHM ($0.08c$). Similar profiles have been observed in optically selected TDEs such as TDE\,2019qiz \citep{Nicholl_2020} and TDE\,2023vto \citep{kumar_2024}, consistent with emission from an expanding, optically thick outflow where line broadening is dominated by bulk gas motion.  

\subsubsection{Broad H$\alpha$ and He (IDs 5, 26, 58, 66)}

Four objects (IDs 5, 26, 58, 66) show both He\,\textsc{ii} $\lambda$4686 and H$\alpha$ broad lines, all with multiple LRIS spectra. A common feature is extremely broad \ion{He}{II} $\lambda 4686$ (${\rm FWHM}\gtrsim 2\times 10^{4}\,{\rm km\,s^{-1}}$). The best-fit Gaussian centers are consistent with the rest-frame wavelength, indicating that the line broadening is dominated by electron scattering rather than bulk kinematic motion. This suggests that in X-ray-selected TDEs, the \ion{He}{II} line photosphere is likely located closer to the central ionizing source, resulting in higher electron scattering optical depth and consequently greater line width.

Two objects (IDs 58, 66) exhibit complex  H$\alpha$ profiles requiring multiple Gaussian components. In SRGe\,J175023.7+712857 (ID 58), the H$\alpha$ profile resembles known double-peaked TDEs such as TDE\,2018hyz \citep{Hung2020, Short2020}, TDE\,2020zso \citep{Wevers2022}, and TDE\,2020nov \citep{Earl2025}. The bluest and reddest peaks are $\approx 4000\,{\rm km\,s^{-1}}$ shifted from rest-frame H$\alpha$, consistent with an accretion disk origin. In SRGe J213527.2-181635 (ID 66), both the +11 day and +33 day spectra show a blue component peaked $\approx 4000\,{\rm km\,s^{-1}}$ blueward of rest-frame  H$\alpha$, with no corresponding red component. Notably, both events show simple \ion{He}{II} $\lambda$4686 profiles well-described by single Gaussians with ${\rm FWHM}\sim 2\times 10^4\,{\rm km\,s^{-1}}$, suggesting different dominant line production mechanisms for \ion{He}{II} $\lambda$4686 and H$\alpha$.

In SRGe J131014.7+444319 (ID 26), the broad  He\,\textsc{ii} $\lambda$4686 and H$\alpha$ have comparable widths. 
The $+29$\,d spectrum shows a luminosity ratio  $L_{\rm He \,II}/ L_{\rm H\alpha}\sim2.9$, which declined to $\sim 1.1$ at $+67$\,d. This evolution is consistent with the optically thick reprocessing picture of \citet{Roth_2016}, where H$\alpha$ line strength is suppressed when the outer envelope radius is small, as expected in this X-ray-selected TDE.

In SRGe J011603.1+072258 (ID 5), the H$\alpha$ component (${\rm FWHM}\sim 4000\,{\rm km\,s^{-1}}$) is much narrower than the He\,\textsc{ii} $\lambda$4686 line. We speculate that the H$\alpha$ originates from a region farther from the black hole, similar to SRGe J153503.3+455054 discussed in \S\ref{subsubsec:h_no_he}.

\subsection{Host properties}

In this section, we discuss the host galaxy properties of our gold TDE sample. The silver sample is shown in figures for reference but is not the focus of our discussion.

\subsubsection{Host Colors versus Stellar Mass}
\label{subsubsec:color_mass}

\begin{figure}[htbp]
    \centering
    \includegraphics[width=\columnwidth]{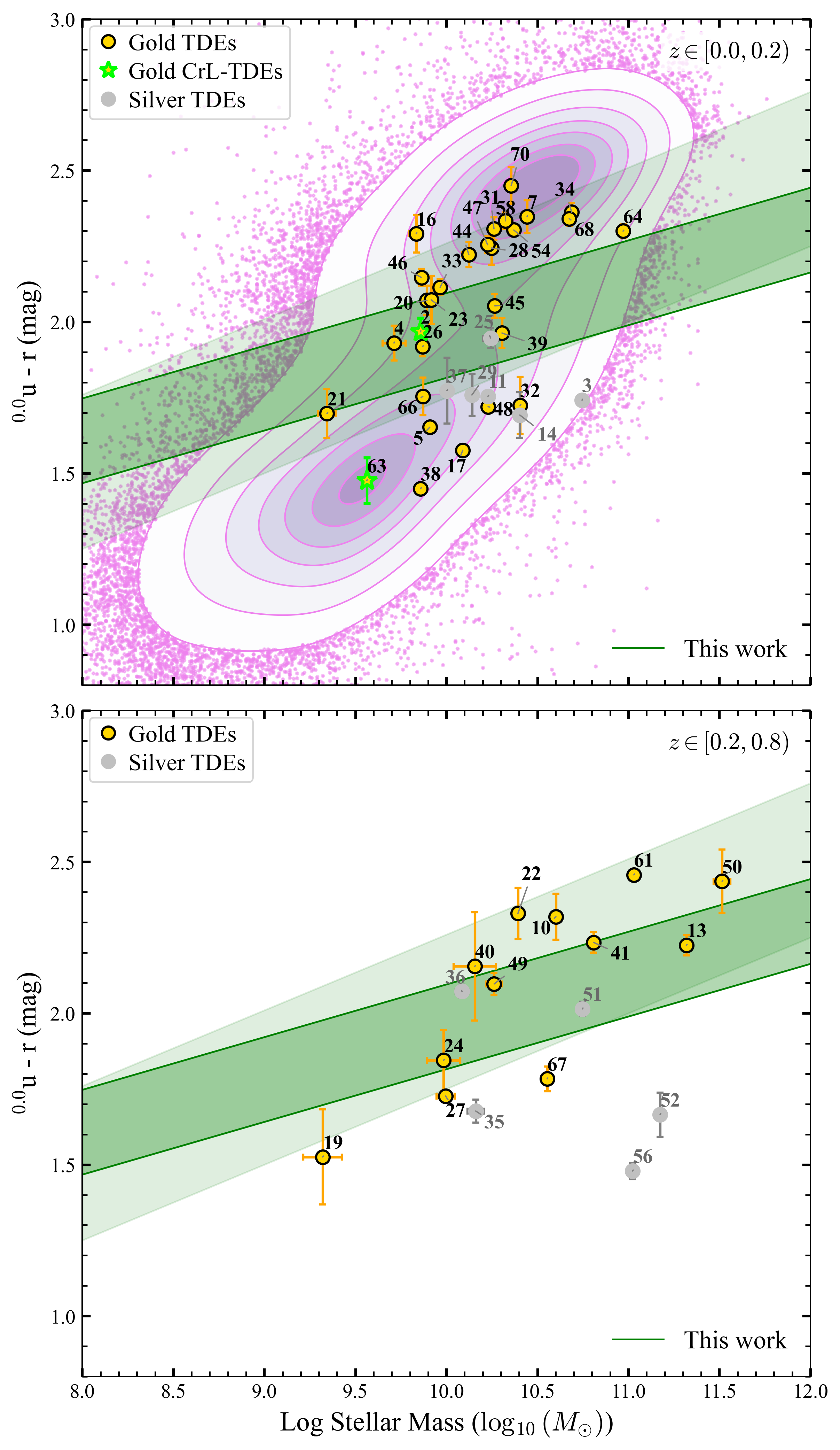}
    \caption{Color–mass diagram for the final TDE host sample. The light-green band indicates the green-valley region defined by \citet{Schawinski_2014}. The darker-green band shows our empirically derived GV definition based on our SDSS comparison sample. Background contours are derived from our SDSS comparison galaxies (\S\ref{subsec:SDSS_comparison_sample}), shown at the 6.7--93.3\% density levels.
}  
\label{fig:u-r color}
\end{figure}

Galaxy colors provide a simple diagnostic of star formation activity and quenching, allowing us to test whether TDE hosts preferentially reside in star-forming, quiescent, or transitional (green valley; GV) galaxies. Figure~\ref{fig:u-r color} presents the Galactic extinction corrected rest-frame $u-r$ color as a function of stellar mass for our final TDE host sample. Both quantities are derived from the \texttt{Prospector} SED fits. Gold (yellow) and silver (grey) subsamples are distinguished, with CrL-TDEs (in the gold sample) highlighted separately in green. The background contours show the distribution of our volume limited SDSS comparison galaxies (see \S\ref{subsec:SDSS_comparison_sample}).

We define the GV empirically by identifying the density minimum in the color distribution between the ``blue cloud'' and ``red sequence'' galaxy populations in our comparison sample. 
A linear fit yields:
\begin{align}
    (u-r)_{\rm GV} = 0.215 + 0.174 \times \log(M_*/M_\odot),
\end{align}
with a half-width $\delta = 0.140$\,mag estimated from the residual distribution. This approach directly traces the low-density transitional region.

To assess whether TDE hosts preferentially occupy the green valley, we perform a one-sided binomial test on our gold sample under the null hypothesis that TDE hosts are randomly drawn from the volume-limited SDSS comparison sample. The enrichment $p$-value is defined as
\begin{align}
    p_{\rm enrich} = P(X \ge k \mid n, f_{\rm SDSS}),
\end{align}
where \(k\) is the number of TDE hosts in the green valley, \(n\) is the total sample size, and \(f_{\rm SDSS}\) is the green-valley fraction in the volume-limited SDSS comparison sample. We also report the fold enrichment ${\rm Fold} = f_{\rm TDE}/f_{\rm SDSS}$. We adopt a stringent significance threshold of $p_{\rm enrich} < 0.001$ to claim genuine GV preference, given our modest sample size. 

Quantitatively, 26.8\% (11/41) of our gold sample fall within the tuned green valley (GV), with a 1.84-fold enrichment over our SDSS comparison sample. However, this enrichment does not meet the statistical significance for either the entire sample ($p = 0.0295> 0.001$) or the low redshift bin ($p = 0.1191> 0.001$), and we therefore do not find strong evidence for GV preference in X-ray-selected TDEs. The majority of the remaining TDEs are found in red sequence galaxies (46.3\%, 19/41), with a smaller fraction in blue sequence galaxies (26.8\%, 11/41). While the red sequence shows a modest 1.24-fold enhancement that is not statistically significant ($p = 0.154$), we find no enrichment of TDEs in blue sequence galaxies relative to the SDSS comparison sample (0.56-fold). These results are consistent across both the full sample and the $z=0.0-0.2$ bin.

Our gold sample spans a broader $u-r$ color and stellar mass range than previous studies \citep{Sazonov2021, Grotova2025}, aligning more closely with the ZTF optical TDE sample \citep{Yao2023}, indicating a more diverse host population.

\subsubsection{Lick indices: H$\delta_A$ versus H$\alpha$ EW} \label{subsubsec:qbs}

Previous work has shown that optically selected TDEs are frequently hosted by rare QBS or E+A galaxies with strong Balmer absorption and weak H$\alpha$ emission. Specifically, using a sample of 8 optically-selected TDEs, \citet{French2016} found that quiescent galaxies with at least moderately strong Balmer absorption are over-represented among the TDE hosts by a factor of $33_{-11}^{+7}$. \citet{Graur2018} considered 35 TDE candidates selected by both X-ray and optical surveys, finding that the over-representation factor is $18^{+8}_{-7}$. More recently, using a sample of 19 ZTF TDEs, \citet{Hammerstein_2021} found an over-representation factor of 16. The eROSITA selected sample (13 objects) presented by \citet{Sazonov2021} suggests a much weaker association with this class of hosts.

\begin{figure}[ht]
    \centering
    \includegraphics[width=\columnwidth]{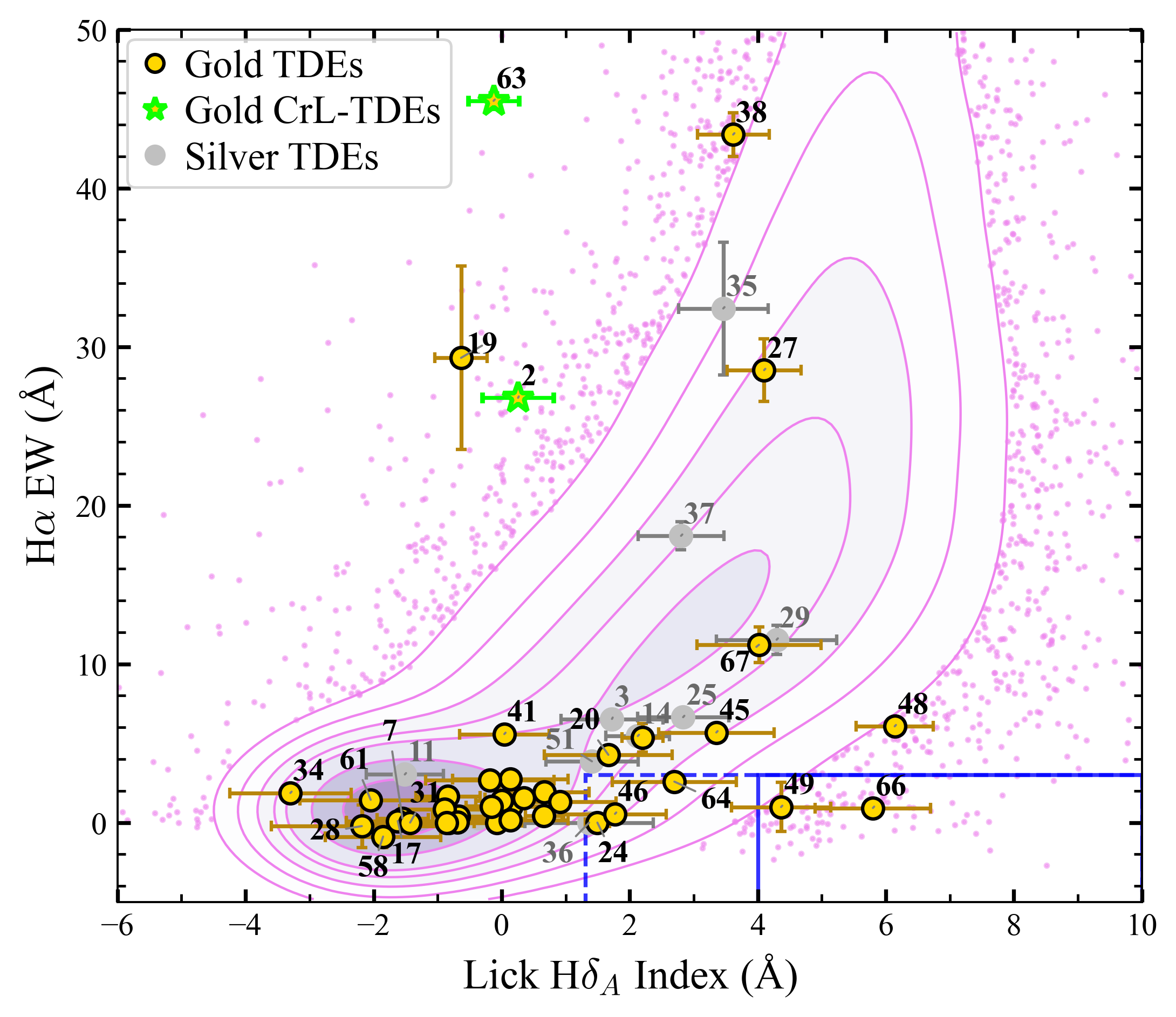}
    \caption{
    Lick H$\delta_A$ absorption index versus H$\alpha$ emission equivalent width for our TDE host sample. The background shows the SDSS comparison sample with contours at $\sigma$-equivalent probability levels (6.7--93.3\%). The solid and dashed boxes mark the E+A and QBS regions, respectively. To avoid label overlap, our gold and silver hosts near the origin are not individually labeled since many are clustered there.}
    \label{fig:Lick}
\end{figure}

Figure~\ref{fig:Lick} shows the Lick H$\delta_A$ absorption index versus the H$\alpha$ EW for our TDE host sample. Line indices and uncertainties are measured with \texttt{pPXF} (\S\ref{subsec:ppxf}). Following \citet{French2016} and \citet{Hammerstein_2021}, we define the E+A region by $H\delta_A-\sigma(H\delta_A)>4.0$ and $\mathrm{EW(H\alpha_{em})}<3.0$ (solid box), and the QBS region by $H\delta_A>1.31$ and $\mathrm{EW(H\alpha_{em})}<3.0$ (dashed box).

First, we test the hypothesis that the distribution of our 40 gold TDE host galaxies in the Lick H$\delta_A$ vs. EW(H$\alpha$) diagram is consistent with the comparison sample using a 2D Kolmogorov--Smirnov (KS) test \citep{Fasano&Franceschini_1987}. We do not include SRGe~J163831.9+534018 (ID 50) in this analysis, as H$\alpha$ is not available due to limited spectral coverage. The resulting $p= 2.39\times10^{-9}$ strongly rejects the null hypothesis. This is not unexpected, as the EW(H$\alpha$) values of our gold TDE hosts are generally lower than the comparison sample, indicating suppressed star formation rates.

Next, we test whether our gold TDE hosts are overrepresented in the QBS and E+A regions using the binomial test described in \S\ref{subsubsec:color_mass}.  For the E+A region, 2 of 40 galaxies (5.0\%) fall within the boundary, with ${\rm Fold} = 10.17$ and $p_{\rm enrich}=1.66\times 10^{-2}$. This suggests enrichment but does not reach our statistical significance threshold. 
For the QBS region, 5 of 40 galaxies (12.5\%) fall within the boundary, with ${\rm Fold}=5.27$ and $p_{\rm enrich}=2.48\times 10^{-3}$. This indicates QBS hosts enrichment but again does not reach our statistical significance threshold. Unlike the strong QBS preference seen previous works \citep{French2016, Graur2018, Hammerstein_2021}, our X-ray-selected sample shows more modest enrichment.

\subsubsection{Recent SFR versus Stellar Mass} \label{subsubsec:sfr_Mgal}

Unlike single-band tracers (e.g., H$\alpha$, UV), our SFR$_{100}$ are derived from \texttt{Prospector}’s nonparametric SFHs, providing a self-consistent 100\,Myr average. Compared to the color--mass diagram (\S\ref{subsubsec:color_mass}), the SFR$_{100}$--$M_\ast$ relation offers a complementary view of star formation in X-ray selected TDE hosts.

\begin{figure}[ht]
    \centering
    \includegraphics[width=\columnwidth]{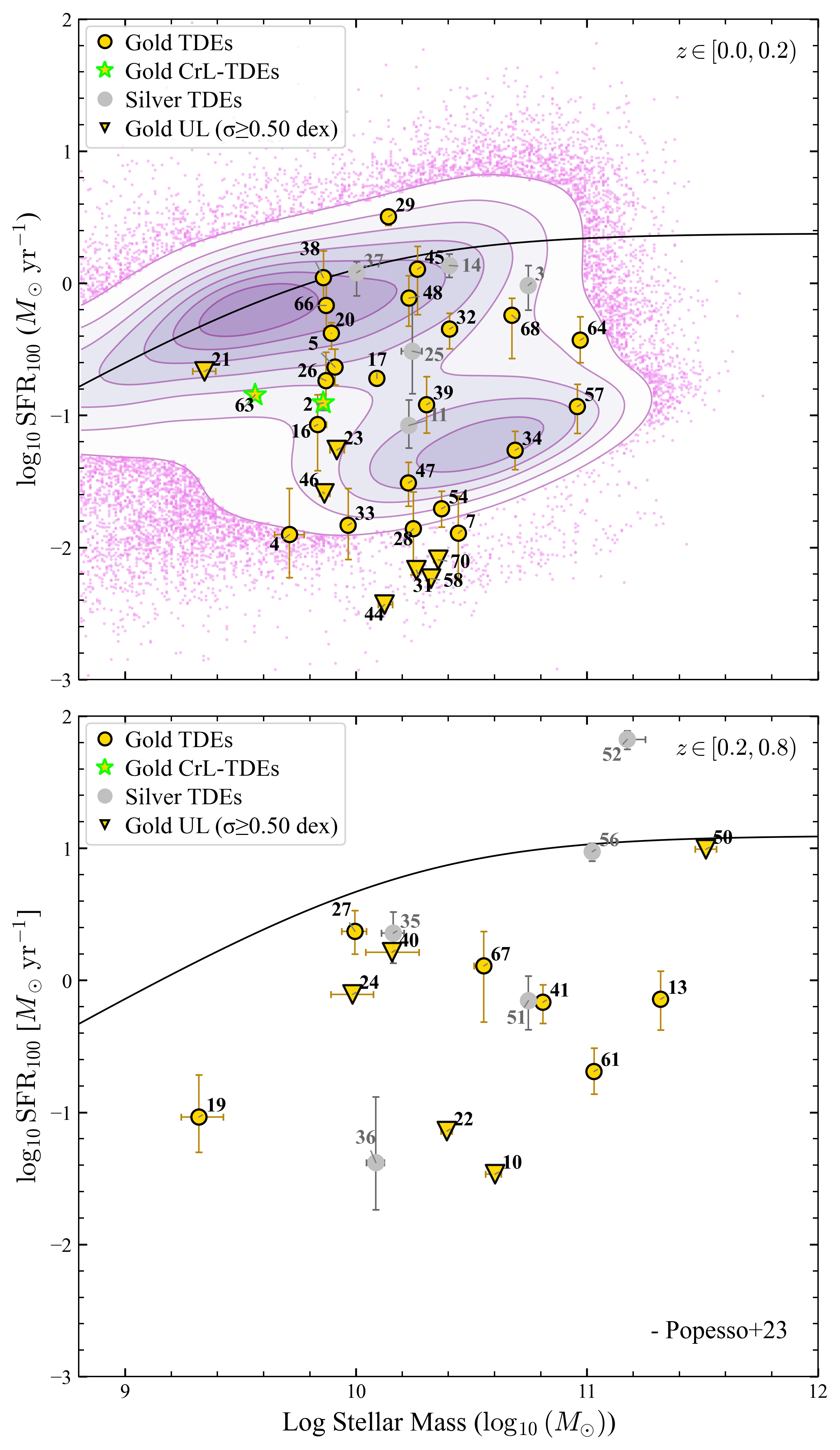}
    \caption{
    Star formation rate averaged over the past 100 Myr (SFR$_{100}$) versus stellar mass for our TDE hosts, split into two redshift bins ($0.0 \leq z < 0.2$ and $0.2 \leq z < 0.8$). The pink background shows the volume-limited SDSS comparison sample described with SFR measurement. The solid black line marks the redshift-dependent star-forming main sequence from \citet{Popesso2023}. 
    } 
    \label{fig:SFR_mass}
\end{figure}

Figure~\ref{fig:SFR_mass} shows the SFR averaged over the past 100 Myr (SFR$_{100}$) versus stellar mass from our Prospector fits. The background distribution shows the volume-limited SDSS comparison sample, and the solid black line indicates the star-forming main sequence from \citet{Popesso2023}. Following \citet{Csizi_2024}, we split into two redshift bins ($0.0 \leq z < 0.2$ and $0.2 \leq z < 0.8$), which affects the main sequence. 


Quantitatively, across both redshift bins, the vast majority of hosts fall below the star-forming main sequence (defined as lying below the main sequence of \citealt{Popesso2023}). In the low-redshift bin ($0.0 \leq z < 0.2$), 96.3\% (26/27) of gold hosts are below the main sequence. In the high-redshift bin ($0.2 \leq z < 0.8$), 100.0\% (10/10) of gold hosts are below the main sequence. Most gold hosts have SFR$_{100}$ values typically between $10^{-3}$ and $1\,M_\odot\,\mathrm{yr}^{-1}$, consistent with suppressed or quenching star formation. 
SRGe J170139.0-085911 (ID 52) appears as an outlier above the sequence, though its Prospector fit may likely overestimate the SFR. 

X-ray-selected TDE hosts predominantly avoid the star-forming main sequence. 
This result is consistent with the color–mass and line-index diagnostics, which also place most hosts in transitional or weakly star-forming regimes rather than strongly star-forming disks. 
The preference for hosts below the main sequence is expected, as dustier star-forming galaxies can strongly absorb soft X-ray emission, making TDEs harder to detect \citep{Panagiotou_2023}.

\section{Conclusion} \label{sec:conclusion}
We present uniform Keck-I/LRIS observations, host galaxy modeling, as well as light curve and spectroscopic analysis for 70 TDE candidates selected from eROSITA all-sky surveys in the Eastern Galactic hemisphere. We develop a robust method for removing AGN interlopers, yielding a final sample of 52 TDEs with redshifts of $0.018 \leq z\leq0.714$. The final TDE sample includes 41 high-confidence (gold) events and 11 lower-confidence (silver) events, representing the largest systematically selected X-ray TDE sample to date. The main results for our gold sample are:
\begin{itemize}[label=\textbullet]
    \item The vast majority of X-ray-selected TDEs are intrinsically X-ray-bright, with 38/41 (93\%) exhibiting $L_{\rm X,peak}>L_{\rm opt, peak}$ (see Figure~\ref{fig:Lx_Lopt}). 
    \item X-ray emission typically peaks 3--276 days (68\% range) after optical peak in events with detected optical flares (Figure~\ref{fig:tpeak_diff}), confirming that delayed X-ray brightening is common to the TDE population.
    \item We identify TDE-associated spectral features in eight events: two coronal-line TDEs (\S\ref{subsubsec:crl-tde}) and six with prominent broad emission lines (\S\ref{subsubsec:tde-broad}). The uniformly broad \ion{He}{II} $\lambda$4686 emission (FWHM $\sim 2\times 10^4\,{\rm km\,s^{-1}}$) indicates high electron scattering optical depth, while the diverse H$\alpha$ line profiles suggest multiple emission mechanisms (\S\ref{subsec:broad_lines}).
    \item X-ray-selected TDE hosts show modest overrepresentation in green valley ($1.8\times$; \S\ref{subsubsec:color_mass}) and quiescent Balmer-strong ($5.3\times$; \S\ref{subsubsec:qbs}) galaxies, which are significantly weaker than optical and pre-eROSITA X-ray samples. This demonstrates that eROSITA TDE hosts span a broader range of star formation histories than previous samples of TDE hosts.
    \item  Our TDE hosts generally exhibit lower star formation rates relative to the star-forming main sequence (\S\ref{subsubsec:sfr_Mgal}). This might be explained by the fact that star-forming galaxies contain more dust that strongly absorb soft X-rays, making TDE selection in the soft X-ray band challenging.
\end{itemize}

This work represents a substantial and well-characterized subsample of eROSITA-RU-selected TDEs. A forthcoming study will present the complete eROSITA-RU TDE sample with uniform analysis. The observed diversity in both TDE properties and host galaxy characteristics presented here should guide future theoretical models of TDE emission mechanisms and their dependence on environment.

\begin{acknowledgements}
This work uses data obtained with eROSITA telescope onboard SRG observatory. The SRG observatory was built by Roskosmos with the participation of the Deutsches Zentrum für Luft- und Raumfahrt (DLR). The SRG/eROSITA X-ray telescope was built by a consortium of German Institutes led by MPE, and supported by DLR. The SRG spacecraft was designed, built, launched and is operated by the Lavochkin Association and its subcontractors. The science data were downlinked via the Deep Space Network Antennae in Bear Lakes, Ussurijsk, and Baykonur, funded by Roskosmos. The eROSITA data used in this work were processed using the eSASS software system developed by the German eROSITA consortium and proprietary data reduction and analysis software developed by the Russian eROSITA Consortium. MG and SS acknowledge support by the Ministry of Science and Higher Education grant 075-15-2024-647.

Some of the data presented herein were obtained at Keck Observatory, which is a private 501(c)3 non-profit organization operated as a scientific partnership among the California Institute of Technology, the University of California, and the National Aeronautics and Space Administration. The Observatory was made possible by the generous financial support of the W. M. Keck Foundation. The authors wish to recognize and acknowledge the very significant cultural role and reverence that the summit of Maunakea has always had within the Native Hawaiian community. We are most fortunate to have the opportunity to conduct observations from this mountain. 

\end{acknowledgements}

\bibliographystyle{aa}
\bibliography{bibliography}

@ARTICLE{2025A&A...693A..62G,
       author = {{Grotova}, I. and {Rau}, A. and {Salvato}, M. and {Buchner}, J. and {Goodwin}, A.~J. and {Liu}, Z. and {Malyali}, A. and {Merloni}, A. and {Tub{\'\i}n-Arenas}, D. and {Homan}, D. and {Krumpe}, M. and {Nandra}, K. and {Shirley}, R. and {Anderson}, G.~E. and {Arcodia}, R. and {Bahic}, S. and {Baldini}, P. and {Buckley}, D.~A.~H. and {Ciroi}, S. and {Kawka}, A. and {Masterson}, M. and {Miller-Jones}, J.~C.~A. and {Di Mille}, F.},
        title = "{eRO-ExTra: eROSITA extragalactic non-AGN X-ray transients and variables in eRASS1 and eRASS2}",
      journal = {\aap},
         year = 2025,
        month = jan,
       volume = {693},
          eid = {A62},
        pages = {A62},
          doi = {10.1051/0004-6361/202451253},
archivePrefix = {arXiv},
       eprint = {2501.04208},
 primaryClass = {astro-ph.HE},
       adsurl = {https://ui.adsabs.harvard.edu/abs/2025A&A...693A..62G}
}

@ARTICLE{2020ATel13499....1K,
       author = {{Khabibullin}, I. and {Medvedev}, P. and {Churazov}, E. and {Gilfanov}, M. and {Sazonov}, S. and {Sunyaev}, R. and {Burenin}, R. and {Pavlinsky}, M. and {Russian Srg/Erosita Consortium} and {Srg/Art-Xc Team}},
        title = "{Bright X-ray source SRGet J134954.70+432859.5 in the direction of galaxy SDSS J134954.68+432856.0}",
      journal = {The Astronomer's Telegram},
         year = 2020,
        month = feb,
       volume = {13499},
        pages = {1},
       adsurl = {https://ui.adsabs.harvard.edu/abs/2020ATel13499....1K}
}

@ARTICLE{Earl2025,
       author = {{Earl}, Nicholas and {French}, K. Decker and {Ramirez-Ruiz}, Enrico and {Auchettl}, Katie and {Raimundo}, Sandra I. and {Davis}, Kyle W. and {Masterson}, Megan and {Arcavi}, Iair and {Lu}, Wenbin and {Baldassare}, Vivienne F. and {Coulter}, David A. and {de Boer}, Thomas and {Drout}, Maria R. and {Dykaar}, Hannah and {Foley}, Ryan J. and {Gall}, Christa and {Gao}, Hua and {Huber}, Mark E. and {Jones}, David O. and {Langeroodi}, Danial and {Lin}, Chien-Cheng and {Magnier}, Eugene A. and {Mockler}, Brenna and {Shepherd}, Margaret and {Verrico}, Margaret E.},
        title = "{AT 2020nov: Evidence for Disk Reprocessing in a Rare Tidal Disruption Event}",
      journal = {\apj},
         year = 2025,
        month = apr,
       volume = {983},
       number = {1},
          eid = {28},
        pages = {28},
          doi = {10.3847/1538-4357/adb974},
archivePrefix = {arXiv},
       eprint = {2412.12991},
 primaryClass = {astro-ph.HE},
       adsurl = {https://ui.adsabs.harvard.edu/abs/2025ApJ...983...28E}
}

@ARTICLE{Wevers2022,
       author = {{Wevers}, T. and {Nicholl}, M. and {Guolo}, M. and {Charalampopoulos}, P. and {Gromadzki}, M. and {Reynolds}, T.~M. and {Kankare}, E. and {Leloudas}, G. and {Anderson}, J.~P. and {Arcavi}, I. and {Cannizzaro}, G. and {Chen}, T.-W. and {Ihanec}, N. and {Inserra}, C. and {Guti{\'e}rrez}, C.~P. and {Jonker}, P.~G. and {Lawrence}, A. and {Magee}, M.~R. and {M{\"u}ller-Bravo}, T.~E. and {Onori}, F. and {Ridley}, E. and {Schulze}, S. and {Short}, P. and {Hiramatsu}, D. and {Newsome}, M. and {Terwel}, J.~H. and {Yang}, S. and {Young}, D.},
        title = "{An elliptical accretion disk following the tidal disruption event AT 2020zso}",
      journal = {\aap},
         year = 2022,
        month = oct,
       volume = {666},
          eid = {A6},
        pages = {A6},
          doi = {10.1051/0004-6361/202142616},
archivePrefix = {arXiv},
       eprint = {2202.08268},
 primaryClass = {astro-ph.HE},
       adsurl = {https://ui.adsabs.harvard.edu/abs/2022A&A...666A...6W}
}

@ARTICLE{Somalwar2025_vlass2,
       author = {{Somalwar}, Jean J. and {Ravi}, Vikram and {Lu}, Wenbin},
        title = "{VLASS Tidal Disruption Events with Optical Flares. II. Discovery of Two TDEs with Intermediate Width Balmer Emission Lines and Connections to the Ambiguous Extreme Coronal Line Emitters}",
      journal = {\apj},
         year = 2025,
        month = apr,
       volume = {983},
       number = {2},
          eid = {159},
        pages = {159},
          doi = {10.3847/1538-4357/adc002},
archivePrefix = {arXiv},
       eprint = {2310.03795},
 primaryClass = {astro-ph.HE},
       adsurl = {https://ui.adsabs.harvard.edu/abs/2025ApJ...983..159S}
}

@ARTICLE{Short2020,
       author = {{Short}, P. and {Nicholl}, M. and {Lawrence}, A. and {Gomez}, S. and {Arcavi}, I. and {Wevers}, T. and {Leloudas}, G. and {Schulze}, S. and {Anderson}, J.~P. and {Berger}, E. and {Blanchard}, P.~K. and {Burke}, J. and {Castro Segura}, N. and {Charalampopoulos}, P. and {Chornock}, R. and {Galbany}, L. and {Gromadzki}, M. and {Herzog}, L.~J. and {Hiramatsu}, D. and {Horne}, Keith and {Hosseinzadeh}, G. and {Howell}, D. Andrew and {Ihanec}, N. and {Inserra}, C. and {Kankare}, E. and {Maguire}, K. and {McCully}, C. and {M{\"u}ller Bravo}, T.~E. and {Onori}, F. and {Sollerman}, J. and {Young}, D.~R.},
        title = "{The tidal disruption event AT 2018hyz - I. Double-peaked emission lines and a flat Balmer decrement}",
      journal = {\mnras},
         year = 2020,
        month = nov,
       volume = {498},
       number = {3},
        pages = {4119-4133},
          doi = {10.1093/mnras/staa2065},
archivePrefix = {arXiv},
       eprint = {2003.05470},
 primaryClass = {astro-ph.GA},
       adsurl = {https://ui.adsabs.harvard.edu/abs/2020MNRAS.498.4119S}
}

@ARTICLE{Hung2020,
       author = {{Hung}, Tiara and {Foley}, Ryan J. and {Ramirez-Ruiz}, Enrico and {Dai}, Jane L. and {Auchettl}, Katie and {Kilpatrick}, Charles D. and {Mockler}, Brenna and {Brown}, Jonathan S. and {Coulter}, David A. and {Dimitriadis}, Georgios and {Holoien}, Thomas W.-S. and {Law-Smith}, Jamie A.~P. and {Piro}, Anthony L. and {Rest}, Armin and {Rojas-Bravo}, C{\'e}sar and {Siebert}, Matthew R.},
        title = "{Double-peaked Balmer Emission Indicating Prompt Accretion Disk Formation in an X-Ray Faint Tidal Disruption Event}",
      journal = {\apj},
         year = 2020,
        month = nov,
       volume = {903},
       number = {1},
          eid = {31},
        pages = {31},
          doi = {10.3847/1538-4357/abb606},
archivePrefix = {arXiv},
       eprint = {2003.09427},
 primaryClass = {astro-ph.HE},
       adsurl = {https://ui.adsabs.harvard.edu/abs/2020ApJ...903...31H}
}

@ARTICLE{Yao2019,
       author = {{Yao}, Yuhan and {Miller}, Adam A. and {Kulkarni}, S.~R. and {Bulla}, Mattia and {Masci}, Frank J. and {Goldstein}, Daniel A. and {Goobar}, Ariel and {Nugent}, Peter and {Dugas}, Alison and {Blagorodnova}, Nadia and {Neill}, James D. and {Rigault}, Mickael and {Sollerman}, Jesper and {Nordin}, J. and {Bellm}, Eric C. and {Cenko}, S. Bradley and {De}, Kishalay and {Dhawan}, Suhail and {Feindt}, Ulrich and {Fremling}, C. and {Gatkine}, Pradip and {Graham}, Matthew J. and {Graham}, Melissa L. and {Ho}, Anna Y.~Q. and {Hung}, T. and {Kasliwal}, Mansi M. and {Kupfer}, Thomas and {Laher}, Russ R. and {Perley}, Daniel A. and {Rusholme}, Ben and {Shupe}, David L. and {Soumagnac}, Maayane T. and {Taggart}, K. and {Walters}, Richard and {Yan}, Lin},
        title = "{ZTF Early Observations of Type Ia Supernovae. I. Properties of the 2018 Sample}",
      journal = {\apj},
         year = 2019,
        month = dec,
       volume = {886},
       number = {2},
          eid = {152},
        pages = {152},
          doi = {10.3847/1538-4357/ab4cf5},
archivePrefix = {arXiv},
       eprint = {1910.02967},
 primaryClass = {astro-ph.HE},
       adsurl = {https://ui.adsabs.harvard.edu/abs/2019ApJ...886..152Y}
}

@ARTICLE{Wevers2024,
       author = {{Wevers}, T. and {Guolo}, M. and {Pasham}, D.~R. and {Coughlin}, E.~R. and {Tombesi}, F. and {Yao}, Y. and {Gezari}, S.},
        title = "{Delayed X-Ray Brightening Accompanied by Variable Ionized Absorption Following a Tidal Disruption Event}",
      journal = {\apj},
         year = 2024,
        month = mar,
       volume = {963},
       number = {1},
          eid = {75},
        pages = {75},
          doi = {10.3847/1538-4357/ad1878},
archivePrefix = {arXiv},
       eprint = {2311.09371},
 primaryClass = {astro-ph.HE},
       adsurl = {https://ui.adsabs.harvard.edu/abs/2024ApJ...963...75W}
}

@ARTICLE{Yao2020_19dge,
       author = {{Yao}, Yuhan and {De}, Kishalay and {Kasliwal}, Mansi M. and {Ho}, Anna Y.~Q. and {Schulze}, Steve and {Li}, Zhihui and {Kulkarni}, S.~R. and {Fruchter}, Andrew and {Rubin}, David and {Perley}, Daniel A. and {Fuller}, Jim and {Piro}, Anthony L. and {Fremling}, C. and {Bellm}, Eric C. and {Burruss}, Rick and {Duev}, Dmitry A. and {Feeney}, Michael and {Gal-Yam}, Avishay and {Golkhou}, V. Zach and {Graham}, Matthew J. and {Helou}, George and {Kupfer}, Thomas and {Laher}, Russ R. and {Masci}, Frank J. and {Miller}, Adam A. and {Rusholme}, Ben and {Shupe}, David L. and {Smith}, Roger and {Sollerman}, Jesper and {Soumagnac}, Maayane T. and {Zolkower}, Jeffry},
        title = "{SN2019dge: A Helium-rich Ultra-stripped Envelope Supernova}",
      journal = {\apj},
         year = 2020,
        month = sep,
       volume = {900},
       number = {1},
          eid = {46},
        pages = {46},
          doi = {10.3847/1538-4357/abaa3d},
archivePrefix = {arXiv},
       eprint = {2005.12922},
 primaryClass = {astro-ph.HE},
       adsurl = {https://ui.adsabs.harvard.edu/abs/2020ApJ...900...46Y}
}

@ARTICLE{2020MNRAS.497L...1W,
       author = {{Wevers}, Thomas},
        title = "{Fainter harder brighter softer: a correlation between {\ensuremath{\alpha}}$_{ox}$, X-ray spectral state, and Eddington ratio in tidal disruption events}",
      journal = {\mnras},
         year = 2020,
        month = sep,
       volume = {497},
       number = {1},
        pages = {L1-L6},
          doi = {10.1093/mnrasl/slaa097},
archivePrefix = {arXiv},
       eprint = {2006.06684},
 primaryClass = {astro-ph.HE},
       adsurl = {https://ui.adsabs.harvard.edu/abs/2020MNRAS.497L...1W}
}

@ARTICLE{Bellm2019,
       author = {{Bellm}, Eric C. and {Kulkarni}, Shrinivas R. and {Graham}, Matthew J. and {Dekany}, Richard and {Smith}, Roger M. and {Riddle}, Reed and {Masci}, Frank J. and {Helou}, George and {Prince}, Thomas A. and {Adams}, Scott M. and {Barbarino}, C. and {Barlow}, Tom and {Bauer}, James and {Beck}, Ron and {Belicki}, Justin and {Biswas}, Rahul and {Blagorodnova}, Nadejda and {Bodewits}, Dennis and {Bolin}, Bryce and {Brinnel}, Valery and {Brooke}, Tim and {Bue}, Brian and {Bulla}, Mattia and {Burruss}, Rick and {Cenko}, S. Bradley and {Chang}, Chan-Kao and {Connolly}, Andrew and {Coughlin}, Michael and {Cromer}, John and {Cunningham}, Virginia and {De}, Kishalay and {Delacroix}, Alex and {Desai}, Vandana and {Duev}, Dmitry A. and {Eadie}, Gwendolyn and {Farnham}, Tony L. and {Feeney}, Michael and {Feindt}, Ulrich and {Flynn}, David and {Franckowiak}, Anna and {Frederick}, S. and {Fremling}, C. and {Gal-Yam}, Avishay and {Gezari}, Suvi and {Giomi}, Matteo and {Goldstein}, Daniel A. and {Golkhou}, V. Zach and {Goobar}, Ariel and {Groom}, Steven and {Hacopians}, Eugean and {Hale}, David and {Henning}, John and {Ho}, Anna Y.~Q. and {Hover}, David and {Howell}, Justin and {Hung}, Tiara and {Huppenkothen}, Daniela and {Imel}, David and {Ip}, Wing-Huen and {Ivezi{\'c}}, {\v{Z}}eljko and {Jackson}, Edward and {Jones}, Lynne and {Juric}, Mario and {Kasliwal}, Mansi M. and {Kaspi}, S. and {Kaye}, Stephen and {Kelley}, Michael S.~P. and {Kowalski}, Marek and {Kramer}, Emily and {Kupfer}, Thomas and {Landry}, Walter and {Laher}, Russ R. and {Lee}, Chien-De and {Lin}, Hsing Wen and {Lin}, Zhong-Yi and {Lunnan}, Ragnhild and {Giomi}, Matteo and {Mahabal}, Ashish and {Mao}, Peter and {Miller}, Adam A. and {Monkewitz}, Serge and {Murphy}, Patrick and {Ngeow}, Chow-Choong and {Nordin}, Jakob and {Nugent}, Peter and {Ofek}, Eran and {Patterson}, Maria T. and {Penprase}, Bryan and {Porter}, Michael and {Rauch}, Ludwig and {Rebbapragada}, Umaa and {Reiley}, Dan and {Rigault}, Mickael and {Rodriguez}, Hector and {van Roestel}, Jan and {Rusholme}, Ben and {van Santen}, Jakob and {Schulze}, S. and {Shupe}, David L. and {Singer}, Leo P. and {Soumagnac}, Maayane T. and {Stein}, Robert and {Surace}, Jason and {Sollerman}, Jesper and {Szkody}, Paula and {Taddia}, F. and {Terek}, Scott and {Van Sistine}, Angela and {van Velzen}, Sjoert and {Vestrand}, W. Thomas and {Walters}, Richard and {Ward}, Charlotte and {Ye}, Quan-Zhi and {Yu}, Po-Chieh and {Yan}, Lin and {Zolkower}, Jeffry},
        title = "{The Zwicky Transient Facility: System Overview, Performance, and First Results}",
      journal = {\pasp},
         year = 2019,
        month = jan,
       volume = {131},
       number = {995},
        pages = {018002},
          doi = {10.1088/1538-3873/aaecbe},
archivePrefix = {arXiv},
       eprint = {1902.01932},
 primaryClass = {astro-ph.IM},
       adsurl = {https://ui.adsabs.harvard.edu/abs/2019PASP..131a8002B}
}

@ARTICLE{Khabibullin2014,
       author = {{Khabibullin}, I. and {Sazonov}, S. and {Sunyaev}, R.},
        title = "{SRG/eROSITA prospects for the detection of stellar tidal disruption flares}",
      journal = {\mnras},
         year = 2014,
        month = jan,
       volume = {437},
       number = {1},
        pages = {327-337},
          doi = {10.1093/mnras/stt1889},
archivePrefix = {arXiv},
       eprint = {1304.3376},
 primaryClass = {astro-ph.HE},
       adsurl = {https://ui.adsabs.harvard.edu/abs/2014MNRAS.437..327K}
}

@ARTICLE{Onori2022,
       author = {{Onori}, F. and {Cannizzaro}, G. and {Jonker}, P.~G. and {Kim}, M. and {Nicholl}, M. and {Mattila}, S. and {Reynolds}, T.~M. and {Fraser}, M. and {Wevers}, T. and {Brocato}, E. and {Anderson}, J.~P. and {Carini}, R. and {Charalampopoulos}, P. and {Clark}, P. and {Gromadzki}, M. and {Guti{\'e}rrez}, C.~P. and {Ihanec}, N. and {Inserra}, C. and {Lawrence}, A. and {Leloudas}, G. and {Lundqvist}, P. and {M{\"u}ller-Bravo}, T.~E. and {Piranomonte}, S. and {Pursiainen}, M. and {Rybicki}, K.~A. and {Somero}, A. and {Young}, D.~R. and {Chambers}, K.~C. and {Gao}, H. and {de Boer}, T.~J.~L. and {Magnier}, E.~A.},
        title = "{The nuclear transient AT 2017gge: a tidal disruption event in a dusty and gas-rich environment and the awakening of a dormant SMBH}",
      journal = {\mnras},
         year = 2022,
        month = nov,
       volume = {517},
       number = {1},
        pages = {76-98},
          doi = {10.1093/mnras/stac2673},
archivePrefix = {arXiv},
       eprint = {2206.00049},
 primaryClass = {astro-ph.HE},
       adsurl = {https://ui.adsabs.harvard.edu/abs/2022MNRAS.517...76O}
}

@ARTICLE{Flewelling2020,
       author = {{Flewelling}, H.~A. and {Magnier}, E.~A. and {Chambers}, K.~C. and {Heasley}, J.~N. and {Holmberg}, C. and {Huber}, M.~E. and {Sweeney}, W. and {Waters}, C.~Z. and {Calamida}, A. and {Casertano}, S. and {Chen}, X. and {Farrow}, D. and {Hasinger}, G. and {Henderson}, R. and {Long}, K.~S. and {Metcalfe}, N. and {Narayan}, G. and {Nieto-Santisteban}, M.~A. and {Norberg}, P. and {Rest}, A. and {Saglia}, R.~P. and {Szalay}, A. and {Thakar}, A.~R. and {Tonry}, J.~L. and {Valenti}, J. and {Werner}, S. and {White}, R. and {Denneau}, L. and {Draper}, P.~W. and {Hodapp}, K.~W. and {Jedicke}, R. and {Kaiser}, N. and {Kudritzki}, R.~P. and {Price}, P.~A. and {Wainscoat}, R.~J. and {Chastel}, S. and {McLean}, B. and {Postman}, M. and {Shiao}, B.},
        title = "{The Pan-STARRS1 Database and Data Products}",
      journal = {\apjs},
         year = 2020,
        month = nov,
       volume = {251},
       number = {1},
          eid = {7},
        pages = {7},
          doi = {10.3847/1538-4365/abb82d},
archivePrefix = {arXiv},
       eprint = {1612.05243},
 primaryClass = {astro-ph.IM},
       adsurl = {https://ui.adsabs.harvard.edu/abs/2020ApJS..251....7F}
}

@ARTICLE{Singh2013,
       author = {{Singh}, R. and {van de Ven}, G. and {Jahnke}, K. and {Lyubenova}, M. and {Falc{\'o}n-Barroso}, J. and {Alves}, J. and {Cid Fernandes}, R. and {Galbany}, L. and {Garc{\'\i}a-Benito}, R. and {Husemann}, B. and {Kennicutt}, R.~C. and {Marino}, R.~A. and {M{\'a}rquez}, I. and {Masegosa}, J. and {Mast}, D. and {Pasquali}, A. and {S{\'a}nchez}, S.~F. and {Walcher}, J. and {Wild}, V. and {Wisotzki}, L. and {Ziegler}, B.},
        title = "{The nature of LINER galaxies:. Ubiquitous hot old stars and rare accreting black holes}",
      journal = {\aap},
         year = 2013,
        month = oct,
       volume = {558},
          eid = {A43},
        pages = {A43},
          doi = {10.1051/0004-6361/201322062},
archivePrefix = {arXiv},
       eprint = {1308.4271},
 primaryClass = {astro-ph.GA},
       adsurl = {https://ui.adsabs.harvard.edu/abs/2013A&A...558A..43S}
}

@INPROCEEDINGS{Filippenko1996,
       author = {{Filippenko}, A.~V.},
        title = "{An Introduction to LINERs}",
    booktitle = {The Physics of Liners in View of Recent Observations},
         year = 1996,
       editor = {{Eracleous}, M. and {Koratkar}, A. and {Leitherer}, C. and {Ho}, L.},
       series = {Astronomical Society of the Pacific Conference Series},
       volume = {103},
        month = jan,
        pages = {17},
       adsurl = {https://ui.adsabs.harvard.edu/abs/1996ASPC..103...17F}
}

@ARTICLE{Ho1993,
       author = {{Ho}, Luis C. and {Filippenko}, Alexei V. and {Sargent}, Wallace L.~W.},
        title = "{A Reevaluation of the Excitation Mechanism of LINERs}",
      journal = {\apj},
         year = 1993,
        month = nov,
       volume = {417},
        pages = {63},
          doi = {10.1086/173291},
       adsurl = {https://ui.adsabs.harvard.edu/abs/1993ApJ...417...63H}
}

@ARTICLE{Heckman1980,
       author = {{Heckman}, T.~M.},
        title = "{An Optical and Radio Survey of the Nuclei of Bright Galaxies - Activity in the Normal Galactic Nuclei}",
      journal = {\aap},
         year = 1980,
        month = jul,
       volume = {87},
        pages = {152},
       adsurl = {https://ui.adsabs.harvard.edu/abs/1980A&A....87..152H}
}

@ARTICLE{Hannah2025,
       author = {{Hannah}, Christian H. and {Stone}, Nicholas C. and {Seth}, Anil C. and {van Velzen}, Sjoert},
        title = "{Counting the Unseen. II. Tidal Disruption Event Rates in Nearby Galaxies with REPTiDE}",
      journal = {\apj},
         year = 2025,
        month = jul,
       volume = {988},
       number = {1},
          eid = {29},
        pages = {29},
          doi = {10.3847/1538-4357/addd1b},
archivePrefix = {arXiv},
       eprint = {2412.19935},
 primaryClass = {astro-ph.GA},
       adsurl = {https://ui.adsabs.harvard.edu/abs/2025ApJ...988...29H}
}

@ARTICLE{Ryu2023,
       author = {{Ryu}, Taeho and {Krolik}, Julian and {Piran}, Tsvi and {Noble}, Scott C. and {Avara}, Mark},
        title = "{Shocks Power Tidal Disruption Events}",
      journal = {\apj},
         year = 2023,
        month = nov,
       volume = {957},
       number = {1},
          eid = {12},
        pages = {12},
          doi = {10.3847/1538-4357/acf5de},
archivePrefix = {arXiv},
       eprint = {2305.05333},
 primaryClass = {astro-ph.HE},
       adsurl = {https://ui.adsabs.harvard.edu/abs/2023ApJ...957...12R}
}

@ARTICLE{Price2024,
       author = {{Price}, Daniel J. and {Liptai}, David and {Mandel}, Ilya and {Shepherd}, Joanna and {Lodato}, Giuseppe and {Levin}, Yuri},
        title = "{Eddington Envelopes: The Fate of Stars on Parabolic Orbits Tidally Disrupted by Supermassive Black Holes}",
      journal = {\apjl},
         year = 2024,
        month = aug,
       volume = {971},
       number = {2},
          eid = {L46},
        pages = {L46},
          doi = {10.3847/2041-8213/ad6862},
archivePrefix = {arXiv},
       eprint = {2404.09381},
 primaryClass = {astro-ph.HE},
       adsurl = {https://ui.adsabs.harvard.edu/abs/2024ApJ...971L..46P}
}

@ARTICLE{Steinberg2024,
       author = {{Steinberg}, Elad and {Stone}, Nicholas C.},
        title = "{Stream-disk shocks as the origins of peak light in tidal disruption events}",
      journal = {\nat},
         year = 2024,
        month = jan,
       volume = {625},
       number = {7995},
        pages = {463-467},
          doi = {10.1038/s41586-023-06875-y},
archivePrefix = {arXiv},
       eprint = {2206.10641},
 primaryClass = {astro-ph.HE},
       adsurl = {https://ui.adsabs.harvard.edu/abs/2024Natur.625..463S}
}

@ARTICLE{Andalman2022,
       author = {{Andalman}, Zachary L. and {Liska}, Matthew T.~P. and {Tchekhovskoy}, Alexander and {Coughlin}, Eric R. and {Stone}, Nicholas},
        title = "{Tidal disruption discs formed and fed by stream-stream and stream-disc interactions in global GRHD simulations}",
      journal = {\mnras},
         year = 2022,
        month = feb,
       volume = {510},
       number = {2},
        pages = {1627-1648},
          doi = {10.1093/mnras/stab3444},
archivePrefix = {arXiv},
       eprint = {2008.04922},
 primaryClass = {astro-ph.HE},
       adsurl = {https://ui.adsabs.harvard.edu/abs/2022MNRAS.510.1627A}
}

@ARTICLE{Graham2019,
       author = {{Graham}, Matthew J. and {Kulkarni}, S.~R. and {Bellm}, Eric C. and {Adams}, Scott M. and {Barbarino}, Cristina and {Blagorodnova}, Nadejda and {Bodewits}, Dennis and {Bolin}, Bryce and {Brady}, Patrick R. and {Cenko}, S. Bradley and {Chang}, Chan-Kao and {Coughlin}, Michael W. and {De}, Kishalay and {Eadie}, Gwendolyn and {Farnham}, Tony L. and {Feindt}, Ulrich and {Franckowiak}, Anna and {Fremling}, Christoffer and {Gezari}, Suvi and {Ghosh}, Shaon and {Goldstein}, Daniel A. and {Golkhou}, V. Zach and {Goobar}, Ariel and {Ho}, Anna Y.~Q. and {Huppenkothen}, Daniela and {Ivezi{\'c}}, {\v{Z}}eljko and {Jones}, R. Lynne and {Juric}, Mario and {Kaplan}, David L. and {Kasliwal}, Mansi M. and {Kelley}, Michael S.~P. and {Kupfer}, Thomas and {Lee}, Chien-De and {Lin}, Hsing Wen and {Lunnan}, Ragnhild and {Mahabal}, Ashish A. and {Miller}, Adam A. and {Ngeow}, Chow-Choong and {Nugent}, Peter and {Ofek}, Eran O. and {Prince}, Thomas A. and {Rauch}, Ludwig and {van Roestel}, Jan and {Schulze}, Steve and {Singer}, Leo P. and {Sollerman}, Jesper and {Taddia}, Francesco and {Yan}, Lin and {Ye}, Quan-Zhi and {Yu}, Po-Chieh and {Barlow}, Tom and {Bauer}, James and {Beck}, Ron and {Belicki}, Justin and {Biswas}, Rahul and {Brinnel}, Valery and {Brooke}, Tim and {Bue}, Brian and {Bulla}, Mattia and {Burruss}, Rick and {Connolly}, Andrew and {Cromer}, John and {Cunningham}, Virginia and {Dekany}, Richard and {Delacroix}, Alex and {Desai}, Vandana and {Duev}, Dmitry A. and {Feeney}, Michael and {Flynn}, David and {Frederick}, Sara and {Gal-Yam}, Avishay and {Giomi}, Matteo and {Groom}, Steven and {Hacopians}, Eugean and {Hale}, David and {Helou}, George and {Henning}, John and {Hover}, David and {Hillenbrand}, Lynne A. and {Howell}, Justin and {Hung}, Tiara and {Imel}, David and {Ip}, Wing-Huen and {Jackson}, Edward and {Kaspi}, Shai and {Kaye}, Stephen and {Kowalski}, Marek and {Kramer}, Emily and {Kuhn}, Michael and {Landry}, Walter and {Laher}, Russ R. and {Mao}, Peter and {Masci}, Frank J. and {Monkewitz}, Serge and {Murphy}, Patrick and {Nordin}, Jakob and {Patterson}, Maria T. and {Penprase}, Bryan and {Porter}, Michael and {Rebbapragada}, Umaa and {Reiley}, Dan and {Riddle}, Reed and {Rigault}, Mickael and {Rodriguez}, Hector and {Rusholme}, Ben and {van Santen}, Jakob and {Shupe}, David L. and {Smith}, Roger M. and {Soumagnac}, Maayane T. and {Stein}, Robert and {Surace}, Jason and {Szkody}, Paula and {Terek}, Scott and {Van Sistine}, Angela and {van Velzen}, Sjoert and {Vestrand}, W. Thomas and {Walters}, Richard and {Ward}, Charlotte and {Zhang}, Chaoran and {Zolkower}, Jeffry},
        title = "{The Zwicky Transient Facility: Science Objectives}",
      journal = {\pasp},
         year = 2019,
        month = jul,
       volume = {131},
       number = {1001},
        pages = {078001},
          doi = {10.1088/1538-3873/ab006c},
archivePrefix = {arXiv},
       eprint = {1902.01945},
 primaryClass = {astro-ph.IM},
       adsurl = {https://ui.adsabs.harvard.edu/abs/2019PASP..131g8001G}
}

@ARTICLE{Masci2023,
       author = {{Masci}, Frank J. and {Laher}, Russ R. and {Rusholme}, Benjamin and {Shupe}, David and {Paladini}, Roberta and {Groom}, Steve and {Wold}, Avery and {Miller}, Adam A. and {Drake}, Andrew},
        title = "{A New Forced Photometry Service for the Zwicky Transient Facility}",
      journal = {arXiv e-prints},
         year = 2023,
        month = may,
          eid = {arXiv:2305.16279},
        pages = {arXiv:2305.16279},
          doi = {10.48550/arXiv.2305.16279},
archivePrefix = {arXiv},
       eprint = {2305.16279},
 primaryClass = {astro-ph.IM},
       adsurl = {https://ui.adsabs.harvard.edu/abs/2023arXiv230516279M}
}

@ARTICLE{Masci2019,
   author = {{Masci}, F.~J. and {Laher}, R.~R. and {Rusholme}, B. and {Shupe}, D.~L. and 
	{Groom}, S. and {Surace}, J. and {Jackson}, E. and {Monkewitz}, S. and 
	{Beck}, R. and {Flynn}, D. and {Terek}, S. and {Landry}, W. and 
	{Hacopians}, E. and {Desai}, V. and {Howell}, J. and {Brooke}, T. and 
	{Imel}, D. and {Wachter}, S. and {Ye}, Q.-Z. and {Lin}, H.-W. and 
	{Cenko}, S.~B. and {Cunningham}, V. and {Rebbapragada}, U. and 
	{Bue}, B. and {Miller}, A.~A. and {Mahabal}, A. and {Bellm}, E.~C. and 
	{Patterson}, M.~T. and {Juri{\'c}}, M. and {Golkhou}, V.~Z. and 
	{Ofek}, E.~O. and {Walters}, R. and {Graham}, M. and {Kasliwal}, M.~M. and 
	{Dekany}, R.~G. and {Kupfer}, T. and {Burdge}, K. and {Cannella}, C.~B. and 
	{Barlow}, T. and {Van Sistine}, A. and {Giomi}, M. and {Fremling}, C. and 
	{Blagorodnova}, N. and {Levitan}, D. and {Riddle}, R. and {Smith}, R.~M. and 
	{Helou}, G. and {Prince}, T.~A. and {Kulkarni}, S.~R.},
    title = "{The Zwicky Transient Facility: Data Processing, Products, and Archive}",
  journal = {\pasp},
     year = 2019,
    month = jan,
   volume = 131,
   number = 1,
    pages = {018003},
      doi = {10.1088/1538-3873/aae8ac},
   adsurl = {http://adsabs.harvard.edu/abs/2019PASP..131a8003M}
}

@ARTICLE{Tonry2018,
       author = {{Tonry}, J.~L. and {Denneau}, L. and {Heinze}, A.~N. and {Stalder}, B. and
         {Smith}, K.~W. and {Smartt}, S.~J. and {Stubbs}, C.~W. and {Weiland
        }, H.~J. and {Rest}, A.},
        title = "{ATLAS: A High-cadence All-sky Survey System}",
      journal = {\pasp},
         year = 2018,
        month = jun,
       volume = {130},
       number = {988},
        pages = {064505},
          doi = {10.1088/1538-3873/aabadf},
archivePrefix = {arXiv},
       eprint = {1802.00879},
 primaryClass = {astro-ph.IM},
       adsurl = {https://ui.adsabs.harvard.edu/abs/2018PASP..130f4505T}
}

@ARTICLE{Shingles2021,
       author = {{Shingles}, L. and {Smith}, K.~W. and {Young}, D.~R. and {Smartt}, S.~J. and {Tonry}, J. and {Denneau}, L. and {Heinze}, A. and {Weiland}, H. and {Flewelling}, H. and {Stalder}, B. and {Clocchiatti}, A. and {F{\"o}rster}, F. and {Pignata}, G. and {Rest}, A. and {Anderson}, J. and {Stubbs}, C. and {Erasmus}, N.},
        title = "{Release of the ATLAS Forced Photometry server for public use}",
      journal = {Transient Name Server AstroNote},
         year = 2021,
        month = jan,
       volume = {7},
        pages = {1-7},
       adsurl = {https://ui.adsabs.harvard.edu/abs/2021TNSAN...7....1S}
}

@article{Hung2019,
   title={Discovery of Highly Blueshifted Broad Balmer and Metastable Helium Absorption Lines in a Tidal Disruption Event},
   volume={879},
   ISSN={1538-4357},
   url={http://dx.doi.org/10.3847/1538-4357/ab24de},
   DOI={10.3847/1538-4357/ab24de},
   number={2},
   journal={The Astrophysical Journal},
   publisher={American Astronomical Society},
   author={Hung, T. and Cenko, S. B. and Roth, Nathaniel and Gezari, S. and Veilleux, S. and Velzen, Sjoert van and Gaskell, C. Martin and Foley, Ryan J. and Blagorodnova, N. and Yan, Lin and Graham, M. J. and Brown, J. S. and Siebert, M. R. and Frederick, Sara and Ward, Charlotte and Gatkine, Pradip and Gal-Yam, Avishay and Yang, Yi and Schulze, S. and Dimitriadis, G. and Kupfer, Thomas and Shupe, David L. and Rusholme, Ben and Masci, Frank J. and Riddle, Reed and Soumagnac, Maayane T. and Roestel, J. van and Dekany, Richard},
   year={2019},
   month=jul, pages={119} }

@ARTICLE{Smith2020,
       author = {{Smith}, K.~W. and {Smartt}, S.~J. and {Young}, D.~R. and {Tonry}, J.~L. and {Denneau}, L. and {Flewelling}, H. and {Heinze}, A.~N. and {Weiland}, H.~J. and {Stalder}, B. and {Rest}, A. and {Stubbs}, C.~W. and {Anderson}, J.~P. and {Chen}, T. -W. and {Clark}, P. and {Do}, A. and {F{\"o}rster}, F. and {Fulton}, M. and {Gillanders}, J. and {McBrien}, O.~R. and {O'Neill}, D. and {Srivastav}, S. and {Wright}, D.~E.},
        title = "{Design and Operation of the ATLAS Transient Science Server}",
      journal = {\pasp},
         year = 2020,
        month = aug,
       volume = {132},
       number = {1014},
          eid = {085002},
        pages = {085002},
          doi = {10.1088/1538-3873/ab936e},
archivePrefix = {arXiv},
       eprint = {2003.09052},
 primaryClass = {astro-ph.IM},
       adsurl = {https://ui.adsabs.harvard.edu/abs/2020PASP..132h5002S}
}

@misc{kumar_2024,
      title={AT2023vto: An Exceptionally Luminous Helium Tidal Disruption Event from a Massive Star}, 
      author={Harsh Kumar and Edo Berger and Daichi Hiramatsu and Sebastian Gomez and Peter K. Blanchard and Yvette Cendes and K. Azalee Bostroem and Joseph Farah and Estefania Padilla Gonzalez and Andrew Howell and Curtis McCully and Megan Newsome and Giacomo Terreran},
      year={2024},
      eprint={2408.01482},
      archivePrefix={arXiv},
      primaryClass={astro-ph.HE},
      url={https://arxiv.org/abs/2408.01482}, 
}

@ARTICLE{Roth_2018,
       author = {{Roth}, Nathaniel and {Kasen}, Daniel},
        title = "{What Sets the Line Profiles in Tidal Disruption Events?}",
      journal = {\apj},
         year = 2018,
        month = mar,
       volume = {855},
       number = {1},
          eid = {54},
        pages = {54},
          doi = {10.3847/1538-4357/aaaec6},
archivePrefix = {arXiv},
       eprint = {1707.02993},
 primaryClass = {astro-ph.HE},
       adsurl = {https://ui.adsabs.harvard.edu/abs/2018ApJ...855...54R}
}

@article{Charalampopoulos_2022,
   title={A detailed spectroscopic study of tidal disruption events},
   volume={659},
   ISSN={1432-0746},
   url={http://dx.doi.org/10.1051/0004-6361/202142122},
   DOI={10.1051/0004-6361/202142122},
   journal={Astronomy \& Astrophysics},
   publisher={EDP Sciences},
   author={Charalampopoulos, P. and Leloudas, G. and Malesani, D. B. and Wevers, T. and Arcavi, I. and Nicholl, M. and Pursiainen, M. and Lawrence, A. and Anderson, J. P. and Benetti, S. and Cannizzaro, G. and Chen, T.-W. and Galbany, L. and Gromadzki, M. and Gutiérrez, C. P. and Inserra, C. and Jonker, P. G. and Müller-Bravo, T. E. and Onori, F. and Short, P. and Sollerman, J. and Young, D. R.},
   year={2022},
   month=mar, pages={A34} }

@article{Panagiotou_2023,
   title={A Luminous Dust-obscured Tidal Disruption Event Candidate in a Star-forming Galaxy at 42 Mpc},
   volume={948},
   ISSN={2041-8213},
   url={http://dx.doi.org/10.3847/2041-8213/acc02f},
   DOI={10.3847/2041-8213/acc02f},
   number={1},
   journal={The Astrophysical Journal Letters},
   publisher={American Astronomical Society},
   author={Panagiotou, Christos and De, Kishalay and Masterson, Megan and Kara, Erin and Calzadilla, Michael and Eilers, Anna-Christina and Frostig, Danielle and Karambelkar, Viraj and Kasliwal, Mansi and Lourie, Nathan and Meisner, Aaron M. and Simcoe, Robert A. and Stein, Robert and Zolkower, Jeffry},
   year={2023},
   month=apr, pages={L5} }

@dataset{2014yCat.2328....0C,
       author = {{Cutri}, R.~M. and {Wright}, E.~L. and {Conrow}, T. and {Fowler}, J.~W. and {Eisenhardt}, P.~R.~M. and {Grillmair}, C. and {Kirkpatrick}, J.~D. and {Masci}, F. and {McCallon}, H.~L. and {Wheelock}, S.~L. and {Fajardo-Acosta}, S. and {Yan}, L. and {Benford}, D. and {Harbut}, M. and {Jarrett}, T. and {Lake}, S. and {Leisawitz}, D. and {Ressler}, M.~E. and {Stanford}, S.~A. and {Tsai}, C. -W. and {Liu}, F. and {Helou}, G. and {Mainzer}, A. and {Gettngs}, D. and {Gonzalez}, A. and {Hoffman}, D. and {Marsh}, K.~A. and {Padgett}, D. and {Skrutskie}, M.~F. and {Beck}, R. and {Papin}, M. and {Wittman}, M.},
        title = "{VizieR Online Data Catalog: AllWISE Data Release (Cutri+ 2013)}",
 howpublished = {VizieR On-line Data Catalog: II/328.  Originally published in: IPAC/Caltech (2013)},
         year = 2021,
        month = feb,
          eid = {II/328},
       adsurl = {https://ui.adsabs.harvard.edu/abs/2014yCat.2328....0C}
}

@ARTICLE{2023A&A...674A...1G,
       author = {{Gaia Collaboration} and {Vallenari}, A. and {Brown}, A.~G.~A. and {Prusti}, T. and {de Bruijne}, J.~H.~J. and {Arenou}, F. and {Babusiaux}, C. and {Biermann}, M. and {Creevey}, O.~L. and {Ducourant}, C. and {Evans}, D.~W. and {Eyer}, L. and {Guerra}, R. and {Hutton}, A. and {Jordi}, C. and {Klioner}, S.~A. and {Lammers}, U.~L. and {Lindegren}, L. and {Luri}, X. and {Mignard}, F. and {Panem}, C. and {Pourbaix}, D. and {Randich}, S. and {Sartoretti}, P. and {Soubiran}, C. and {Tanga}, P. and {Walton}, N.~A. and {Bailer-Jones}, C.~A.~L. and {Bastian}, U. and {Drimmel}, R. and {Jansen}, F. and {Katz}, D. and {Lattanzi}, M.~G. and {van Leeuwen}, F. and {Bakker}, J. and {Cacciari}, C. and {Casta{\~n}eda}, J. and {De Angeli}, F. and {Fabricius}, C. and {Fouesneau}, M. and {Fr{\'e}mat}, Y. and {Galluccio}, L. and {Guerrier}, A. and {Heiter}, U. and {Masana}, E. and {Messineo}, R. and {Mowlavi}, N. and {Nicolas}, C. and {Nienartowicz}, K. and {Pailler}, F. and {Panuzzo}, P. and {Riclet}, F. and {Roux}, W. and {Seabroke}, G.~M. and {Sordo}, R. and {Th{\'e}venin}, F. and {Gracia-Abril}, G. and {Portell}, J. and {Teyssier}, D. and {Altmann}, M. and {Andrae}, R. and {Audard}, M. and {Bellas-Velidis}, I. and {Benson}, K. and {Berthier}, J. and {Blomme}, R. and {Burgess}, P.~W. and {Busonero}, D. and {Busso}, G. and {C{\'a}novas}, H. and {Carry}, B. and {Cellino}, A. and {Cheek}, N. and {Clementini}, G. and {Damerdji}, Y. and {Davidson}, M. and {de Teodoro}, P. and {Nu{\~n}ez Campos}, M. and {Delchambre}, L. and {Dell'Oro}, A. and {Esquej}, P. and {Fern{\'a}ndez-Hern{\'a}ndez}, J. and {Fraile}, E. and {Garabato}, D. and {Garc{\'\i}a-Lario}, P. and {Gosset}, E. and {Haigron}, R. and {Halbwachs}, J. -L. and {Hambly}, N.~C. and {Harrison}, D.~L. and {Hern{\'a}ndez}, J. and {Hestroffer}, D. and {Hodgkin}, S.~T. and {Holl}, B. and {Jan{\ss}en}, K. and {Jevardat de Fombelle}, G. and {Jordan}, S. and {Krone-Martins}, A. and {Lanzafame}, A.~C. and {L{\"o}ffler}, W. and {Marchal}, O. and {Marrese}, P.~M. and {Moitinho}, A. and {Muinonen}, K. and {Osborne}, P. and {Pancino}, E. and {Pauwels}, T. and {Recio-Blanco}, A. and {Reyl{\'e}}, C. and {Riello}, M. and {Rimoldini}, L. and {Roegiers}, T. and {Rybizki}, J. and {Sarro}, L.~M. and {Siopis}, C. and {Smith}, M. and {Sozzetti}, A. and {Utrilla}, E. and {van Leeuwen}, M. and {Abbas}, U. and {{\'A}brah{\'a}m}, P. and {Abreu Aramburu}, A. and {Aerts}, C. and {Aguado}, J.~J. and {Ajaj}, M. and {Aldea-Montero}, F. and {Altavilla}, G. and {{\'A}lvarez}, M.~A. and {Alves}, J. and {Anders}, F. and {Anderson}, R.~I. and {Anglada Varela}, E. and {Antoja}, T. and {Baines}, D. and {Baker}, S.~G. and {Balaguer-N{\'u}{\~n}ez}, L. and {Balbinot}, E. and {Balog}, Z. and {Barache}, C. and {Barbato}, D. and {Barros}, M. and {Barstow}, M.~A. and {Bartolom{\'e}}, S. and {Bassilana}, J. -L. and {Bauchet}, N. and {Becciani}, U. and {Bellazzini}, M. and {Berihuete}, A. and {Bernet}, M. and {Bertone}, S. and {Bianchi}, L. and {Binnenfeld}, A. and {Blanco-Cuaresma}, S. and {Blazere}, A. and {Boch}, T. and {Bombrun}, A. and {Bossini}, D. and {Bouquillon}, S. and {Bragaglia}, A. and {Bramante}, L. and {Breedt}, E. and {Bressan}, A. and {Brouillet}, N. and {Brugaletta}, E. and {Bucciarelli}, B. and {Burlacu}, A. and {Butkevich}, A.~G. and {Buzzi}, R. and {Caffau}, E. and {Cancelliere}, R. and {Cantat-Gaudin}, T. and {Carballo}, R. and {Carlucci}, T. and {Carnerero}, M.~I. and {Carrasco}, J.~M. and {Casamiquela}, L. and {Castellani}, M. and {Castro-Ginard}, A. and {Chaoul}, L. and {Charlot}, P. and {Chemin}, L. and {Chiaramida}, V. and {Chiavassa}, A. and {Chornay}, N. and {Comoretto}, G. and {Contursi}, G. and {Cooper}, W.~J. and {Cornez}, T. and {Cowell}, S. and {Crifo}, F. and {Cropper}, M. and {Crosta}, M. and {Crowley}, C. and {Dafonte}, C. and {Dapergolas}, A. and {David}, M. and {David}, P. and {de Laverny}, P. and {De Luise}, F. and {De March}, R.},
        title = "{Gaia Data Release 3. Summary of the content and survey properties}",
      journal = {\aap},
         year = 2023,
        month = jun,
       volume = {674},
          eid = {A1},
        pages = {A1},
          doi = {10.1051/0004-6361/202243940},
archivePrefix = {arXiv},
       eprint = {2208.00211},
 primaryClass = {astro-ph.GA},
       adsurl = {https://ui.adsabs.harvard.edu/abs/2023A&A...674A...1G}
}

@article{Nicholl2019,
    author = {Nicholl, M and Blanchard, P K and Berger, E and Gomez, S and Margutti, R and Alexander, K D and Guillochon, J and Leja, J and Chornock, R and Snios, B and Auchettl, K and Bruce, A G and Challis, P and D’Orazio, D J and Drout, M R and Eftekhari, T and Foley, R J and Graur, O and Kilpatrick, C D and Lawrence, A and Piro, A L and Rojas-Bravo, C and Ross, N P and Short, P and Smartt, S J and Smith, K W and Stalder, B},
    title = {The tidal disruption event AT2017eqx: spectroscopic evolution from hydrogen rich to poor suggests an atmosphere and outflow},
    journal = {Monthly Notices of the Royal Astronomical Society},
    volume = {488},
    number = {2},
    pages = {1878-1893},
    year = {2019},
    month = {07},
    issn = {0035-8711},
    doi = {10.1093/mnras/stz1837},
    url = {https://doi.org/10.1093/mnras/stz1837},
    eprint = {https://academic.oup.com/mnras/article-pdf/488/2/1878/28954577/stz1837.pdf},
}

@ARTICLE{2022AstL...48..735M,
       author = {{Medvedev}, P.~S. and {Gilfanov}, M.~R. and {Sazonov}, S. Yu. and {Sunyaev}, R.~A. and {Khorunzhev}, G.~A.},
        title = "{Highly Variable Active Galactic Nuclei in the SRG/eROSITA Sky Survey: I. The Constriction of a Sample and the Catalog of Objects Detected in a Low State}",
      journal = {Astronomy Letters},
         year = 2022,
        month = dec,
       volume = {48},
       number = {12},
        pages = {735-754},
          doi = {10.1134/S1063773722120015},
archivePrefix = {arXiv},
       eprint = {2309.11266},
 primaryClass = {astro-ph.HE},
       adsurl = {https://ui.adsabs.harvard.edu/abs/2022AstL...48..735M}
}

@ARTICLE{2024AstL...50..744K,
       author = {{Khorunzhev}, G.~A. and {Sazonov}, S. Yu. and {Medvedev}, P.~S. and {Gilfanov}, M.~R. and {Dodin}, A.~V. and {Moiseev}, A.~V. and {Zaznobin}, I.~A. and {Moskaleva}, A.~V. and {Oparin}, D.~V. and {Burlak}, M.~A. and {Vozyakova}, O.~V. and {Eselevich}, M.~V. and {Sunyaev}, R.~A.},
        title = "{Highly Variable Active Galactic Nuclei in the SRG/eROSITA Sky Survey: II. Spectroscopic Identification of Sources with Russian Telescopes}",
      journal = {Astronomy Letters},
         year = 2024,
        month = dec,
       volume = {50},
       number = {12},
        pages = {744-755},
          doi = {10.1134/S1063773725700082},
       adsurl = {https://ui.adsabs.harvard.edu/abs/2024AstL...50..744K}
}

@ARTICLE{2020ATel13494....1K,
       author = {{Khabibullin}, I. and {Sunyaev}, R. and {Churazov}, E. and {Gilfanov}, M. and {Medvedev}, P. and {Sazonov}, S.},
        title = "{A bright X-ray TDE candidate SRGet J143359.25+400638.5 in SDSS J143359.16+400636.0}",
      journal = {The Astronomer's Telegram},
         year = 2020,
        month = feb,
       volume = {13494},
        pages = {1},
       adsurl = {https://ui.adsabs.harvard.edu/abs/2020ATel13494....1K}
}

@ARTICLE{1998ApJ...500..525S,
       author = {{Schlegel}, David J. and {Finkbeiner}, Douglas P. and {Davis}, Marc},
        title = "{Maps of Dust Infrared Emission for Use in Estimation of Reddening and Cosmic Microwave Background Radiation Foregrounds}",
      journal = {\apj},
         year = 1998,
        month = jun,
       volume = {500},
       number = {2},
        pages = {525-553},
          doi = {10.1086/305772},
archivePrefix = {arXiv},
       eprint = {astro-ph/9710327},
 primaryClass = {astro-ph},
       adsurl = {https://ui.adsabs.harvard.edu/abs/1998ApJ...500..525S}
}

@ARTICLE{DeColle2020,
       author = {{De Colle}, Fabio and {Lu}, Wenbin},
        title = "{Jets from Tidal Disruption Events}",
      journal = {\nar},
         year = 2020,
        month = sep,
       volume = {89},
          eid = {101538},
        pages = {101538},
          doi = {10.1016/j.newar.2020.101538},
archivePrefix = {arXiv},
       eprint = {1911.01442},
 primaryClass = {astro-ph.HE},
       adsurl = {https://ui.adsabs.harvard.edu/abs/2020NewAR..8901538D}
}

@INPROCEEDINGS{Phinney1989,
       author = {{Phinney}, E.~S.},
        title = "{Manifestations of a Massive Black Hole in the Galactic Center}",
    booktitle = {The Center of the Galaxy},
         year = 1989,
       editor = {{Morris}, Mark},
       series = {IAU Symposium},
       volume = {136},
        month = jan,
        pages = {543},
       adsurl = {https://ui.adsabs.harvard.edu/abs/1989IAUS..136..543P}
}

@ARTICLE{Rees1988,
       author = {{Rees}, Martin J.},
        title = "{Tidal disruption of stars by black holes of {}10$^{6}$-{}10$^{8}$ solar masses in nearby galaxies}",
      journal = {\nat},
         year = 1988,
        month = jun,
       volume = {333},
       number = {6173},
        pages = {523-528},
          doi = {10.1038/333523a0},
       adsurl = {https://ui.adsabs.harvard.edu/abs/1988Natur.333..523R}
}

@ARTICLE{Malyali2023,
       author = {{Malyali}, A. and {Liu}, Z. and {Merloni}, A. and {Rau}, A. and {Buchner}, J. and {Ciroi}, S. and {Di Mille}, F. and {Grotova}, I. and {Dwelly}, T. and {Nandra}, K. and {Salvato}, M. and {Homan}, D. and {Krumpe}, M.},
        title = "{eRASSt J074426.3 + 291606: prompt accretion disc formation in a 'faint and slow' tidal disruption event}",
      journal = {\mnras},
         year = 2023,
        month = apr,
       volume = {520},
       number = {3},
        pages = {4209-4225},
          doi = {10.1093/mnras/stad046},
archivePrefix = {arXiv},
       eprint = {2301.05484},
 primaryClass = {astro-ph.HE},
       adsurl = {https://ui.adsabs.harvard.edu/abs/2023MNRAS.520.4209M}
}

@ARTICLE{Liu2023_pTDE,
       author = {{Liu}, Z. and {Malyali}, A. and {Krumpe}, M. and {Homan}, D. and {Goodwin}, A.~J. and {Grotova}, I. and {Kawka}, A. and {Rau}, A. and {Merloni}, A. and {Anderson}, G.~E. and {Miller-Jones}, J.~C.~A. and {Markowitz}, A.~G. and {Ciroi}, S. and {Di Mille}, F. and {Schramm}, M. and {Tang}, S. and {Buckley}, D.~A.~H. and {Gromadzki}, M. and {Jin}, C. and {Buchner}, J.},
        title = "{Deciphering the extreme X-ray variability of the nuclear transient eRASSt J045650.3{\ensuremath{-}}203750. A likely repeating partial tidal disruption event}",
      journal = {\aap},
         year = 2023,
        month = jan,
       volume = {669},
          eid = {A75},
        pages = {A75},
          doi = {10.1051/0004-6361/202244805},
archivePrefix = {arXiv},
       eprint = {2208.12452},
 primaryClass = {astro-ph.HE},
       adsurl = {https://ui.adsabs.harvard.edu/abs/2023A&A...669A..75L}
}

@ARTICLE{Malyali2024,
       author = {{Malyali}, A. and {Rau}, A. and {Bonnerot}, C. and {Goodwin}, A.~J. and {Liu}, Z. and {Anderson}, G.~E. and {Brink}, J. and {Buckley}, D.~A.~H. and {Merloni}, A. and {Miller-Jones}, J.~C.~A. and {Grotova}, I. and {Kawka}, A.},
        title = "{Transient fading X-ray emission detected during the optical rise of a tidal disruption event}",
      journal = {\mnras},
         year = 2024,
        month = jun,
       volume = {531},
       number = {1},
        pages = {1256-1275},
          doi = {10.1093/mnras/stae927},
archivePrefix = {arXiv},
       eprint = {2309.16336},
 primaryClass = {astro-ph.HE},
       adsurl = {https://ui.adsabs.harvard.edu/abs/2024MNRAS.531.1256M}
}

@ARTICLE{Popesso2023,
       author = {{Popesso}, P. and {Concas}, A. and {Cresci}, G. and {Belli}, S. and {Rodighiero}, G. and {Inami}, H. and {Dickinson}, M. and {Ilbert}, O. and {Pannella}, M. and {Elbaz}, D.},
        title = "{The main sequence of star-forming galaxies across cosmic times}",
      journal = {\mnras},
         year = 2023,
        month = feb,
       volume = {519},
       number = {1},
        pages = {1526-1544},
          doi = {10.1093/mnras/stac3214},
archivePrefix = {arXiv},
       eprint = {2203.10487},
 primaryClass = {astro-ph.GA},
       adsurl = {https://ui.adsabs.harvard.edu/abs/2023MNRAS.519.1526P}
}

@ARTICLE{Graur2018,
       author = {{Graur}, Or and {French}, K. Decker and {Zahid}, H. Jabran and {Guillochon}, James and {Mandel}, Kaisey S. and {Auchettl}, Katie and {Zabludoff}, Ann I.},
        title = "{A Dependence of the Tidal Disruption Event Rate on Global Stellar Surface Mass Density and Stellar Velocity Dispersion}",
      journal = {\apj},
         year = 2018,
        month = jan,
       volume = {853},
       number = {1},
          eid = {39},
        pages = {39},
          doi = {10.3847/1538-4357/aaa3fd},
archivePrefix = {arXiv},
       eprint = {1707.02986},
 primaryClass = {astro-ph.HE},
       adsurl = {https://ui.adsabs.harvard.edu/abs/2018ApJ...853...39G}
}

@article{Csizi_2024,
   title={The PAU Survey: Galaxy stellar population properties estimates with narrowband data},
   volume={689},
   ISSN={1432-0746},
   url={http://dx.doi.org/10.1051/0004-6361/202449838},
   DOI={10.1051/0004-6361/202449838},
   journal={Astronomy \& Astrophysics},
   publisher={EDP Sciences},
   author={Csizi, B. and Tortorelli, L. and Siudek, M. and Grün, D. and Renard, P. and Tallada-Crespí, P. and Sánchez, E. and Miquel, R. and Padilla, C. and García-Bellido, J. and Gaztañaga, E. and Casas, R. and Serrano, S. and De Vicente, J. and Fernandez, E. and Eriksen, M. and Manzoni, G. and Baugh, C. M. and Carretero, J. and Castander, F. J.},
   year={2024},
   month=aug, pages={A37} }

@article{Hammerstein_2021,
   title={Tidal Disruption Event Hosts Are Green and Centrally Concentrated: Signatures of a Post-merger System},
   volume={908},
   ISSN={2041-8213},
   url={http://dx.doi.org/10.3847/2041-8213/abdcb4},
   DOI={10.3847/2041-8213/abdcb4},
   number={1},
   journal={The Astrophysical Journal Letters},
   publisher={American Astronomical Society},
   author={Hammerstein, Erica and Gezari, Suvi and van Velzen, Sjoert and Cenko, S. Bradley and Roth, Nathaniel and Ward, Charlotte and Frederick, Sara and Hung, Tiara and Graham, Matthew and Foley, Ryan J. and Bellm, Eric C. and Cannella, Christopher and Drake, Andrew J. and Kupfer, Thomas and Laher, Russ R. and Mahabal, Ashish A. and Masci, Frank J. and Riddle, Reed and Rojas-Bravo, César and Smith, Roger},
   year={2021},
   month=feb, pages={L20} }

@ARTICLE{French2016,
       author = {{French}, K. Decker and {Arcavi}, Iair and {Zabludoff}, Ann},
        title = "{Tidal Disruption Events Prefer Unusual Host Galaxies}",
      journal = {\apjl},
         year = 2016,
        month = feb,
       volume = {818},
       number = {1},
          eid = {L21},
        pages = {L21},
          doi = {10.3847/2041-8205/818/1/L21},
archivePrefix = {arXiv},
       eprint = {1601.04705},
 primaryClass = {astro-ph.GA},
       adsurl = {https://ui.adsabs.harvard.edu/abs/2016ApJ...818L..21F}
}

@ARTICLE{Baldwin1981,
       author = {{Baldwin}, J.~A. and {Phillips}, M.~M. and {Terlevich}, R.},
        title = "{Classification parameters for the emission-line spectra of extragalactic objects.}",
      journal = {\pasp},
         year = 1981,
        month = feb,
       volume = {93},
        pages = {5-19},
          doi = {10.1086/130766},
       adsurl = {https://ui.adsabs.harvard.edu/abs/1981PASP...93....5B}
}

@ARTICLE{Mondal2025,
       author = {{Mondal}, Samaresh and {French}, K. Decker},
        title = "{Utilizing Maximum Variability to Discern TDE Emission from AGN Flares}",
      journal = {arXiv e-prints},
         year = 2025,
        month = aug,
          eid = {arXiv:2508.03889},
        pages = {arXiv:2508.03889},
          doi = {10.48550/arXiv.2508.03889},
archivePrefix = {arXiv},
       eprint = {2508.03889},
 primaryClass = {astro-ph.HE},
       adsurl = {https://ui.adsabs.harvard.edu/abs/2025arXiv250803889M}
}

@ARTICLE{Martin2005,
       author = {{Martin}, D. Christopher and {Fanson}, James and {Schiminovich}, David and {Morrissey}, Patrick and {Friedman}, Peter G. and {Barlow}, Tom A. and {Conrow}, Tim and {Grange}, Robert and {Jelinsky}, Patrick N. and {Milliard}, Bruno and {Siegmund}, Oswald H.~W. and {Bianchi}, Luciana and {Byun}, Yong-Ik and {Donas}, Jose and {Forster}, Karl and {Heckman}, Timothy M. and {Lee}, Young-Wook and {Madore}, Barry F. and {Malina}, Roger F. and {Neff}, Susan G. and {Rich}, R. Michael and {Small}, Todd and {Surber}, Frank and {Szalay}, Alex S. and {Welsh}, Barry and {Wyder}, Ted K.},
        title = "{The Galaxy Evolution Explorer: A Space Ultraviolet Survey Mission}",
      journal = {\apjl},
         year = 2005,
        month = jan,
       volume = {619},
       number = {1},
        pages = {L1-L6},
          doi = {10.1086/426387},
archivePrefix = {arXiv},
       eprint = {astro-ph/0411302},
 primaryClass = {astro-ph},
       adsurl = {https://ui.adsabs.harvard.edu/abs/2005ApJ...619L...1M}
}

@ARTICLE{Grotova2025,
       author = {{Grotova}, I. and {Rau}, A. and {Baldini}, P. and {Goodwin}, A.~J. and {Liu}, Z. and {Merloni}, A. and {Salvato}, M. and {Anderson}, G.~E. and {Arcodia}, R. and {Buchner}, J. and {Krumpe}, M. and {Malyali}, A. and {Masterson}, M. and {Miller-Jones}, J.~C.~A. and {Nandra}, K. and {Shirley}, R.},
        title = "{The population of tidal disruption events discovered with eROSITA}",
      journal = {\aap},
         year = 2025,
        month = may,
       volume = {697},
          eid = {A159},
        pages = {A159},
          doi = {10.1051/0004-6361/202553669},
archivePrefix = {arXiv},
       eprint = {2504.08424},
 primaryClass = {astro-ph.HE},
       adsurl = {https://ui.adsabs.harvard.edu/abs/2025A&A...697A.159G}
}

@ARTICLE{2008ApJ...685..160N,
       author = {{Nenkova}, Maia and {Sirocky}, Matthew M. and {Nikutta}, Robert and {Ivezi{\'c}}, {\v{Z}}eljko and {Elitzur}, Moshe},
        title = "{AGN Dusty Tori. II. Observational Implications of Clumpiness}",
      journal = {\apj},
         year = 2008,
        month = sep,
       volume = {685},
       number = {1},
        pages = {160-180},
          doi = {10.1086/590483},
archivePrefix = {arXiv},
       eprint = {0806.0512},
 primaryClass = {astro-ph},
       adsurl = {https://ui.adsabs.harvard.edu/abs/2008ApJ...685..160N}
}

@ARTICLE{2007ApJ...657..810D,
       author = {{Draine}, B.~T. and {Li}, Aigen},
        title = "{Infrared Emission from Interstellar Dust. IV. The Silicate-Graphite-PAH Model in the Post-Spitzer Era}",
      journal = {\apj},
         year = 2007,
        month = mar,
       volume = {657},
       number = {2},
        pages = {810-837},
          doi = {10.1086/511055},
archivePrefix = {arXiv},
       eprint = {astro-ph/0608003},
 primaryClass = {astro-ph},
       adsurl = {https://ui.adsabs.harvard.edu/abs/2007ApJ...657..810D}
}

@ARTICLE{2000ApJ...539..718C,
       author = {{Charlot}, St{\'e}phane and {Fall}, S. Michael},
        title = "{A Simple Model for the Absorption of Starlight by Dust in Galaxies}",
      journal = {\apj},
         year = 2000,
        month = aug,
       volume = {539},
       number = {2},
        pages = {718-731},
          doi = {10.1086/309250},
archivePrefix = {arXiv},
       eprint = {astro-ph/0003128},
 primaryClass = {astro-ph},
       adsurl = {https://ui.adsabs.harvard.edu/abs/2000ApJ...539..718C}
}

@ARTICLE{2020A&C....3300411N,
       author = {{Nikutta}, R. and {Fitzpatrick}, M. and {Scott}, A. and {Weaver}, B.~A.},
        title = "{Data Lab-A community science platform}",
      journal = {Astronomy and Computing},
         year = 2020,
        month = oct,
       volume = {33},
          eid = {100411},
        pages = {100411},
          doi = {10.1016/j.ascom.2020.100411},
       adsurl = {https://ui.adsabs.harvard.edu/abs/2020A&C....3300411N}
}

@ARTICLE{DESI2025_dr1,
       author = {{DESI Collaboration} and {Abdul-Karim}, M. and {Adame}, A.~G. and {Aguado}, D. and {Aguilar}, J. and {Ahlen}, S. and {Alam}, S. and {Aldering}, G. and {Alexander}, D.~M. and {Alfarsy}, R. and {Allen}, L. and {Allende Prieto}, C. and {Alves}, O. and {Anand}, A. and {Andrade}, U. and {Armengaud}, E. and {Avila}, S. and {Aviles}, A. and {Awan}, H. and {Bailey}, S. and {Baleato Lizancos}, A. and {Ballester}, O. and {Bault}, A. and {Bautista}, J. and {BenZvi}, S. and {Beraldo e Silva}, L. and {Bermejo-Climent}, J.~R. and {Beutler}, F. and {Bianchi}, D. and {Blake}, C. and {Blum}, R. and {Bolton}, A.~S. and {Bonici}, M. and {Brieden}, S. and {Brodzeller}, A. and {Brooks}, D. and {Buckley-Geer}, E. and {Burtin}, E. and {Canning}, R. and {Carnero Rosell}, A. and {Carr}, A. and {Carrilho}, P. and {Casas}, L. and {Castander}, F.~J. and {Cereskaite}, R. and {Cervantes-Cota}, J.~L. and {Chaussidon}, E. and {Chaves-Montero}, J. and {Chen}, S. and {Chen}, X. and {Claybaugh}, T. and {Cole}, S. and {Cooper}, A.~P. and {Cousinou}, M. -C. and {Cuceu}, A. and {Davis}, T.~M. and {Dawson}, K.~S. and {de Belsunce}, R. and {de la Cruz}, R. and {de la Macorra}, A. and {de Mattia}, A. and {Deiosso}, N. and {Della Costa}, J. and {Demina}, R. and {Demirbozan}, U. and {DeRose}, J. and {Dey}, A. and {Dey}, B. and {Ding}, J. and {Ding}, Z. and {Doel}, P. and {Douglass}, K. and {Dowicz}, M. and {Ebina}, H. and {Edelstein}, J. and {Eisenstein}, D.~J. and {Elbers}, W. and {Emas}, N. and {Escoffier}, S. and {Fagrelius}, P. and {Fan}, X. and {Fanning}, K. and {Fawcett}, V.~A. and {Fern\textbackslash'andez-Garc\textbackslash'ia}, E. and {Ferraro}, S. and {Findlay}, N. and {Font-Ribera}, A. and {Forero-Romero}, J.~E. and {Forero-S\textbackslash'anchez}, D. and {Frenk}, C.~S. and {G\textbackslash''ansicke}, B.~T. and {Galbany}, L. and {Garc\textbackslash'ia-Bellido}, J. and {Garcia-Quintero}, C. and {Garrison}, L.~H. and {{Gazta\~naga}}, E. and {Gil-Mar\textbackslash'in}, H. and {Gnedin}, O.~Y. and {Gontcho}, S. Gontcho A and {Gonzalez-Morales}, A.~X. and {Gonzalez-Perez}, V. and {Gordon}, C. and {Graur}, O. and {Green}, D. and {Gruen}, D. and {Gsponer}, R. and {Guandalin}, C. and {Gutierrez}, G. and {Guy}, J. and {Hahn}, C. and {Han}, J.~J. and {Han}, J. and {He}, S. and {Herrera-Alcantar}, H.~K. and {Honscheid}, K. and {Hou}, J. and {Howlett}, C. and {Huterer}, D. and {{Ir\v{s}i\v{c}}}, V. and {Ishak}, M. and {Jacques}, A. and {Jimenez}, J. and {Jing}, Y.~P. and {Joachimi}, B. and {Joudaki}, S. and {Joyce}, R. and {Jullo}, E. and {Juneau}, S. and {{Kara\c{c}ayl{\i}}}, N.~G. and {Karim}, T. and {Kehoe}, R. and {Kent}, S. and {Khederlarian}, A. and {Kirkby}, D. and {Kisner}, T. and {Kitaura}, F. -S. and {Kizhuprakkat}, N. and {Kong}, H. and {Koposov}, S.~E. and {Kremin}, A. and {Krolewski}, A. and {Lahav}, O. and {Lai}, Y. and {Lamman}, C. and {Lan}, T. -W. and {Landriau}, M. and {Lang}, D. and {Lange}, J.~U. and {Lasker}, J. and {Le Goff}, J.~M. and {Le Guillou}, L. and {Leauthaud}, A. and {Levi}, M.~E. and {Li}, S. and {Li}, T.~S. and {Lodha}, K. and {Lokken}, M. and {Luo}, Y. and {Magneville}, C. and {Manera}, M. and {Manser}, C.~J. and {Margala}, D. and {Martini}, P. and {Maus}, M. and {McCullough}, J. and {McDonald}, P. and {Medina}, G.~E. and {Medina-Varela}, L. and {Meisner}, A. and {{Mena-Fern\'andez}}, J. and {Menegas}, A. and {Mezcua}, M. and {Miquel}, R. and {Montero-Camacho}, P. and {Moon}, J. and {Moustakas}, J. and {{Mu\~{n}oz-Guti\'errez}}, A. and {{Mu\~{n}oz-Santos}}, D. and {Myers}, A.~D. and {Myles}, J. and {Nadathur}, S. and {Najita}, J. and {Napolitano}, L. and {Newman}, J.~A. and {Nikakhtar}, F. and {Nikutta}, R. and {Niz}, G. and {Noriega}, H.~E. and {Padmanabhan}, N. and {Paillas}, E. and {Palanque-Delabrouille}, N. and {Palmese}, A. and {Pan}, J. and {Pan}, Z. and {Parkinson}, D. and {Peacock}, J. and {Percival}, W.~J. and {{P\'erez-Fern\'andez}}, A. and {{P\'erez-R\`afols}}, I. and {Peterson}, P.},
        title = "{Data Release 1 of the Dark Energy Spectroscopic Instrument}",
      journal = {arXiv e-prints},
         year = 2025,
        month = mar,
          eid = {arXiv:2503.14745},
        pages = {arXiv:2503.14745},
          doi = {10.48550/arXiv.2503.14745},
archivePrefix = {arXiv},
       eprint = {2503.14745},
 primaryClass = {astro-ph.CO},
       adsurl = {https://ui.adsabs.harvard.edu/abs/2025arXiv250314745D}
}

@ARTICLE{Oke1995,
       author = {{Oke}, J.~B. and {Cohen}, J.~G. and {Carr}, M. and {Cromer}, J. and
         {Dingizian}, A. and {Harris}, F.~H. and {Labrecque}, S. and
         {Lucinio}, R. and {Schaal}, W. and {Epps}, H. and {Miller}, J.},
        title = "{The Keck Low-Resolution Imaging Spectrometer}",
      journal = {\pasp},
         year = "1995",
        month = "Apr",
       volume = {107},
        pages = {375},
          doi = {10.1086/133562},
       adsurl = {https://ui.adsabs.harvard.edu/abs/1995PASP..107..375O}
}

@ARTICLE{Million2016,
       author = {{Million}, Chase and {Fleming}, Scott W. and {Shiao}, Bernie and {Seibert}, Mark and {Loyd}, Parke and {Tucker}, Michael and {Smith}, Myron and {Thompson}, Randy and {White}, Richard L.},
        title = "{gPhoton: The GALEX Photon Data Archive}",
      journal = {\apj},
         year = 2016,
        month = dec,
       volume = {833},
       number = {2},
          eid = {292},
        pages = {292},
          doi = {10.3847/1538-4357/833/2/292},
archivePrefix = {arXiv},
       eprint = {1609.09492},
 primaryClass = {astro-ph.IM},
       adsurl = {https://ui.adsabs.harvard.edu/abs/2016ApJ...833..292M}
}

@ARTICLE{2006AJ....131.1163S,
       author = {{Skrutskie}, M.~F. and {Cutri}, R.~M. and {Stiening}, R. and {Weinberg}, M.~D. and {Schneider}, S. and {Carpenter}, J.~M. and {Beichman}, C. and {Capps}, R. and {Chester}, T. and {Elias}, J. and {Huchra}, J. and {Liebert}, J. and {Lonsdale}, C. and {Monet}, D.~G. and {Price}, S. and {Seitzer}, P. and {Jarrett}, T. and {Kirkpatrick}, J.~D. and {Gizis}, J.~E. and {Howard}, E. and {Evans}, T. and {Fowler}, J. and {Fullmer}, L. and {Hurt}, R. and {Light}, R. and {Kopan}, E.~L. and {Marsh}, K.~A. and {McCallon}, H.~L. and {Tam}, R. and {Van Dyk}, S. and {Wheelock}, S.},
        title = "{The Two Micron All Sky Survey (2MASS)}",
      journal = {\aj},
         year = 2006,
        month = feb,
       volume = {131},
       number = {2},
        pages = {1163-1183},
          doi = {10.1086/498708},
       adsurl = {https://ui.adsabs.harvard.edu/abs/2006AJ....131.1163S}
}

@MISC{2013wise.rept....1C,
       author = {{Cutri}, R.~M. and {Wright}, E.~L. and {Conrow}, T. and {Fowler}, J.~W. and {Eisenhardt}, P.~R.~M. and {Grillmair}, C. and {Kirkpatrick}, J.~D. and {Masci}, F. and {McCallon}, H.~L. and {Wheelock}, S.~L. and {Fajardo-Acosta}, S. and {Yan}, L. and {Benford}, D. and {Harbut}, M. and {Jarrett}, T. and {Lake}, S. and {Leisawitz}, D. and {Ressler}, M.~E. and {Stanford}, S.~A. and {Tsai}, C.~W. and {Liu}, F. and {Helou}, G. and {Mainzer}, A. and {Gettings}, D. and {Gonzalez}, A. and {Hoffman}, D. and {Marsh}, K.~A. and {Padgett}, D. and {Skrutskie}, M.~F. and {Beck}, R.~P. and {Papin}, M. and {Wittman}, M.},
        title = "{Explanatory Supplement to the AllWISE Data Release Products}",
 howpublished = {Explanatory Supplement to the AllWISE Data Release Products, by R. M. Cutri et al.},
         year = 2013,
        month = nov,
        pages = {1},
       adsurl = {https://ui.adsabs.harvard.edu/abs/2013wise.rept....1C}
}

@ARTICLE{2021ApJS..254...22J,
       author = {{Johnson}, Benjamin D. and {Leja}, Joel and {Conroy}, Charlie and {Speagle}, Joshua S.},
        title = "{Stellar Population Inference with Prospector}",
      journal = {\apjs},
         year = 2021,
        month = jun,
       volume = {254},
       number = {2},
          eid = {22},
        pages = {22},
          doi = {10.3847/1538-4365/abef67},
archivePrefix = {arXiv},
       eprint = {2012.01426},
 primaryClass = {astro-ph.GA},
       adsurl = {https://ui.adsabs.harvard.edu/abs/2021ApJS..254...22J}
}

@software{2010ascl.soft10043C,
       author = {{Conroy}, Charlie and {Gunn}, James E.},
        title = "{FSPS: Flexible Stellar Population Synthesis}",
 howpublished = {Astrophysics Source Code Library, record ascl:1010.043},
         year = 2010,
        month = oct,
          eid = {ascl:1010.043},
       adsurl = {https://ui.adsabs.harvard.edu/abs/2010ascl.soft10043C}
}

@ARTICLE{2005MNRAS.362...41G,
       author = {{Gallazzi}, Anna and {Charlot}, St{\'e}phane and {Brinchmann}, Jarle and {White}, Simon D.~M. and {Tremonti}, Christy A.},
        title = "{The ages and metallicities of galaxies in the local universe}",
      journal = {\mnras},
         year = 2005,
        month = sep,
       volume = {362},
       number = {1},
        pages = {41-58},
          doi = {10.1111/j.1365-2966.2005.09321.x},
archivePrefix = {arXiv},
       eprint = {astro-ph/0506539},
 primaryClass = {astro-ph},
       adsurl = {https://ui.adsabs.harvard.edu/abs/2005MNRAS.362...41G}
}

@ARTICLE{2020MNRAS.493.3132S,
       author = {{Speagle}, Joshua S.},
        title = "{DYNESTY: a dynamic nested sampling package for estimating Bayesian posteriors and evidences}",
      journal = {\mnras},
         year = 2020,
        month = apr,
       volume = {493},
       number = {3},
        pages = {3132-3158},
          doi = {10.1093/mnras/staa278},
archivePrefix = {arXiv},
       eprint = {1904.02180},
 primaryClass = {astro-ph.IM},
       adsurl = {https://ui.adsabs.harvard.edu/abs/2020MNRAS.493.3132S}
}

@ARTICLE{2013ApJ...775L..16K,
       author = {{Kriek}, Mariska and {Conroy}, Charlie},
        title = "{The Dust Attenuation Law in Distant Galaxies: Evidence for Variation with Spectral Type}",
      journal = {\apjl},
         year = 2013,
        month = sep,
       volume = {775},
       number = {1},
          eid = {L16},
        pages = {L16},
          doi = {10.1088/2041-8205/775/1/L16},
archivePrefix = {arXiv},
       eprint = {1308.1099},
 primaryClass = {astro-ph.CO},
       adsurl = {https://ui.adsabs.harvard.edu/abs/2013ApJ...775L..16K}
}

@ARTICLE{2016arXiv161205560C,
       author = {{Chambers}, K.~C. and {Magnier}, E.~A. and {Metcalfe}, N. and {Flewelling}, H.~A. and {Huber}, M.~E. and {Waters}, C.~Z. and {Denneau}, L. and {Draper}, P.~W. and {Farrow}, D. and {Finkbeiner}, D.~P. and {Holmberg}, C. and {Koppenhoefer}, J. and {Price}, P.~A. and {Rest}, A. and {Saglia}, R.~P. and {Schlafly}, E.~F. and {Smartt}, S.~J. and {Sweeney}, W. and {Wainscoat}, R.~J. and {Burgett}, W.~S. and {Chastel}, S. and {Grav}, T. and {Heasley}, J.~N. and {Hodapp}, K.~W. and {Jedicke}, R. and {Kaiser}, N. and {Kudritzki}, R. -P. and {Luppino}, G.~A. and {Lupton}, R.~H. and {Monet}, D.~G. and {Morgan}, J.~S. and {Onaka}, P.~M. and {Shiao}, B. and {Stubbs}, C.~W. and {Tonry}, J.~L. and {White}, R. and {Ba{\~n}ados}, E. and {Bell}, E.~F. and {Bender}, R. and {Bernard}, E.~J. and {Boegner}, M. and {Boffi}, F. and {Botticella}, M.~T. and {Calamida}, A. and {Casertano}, S. and {Chen}, W. -P. and {Chen}, X. and {Cole}, S. and {Deacon}, N. and {Frenk}, C. and {Fitzsimmons}, A. and {Gezari}, S. and {Gibbs}, V. and {Goessl}, C. and {Goggia}, T. and {Gourgue}, R. and {Goldman}, B. and {Grant}, P. and {Grebel}, E.~K. and {Hambly}, N.~C. and {Hasinger}, G. and {Heavens}, A.~F. and {Heckman}, T.~M. and {Henderson}, R. and {Henning}, T. and {Holman}, M. and {Hopp}, U. and {Ip}, W. -H. and {Isani}, S. and {Jackson}, M. and {Keyes}, C.~D. and {Koekemoer}, A.~M. and {Kotak}, R. and {Le}, D. and {Liska}, D. and {Long}, K.~S. and {Lucey}, J.~R. and {Liu}, M. and {Martin}, N.~F. and {Masci}, G. and {McLean}, B. and {Mindel}, E. and {Misra}, P. and {Morganson}, E. and {Murphy}, D.~N.~A. and {Obaika}, A. and {Narayan}, G. and {Nieto-Santisteban}, M.~A. and {Norberg}, P. and {Peacock}, J.~A. and {Pier}, E.~A. and {Postman}, M. and {Primak}, N. and {Rae}, C. and {Rai}, A. and {Riess}, A. and {Riffeser}, A. and {Rix}, H.~W. and {R{\"o}ser}, S. and {Russel}, R. and {Rutz}, L. and {Schilbach}, E. and {Schultz}, A.~S.~B. and {Scolnic}, D. and {Strolger}, L. and {Szalay}, A. and {Seitz}, S. and {Small}, E. and {Smith}, K.~W. and {Soderblom}, D.~R. and {Taylor}, P. and {Thomson}, R. and {Taylor}, A.~N. and {Thakar}, A.~R. and {Thiel}, J. and {Thilker}, D. and {Unger}, D. and {Urata}, Y. and {Valenti}, J. and {Wagner}, J. and {Walder}, T. and {Walter}, F. and {Watters}, S.~P. and {Werner}, S. and {Wood-Vasey}, W.~M. and {Wyse}, R.},
        title = "{The Pan-STARRS1 Surveys}",
      journal = {arXiv e-prints},
         year = 2016,
        month = dec,
          eid = {arXiv:1612.05560},
        pages = {arXiv:1612.05560},
          doi = {10.48550/arXiv.1612.05560},
archivePrefix = {arXiv},
       eprint = {1612.05560},
 primaryClass = {astro-ph.IM},
       adsurl = {https://ui.adsabs.harvard.edu/abs/2016arXiv161205560C}
}

@ARTICLE{2010AJ....140.1868W,
       author = {{Wright}, Edward L. and {Eisenhardt}, Peter R.~M. and {Mainzer}, Amy K. and {Ressler}, Michael E. and {Cutri}, Roc M. and {Jarrett}, Thomas and {Kirkpatrick}, J. Davy and {Padgett}, Deborah and {McMillan}, Robert S. and {Skrutskie}, Michael and {Stanford}, S.~A. and {Cohen}, Martin and {Walker}, Russell G. and {Mather}, John C. and {Leisawitz}, David and {Gautier}, III, Thomas N. and {McLean}, Ian and {Benford}, Dominic and {Lonsdale}, Carol J. and {Blain}, Andrew and {Mendez}, Bryan and {Irace}, William R. and {Duval}, Valerie and {Liu}, Fengchuan and {Royer}, Don and {Heinrichsen}, Ingolf and {Howard}, Joan and {Shannon}, Mark and {Kendall}, Martha and {Walsh}, Amy L. and {Larsen}, Mark and {Cardon}, Joel G. and {Schick}, Scott and {Schwalm}, Mark and {Abid}, Mohamed and {Fabinsky}, Beth and {Naes}, Larry and {Tsai}, Chao-Wei},
        title = "{The Wide-field Infrared Survey Explorer (WISE): Mission Description and Initial On-orbit Performance}",
      journal = {\aj},
         year = 2010,
        month = dec,
       volume = {140},
       number = {6},
        pages = {1868-1881},
          doi = {10.1088/0004-6256/140/6/1868},
archivePrefix = {arXiv},
       eprint = {1008.0031},
 primaryClass = {astro-ph.IM},
       adsurl = {https://ui.adsabs.harvard.edu/abs/2010AJ....140.1868W}
}

@ARTICLE{Cappellari2023,
    author = {{Cappellari}, M.},
    title = "{Full spectrum fitting with photometry in PPXF: stellar population
        versus dynamical masses, non-parametric star formation history and
        metallicity for 3200 LEGA-C galaxies at redshift $z\approx0.8$}",
    journal = {MNRAS},
    eprint = {2208.14974},
    year = 2023,
    volume = 526,
    pages = {3273-3300},
    doi = {10.1093/mnras/stad2597}
}

@ARTICLE{Khorunzhev2022,
       author = {{Khorunzhev}, G.~A. and {Sazonov}, S. Yu. and {Medvedev}, P.~S. and {Gilfanov}, M.~R. and {Atapin}, K.~E. and {Belinski}, A.~A. and {Vozyakova}, O.~V. and {Dodin}, A.~V. and {Safonov}, B.~S. and {Tatarnikov}, A.~M. and {Bikmaev}, I.~F. and {Burenin}, R.~A. and {Dodonov}, S.~N. and {Eselevich}, M.~V. and {Zaznobin}, I.~A. and {Krivonos}, R.~A. and {Uklein}, R.~I. and {Postnov}, K.~A. and {Sunyaev}, R.~A.},
        title = "{Search for Tidal Disruption Events Based on the SRG/eROSITA Survey with Subsequent Optical Spectroscopy}",
      journal = {Astronomy Letters},
         year = 2022,
        month = dec,
       volume = {48},
       number = {12},
        pages = {767-789},
          doi = {10.1134/S1063773723010036},
       adsurl = {https://ui.adsabs.harvard.edu/abs/2022AstL...48..767K}
}

@ARTICLE{Nicholl_2020,
       author = {{Nicholl}, M. and {Wevers}, T. and {Oates}, S.~R. and {Alexander}, K.~D. and {Leloudas}, G. and {Onori}, F. and {Jerkstrand}, A. and {Gomez}, S. and {Campana}, S. and {Arcavi}, I. and {Charalampopoulos}, P. and {Gromadzki}, M. and {Ihanec}, N. and {Jonker}, P.~G. and {Lawrence}, A. and {Mandel}, I. and {Schulze}, S. and {Short}, P. and {Burke}, J. and {McCully}, C. and {Hiramatsu}, D. and {Howell}, D.~A. and {Pellegrino}, C. and {Abbot}, H. and {Anderson}, J.~P. and {Berger}, E. and {Blanchard}, P.~K. and {Cannizzaro}, G. and {Chen}, T.-W. and {Dennefeld}, M. and {Galbany}, L. and {Gonz{\'a}lez-Gait{\'a}n}, S. and {Hosseinzadeh}, G. and {Inserra}, C. and {Irani}, I. and {Kuin}, P. and {M{\"u}ller-Bravo}, T. and {Pineda}, J. and {Ross}, N.~P. and {Roy}, R. and {Smartt}, S.~J. and {Smith}, K.~W. and {Tucker}, B. and {Wyrzykowski}, {\L}. and {Young}, D.~R.},
        title = "{An outflow powers the optical rise of the nearby, fast-evolving tidal disruption event AT2019qiz}",
      journal = {\mnras},
         year = 2020,
        month = nov,
       volume = {499},
       number = {1},
        pages = {482-504},
          doi = {10.1093/mnras/staa2824},
archivePrefix = {arXiv},
       eprint = {2006.02454},
 primaryClass = {astro-ph.HE},
       adsurl = {https://ui.adsabs.harvard.edu/abs/2020MNRAS.499..482N}
}

@ARTICLE{Arcavi_2014,
       author = {{Arcavi}, Iair and {Gal-Yam}, Avishay and {Sullivan}, Mark and {Pan}, Yen-Chen and {Cenko}, S. Bradley and {Horesh}, Assaf and {Ofek}, Eran O. and {De Cia}, Annalisa and {Yan}, Lin and {Yang}, Chen-Wei and {Howell}, D.~A. and {Tal}, David and {Kulkarni}, Shrinivas R. and {Tendulkar}, Shriharsh P. and {Tang}, Sumin and {Xu}, Dong and {Sternberg}, Assaf and {Cohen}, Judith G. and {Bloom}, Joshua S. and {Nugent}, Peter E. and {Kasliwal}, Mansi M. and {Perley}, Daniel A. and {Quimby}, Robert M. and {Miller}, Adam A. and {Theissen}, Christopher A. and {Laher}, Russ R.},
        title = "{A Continuum of H- to He-rich Tidal Disruption Candidates With a Preference for E+A Galaxies}",
      journal = {\apj},
         year = 2014,
        month = sep,
       volume = {793},
       number = {1},
          eid = {38},
        pages = {38},
          doi = {10.1088/0004-637X/793/1/38},
archivePrefix = {arXiv},
       eprint = {1405.1415},
 primaryClass = {astro-ph.HE},
       adsurl = {https://ui.adsabs.harvard.edu/abs/2014ApJ...793...38A}
}

@ARTICLE{Yao2024_22cmc,
       author = {{Yao}, Yuhan and {Lu}, Wenbin and {Harrison}, Fiona and {Kulkarni}, S.~R. and {Gezari}, Suvi and {Guolo}, Muryel and {Cenko}, S. Bradley and {Ho}, Anna Y.~Q.},
        title = "{The On-axis Jetted Tidal Disruption Event AT2022cmc: X-Ray Observations and Broadband Spectral Modeling}",
      journal = {\apj},
         year = 2024,
        month = apr,
       volume = {965},
       number = {1},
          eid = {39},
        pages = {39},
          doi = {10.3847/1538-4357/ad2b6b},
archivePrefix = {arXiv},
       eprint = {2308.09834},
 primaryClass = {astro-ph.HE},
       adsurl = {https://ui.adsabs.harvard.edu/abs/2024ApJ...965...39Y}
}

@ARTICLE{Brandt1995,
       author = {{Brandt}, W.~N. and {Pounds}, K.~A. and {Fink}, H.},
        title = "{The unusual X-ray and optical properties of the ultrasoft active galactic nucleus Zwicky 159.034 (RE J1237+264)}",
      journal = {\mnras},
         year = 1995,
        month = apr,
       volume = {273},
       number = {3},
        pages = {L47-L52},
          doi = {10.1093/mnras/273.1.L47},
archivePrefix = {arXiv},
       eprint = {astro-ph/9501108},
 primaryClass = {astro-ph},
       adsurl = {https://ui.adsabs.harvard.edu/abs/1995MNRAS.273L..47B}
}

@ARTICLE{Grupe1995,
       author = {{Grupe}, D. and {Beuerman}, K. and {Mannheim}, K. and {Thomas}, H. -C. and {Fink}, H.~H. and {de Martino}, D.},
        title = "{Discovery of an ultrasoft transient ROSAT AGN: WPVS 007.}",
      journal = {\aap},
         year = 1995,
        month = aug,
       volume = {300},
        pages = {L21},
          doi = {10.48550/arXiv.astro-ph/9506087},
archivePrefix = {arXiv},
       eprint = {astro-ph/9506087},
 primaryClass = {astro-ph},
       adsurl = {https://ui.adsabs.harvard.edu/abs/1995A&A...300L..21G}
}

@ARTICLE{Jiang2021_tde_echo,
       author = {{Jiang}, Ning and {Wang}, Tinggui and {Hu}, Xueyang and {Sun}, Luming and {Dou}, Liming and {Xiao}, Lin},
        title = "{Infrared Echoes of Optical Tidal Disruption Events: {\ensuremath{\sim}}1\% Dust-covering Factor or Less at Subparsec Scale}",
      journal = {\apj},
         year = 2021,
        month = apr,
       volume = {911},
       number = {1},
          eid = {31},
        pages = {31},
          doi = {10.3847/1538-4357/abe772},
archivePrefix = {arXiv},
       eprint = {2102.08044},
 primaryClass = {astro-ph.GA},
       adsurl = {https://ui.adsabs.harvard.edu/abs/2021ApJ...911...31J}
}

@PHDTHESIS{Yao2023_thesis,
       author = {{Yao}, Yuhan},
        title = "{High Energy Transients Powered by Black Holes}",
       school = {California Institute of Technology, Division of Physics, Mathematics and
        Astronomy},
         year = 2023,
        month = jan,
       adsurl = {https://ui.adsabs.harvard.edu/abs/2023PhDT.........2Y}
}

@ARTICLE{Gezari2021,
       author = {{Gezari}, Suvi},
        title = "{Tidal Disruption Events}",
      journal = {\araa},
         year = 2021,
        month = sep,
       volume = {59},
        pages = {21-58},
          doi = {10.1146/annurev-astro-111720-030029},
archivePrefix = {arXiv},
       eprint = {2104.14580},
 primaryClass = {astro-ph.HE},
       adsurl = {https://ui.adsabs.harvard.edu/abs/2021ARA&A..59...21G}
}

@ARTICLE{Komossa2015,
       author = {{Komossa}, S.},
        title = "{Tidal disruption of stars by supermassive black holes: Status of observations}",
      journal = {Journal of High Energy Astrophysics},
         year = 2015,
        month = sep,
       volume = {7},
        pages = {148-157},
          doi = {10.1016/j.jheap.2015.04.006},
archivePrefix = {arXiv},
       eprint = {1505.01093},
 primaryClass = {astro-ph.HE},
       adsurl = {https://ui.adsabs.harvard.edu/abs/2015JHEAp...7..148K}
}

@ARTICLE{2012ApJ...753...30S,
       author = {{Stern}, Daniel and {Assef}, Roberto J. and {Benford}, Dominic J. and {Blain}, Andrew and {Cutri}, Roc and {Dey}, Arjun and {Eisenhardt}, Peter and {Griffith}, Roger L. and {Jarrett}, T.~H. and {Lake}, Sean and {Masci}, Frank and {Petty}, Sara and {Stanford}, S.~A. and {Tsai}, Chao-Wei and {Wright}, E.~L. and {Yan}, Lin and {Harrison}, Fiona and {Madsen}, Kristin},
        title = "{Mid-infrared Selection of Active Galactic Nuclei with the Wide-Field Infrared Survey Explorer. I. Characterizing WISE-selected Active Galactic Nuclei in COSMOS}",
      journal = {\apj},
         year = 2012,
        month = jul,
       volume = {753},
       number = {1},
          eid = {30},
        pages = {30},
          doi = {10.1088/0004-637X/753/1/30},
archivePrefix = {arXiv},
       eprint = {1205.0811},
 primaryClass = {astro-ph.CO},
       adsurl = {https://ui.adsabs.harvard.edu/abs/2012ApJ...753...30S}
}

@article{Schawinski_2014,
   title={The green valley is a red herring: Galaxy Zoo reveals two evolutionary pathways towards quenching of star formation in early- and late-type galaxies★},
   volume={440},
   ISSN={0035-8711},
   url={http://dx.doi.org/10.1093/mnras/stu327},
   DOI={10.1093/mnras/stu327},
   number={1},
   journal={Monthly Notices of the Royal Astronomical Society},
   publisher={Oxford University Press (OUP)},
   author={Schawinski, Kevin and Urry, C. Megan and Simmons, Brooke D. and Fortson, Lucy and Kaviraj, Sugata and Keel, William C. and Lintott, Chris J. and Masters, Karen L. and Nichol, Robert C. and Sarzi, Marc and Skibba, Ramin and Treister, Ezequiel and Willett, Kyle W. and Wong, O. Ivy and Yi, Sukyoung K.},
   year={2014},
   month=mar, pages={889–907} }

@ARTICLE{Johnson2021,
       author = {{Johnson}, Benjamin D. and {Leja}, Joel and {Conroy}, Charlie and {Speagle}, Joshua S.},
        title = "{Stellar Population Inference with Prospector}",
      journal = {\apjs},
         year = 2021,
        month = jun,
       volume = {254},
       number = {2},
          eid = {22},
        pages = {22},
          doi = {10.3847/1538-4365/abef67},
archivePrefix = {arXiv},
       eprint = {2012.01426},
 primaryClass = {astro-ph.GA},
       adsurl = {https://ui.adsabs.harvard.edu/abs/2021ApJS..254...22J}
}

@ARTICLE{Tremonti2004,
       author = {{Tremonti}, Christy A. and {Heckman}, Timothy M. and {Kauffmann}, Guinevere and {Brinchmann}, Jarle and {Charlot}, St{\'e}phane and {White}, Simon D.~M. and {Seibert}, Mark and {Peng}, Eric W. and {Schlegel}, David J. and {Uomoto}, Alan and {Fukugita}, Masataka and {Brinkmann}, Jon},
        title = "{The Origin of the Mass-Metallicity Relation: Insights from 53,000 Star-forming Galaxies in the Sloan Digital Sky Survey}",
      journal = {\apj},
         year = 2004,
        month = oct,
       volume = {613},
       number = {2},
        pages = {898-913},
          doi = {10.1086/423264},
archivePrefix = {arXiv},
       eprint = {astro-ph/0405537},
 primaryClass = {astro-ph},
       adsurl = {https://ui.adsabs.harvard.edu/abs/2004ApJ...613..898T}
}

@ARTICLE{Kauffmann2003,
       author = {{Kauffmann}, Guinevere and {Heckman}, Timothy M. and {White}, Simon D.~M. and {Charlot}, St{\'e}phane and {Tremonti}, Christy and {Brinchmann}, Jarle and {Bruzual}, Gustavo and {Peng}, Eric W. and {Seibert}, Mark and {Bernardi}, Mariangela and {Blanton}, Michael and {Brinkmann}, Jon and {Castander}, Francisco and {Cs{\'a}bai}, Istvan and {Fukugita}, Masataka and {Ivezic}, Zeljko and {Munn}, Jeffrey A. and {Nichol}, Robert C. and {Padmanabhan}, Nikhil and {Thakar}, Aniruddha R. and {Weinberg}, David H. and {York}, Donald},
        title = "{Stellar masses and star formation histories for {}10$^{5}$ galaxies from the Sloan Digital Sky Survey}",
      journal = {\mnras},
         year = 2003,
        month = may,
       volume = {341},
       number = {1},
        pages = {33-53},
          doi = {10.1046/j.1365-8711.2003.06291.x},
archivePrefix = {arXiv},
       eprint = {astro-ph/0204055},
 primaryClass = {astro-ph},
       adsurl = {https://ui.adsabs.harvard.edu/abs/2003MNRAS.341...33K}
}

@ARTICLE{Yao2023,
       author = {{Yao}, Yuhan and {Ravi}, Vikram and {Gezari}, Suvi and {van Velzen}, Sjoert and {Lu}, Wenbin and {Schulze}, Steve and {Somalwar}, Jean J. and {Kulkarni}, S.~R. and {Hammerstein}, Erica and {Nicholl}, Matt and {Graham}, Matthew J. and {Perley}, Daniel A. and {Cenko}, S. Bradley and {Stein}, Robert and {Ricarte}, Angelo and {Chadayammuri}, Urmila and {Quataert}, Eliot and {Bellm}, Eric C. and {Bloom}, Joshua S. and {Dekany}, Richard and {Drake}, Andrew J. and {Groom}, Steven L. and {Mahabal}, Ashish A. and {Prince}, Thomas A. and {Riddle}, Reed and {Rusholme}, Ben and {Sharma}, Yashvi and {Sollerman}, Jesper and {Yan}, Lin},
        title = "{Tidal Disruption Event Demographics with the Zwicky Transient Facility: Volumetric Rates, Luminosity Function, and Implications for the Local Black Hole Mass Function}",
      journal = {\apjl},
         year = 2023,
        month = sep,
       volume = {955},
       number = {1},
          eid = {L6},
        pages = {L6},
          doi = {10.3847/2041-8213/acf216},
archivePrefix = {arXiv},
       eprint = {2303.06523},
 primaryClass = {astro-ph.HE},
       adsurl = {https://ui.adsabs.harvard.edu/abs/2023ApJ...955L...6Y}
}

@ARTICLE{Antonucci1993,
       author = {{Antonucci}, Robert},
        title = "{Unified models for active galactic nuclei and quasars.}",
      journal = {\araa},
         year = 1993,
        month = jan,
       volume = {31},
        pages = {473-521},
          doi = {10.1146/annurev.aa.31.090193.002353},
       adsurl = {https://ui.adsabs.harvard.edu/abs/1993ARA&A..31..473A}
}

@ARTICLE{Ho2025,
       author = {{Ho}, Anna Y.~Q. and {Yao}, Yuhan and {Matsumoto}, Tatsuya and {Schroeder}, Genevieve and {Coughlin}, Eric R. and {Perley}, Daniel A. and {Andreoni}, Igor and {Bellm}, Eric C. and {Chen}, Tracy X. and {Chornock}, Ryan and {Covarrubias}, Sofia and {Das}, Kaustav and {Fremling}, Christoffer and {Gilfanov}, Marat and {Hinds}, K.~R. and {Jarvis}, Dan and {Kasliwal}, Mansi M. and {Liu}, Chang and {Lyman}, Joseph D. and {Masci}, Frank J. and {Prince}, Thomas A. and {Ravi}, Vikram and {Rich}, R. Michael and {Riddle}, Reed and {Sevilla}, Cassie and {Smith}, Roger and {Sollerman}, Jesper and {Somalwar}, Jean J. and {Srinivasaragavan}, Gokul P. and {Sunyaev}, Rashid and {Vail}, Jada L. and {Wise}, Jacob L. and {Yun}, Sol Bin},
        title = "{A Luminous Red Optical Flare and Hard X-Ray Emission in the Tidal Disruption Event AT 2024kmq}",
      journal = {\apj},
         year = 2025,
        month = aug,
       volume = {989},
       number = {1},
          eid = {54},
        pages = {54},
          doi = {10.3847/1538-4357/ade8f2},
archivePrefix = {arXiv},
       eprint = {2502.07885},
 primaryClass = {astro-ph.HE},
       adsurl = {https://ui.adsabs.harvard.edu/abs/2025ApJ...989...54H}
}

@ARTICLE{Newsome2024,
       author = {{Newsome}, Megan and {Arcavi}, Iair and {Howell}, D. Andrew and {McCully}, Curtis and {Terreran}, Giacomo and {Hosseinzadeh}, Griffin and {Bostroem}, K. Azalee and {Dgany}, Yael and {Farah}, Joseph and {Faris}, Sara and {Padilla-Gonzalez}, Estefania and {Pellegrino}, Craig and {Andrews}, Moira},
        title = "{Mapping the Inner 0.1 pc of a Supermassive Black Hole Environment with the Tidal Disruption Event and Extreme Coronal-line Emitter AT 2022upj}",
      journal = {\apj},
         year = 2024,
        month = dec,
       volume = {977},
       number = {2},
          eid = {258},
        pages = {258},
          doi = {10.3847/1538-4357/ad8a69},
archivePrefix = {arXiv},
       eprint = {2406.11972},
 primaryClass = {astro-ph.HE},
       adsurl = {https://ui.adsabs.harvard.edu/abs/2024ApJ...977..258N}
}

@ARTICLE{Short2023,
       author = {{Short}, P. and {Lawrence}, A. and {Nicholl}, M. and {Ward}, M. and {Reynolds}, T.~M. and {Mattila}, S. and {Yin}, C. and {Arcavi}, I. and {Carnall}, A. and {Charalampopoulos}, P. and {Gromadzki}, M. and {Jonker}, P.~G. and {Kim}, S. and {Leloudas}, G. and {Mandel}, I. and {Onori}, F. and {Pursiainen}, M. and {Schulze}, S. and {Villforth}, C. and {Wevers}, T.},
        title = "{Delayed appearance and evolution of coronal lines in the TDE AT2019qiz}",
      journal = {\mnras},
         year = 2023,
        month = oct,
       volume = {525},
       number = {1},
        pages = {1568-1587},
          doi = {10.1093/mnras/stad2270},
archivePrefix = {arXiv},
       eprint = {2307.13674},
 primaryClass = {astro-ph.HE},
       adsurl = {https://ui.adsabs.harvard.edu/abs/2023MNRAS.525.1568S}
}

@ARTICLE{Hinkle2024_CrLTDE,
       author = {{Hinkle}, Jason T. and {Shappee}, Benjamin J. and {Holoien}, Thomas W.-S.},
        title = "{Coronal line emitters are tidal disruption events in gas-rich environments}",
      journal = {\mnras},
         year = 2024,
        month = mar,
       volume = {528},
       number = {3},
        pages = {4775-4784},
          doi = {10.1093/mnras/stae022},
archivePrefix = {arXiv},
       eprint = {2303.05525},
 primaryClass = {astro-ph.HE},
       adsurl = {https://ui.adsabs.harvard.edu/abs/2024MNRAS.528.4775H}
}

@ARTICLE{Callow2024,
       author = {{Callow}, J. and {Graur}, O. and {Clark}, P. and {Palmese}, A. and {Aguilar}, J. and {Ahlen}, S. and {BenZvi}, S. and {Brooks}, D. and {Claybaugh}, T. and {de la Macorra}, A. and {Doel}, P. and {Forero-Romero}, J.~E. and {Gazta{\~n}aga}, E. and {Gontcho A Gontcho}, S. and {Lambert}, A. and {Landriau}, M. and {Manera}, M. and {Meisner}, A. and {Miquel}, R. and {Moustakas}, J. and {Nie}, J. and {Poppett}, C. and {Prada}, F. and {Rezaie}, M. and {Rossi}, G. and {Sanchez}, E. and {Silber}, J. and {Tarl{\'e}}, G. and {Weaver}, B.~A. and {Zhou}, Z.},
        title = "{The rate of extreme coronal line emitting galaxies in the Sloan Digital Sky Survey and their relation to tidal disruption events}",
      journal = {\mnras},
         year = 2024,
        month = nov,
       volume = {535},
       number = {1},
        pages = {1095-1122},
          doi = {10.1093/mnras/stae2384},
archivePrefix = {arXiv},
       eprint = {2402.16951},
 primaryClass = {astro-ph.HE},
       adsurl = {https://ui.adsabs.harvard.edu/abs/2024MNRAS.535.1095C}
}

@ARTICLE{Wang2012,
       author = {{Wang}, Ting-Gui and {Zhou}, Hong-Yan and {Komossa}, S. and {Wang}, Hui-Yuan and {Yuan}, Weimin and {Yang}, Chenwei},
        title = "{Extreme Coronal Line Emitters: Tidal Disruption of Stars by Massive Black Holes in Galactic Nuclei?}",
      journal = {\apj},
         year = 2012,
        month = apr,
       volume = {749},
       number = {2},
          eid = {115},
        pages = {115},
          doi = {10.1088/0004-637X/749/2/115},
archivePrefix = {arXiv},
       eprint = {1202.1064},
 primaryClass = {astro-ph.HE},
       adsurl = {https://ui.adsabs.harvard.edu/abs/2012ApJ...749..115W}
}

@ARTICLE{Clark2025,
       author = {{Clark}, Peter and {Callow}, Joseph and {Graur}, Or and {Greenwell}, Claire and {Hu}, Lei and {Aguilar}, Jessica and {Ahlen}, Steven and {Bianchi}, Davide and {Brooks}, David and {Claybaugh}, Todd and {Dawson}, Kyle and {de la Macorra}, Axel and {Doel}, Peter and {Gontcho A Gontcho}, Satya and {Gutierrez}, Gaston and {Honscheid}, Klaus and {Juneau}, Stephanie and {Kehoe}, Robert and {Kisner}, Theodore and {Kremin}, Anthony and {Landriau}, Martin and {Le Guillou}, Laurent and {Meisner}, Aaron and {Miquel}, Ramon and {Moustakas}, John and {P{\'e}rez-R{\`a}fols}, Ignasi and {Sanchez}, Eusebio and {Schubnell}, Michael and {Sprayberry}, David and {Tarl{\'e}}, Gregory and {Weaver}, Benjamin A. and {Zou}, Hu},
        title = "{AT 2018dyk: tidal disruption event or active galactic nucleus? Follow-up observations of an extreme coronal line emitter with the Dark Energy Spectroscopic Instrument}",
      journal = {\mnras},
         year = 2025,
        month = jun,
       volume = {540},
       number = {1},
        pages = {871-906},
          doi = {10.1093/mnras/staf724},
archivePrefix = {arXiv},
       eprint = {2502.04080},
 primaryClass = {astro-ph.HE},
       adsurl = {https://ui.adsabs.harvard.edu/abs/2025MNRAS.540..871C}
}

@ARTICLE{Wang2011,
       author = {{Wang}, Ting-Gui and {Zhou}, Hong-Yan and {Wang}, Li-Fan and {Lu}, Hong-Lin and {Xu}, Dawei},
        title = "{Transient Superstrong Coronal Lines and Broad Bumps in the Galaxy SDSS J074820.67+471214.3}",
      journal = {\apj},
         year = 2011,
        month = oct,
       volume = {740},
       number = {2},
          eid = {85},
        pages = {85},
          doi = {10.1088/0004-637X/740/2/85},
archivePrefix = {arXiv},
       eprint = {1108.2790},
 primaryClass = {astro-ph.CO},
       adsurl = {https://ui.adsabs.harvard.edu/abs/2011ApJ...740...85W}
}

@ARTICLE{Komossa2008,
       author = {{Komossa}, S. and {Zhou}, H. and {Wang}, T. and {Ajello}, M. and {Ge}, J. and {Greiner}, J. and {Lu}, H. and {Salvato}, M. and {Saxton}, R. and {Shan}, H. and {Xu}, D. and {Yuan}, W.},
        title = "{Discovery of Superstrong, Fading, Iron Line Emission and Double-peaked Balmer Lines of the Galaxy SDSS J095209.56+214313.3: The Light Echo of a Huge Flare}",
      journal = {\apjl},
         year = 2008,
        month = may,
       volume = {678},
       number = {1},
        pages = {L13},
          doi = {10.1086/588281},
archivePrefix = {arXiv},
       eprint = {0804.2670},
 primaryClass = {astro-ph},
       adsurl = {https://ui.adsabs.harvard.edu/abs/2008ApJ...678L..13K}
}

@ARTICLE{Cao2024_20ocn,
       author = {{Cao}, Z. and {Jonker}, P.~G. and {Pasham}, D.~R. and {Wen}, S. and {Stone}, N.~C. and {Zabludoff}, A.~I.},
        title = "{Tidal Disruption Event AT2020ocn: Early Time X-Ray Flares Caused by a Possible Disk Alignment Process}",
      journal = {\apj},
         year = 2024,
        month = jul,
       volume = {970},
       number = {1},
          eid = {89},
        pages = {89},
          doi = {10.3847/1538-4357/ad496f},
archivePrefix = {arXiv},
       eprint = {2405.07642},
 primaryClass = {astro-ph.HE},
       adsurl = {https://ui.adsabs.harvard.edu/abs/2024ApJ...970...89C}
}

@ARTICLE{Sazonov2021,
       author = {{Sazonov}, S. and {Gilfanov}, M. and {Medvedev}, P. and {Yao}, Y. and {Khorunzhev}, G. and {Semena}, A. and {Sunyaev}, R. and {Burenin}, R. and {Lyapin}, A. and {Meshcheryakov}, A. and {Uskov}, G. and {Zaznobin}, I. and {Postnov}, K.~A. and {Dodin}, A.~V. and {Belinski}, A.~A. and {Cherepashchuk}, A.~M. and {Eselevich}, M. and {Dodonov}, S.~N. and {Grokhovskaya}, A.~A. and {Kotov}, S.~S. and {Bikmaev}, I.~F. and {Zhuchkov}, R. Ya and {Gumerov}, R.~I. and {van Velzen}, S. and {Kulkarni}, S.},
        title = "{First tidal disruption events discovered by SRG/eROSITA: X-ray/optical properties and X-ray luminosity function at z < 0.6}",
      journal = {\mnras},
         year = 2021,
        month = dec,
       volume = {508},
       number = {3},
        pages = {3820-3847},
          doi = {10.1093/mnras/stab2843},
archivePrefix = {arXiv},
       eprint = {2108.02449},
 primaryClass = {astro-ph.HE},
       adsurl = {https://ui.adsabs.harvard.edu/abs/2021MNRAS.508.3820S}
}

@ARTICLE{Alexander2020,
       author = {{Alexander}, Kate D. and {van Velzen}, Sjoert and {Horesh}, Assaf and {Zauderer}, B. Ashley},
        title = "{Radio Properties of Tidal Disruption Events}",
      journal = {\ssr},
         year = 2020,
        month = jun,
       volume = {216},
       number = {5},
          eid = {81},
        pages = {81},
          doi = {10.1007/s11214-020-00702-w},
archivePrefix = {arXiv},
       eprint = {2006.01159},
 primaryClass = {astro-ph.HE},
       adsurl = {https://ui.adsabs.harvard.edu/abs/2020SSRv..216...81A}
}

@ARTICLE{Predehl2021,
       author = {{Predehl}, P. and {Andritschke}, R. and {Arefiev}, V. and {Babyshkin}, V. and {Batanov}, O. and {Becker}, W. and {B{\"o}hringer}, H. and {Bogomolov}, A. and {Boller}, T. and {Borm}, K. and {Bornemann}, W. and {Br{\"a}uninger}, H. and {Br{\"u}ggen}, M. and {Brunner}, H. and {Brusa}, M. and {Bulbul}, E. and {Buntov}, M. and {Burwitz}, V. and {Burkert}, W. and {Clerc}, N. and {Churazov}, E. and {Coutinho}, D. and {Dauser}, T. and {Dennerl}, K. and {Doroshenko}, V. and {Eder}, J. and {Emberger}, V. and {Eraerds}, T. and {Finoguenov}, A. and {Freyberg}, M. and {Friedrich}, P. and {Friedrich}, S. and {F{\"u}rmetz}, M. and {Georgakakis}, A. and {Gilfanov}, M. and {Granato}, S. and {Grossberger}, C. and {Gueguen}, A. and {Gureev}, P. and {Haberl}, F. and {H{\"a}lker}, O. and {Hartner}, G. and {Hasinger}, G. and {Huber}, H. and {Ji}, L. and {Kienlin}, A. v. and {Kink}, W. and {Korotkov}, F. and {Kreykenbohm}, I. and {Lamer}, G. and {Lomakin}, I. and {Lapshov}, I. and {Liu}, T. and {Maitra}, C. and {Meidinger}, N. and {Menz}, B. and {Merloni}, A. and {Mernik}, T. and {Mican}, B. and {Mohr}, J. and {M{\"u}ller}, S. and {Nandra}, K. and {Nazarov}, V. and {Pacaud}, F. and {Pavlinsky}, M. and {Perinati}, E. and {Pfeffermann}, E. and {Pietschner}, D. and {Ramos-Ceja}, M.~E. and {Rau}, A. and {Reiffers}, J. and {Reiprich}, T.~H. and {Robrade}, J. and {Salvato}, M. and {Sanders}, J. and {Santangelo}, A. and {Sasaki}, M. and {Scheuerle}, H. and {Schmid}, C. and {Schmitt}, J. and {Schwope}, A. and {Shirshakov}, A. and {Steinmetz}, M. and {Stewart}, I. and {Str{\"u}der}, L. and {Sunyaev}, R. and {Tenzer}, C. and {Tiedemann}, L. and {Tr{\"u}mper}, J. and {Voron}, V. and {Weber}, P. and {Wilms}, J. and {Yaroshenko}, V.},
        title = "{The eROSITA X-ray telescope on SRG}",
      journal = {\aap},
         year = 2021,
        month = mar,
       volume = {647},
          eid = {A1},
        pages = {A1},
          doi = {10.1051/0004-6361/202039313},
archivePrefix = {arXiv},
       eprint = {2010.03477},
 primaryClass = {astro-ph.HE},
       adsurl = {https://ui.adsabs.harvard.edu/abs/2021A&A...647A...1P}
}

@ARTICLE{Sunyaev2021,
       author = {{Sunyaev}, R. and {Arefiev}, V. and {Babyshkin}, V. and {Bogomolov}, A. and {Borisov}, K. and {Buntov}, M. and {Brunner}, H. and {Burenin}, R. and {Churazov}, E. and {Coutinho}, D. and {Eder}, J. and {Eismont}, N. and {Freyberg}, M. and {Gilfanov}, M. and {Gureyev}, P. and {Hasinger}, G. and {Khabibullin}, I. and {Kolmykov}, V. and {Komovkin}, S. and {Krivonos}, R. and {Lapshov}, I. and {Levin}, V. and {Lomakin}, I. and {Lutovinov}, A. and {Medvedev}, P. and {Merloni}, A. and {Mernik}, T. and {Mikhailov}, E. and {Molodtsov}, V. and {Mzhelsky}, P. and {M{\"u}ller}, S. and {Nandra}, K. and {Nazarov}, V. and {Pavlinsky}, M. and {Poghodin}, A. and {Predehl}, P. and {Robrade}, J. and {Sazonov}, S. and {Scheuerle}, H. and {Shirshakov}, A. and {Tkachenko}, A. and {Voron}, V.},
        title = "{SRG X-ray orbital observatory. Its telescopes and first scientific results}",
      journal = {\aap},
         year = 2021,
        month = dec,
       volume = {656},
          eid = {A132},
        pages = {A132},
          doi = {10.1051/0004-6361/202141179},
archivePrefix = {arXiv},
       eprint = {2104.13267},
 primaryClass = {astro-ph.HE},
       adsurl = {https://ui.adsabs.harvard.edu/abs/2021A&A...656A.132S}
}

@ARTICLE{Komossa1999,
       author = {{Komossa}, Stefanie and {Greiner}, Jochen},
        title = "{Discovery of a giant and luminous X-ray outburst from the optically inactive galaxy pair RX J1242.6-1119}",
      journal = {\aap},
         year = 1999,
        month = sep,
       volume = {349},
        pages = {L45-L48},
          doi = {10.48550/arXiv.astro-ph/9908216},
archivePrefix = {arXiv},
       eprint = {astro-ph/9908216},
 primaryClass = {astro-ph},
       adsurl = {https://ui.adsabs.harvard.edu/abs/1999A&A...349L..45K}
}

@ARTICLE{Bade1996,
       author = {{Bade}, N. and {Komossa}, S. and {Dahlem}, M.},
        title = "{Detection of an extremely soft X-ray outburst in the HII-like nucleus of NGC 5905.}",
      journal = {\aap},
         year = 1996,
        month = may,
       volume = {309},
        pages = {L35-L38},
       adsurl = {https://ui.adsabs.harvard.edu/abs/1996A&A...309L..35B}
}

@ARTICLE{Stone2020,
       author = {{Stone}, N.~C. and {Vasiliev}, E. and {Kesden}, M. and {Rossi}, E.~M. and {Perets}, H.~B. and {Amaro-Seoane}, P.},
        title = "{Rates of Stellar Tidal Disruption}",
      journal = {\ssr},
         year = 2020,
        month = mar,
       volume = {216},
       number = {3},
          eid = {35},
        pages = {35},
          doi = {10.1007/s11214-020-00651-4},
archivePrefix = {arXiv},
       eprint = {2003.08953},
 primaryClass = {astro-ph.HE},
       adsurl = {https://ui.adsabs.harvard.edu/abs/2020SSRv..216...35S}
}

@ARTICLE{Saxton2020,
       author = {{Saxton}, R. and {Komossa}, S. and {Auchettl}, K. and {Jonker}, P.~G.},
        title = "{X-Ray Properties of TDEs}",
      journal = {\ssr},
         year = 2020,
        month = jul,
       volume = {216},
       number = {5},
          eid = {85},
        pages = {85},
          doi = {10.1007/s11214-020-00708-4},
       adsurl = {https://ui.adsabs.harvard.edu/abs/2020SSRv..216...85S}
}

@ARTICLE{2020ApJS..247...69E,
       author = {{Eisenhardt}, Peter R.~M. and {Marocco}, Federico and {Fowler}, John W. and {Meisner}, Aaron M. and {Kirkpatrick}, J. Davy and {Garcia}, Nelson and {Jarrett}, Thomas H. and {Koontz}, Renata and {Marchese}, Elijah J. and {Stanford}, S. Adam and {Caselden}, Dan and {Cushing}, Michael C. and {Cutri}, Roc M. and {Faherty}, Jacqueline K. and {Gelino}, Christopher R. and {Gonzalez}, Anthony H. and {Mainzer}, Amanda and {Mobasher}, Bahram and {Schlegel}, David J. and {Stern}, Daniel and {Teplitz}, Harry I. and {Wright}, Edward L.},
        title = "{The CatWISE Preliminary Catalog: Motions from WISE and NEOWISE Data}",
      journal = {\apjs},
         year = 2020,
        month = apr,
       volume = {247},
       number = {2},
          eid = {69},
        pages = {69},
          doi = {10.3847/1538-4365/ab7f2a},
archivePrefix = {arXiv},
       eprint = {1908.08902},
 primaryClass = {astro-ph.IM},
       adsurl = {https://ui.adsabs.harvard.edu/abs/2020ApJS..247...69E}
}

@ARTICLE{Guolo2024,
       author = {{Guolo}, Muryel and {Gezari}, Suvi and {Yao}, Yuhan and {van Velzen}, Sjoert and {Hammerstein}, Erica and {Cenko}, S. Bradley and {Tokayer}, Yarone M.},
        title = "{A Systematic Analysis of the X-Ray Emission in Optically Selected Tidal Disruption Events: Observational Evidence for the Unification of the Optically and X-Ray-selected Populations}",
      journal = {\apj},
         year = 2024,
        month = may,
       volume = {966},
       number = {2},
          eid = {160},
        pages = {160},
          doi = {10.3847/1538-4357/ad2f9f},
archivePrefix = {arXiv},
       eprint = {2308.13019},
 primaryClass = {astro-ph.HE},
       adsurl = {https://ui.adsabs.harvard.edu/abs/2024ApJ...966..160G}
}

@article{Yao_2022,
   title={The Tidal Disruption Event AT2021ehb: Evidence of Relativistic Disk Reflection, and Rapid Evolution of the Disk–Corona System},
   volume={937},
   ISSN={1538-4357},
   url={http://dx.doi.org/10.3847/1538-4357/ac898a},
   DOI={10.3847/1538-4357/ac898a},
   number={1},
   journal={The Astrophysical Journal},
   publisher={American Astronomical Society},
   author={Yao, Yuhan and Lu, Wenbin and Guolo, Muryel and Pasham, Dheeraj R. and Gezari, Suvi and Gilfanov, Marat and Gendreau, Keith C. and Harrison, Fiona and Cenko, S. Bradley and Kulkarni, S. R. and Miller, Jon M. and Walton, Dominic J. and García, Javier A. and Velzen, Sjoert van and Alexander, Kate D. and Miller-Jones, James C. A. and Nicholl, Matt and Hammerstein, Erica and Medvedev, Pavel and Stern, Daniel and Ravi, Vikram and Sunyaev, R. and Bloom, Joshua S. and Graham, Matthew J. and Kool, Erik C. and Mahabal, Ashish A. and Masci, Frank J. and Purdum, Josiah and Rusholme, Ben and Sharma, Yashvi and Smith, Roger and Sollerman, Jesper},
   year={2022},
   month=sep, pages={8} }

@ARTICLE{Hammerstein2023,
       author = {{Hammerstein}, Erica and {van Velzen}, Sjoert and {Gezari}, Suvi and {Cenko}, S. Bradley and {Yao}, Yuhan and {Ward}, Charlotte and {Frederick}, Sara and {Villanueva}, Natalia and {Somalwar}, Jean J. and {Graham}, Matthew J. and {Kulkarni}, Shrinivas R. and {Stern}, Daniel and {Andreoni}, Igor and {Bellm}, Eric C. and {Dekany}, Richard and {Dhawan}, Suhail and {Drake}, Andrew J. and {Fremling}, Christoffer and {Gatkine}, Pradip and {Groom}, Steven L. and {Ho}, Anna Y.~Q. and {Kasliwal}, Mansi M. and {Karambelkar}, Viraj and {Kool}, Erik C. and {Masci}, Frank J. and {Medford}, Michael S. and {Perley}, Daniel A. and {Purdum}, Josiah and {van Roestel}, Jan and {Sharma}, Yashvi and {Sollerman}, Jesper and {Taggart}, Kirsty and {Yan}, Lin},
        title = "{The Final Season Reimagined: 30 Tidal Disruption Events from the ZTF-I Survey}",
      journal = {\apj},
         year = 2023,
        month = jan,
       volume = {942},
       number = {1},
          eid = {9},
        pages = {9},
          doi = {10.3847/1538-4357/aca283},
archivePrefix = {arXiv},
       eprint = {2203.01461},
 primaryClass = {astro-ph.HE},
       adsurl = {https://ui.adsabs.harvard.edu/abs/2023ApJ...942....9H}
}

@ARTICLE{Cardelli1989,
       author = {{Cardelli}, Jason A. and {Clayton}, Geoffrey C. and {Mathis}, John S.},
        title = "{The Relationship between Infrared, Optical, and Ultraviolet Extinction}",
      journal = {\apj},
         year = 1989,
        month = oct,
       volume = {345},
        pages = {245},
          doi = {10.1086/167900},
       adsurl = {https://ui.adsabs.harvard.edu/abs/1989ApJ...345..245C}
}

@article{Roth_2016,
   title={THE X-RAY THROUGH OPTICAL FLUXES AND LINE STRENGTHS OF TIDAL DISRUPTION EVENTS},
   volume={827},
   ISSN={1538-4357},
   url={http://dx.doi.org/10.3847/0004-637X/827/1/3},
   DOI={10.3847/0004-637x/827/1/3},
   number={1},
   journal={The Astrophysical Journal},
   publisher={American Astronomical Society},
   author={Roth, Nathaniel and Kasen, Daniel and Guillochon, James and Ramirez-Ruiz, Enrico},
   year={2016},
   month=aug, pages={3} }

@ARTICLE{Roth2020,
       author = {{Roth}, Nathaniel and {Rossi}, Elena Maria and {Krolik}, Julian and {Piran}, Tsvi and {Mockler}, Brenna and {Kasen}, Daniel},
        title = "{Radiative Emission Mechanisms}",
      journal = {\ssr},
         year = 2020,
        month = oct,
       volume = {216},
       number = {7},
          eid = {114},
        pages = {114},
          doi = {10.1007/s11214-020-00735-1},
archivePrefix = {arXiv},
       eprint = {2008.01117},
 primaryClass = {astro-ph.HE},
       adsurl = {https://ui.adsabs.harvard.edu/abs/2020SSRv..216..114R}
}

@ARTICLE{vanVelzen2021,
       author = {{van Velzen}, Sjoert and {Gezari}, Suvi and {Hammerstein}, Erica and {Roth}, Nathaniel and {Frederick}, Sara and {Ward}, Charlotte and {Hung}, Tiara and {Cenko}, S. Bradley and {Stein}, Robert and {Perley}, Daniel A. and {Taggart}, Kirsty and {Foley}, Ryan J. and {Sollerman}, Jesper and {Blagorodnova}, Nadejda and {Andreoni}, Igor and {Bellm}, Eric C. and {Brinnel}, Valery and {De}, Kishalay and {Dekany}, Richard and {Feeney}, Michael and {Fremling}, Christoffer and {Giomi}, Matteo and {Golkhou}, V. Zach and {Graham}, Matthew J. and {Ho}, Anna. Y.~Q. and {Kasliwal}, Mansi M. and {Kilpatrick}, Charles D. and {Kulkarni}, Shrinivas R. and {Kupfer}, Thomas and {Laher}, Russ R. and {Mahabal}, Ashish and {Masci}, Frank J. and {Miller}, Adam A. and {Nordin}, Jakob and {Riddle}, Reed and {Rusholme}, Ben and {van Santen}, Jakob and {Sharma}, Yashvi and {Shupe}, David L. and {Soumagnac}, Maayane T.},
        title = "{Seventeen Tidal Disruption Events from the First Half of ZTF Survey Observations: Entering a New Era of Population Studies}",
      journal = {\apj},
         year = 2021,
        month = feb,
       volume = {908},
       number = {1},
          eid = {4},
        pages = {4},
          doi = {10.3847/1538-4357/abc258},
archivePrefix = {arXiv},
       eprint = {2001.01409},
 primaryClass = {astro-ph.HE},
       adsurl = {https://ui.adsabs.harvard.edu/abs/2021ApJ...908....4V}
}

@ARTICLE{Blanton_2005,
       author = {{Blanton}, Michael R. and {Schlegel}, David J. and {Strauss}, Michael A. and {Brinkmann}, J. and {Finkbeiner}, Douglas and {Fukugita}, Masataka and {Gunn}, James E. and {Hogg}, David W. and {Ivezi{\'c}}, {\v{Z}}eljko and {Knapp}, G.~R. and {Lupton}, Robert H. and {Munn}, Jeffrey A. and {Schneider}, Donald P. and {Tegmark}, Max and {Zehavi}, Idit},
        title = "{New York University Value-Added Galaxy Catalog: A Galaxy Catalog Based on New Public Surveys}",
      journal = {\aj},
         year = 2005,
        month = jun,
       volume = {129},
       number = {6},
        pages = {2562-2578},
          doi = {10.1086/429803},
archivePrefix = {arXiv},
       eprint = {astro-ph/0410166},
 primaryClass = {astro-ph},
       adsurl = {https://ui.adsabs.harvard.edu/abs/2005AJ....129.2562B}
}

@article{Fasano&Franceschini_1987,
    author = {Fasano, G. and Franceschini, A.},
    title = {A multidimensional version of the Kolmogorov–Smirnov test},
    journal = {Monthly Notices of the Royal Astronomical Society},
    volume = {225},
    number = {1},
    pages = {155-170},
    year = {1987},
    month = {03},
    issn = {0035-8711},
    doi = {10.1093/mnras/225.1.155},
    url = {https://doi.org/10.1093/mnras/225.1.155},
    eprint = {https://academic.oup.com/mnras/article-pdf/225/1/155/18522274/mnras225-0155.pdf},
}

@ARTICLE{Brinchmann_2004,
       author = {{Brinchmann}, J. and {Charlot}, S. and {White}, S.~D.~M. and {Tremonti}, C. and {Kauffmann}, G. and {Heckman}, T. and {Brinkmann}, J.},
        title = "{The physical properties of star-forming galaxies in the low-redshift Universe}",
      journal = {\mnras},
         year = 2004,
        month = jul,
       volume = {351},
       number = {4},
        pages = {1151-1179},
          doi = {10.1111/j.1365-2966.2004.07881.x},
archivePrefix = {arXiv},
       eprint = {astro-ph/0311060},
 primaryClass = {astro-ph},
       adsurl = {https://ui.adsabs.harvard.edu/abs/2004MNRAS.351.1151B}
}

@ARTICLE{Mendel_2014,
       author = {{Mendel}, J. Trevor and {Simard}, Luc and {Palmer}, Michael and {Ellison}, Sara L. and {Patton}, David R.},
        title = "{A Catalog of Bulge, Disk, and Total Stellar Mass Estimates for the Sloan Digital Sky Survey}",
      journal = {\apjs},
         year = 2014,
        month = jan,
       volume = {210},
       number = {1},
          eid = {3},
        pages = {3},
          doi = {10.1088/0067-0049/210/1/3},
archivePrefix = {arXiv},
       eprint = {1310.8304},
 primaryClass = {astro-ph.CO},
       adsurl = {https://ui.adsabs.harvard.edu/abs/2014ApJS..210....3M}
}

@ARTICLE{2016MNRAS.463.3409V,
       author = {{Vazdekis}, A. and {Koleva}, M. and {Ricciardelli}, E. and {R{\"o}ck}, B. and {Falc{\'o}n-Barroso}, J.},
        title = "{UV-extended E-MILES stellar population models: young components in massive early-type galaxies}",
      journal = {\mnras},
         year = 2016,
        month = dec,
       volume = {463},
       number = {4},
        pages = {3409-3436},
          doi = {10.1093/mnras/stw2231},
archivePrefix = {arXiv},
       eprint = {1612.01187},
 primaryClass = {astro-ph.GA},
       adsurl = {https://ui.adsabs.harvard.edu/abs/2016MNRAS.463.3409V}
}

@ARTICLE{2002AJ....123..485S,
       author = {{Stoughton}, Chris and {Lupton}, Robert H. and {Bernardi}, Mariangela and {Blanton}, Michael R. and {Burles}, Scott and {Castander}, Francisco J. and {Connolly}, A.~J. and {Eisenstein}, Daniel J. and {Frieman}, Joshua A. and {Hennessy}, G.~S. and {Hindsley}, Robert B. and {Ivezi{\'c}}, {\v{Z}}eljko and {Kent}, Stephen and {Kunszt}, Peter Z. and {Lee}, Brian C. and {Meiksin}, Avery and {Munn}, Jeffrey A. and {Newberg}, Heidi Jo and {Nichol}, R.~C. and {Nicinski}, Tom and {Pier}, Jeffrey R. and {Richards}, Gordon T. and {Richmond}, Michael W. and {Schlegel}, David J. and {Smith}, J. Allyn and {Strauss}, Michael A. and {SubbaRao}, Mark and {Szalay}, Alexander S. and {Thakar}, Aniruddha R. and {Tucker}, Douglas L. and {Vanden Berk}, Daniel E. and {Yanny}, Brian and {Adelman}, Jennifer K. and {Anderson}, Jr., John E. and {Anderson}, Scott F. and {Annis}, James and {Bahcall}, Neta A. and {Bakken}, J.~A. and {Bartelmann}, Matthias and {Bastian}, Steven and {Bauer}, Amanda and {Berman}, Eileen and {B{\"o}hringer}, Hans and {Boroski}, William N. and {Bracker}, Steve and {Briegel}, Charlie and {Briggs}, John W. and {Brinkmann}, J. and {Brunner}, Robert and {Carey}, Larry and {Carr}, Michael A. and {Chen}, Bing and {Christian}, Damian and {Colestock}, Patrick L. and {Crocker}, J.~H. and {Csabai}, Istv{\'a}n and {Czarapata}, Paul C. and {Dalcanton}, Julianne and {Davidsen}, Arthur F. and {Davis}, John Eric and {Dehnen}, Walter and {Dodelson}, Scott and {Doi}, Mamoru and {Dombeck}, Tom and {Donahue}, Megan and {Ellman}, Nancy and {Elms}, Brian R. and {Evans}, Michael L. and {Eyer}, Laurent and {Fan}, Xiaohui and {Federwitz}, Glenn R. and {Friedman}, Scott and {Fukugita}, Masataka and {Gal}, Roy and {Gillespie}, Bruce and {Glazebrook}, Karl and {Gray}, Jim and {Grebel}, Eva K. and {Greenawalt}, Bruce and {Greene}, Gretchen and {Gunn}, James E. and {de Haas}, Ernst and {Haiman}, Zolt{\'a}n and {Haldeman}, Merle and {Hall}, Patrick B. and {Hamabe}, Masaru and {Hansen}, Brad and {Harris}, Frederick H. and {Harris}, Hugh and {Harvanek}, Michael and {Hawley}, Suzanne L. and {Hayes}, J.~J.~E. and {Heckman}, Timothy M. and {Helmi}, Amina and {Henden}, Arne and {Hogan}, Craig J. and {Hogg}, David W. and {Holmgren}, Donald J. and {Holtzman}, Jon and {Huang}, Chih-Hao and {Hull}, Charles and {Ichikawa}, Shin-Ichi and {Ichikawa}, Takashi and {Johnston}, David E. and {Kauffmann}, Guinevere and {Kim}, Rita S.~J. and {Kimball}, Tim and {Kinney}, E. and {Klaene}, Mark and {Kleinman}, S.~J. and {Klypin}, Anatoly and {Knapp}, G.~R. and {Korienek}, John and {Krolik}, Julian and {Kron}, Richard G. and {Krzesi{\'n}ski}, Jurek and {Lamb}, D.~Q. and {Leger}, R. French and {Limmongkol}, Siriluk and {Lindenmeyer}, Carl and {Long}, Daniel C. and {Loomis}, Craig and {Loveday}, Jon and {MacKinnon}, Bryan and {Mannery}, Edward J. and {Mantsch}, P.~M. and {Margon}, Bruce and {McGehee}, Peregrine and {McKay}, Timothy A. and {McLean}, Brian and {Menou}, Kristen and {Merelli}, Aronne and {Mo}, H.~J. and {Monet}, David G. and {Nakamura}, Osamu and {Narayanan}, Vijay K. and {Nash}, Thomas and {Neilsen}, Jr., Eric H. and {Newman}, Peter R. and {Nitta}, Atsuko and {Odenkirchen}, Michael and {Okada}, Norio and {Okamura}, Sadanori and {Ostriker}, Jeremiah P. and {Owen}, Russell and {Pauls}, A. George and {Peoples}, John and {Peterson}, R.~S. and {Petravick}, Donald and {Pope}, Adrian and {Pordes}, Ruth and {Postman}, Marc and {Prosapio}, Angela and {Quinn}, Thomas R. and {Rechenmacher}, Ron and {Rivetta}, Claudio H. and {Rix}, Hans-Walter and {Rockosi}, Constance M. and {Rosner}, Robert and {Ruthmansdorfer}, Kurt and {Sandford}, Dale and {Schneider}, Donald P. and {Scranton}, Ryan and {Sekiguchi}, Maki and {Sergey}, Gary and {Sheth}, Ravi and {Shimasaku}, Kazuhiro and {Smee}, Stephen and {Snedden}, Stephanie A. and {Stebbins}, Albert and {Stubbs}, Christopher and {Szapudi}, Istv{\'a}n and {Szkody}, Paula and {Szokoly}, Gyula P. and {Tabachnik}, Serge and {Tsvetanov}, Zlatan and {Uomoto}, Alan and {Vogeley}, Michael S. and {Voges}, Wolfgang and {Waddell}, Patrick and {Walterbos}, Ren{\'e} and {Wang}, Shu-i. and {Watanabe}, Masaru and {Weinberg}, David H. and {White}, Richard L. and {White}, Simon D.~M. and {Wilhite}, Brian and {Wolfe}, David and {Yasuda}, Naoki and {York}, Donald G. and {Zehavi}, Idit and {Zheng}, Wei},
        title = "{Sloan Digital Sky Survey: Early Data Release}",
      journal = {\aj},
         year = 2002,
        month = jan,
       volume = {123},
       number = {1},
        pages = {485-548},
          doi = {10.1086/324741},
       adsurl = {https://ui.adsabs.harvard.edu/abs/2002AJ....123..485S}
}

@ARTICLE{Dey2019,
       author = {{Dey}, Arjun and {Schlegel}, David J. and {Lang}, Dustin and {Blum}, Robert and {Burleigh}, Kaylan and {Fan}, Xiaohui and {Findlay}, Joseph R. and {Finkbeiner}, Doug and {Herrera}, David and {Juneau}, St{\'e}phanie and {Landriau}, Martin and {Levi}, Michael and {McGreer}, Ian and {Meisner}, Aaron and {Myers}, Adam D. and {Moustakas}, John and {Nugent}, Peter and {Patej}, Anna and {Schlafly}, Edward F. and {Walker}, Alistair R. and {Valdes}, Francisco and {Weaver}, Benjamin A. and {Y{\`e}che}, Christophe and {Zou}, Hu and {Zhou}, Xu and {Abareshi}, Behzad and {Abbott}, T.~M.~C. and {Abolfathi}, Bela and {Aguilera}, C. and {Alam}, Shadab and {Allen}, Lori and {Alvarez}, A. and {Annis}, James and {Ansarinejad}, Behzad and {Aubert}, Marie and {Beechert}, Jacqueline and {Bell}, Eric F. and {BenZvi}, Segev Y. and {Beutler}, Florian and {Bielby}, Richard M. and {Bolton}, Adam S. and {Brice{\~n}o}, C{\'e}sar and {Buckley-Geer}, Elizabeth J. and {Butler}, Karen and {Calamida}, Annalisa and {Carlberg}, Raymond G. and {Carter}, Paul and {Casas}, Ricard and {Castander}, Francisco J. and {Choi}, Yumi and {Comparat}, Johan and {Cukanovaite}, Elena and {Delubac}, Timoth{\'e}e and {DeVries}, Kaitlin and {Dey}, Sharmila and {Dhungana}, Govinda and {Dickinson}, Mark and {Ding}, Zhejie and {Donaldson}, John B. and {Duan}, Yutong and {Duckworth}, Christopher J. and {Eftekharzadeh}, Sarah and {Eisenstein}, Daniel J. and {Etourneau}, Thomas and {Fagrelius}, Parker A. and {Farihi}, Jay and {Fitzpatrick}, Mike and {Font-Ribera}, Andreu and {Fulmer}, Leah and {G{\"a}nsicke}, Boris T. and {Gaztanaga}, Enrique and {George}, Koshy and {Gerdes}, David W. and {Gontcho}, Satya Gontcho A. and {Gorgoni}, Claudio and {Green}, Gregory and {Guy}, Julien and {Harmer}, Diane and {Hernandez}, M. and {Honscheid}, Klaus and {Huang}, Lijuan Wendy and {James}, David J. and {Jannuzi}, Buell T. and {Jiang}, Linhua and {Joyce}, Richard and {Karcher}, Armin and {Karkar}, Sonia and {Kehoe}, Robert and {Kneib}, Jean-Paul and {Kueter-Young}, Andrea and {Lan}, Ting-Wen and {Lauer}, Tod R. and {Le Guillou}, Laurent and {Le Van Suu}, Auguste and {Lee}, Jae Hyeon and {Lesser}, Michael and {Perreault Levasseur}, Laurence and {Li}, Ting S. and {Mann}, Justin L. and {Marshall}, Robert and {Mart{\'\i}nez-V{\'a}zquez}, C.~E. and {Martini}, Paul and {du Mas des Bourboux}, H{\'e}lion and {McManus}, Sean and {Meier}, Tobias Gabriel and {M{\'e}nard}, Brice and {Metcalfe}, Nigel and {Mu{\~n}oz-Guti{\'e}rrez}, Andrea and {Najita}, Joan and {Napier}, Kevin and {Narayan}, Gautham and {Newman}, Jeffrey A. and {Nie}, Jundan and {Nord}, Brian and {Norman}, Dara J. and {Olsen}, Knut A.~G. and {Paat}, Anthony and {Palanque-Delabrouille}, Nathalie and {Peng}, Xiyan and {Poppett}, Claire L. and {Poremba}, Megan R. and {Prakash}, Abhishek and {Rabinowitz}, David and {Raichoor}, Anand and {Rezaie}, Mehdi and {Robertson}, A.~N. and {Roe}, Natalie A. and {Ross}, Ashley J. and {Ross}, Nicholas P. and {Rudnick}, Gregory and {Safonova}, Sasha and {Saha}, Abhijit and {S{\'a}nchez}, F. Javier and {Savary}, Elodie and {Schweiker}, Heidi and {Scott}, Adam and {Seo}, Hee-Jong and {Shan}, Huanyuan and {Silva}, David R. and {Slepian}, Zachary and {Soto}, Christian and {Sprayberry}, David and {Staten}, Ryan and {Stillman}, Coley M. and {Stupak}, Robert J. and {Summers}, David L. and {Sien Tie}, Suk and {Tirado}, H. and {Vargas-Maga{\~n}a}, Mariana and {Vivas}, A. Katherina and {Wechsler}, Risa H. and {Williams}, Doug and {Yang}, Jinyi and {Yang}, Qian and {Yapici}, Tolga and {Zaritsky}, Dennis and {Zenteno}, A. and {Zhang}, Kai and {Zhang}, Tianmeng and {Zhou}, Rongpu and {Zhou}, Zhimin},
        title = "{Overview of the DESI Legacy Imaging Surveys}",
      journal = {\aj},
         year = 2019,
        month = may,
       volume = {157},
       number = {5},
          eid = {168},
        pages = {168},
          doi = {10.3847/1538-3881/ab089d},
archivePrefix = {arXiv},
       eprint = {1804.08657},
 primaryClass = {astro-ph.IM},
       adsurl = {https://ui.adsabs.harvard.edu/abs/2019AJ....157..168D}
}

@ARTICLE{Ueda_2015,
       author = {{Ueda}, Y. and {Hashimoto}, Y. and {Ichikawa}, K. and {Ishino}, Y. and {Kniazev}, A.~Y. and {V{\"a}is{\"a}nen}, P. and {Ricci}, C. and {Berney}, S. and {Gandhi}, P. and {Koss}, M. and {Mushotzky}, R. and {Terashima}, Y. and {Trakhtenbrot}, B. and {Crenshaw}, M.},
        title = "{[O iii] {\ensuremath{\lambda}}5007 and X-Ray Properties of a Complete Sample of Hard X-Ray Selected AGNs in the Local Universe}",
      journal = {\apj},
         year = 2015,
        month = dec,
       volume = {815},
       number = {1},
          eid = {1},
        pages = {1},
          doi = {10.1088/0004-637X/815/1/1},
archivePrefix = {arXiv},
       eprint = {1510.03153},
 primaryClass = {astro-ph.HE},
       adsurl = {https://ui.adsabs.harvard.edu/abs/2015ApJ...815....1U}
}

@ARTICLE{2011MNRAS.413.1687C,
       author = {{Cid Fernandes}, R. and {Stasi{\'n}ska}, G. and {Mateus}, A. and {Vale Asari}, N.},
        title = "{A comprehensive classification of galaxies in the Sloan Digital Sky Survey: how to tell true from fake AGN?}",
      journal = {\mnras},
         year = 2011,
        month = may,
       volume = {413},
       number = {3},
        pages = {1687-1699},
          doi = {10.1111/j.1365-2966.2011.18244.x},
archivePrefix = {arXiv},
       eprint = {1012.4426},
 primaryClass = {astro-ph.CO},
       adsurl = {https://ui.adsabs.harvard.edu/abs/2011MNRAS.413.1687C}
}

@misc{pulatova2025,
      title={Optical Emission-Line Properties of eROSITA-selected SDSS-V Galaxies}, 
      author={Nadiia G. Pulatova and Evgenii Rubtsov and Igor V. Chilingarian and Hans-Walter Rix and Mariia Demianenko and Kirill A. Grishin and Ivan Yu. Katkov and Donald P. Schneider and Catarina Aydar and Johannes Buchner and Mara Salvato and Andrea Merloni and Anton M. Koekemoer and Roberto J. Assef and Claudio Ricci and Dominika Wylezalek and Damir Gasymov and William Nielsen Brandt and Castalia Alenka Negrete Peñaloza and Sean Morrison and Scott F. Anderson and Franz E. Bauer and Hector Javier Ibarra-Medel and Qiaoya Wu},
      year={2025},
      eprint={2507.21632},
      archivePrefix={arXiv},
      primaryClass={astro-ph.GA},
      url={https://arxiv.org/abs/2507.21632}, 
}

@ARTICLE{2006MNRAS.372..961K,
       author = {{Kewley}, Lisa J. and {Groves}, Brent and {Kauffmann}, Guinevere and {Heckman}, Tim},
        title = "{The host galaxies and classification of active galactic nuclei}",
      journal = {\mnras},
         year = 2006,
        month = nov,
       volume = {372},
       number = {3},
        pages = {961-976},
          doi = {10.1111/j.1365-2966.2006.10859.x},
archivePrefix = {arXiv},
       eprint = {astro-ph/0605681},
 primaryClass = {astro-ph},
       adsurl = {https://ui.adsabs.harvard.edu/abs/2006MNRAS.372..961K}
}

@ARTICLE{2003MNRAS.346.1055K,
       author = {{Kauffmann}, Guinevere and {Heckman}, Timothy M. and {Tremonti}, Christy and {Brinchmann}, Jarle and {Charlot}, St{\'e}phane and {White}, Simon D.~M. and {Ridgway}, Susan E. and {Brinkmann}, Jon and {Fukugita}, Masataka and {Hall}, Patrick B. and {Ivezi{\'c}}, {\v{Z}}eljko and {Richards}, Gordon T. and {Schneider}, Donald P.},
        title = "{The host galaxies of active galactic nuclei}",
      journal = {\mnras},
         year = 2003,
        month = dec,
       volume = {346},
       number = {4},
        pages = {1055-1077},
          doi = {10.1111/j.1365-2966.2003.07154.x},
archivePrefix = {arXiv},
       eprint = {astro-ph/0304239},
 primaryClass = {astro-ph},
       adsurl = {https://ui.adsabs.harvard.edu/abs/2003MNRAS.346.1055K}
}

@article{Stern_2012,
   title={Type 1 AGN at lowz- II. The relative strength of narrow lines and the nature of intermediate type AGN: Type 1 low-z AGN. II. NLR relative luminosity},
   volume={426},
   ISSN={0035-8711},
   url={http://dx.doi.org/10.1111/j.1365-2966.2012.21772.x},
   DOI={10.1111/j.1365-2966.2012.21772.x},
   number={4},
   journal={Monthly Notices of the Royal Astronomical Society},
   publisher={Oxford University Press (OUP)},
   author={Stern, Jonathan and Laor, Ari},
   year={2012},
   month=oct, pages={2703–2718} }

@ARTICLE{Beck2021,
       author = {{Beck}, R{\'o}bert and {Szapudi}, Istv{\'a}n and {Flewelling}, Heather and {Holmberg}, Conrad and {Magnier}, Eugene and {Chambers}, Kenneth C.},
        title = "{PS1-STRM: neural network source classification and photometric redshift catalogue for PS1 3{\ensuremath{\pi}} DR1}",
      journal = {\mnras},
         year = 2021,
        month = jan,
       volume = {500},
       number = {2},
        pages = {1633-1644},
          doi = {10.1093/mnras/staa2587},
archivePrefix = {arXiv},
       eprint = {1910.10167},
 primaryClass = {astro-ph.GA},
       adsurl = {https://ui.adsabs.harvard.edu/abs/2021MNRAS.500.1633B}
}

@ARTICLE{Sutherland1992,
       author = {{Sutherland}, Will and {Saunders}, Will},
        title = "{On the likelihood ratio for source identification.}",
      journal = {\mnras},
         year = 1992,
        month = dec,
       volume = {259},
        pages = {413-420},
          doi = {10.1093/mnras/259.3.413},
       adsurl = {https://ui.adsabs.harvard.edu/abs/1992MNRAS.259..413S}
}

@ARTICLE{Guolo2024_22d,
       author = {{Guolo}, Muryel and {Pasham}, Dheeraj R. and {Zaja{\v{c}}ek}, Michal and {Coughlin}, Eric R. and {Gezari}, Suvi and {Sukov{\'a}}, Petra and {Wevers}, Thomas and {Witzany}, Vojt{\v{e}}ch and {Tombesi}, Francesco and {van Velzen}, Sjoert and {Alexander}, Kate D. and {Yao}, Yuhan and {Arcodia}, Riccardo and {Karas}, Vladim{\'\i}r and {Miller-Jones}, James C.~A. and {Remillard}, Ronald and {Gendreau}, Keith and {Ferrara}, Elizabeth C.},
        title = "{X-ray eruptions every 22 days from the nucleus of a nearby galaxy}",
      journal = {Nature Astronomy},
         year = 2024,
        month = mar,
       volume = {8},
        pages = {347-358},
          doi = {10.1038/s41550-023-02178-4},
archivePrefix = {arXiv},
       eprint = {2309.03011},
 primaryClass = {astro-ph.HE},
       adsurl = {https://ui.adsabs.harvard.edu/abs/2024NatAs...8..347G}
}

@ARTICLE{Evans2023,
       author = {{Evans}, P.~A. and {Nixon}, C.~J. and {Campana}, S. and {Charalampopoulos}, P. and {Perley}, D.~A. and {Breeveld}, A.~A. and {Page}, K.~L. and {Oates}, S.~R. and {Eyles-Ferris}, R.~A.~J. and {Malesani}, D.~B. and {Izzo}, L. and {Goad}, M.~R. and {O'Brien}, P.~T. and {Osborne}, J.~P. and {Sbarufatti}, B.},
        title = "{Monthly quasi-periodic eruptions from repeated stellar disruption by a massive black hole}",
      journal = {Nature Astronomy},
         year = 2023,
        month = nov,
       volume = {7},
        pages = {1368-1375},
          doi = {10.1038/s41550-023-02073-y},
archivePrefix = {arXiv},
       eprint = {2309.02500},
 primaryClass = {astro-ph.HE},
       adsurl = {https://ui.adsabs.harvard.edu/abs/2023NatAs...7.1368E}
}

@article{Kass1995_BIC,
    author = "Kass, Robert E. and Raftery, Adrian E.",
    title = "{Bayes Factors}",
    doi = "10.1080/01621459.1995.10476572",
    journal = "J. Am. Statist. Assoc.",
    volume = "90",
    number = "430",
    pages = "773--795",
    year = "1995"
}

@ARTICLE{Lorah2019_BIC,
  title    = "Value of sample size for computation of the Bayesian information
              criterion ({BIC}) in multilevel modeling",
  author   = "Lorah, Julie and Womack, Andrew",
  journal  = "Behavior Research Methods",
  volume   =  51,
  number   =  1,
  pages    = "440--450",
  month    =  feb,
  year     =  2019
}

\begin{appendix}

\section{Supplementary Figures} \label{sec:sup_fig}

Figures~\ref{fig:tde_host_galaxies_part1} and \ref{fig:tde_host_galaxies_part2} show the LS color images of the host galaxies of the 70 TDE candidates. 

Figures~\ref{fig:opt_diff_lc}, \ref{fig:opt_diff_lc_2}, \ref{fig:opt_diff_lc_3}, and \ref{fig:opt_diff_lc_4} show the forced photometry light curves of the host galaxy nuclei of the 70 TDE candidates. Note that two optical spectra obtained in June 2025 (IDs 55, 58) are beyond the time range of the x-axis. For ID 47, since the optical flare was detected during ZTF reference image building, we perform baseline correction following \citet{Yao2019}. 

Figures~\ref{fig:sed_fits} and \ref{fig:sed_fits_2} show the \texttt{Prospector} SED fitting results of the host galaxies of our gold and silver samples.

\begin{figure*}[htbp]
    \centering
    \includegraphics[width=0.18\textwidth]{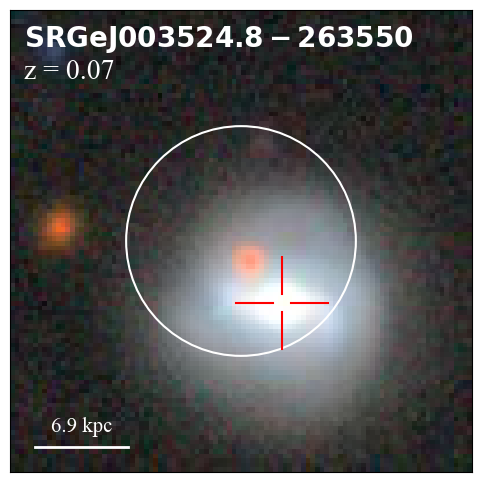}
    \includegraphics[width=0.18\textwidth]{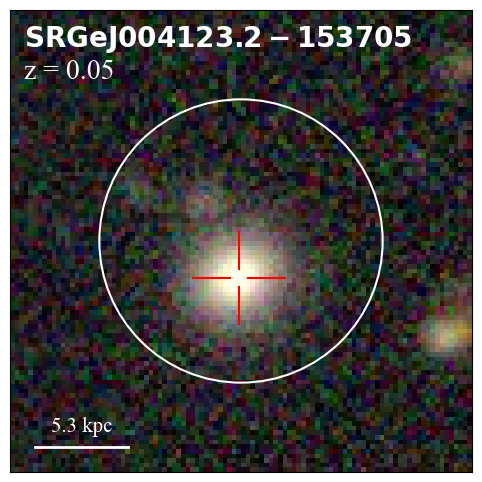}
    \includegraphics[width=0.18\textwidth]{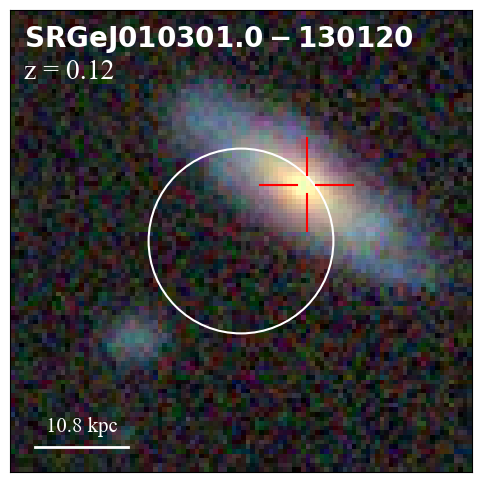}
    \includegraphics[width=0.18\textwidth]{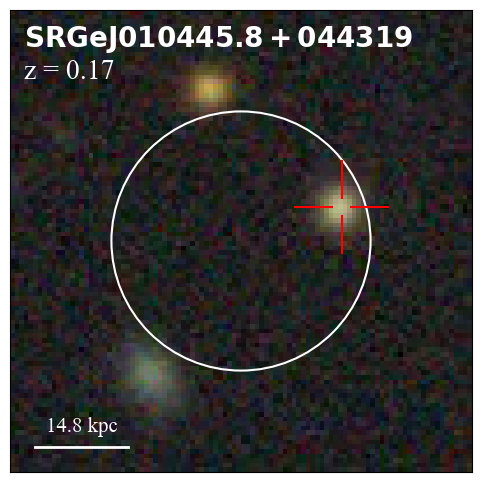}
    \includegraphics[width=0.18\textwidth]{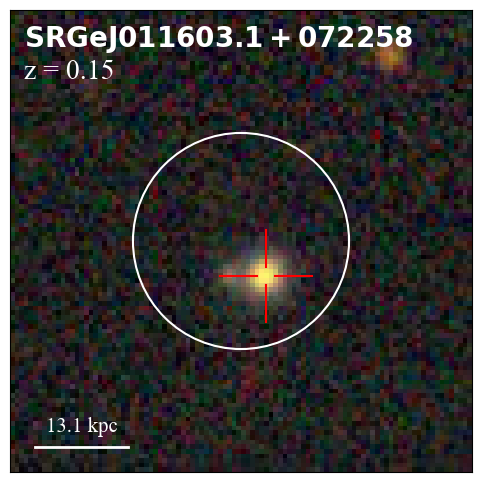}
    
    \includegraphics[width=0.18\textwidth]{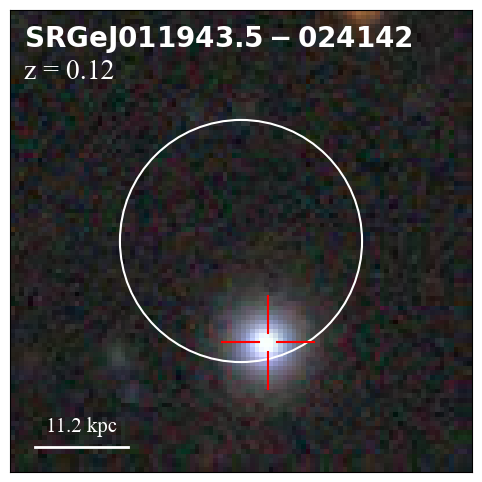}
    \includegraphics[width=0.18\textwidth]{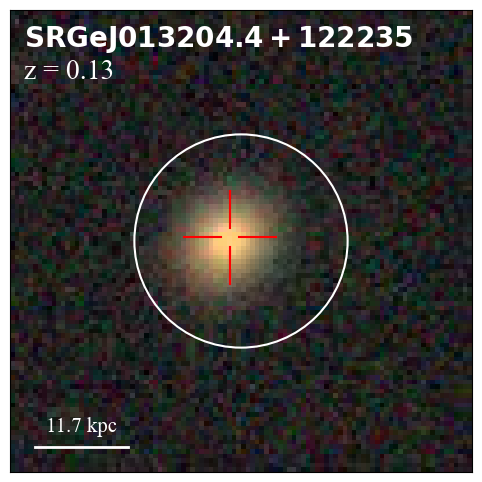}
    \includegraphics[width=0.18\textwidth]{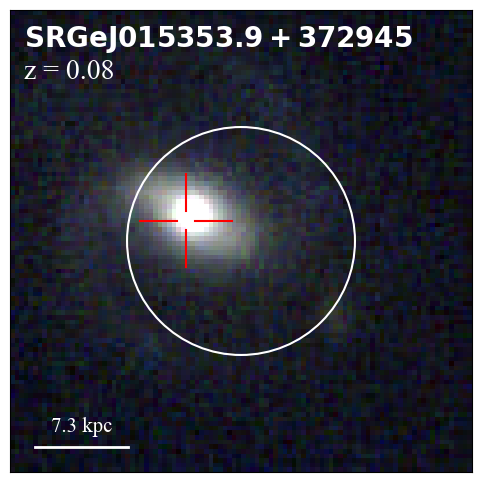}
    \includegraphics[width=0.18\textwidth]{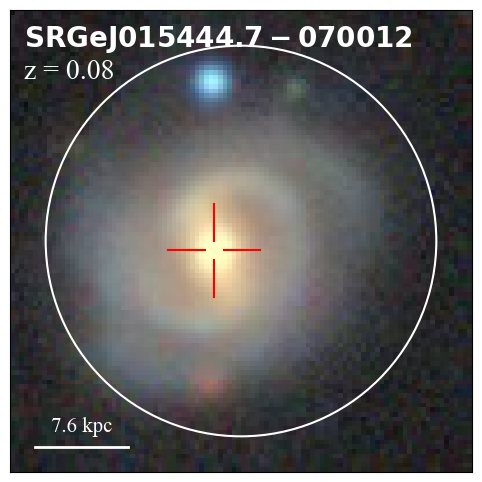}
    \includegraphics[width=0.18\textwidth]{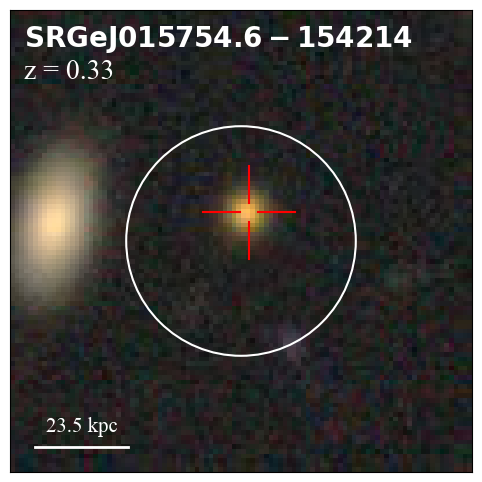}
    
    \includegraphics[width=0.18\textwidth]{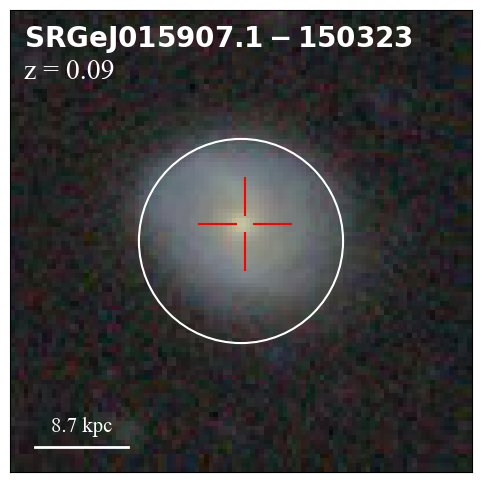}
    \includegraphics[width=0.18\textwidth]{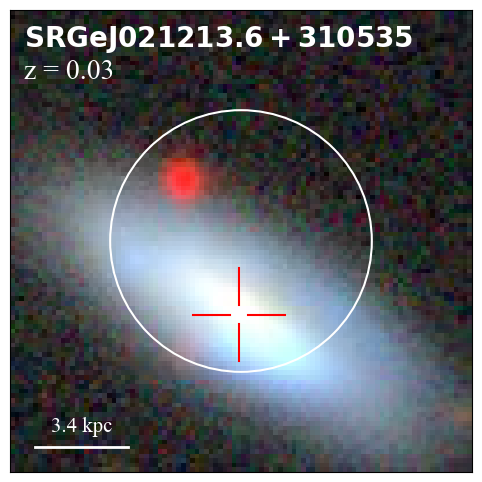}
    \includegraphics[width=0.18\textwidth]{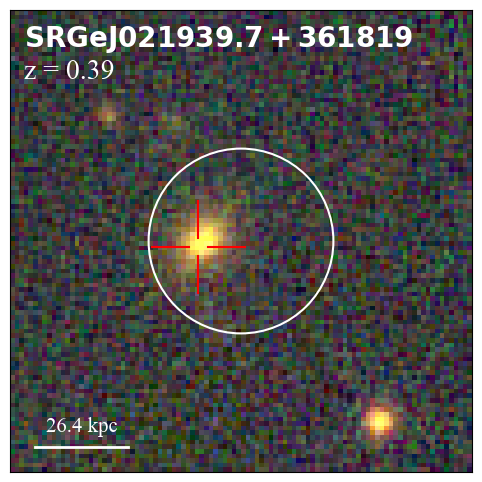}
    \includegraphics[width=0.18\textwidth]{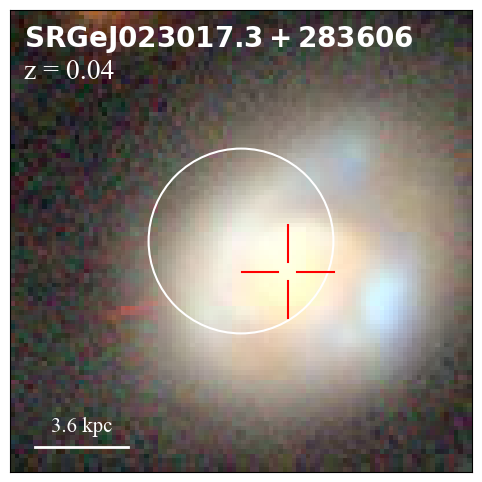}
    \includegraphics[width=0.18\textwidth]{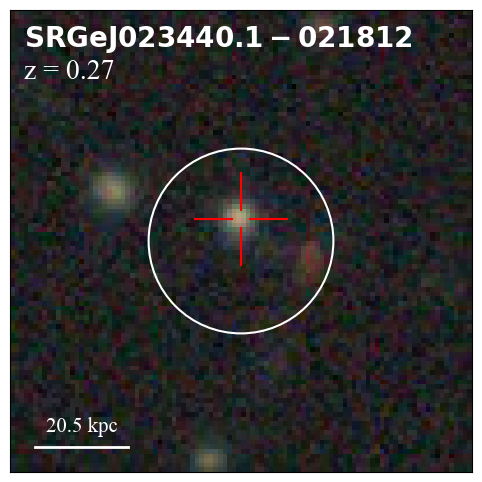}
    
    \includegraphics[width=0.18\textwidth]{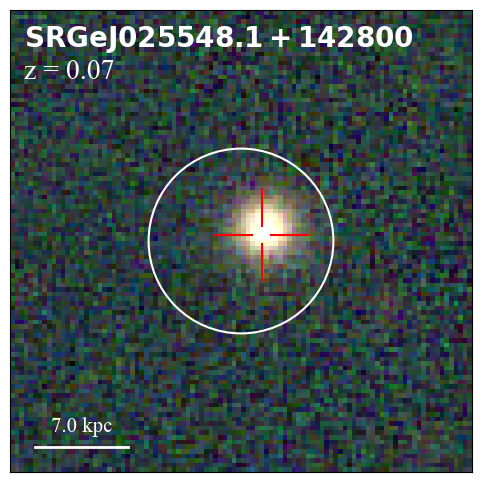}
    \includegraphics[width=0.18\textwidth]{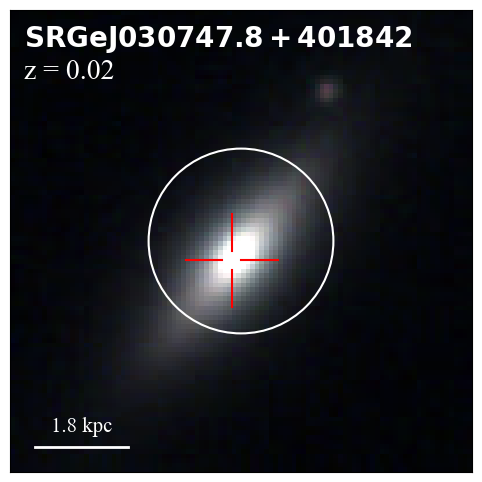}
    \includegraphics[width=0.18\textwidth]{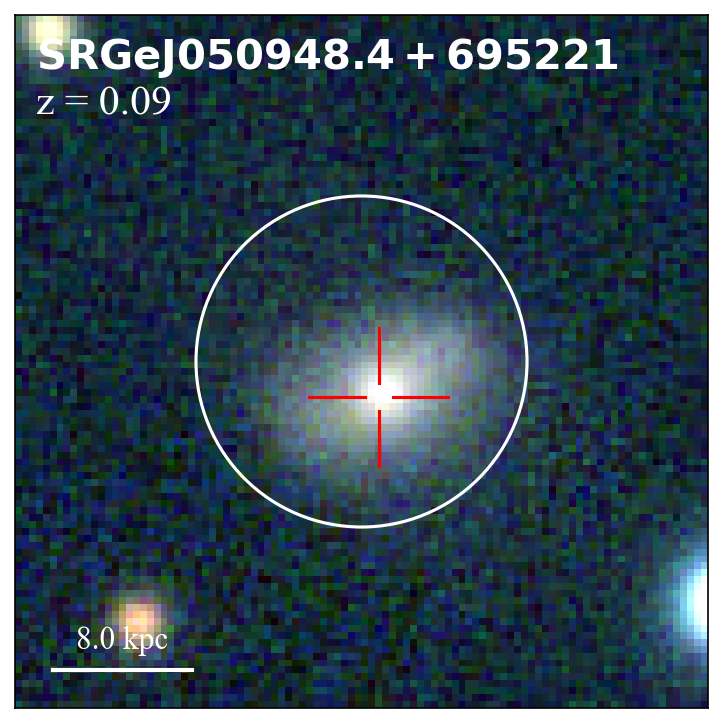}
    \includegraphics[width=0.18\textwidth]{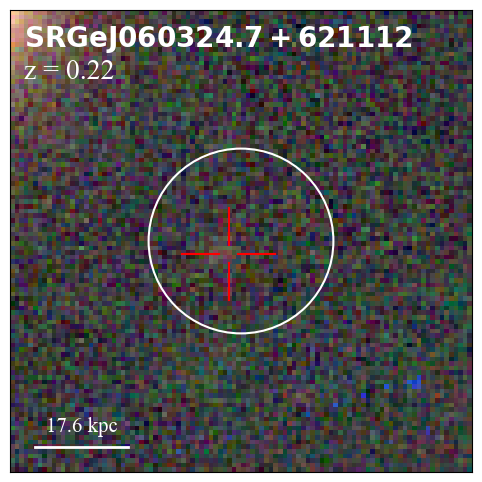}
    \includegraphics[width=0.18\textwidth]{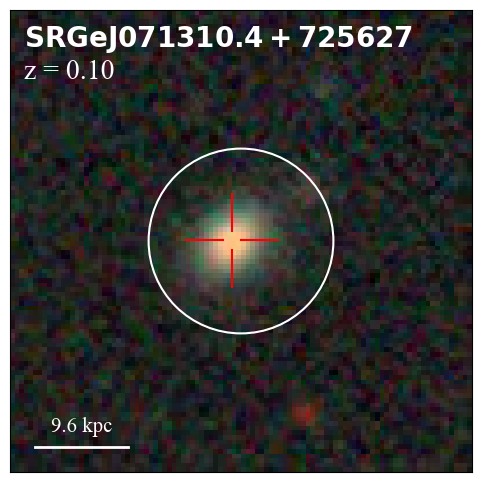}
    
    \includegraphics[width=0.18\textwidth]{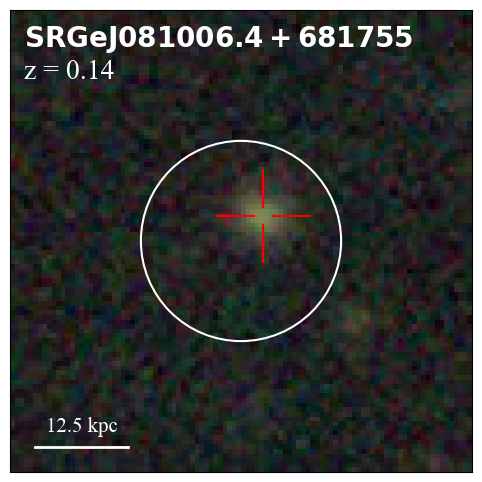}
    \includegraphics[width=0.18\textwidth]{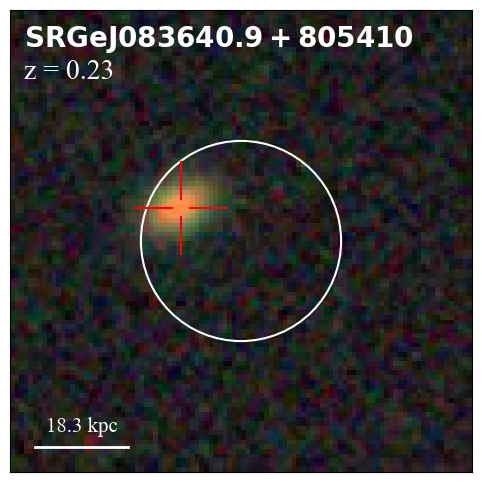}
    \includegraphics[width=0.18\textwidth]{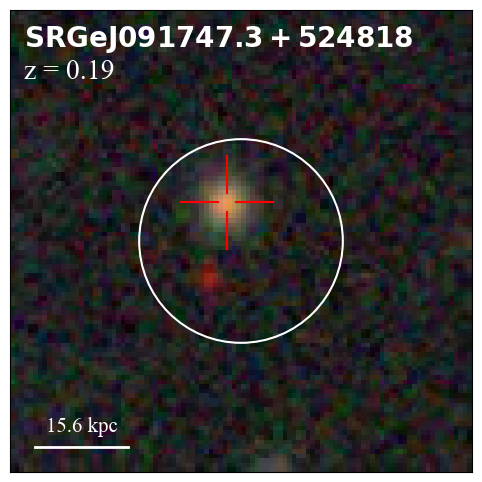}
    \includegraphics[width=0.18\textwidth]{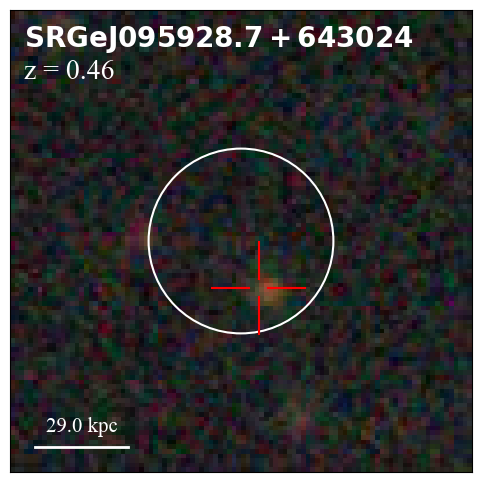}
    \includegraphics[width=0.18\textwidth]{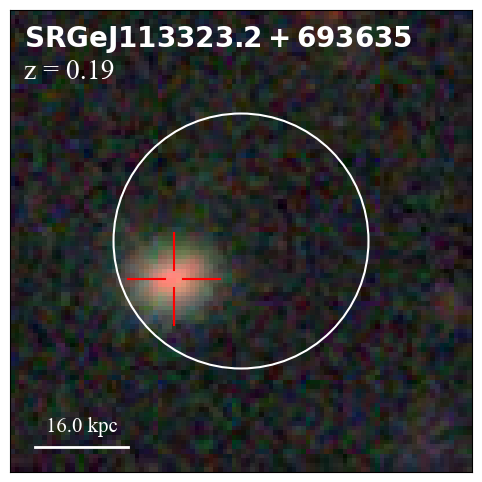}
    
    \includegraphics[width=0.18\textwidth]{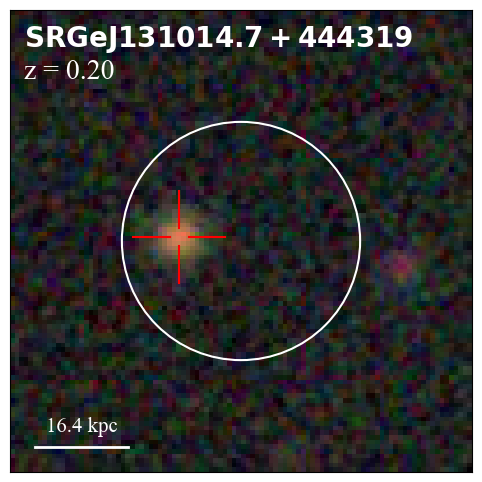}
    \includegraphics[width=0.18\textwidth]{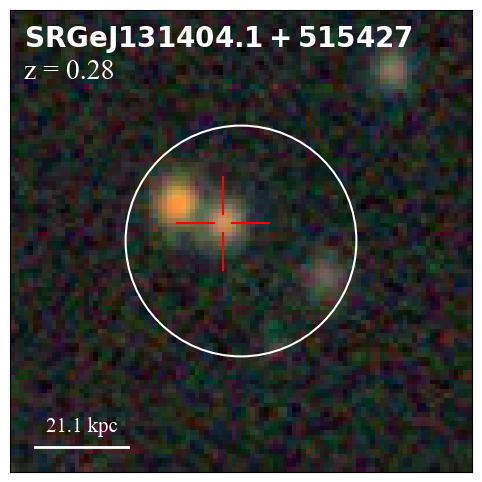}
    \includegraphics[width=0.18\textwidth]{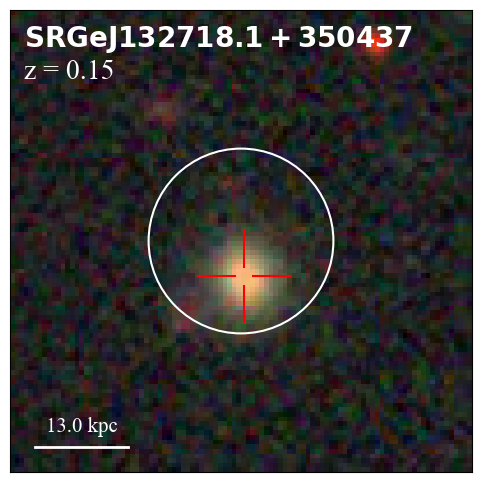}
    \includegraphics[width=0.18\textwidth]{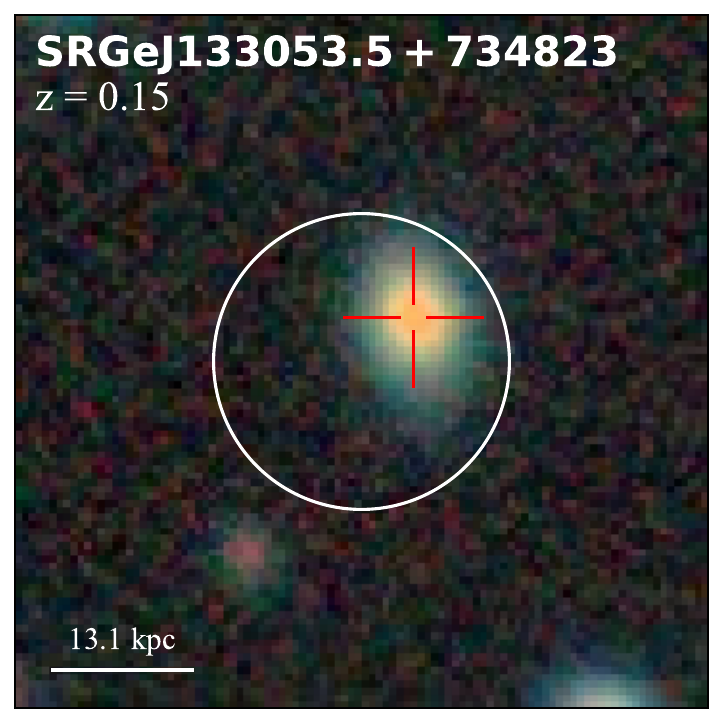}
    \includegraphics[width=0.18\textwidth]{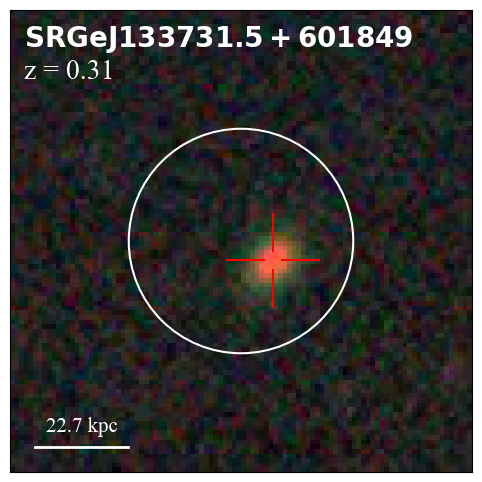}
    
    \includegraphics[width=0.18\textwidth]{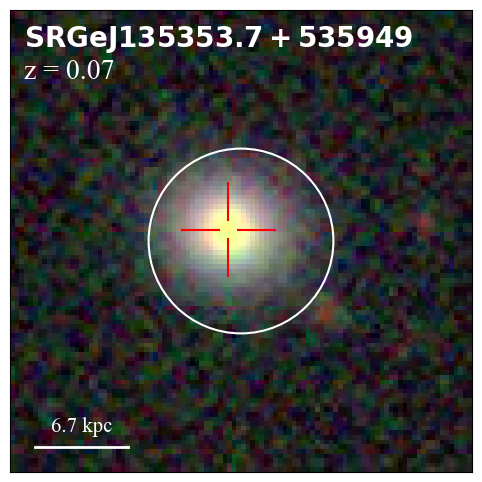}
    \includegraphics[width=0.18\textwidth]{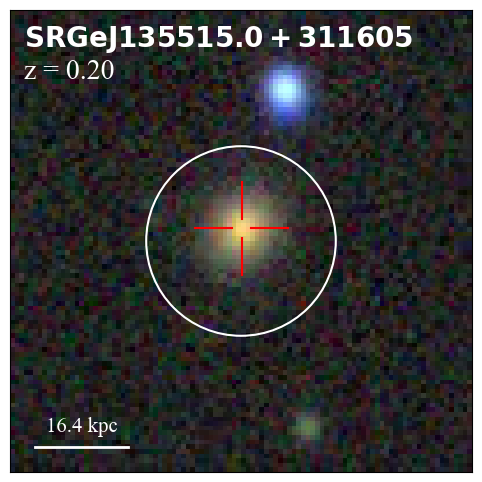}
    \includegraphics[width=0.18\textwidth]{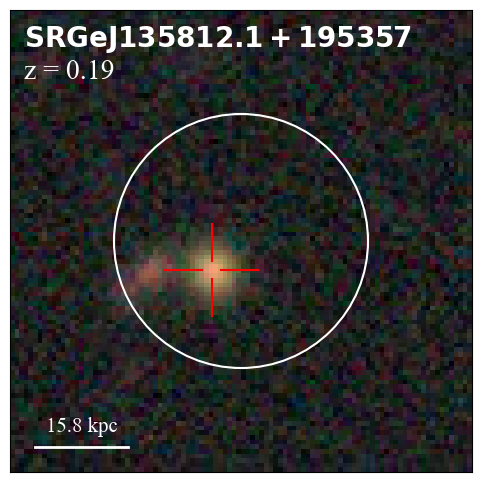}
    \includegraphics[width=0.18\textwidth]{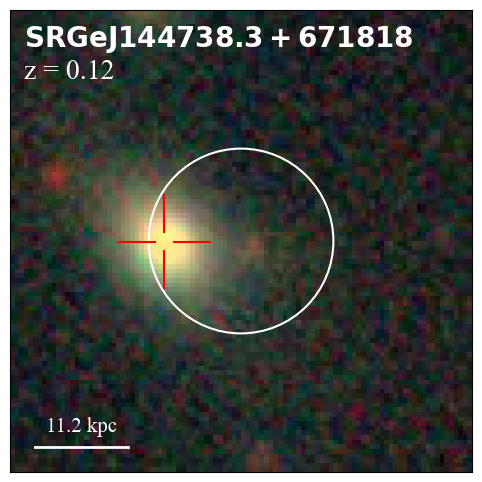}
    \includegraphics[width=0.18\textwidth]{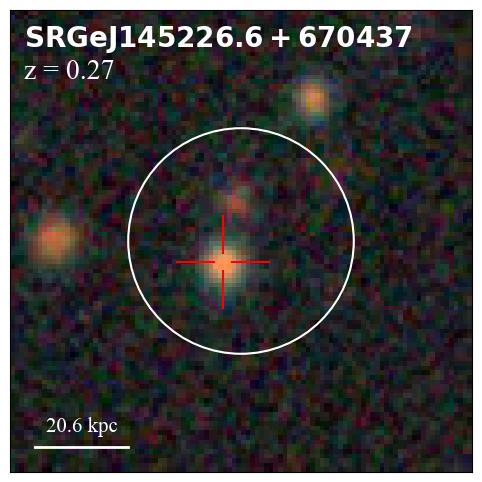}
    
    \caption{Legacy Survey DR9 co-added and PS1 color cutouts (100 $\times$ 100 pixels) of the eROSITA TDE host-galaxy associations. The eROSITA 98\% confidence position is shown as a white circle, and the optical counterpart is marked with an open red cross.\label{fig:tde_host_galaxies_part1}}
\end{figure*}

\begin{figure*}[htbp]
    \centering
    
    \includegraphics[width=0.18\textwidth]{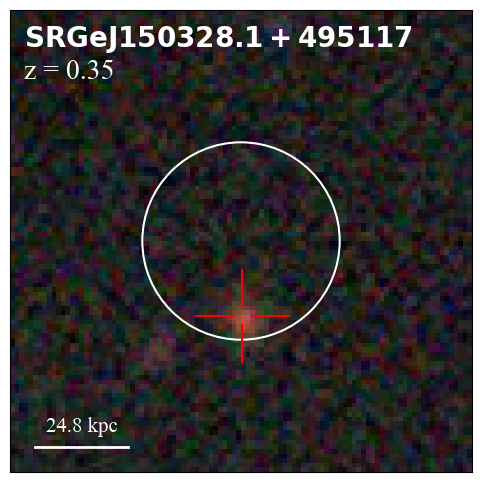}
    \includegraphics[width=0.18\textwidth]{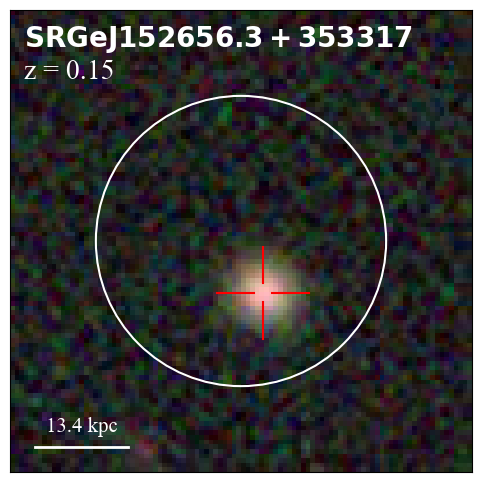}
    \includegraphics[width=0.18\textwidth]{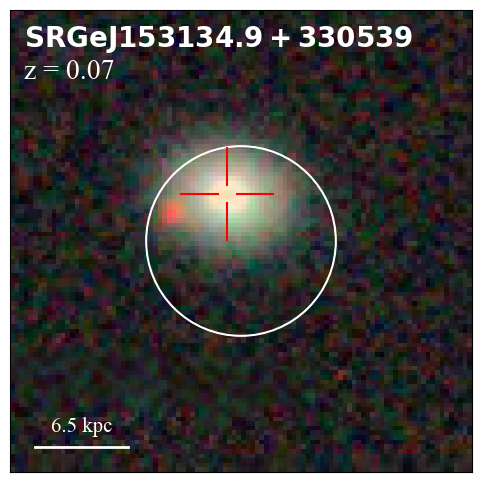}
    \includegraphics[width=0.18\textwidth]{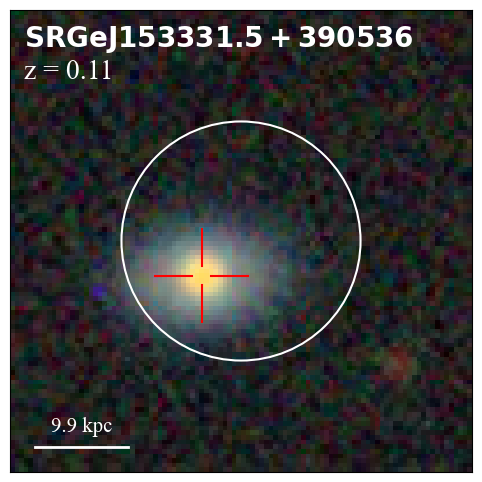}
    \includegraphics[width=0.18\textwidth]{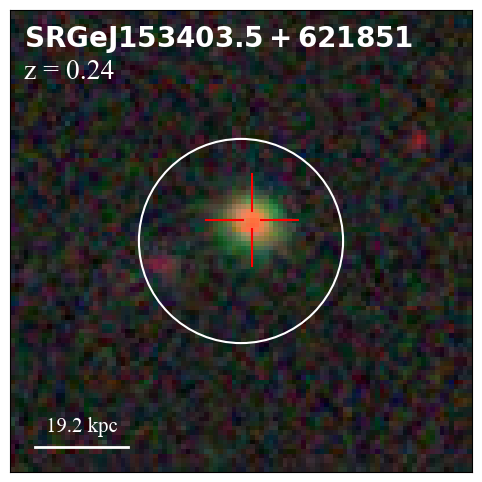}
    
    \includegraphics[width=0.18\textwidth]{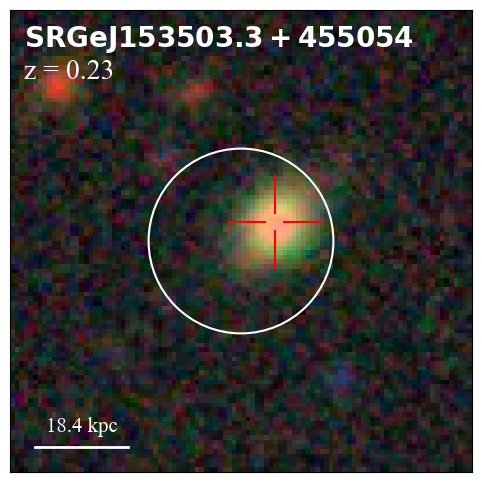}
    \includegraphics[width=0.18\textwidth]{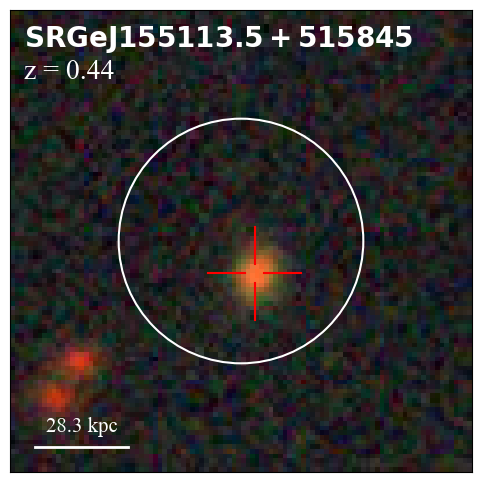}
    \includegraphics[width=0.18\textwidth]{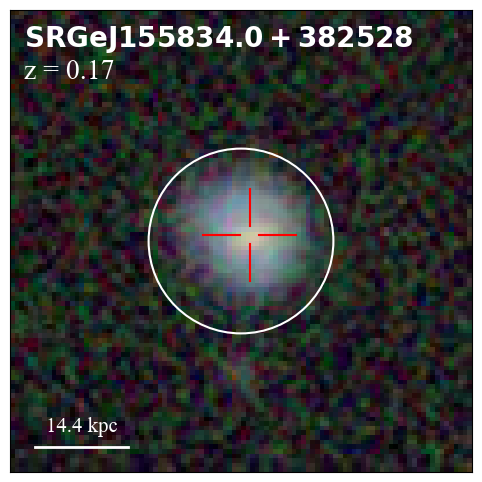}
    \includegraphics[width=0.18\textwidth]{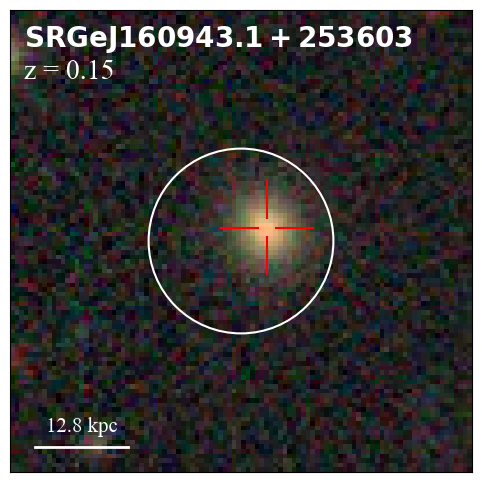}
    \includegraphics[width=0.18\textwidth]{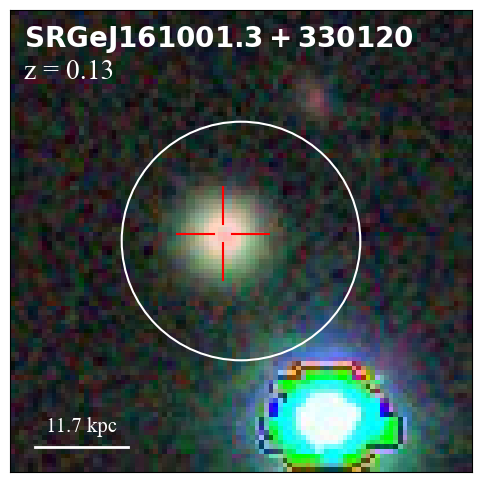}
    
    \includegraphics[width=0.18\textwidth]{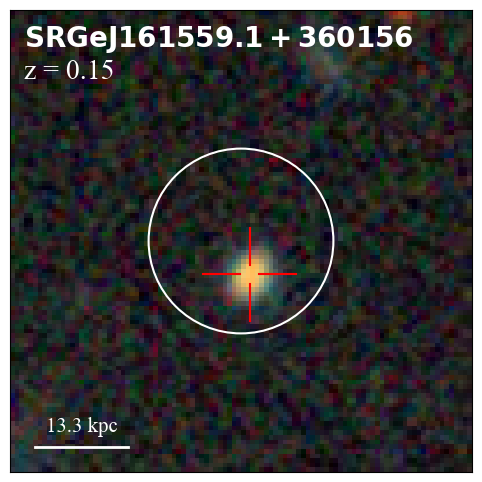}
    \includegraphics[width=0.18\textwidth]{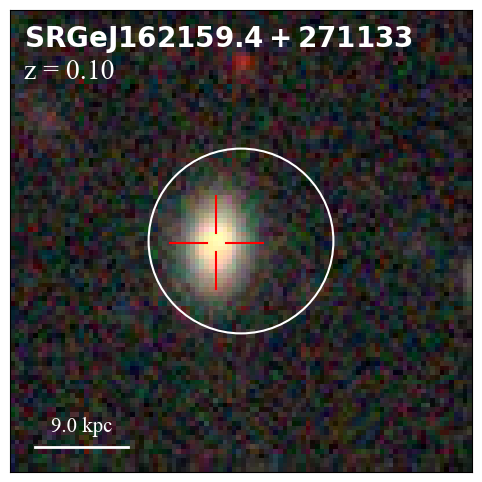}
    \includegraphics[width=0.18\textwidth]{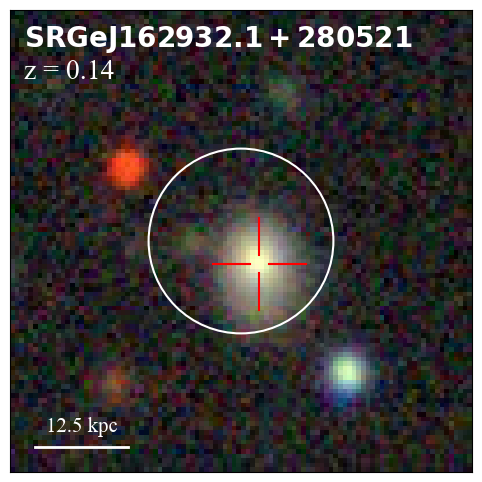}
    \includegraphics[width=0.18\textwidth]{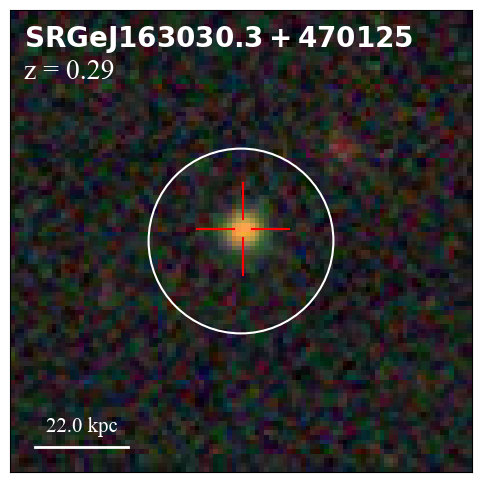}
    \includegraphics[width=0.18\textwidth]{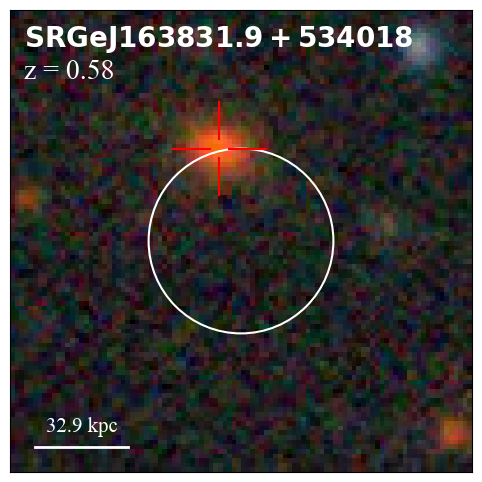}
    
    \includegraphics[width=0.18\textwidth]{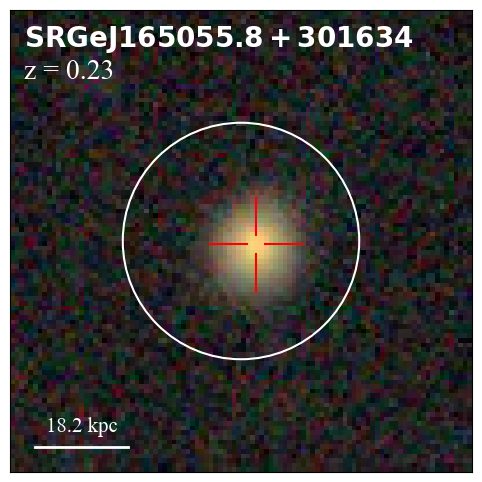}
    \includegraphics[width=0.18\textwidth]{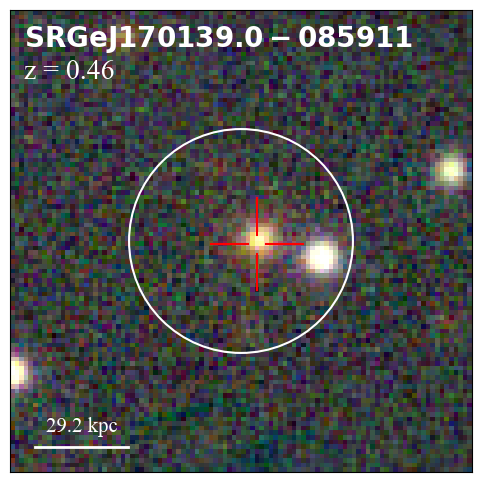}
    \includegraphics[width=0.18\textwidth]{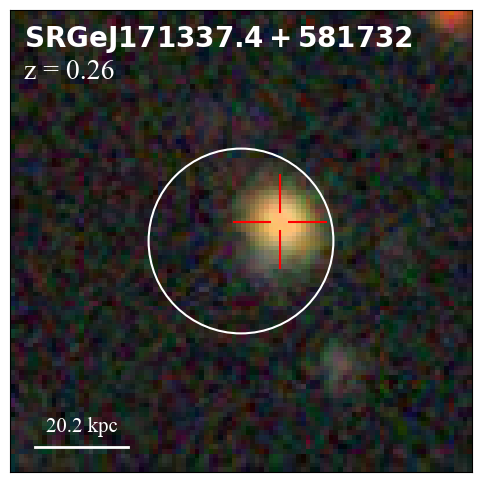}
    \includegraphics[width=0.18\textwidth]{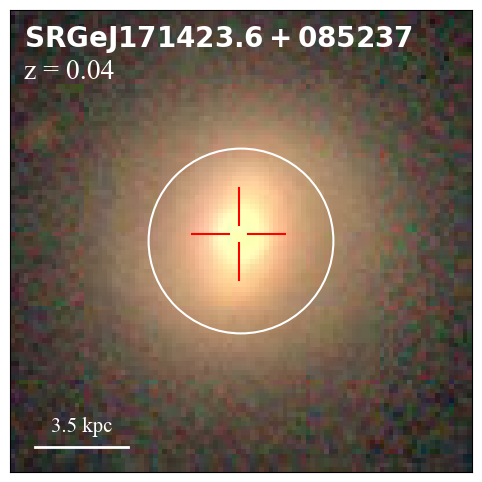}
    \includegraphics[width=0.18\textwidth]{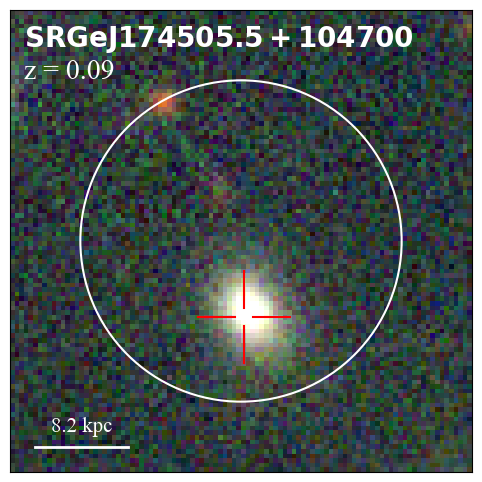}
    
    \includegraphics[width=0.18\textwidth]{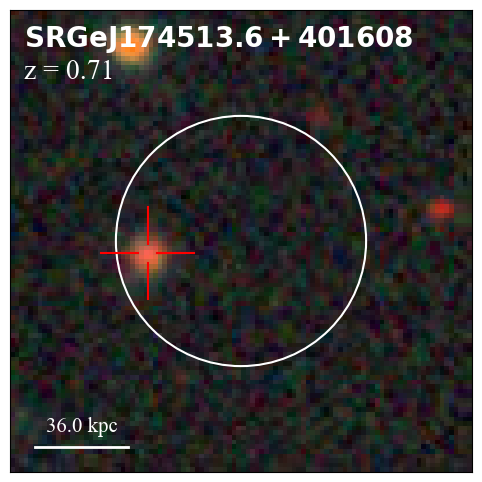}
    \includegraphics[width=0.18\textwidth]{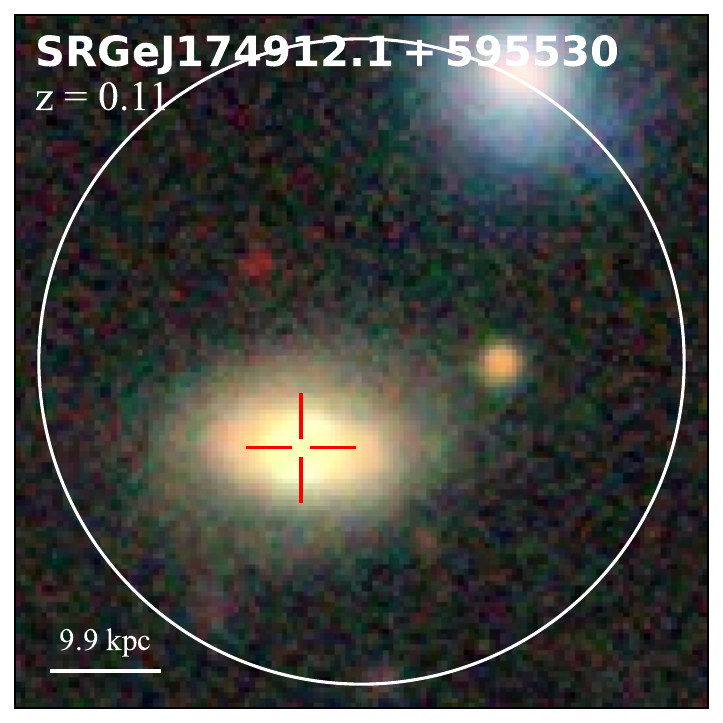}
    \includegraphics[width=0.18\textwidth]{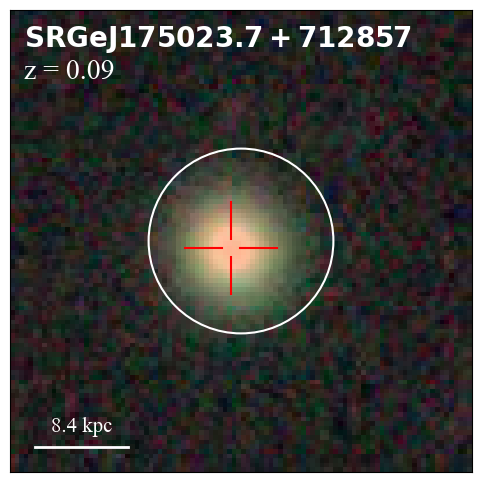}
    \includegraphics[width=0.18\textwidth]{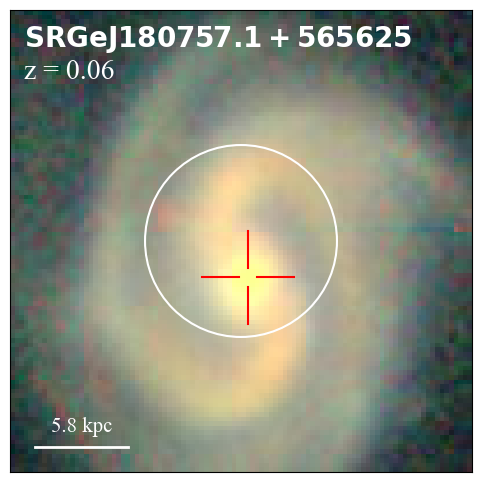}
    \includegraphics[width=0.18\textwidth]{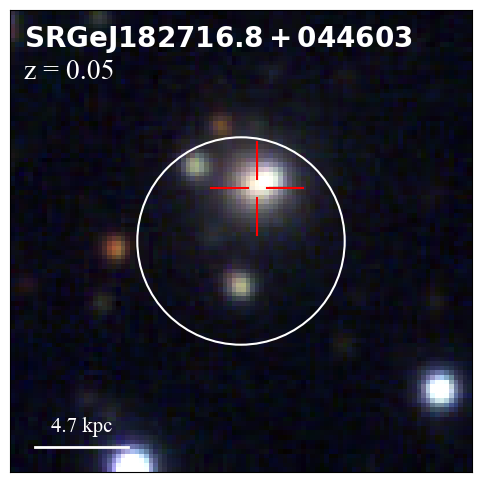}
    
    \includegraphics[width=0.18\textwidth]{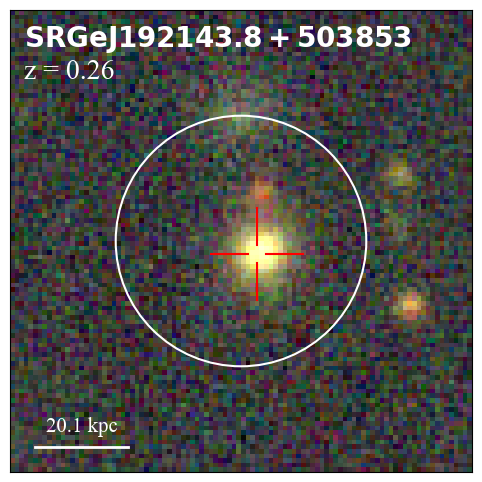}
    \includegraphics[width=0.18\textwidth]{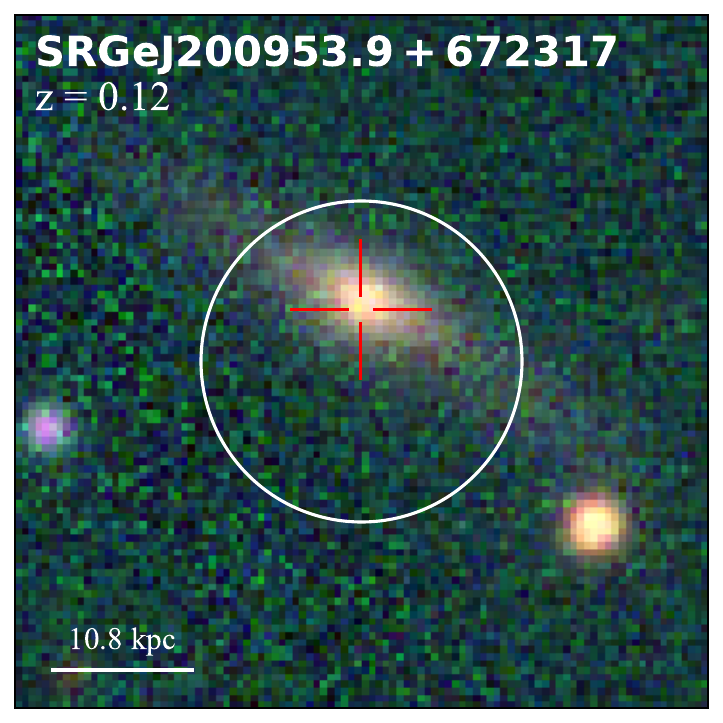}
    \includegraphics[width=0.18\textwidth]{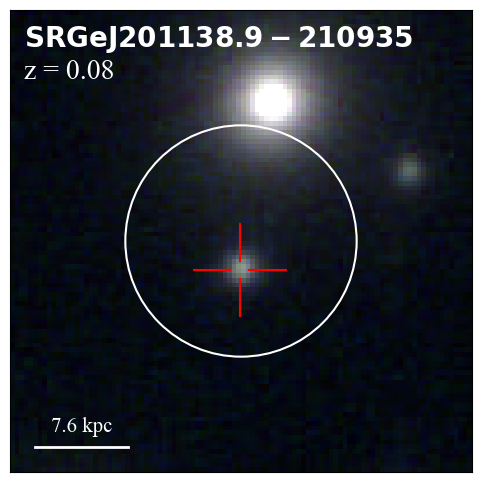}
    \includegraphics[width=0.18\textwidth]{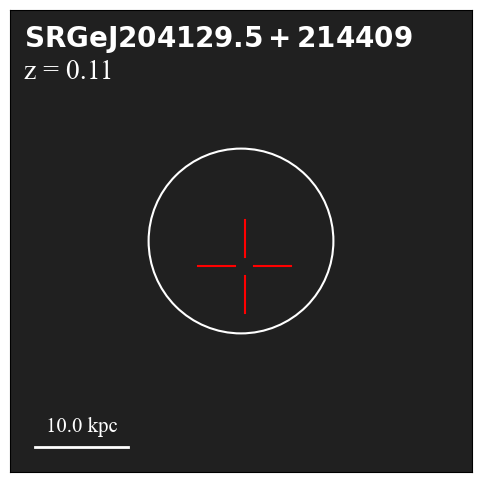}
    \includegraphics[width=0.18\textwidth]{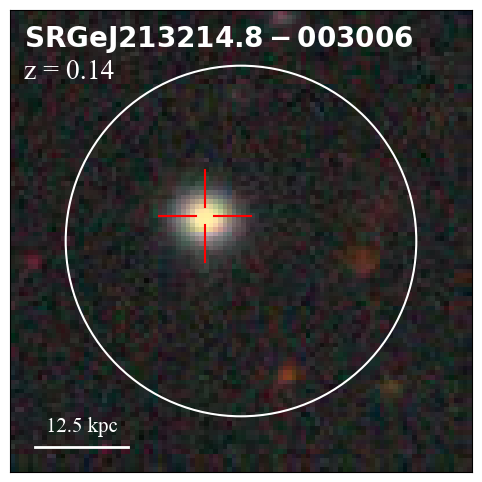}
    
    \includegraphics[width=0.18\textwidth]{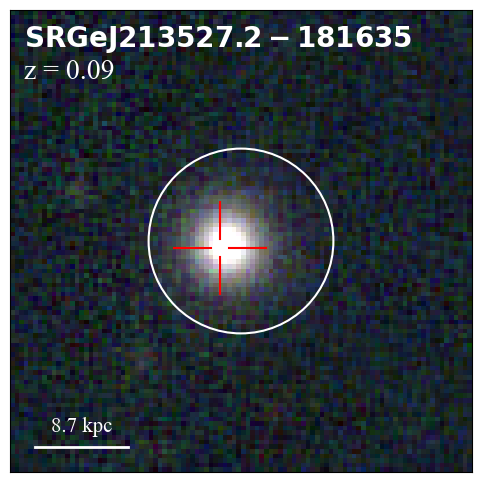}
    \includegraphics[width=0.18\textwidth]{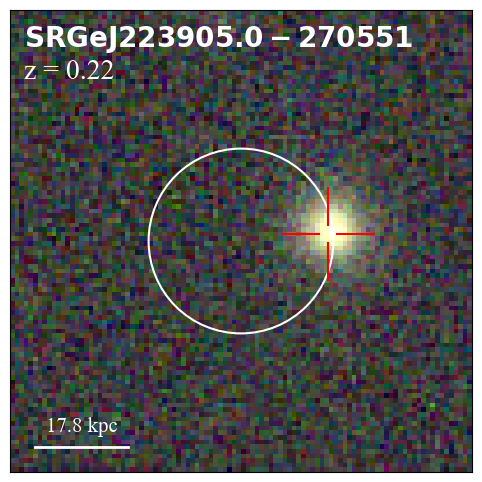}
    \includegraphics[width=0.18\textwidth]{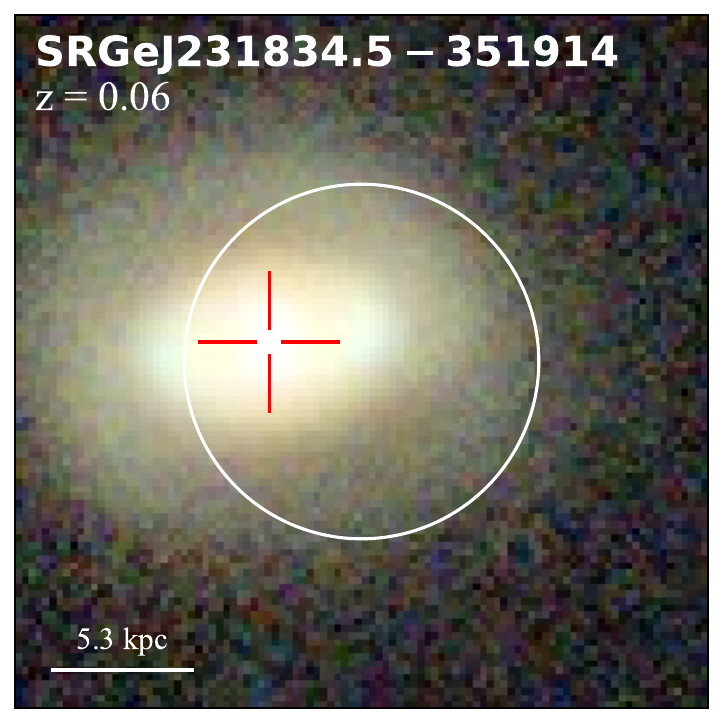}
    \includegraphics[width=0.18\textwidth]{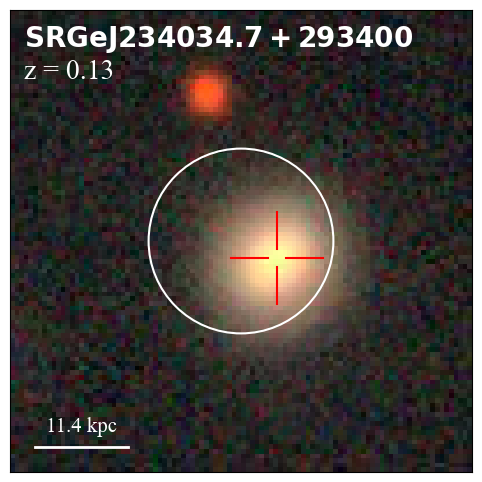}
    \includegraphics[width=0.18\textwidth]{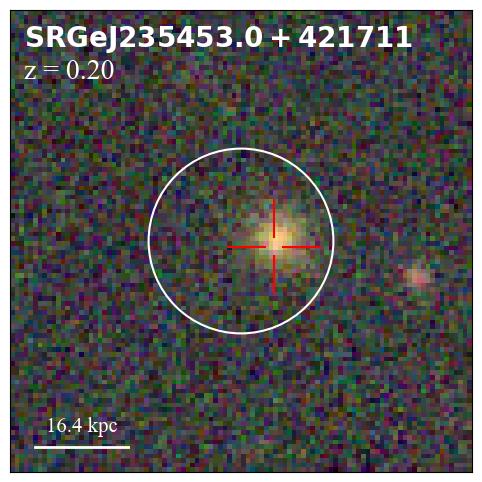}
    
    \caption{
    Continued figure of Fig.~\ref{fig:tde_host_galaxies_part1}. \label{fig:tde_host_galaxies_part2}}
\end{figure*}

\begin{figure*}
    \centering
    \includegraphics[width=0.33\textwidth]{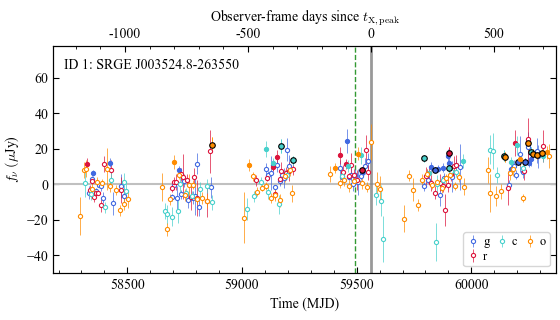}
    \includegraphics[width=0.33\textwidth]{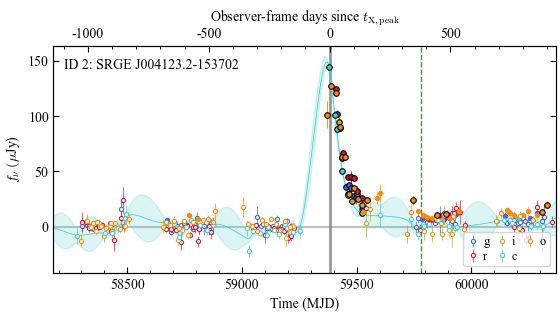}
    \includegraphics[width=0.33\textwidth]{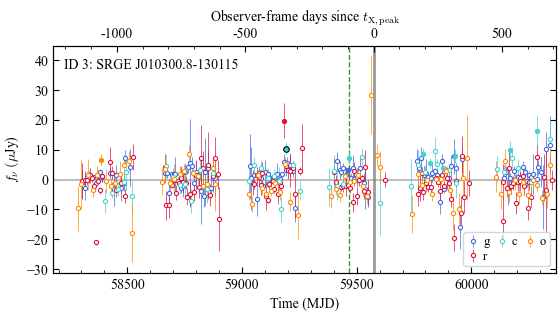}\\
    \includegraphics[width=0.33\textwidth]{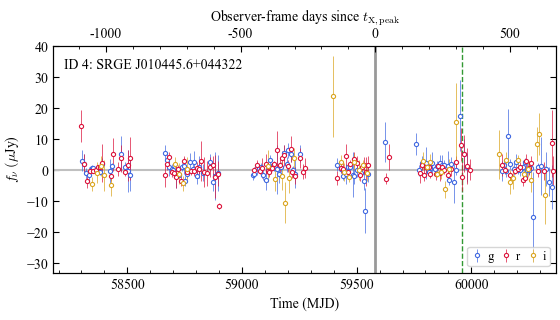}
    \includegraphics[width=0.33\textwidth]{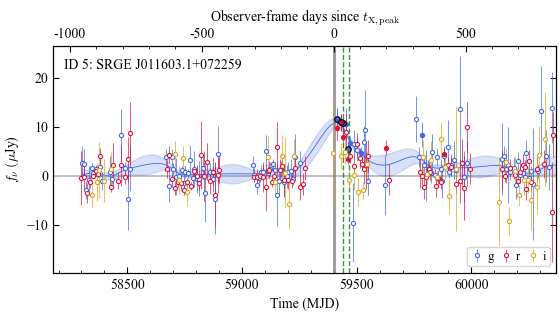}
    \includegraphics[width=0.33\textwidth]{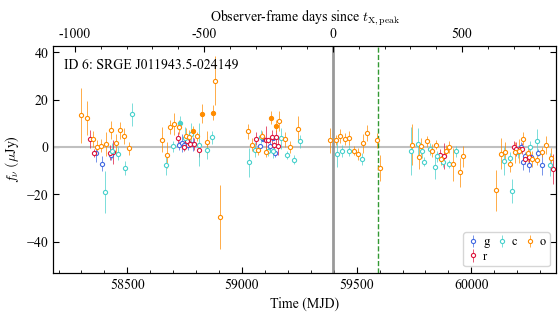}\\
    \includegraphics[width=0.33\textwidth]{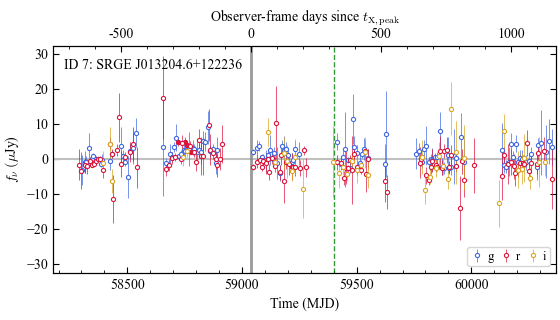}
    \includegraphics[width=0.33\textwidth]{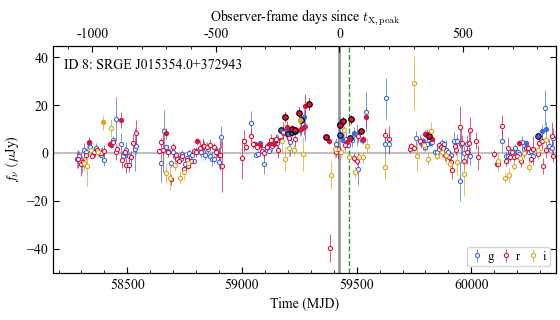}
    \includegraphics[width=0.33\textwidth]{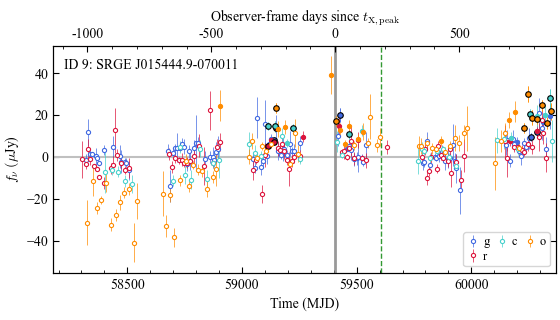}\\
    \includegraphics[width=0.33\textwidth]{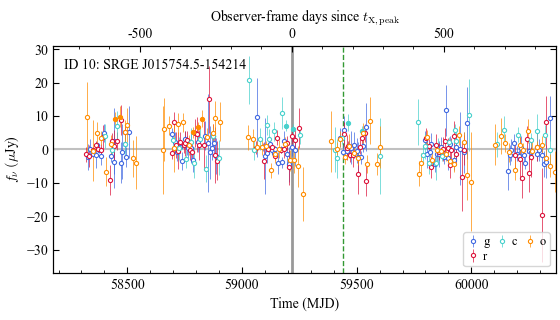}
    \includegraphics[width=0.33\textwidth]{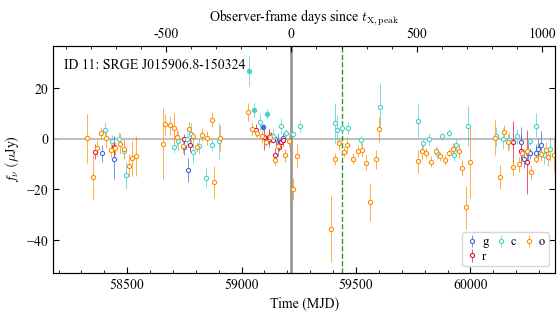}
    \includegraphics[width=0.33\textwidth]{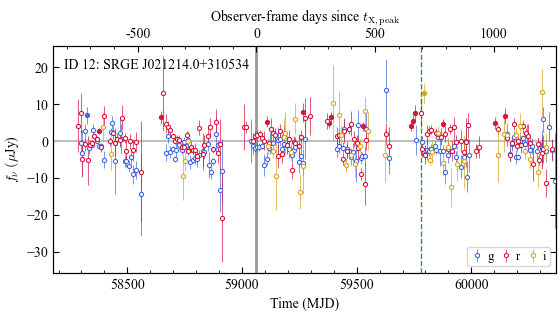}\\
    \includegraphics[width=0.33\textwidth]{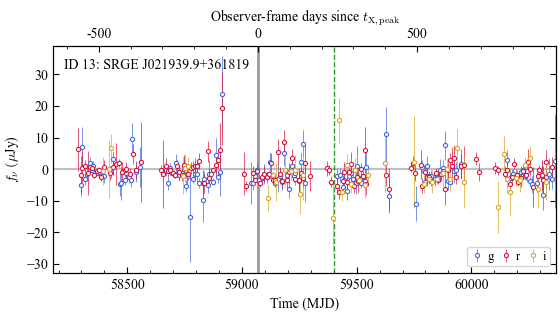}
    \includegraphics[width=0.33\textwidth]{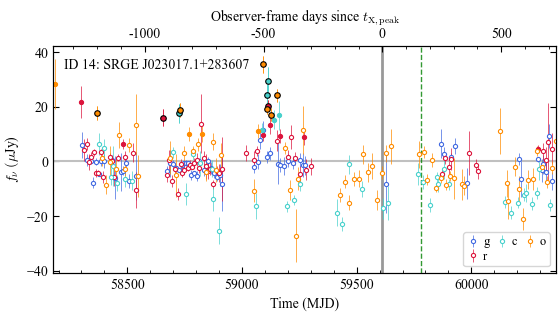}
    \includegraphics[width=0.33\textwidth]{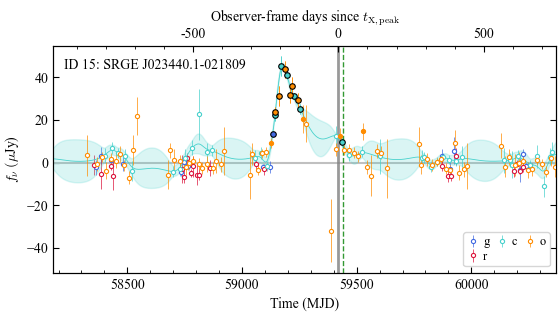}\\
    \includegraphics[width=0.33\textwidth]{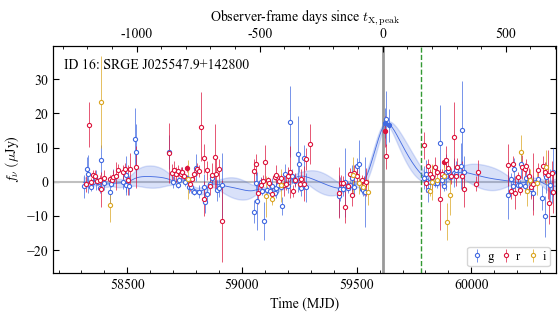}
    \includegraphics[width=0.33\textwidth]{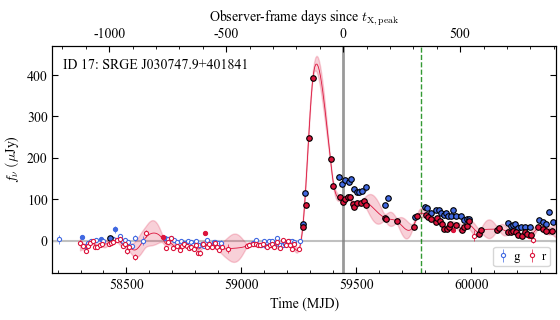}
    \includegraphics[width=0.33\textwidth]{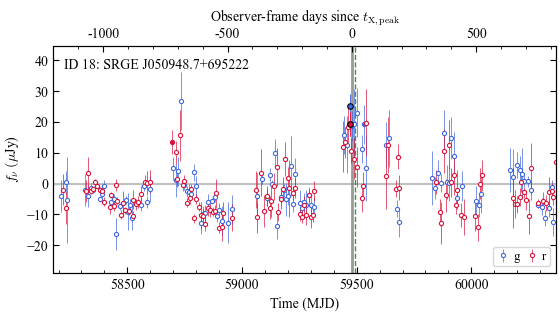}\\
    \caption{ZTF and ATLAS differential photometry performed at the centroids of the host galaxies. Data points with $<3\sigma$ significance are shown as hollow circles, $>3\sigma$ as solid circles, and $>5\sigma$ are further highlighted with black edges. Spectroscopic observation epochs are marked with vertical dashed green lines. Epochs of X-ray peak are marked with vertical solid gray lines. 
    Data are binned in 10-day intervals for clarity. For TDEs with detected optical flares, we show Gaussian process model fits following the procedures described in Appendix B.4 of \citet{Yao2020_19dge}. Models are fitted to the band with either the smallest photometric uncertainties or the best temporal coverage. The x-axis spans from 2018 March 1 to 2024 March 1 for all sources.  \label{fig:opt_diff_lc}}
\end{figure*}

\begin{figure*}
    \centering
    \includegraphics[width=0.33\textwidth]{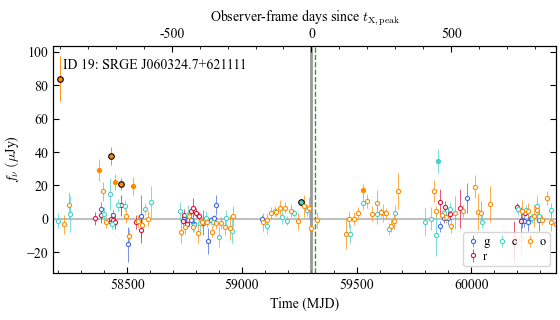}
    \includegraphics[width=0.33\textwidth]{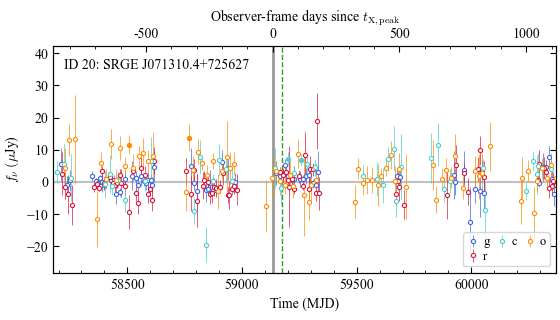}
    \includegraphics[width=0.33\textwidth]{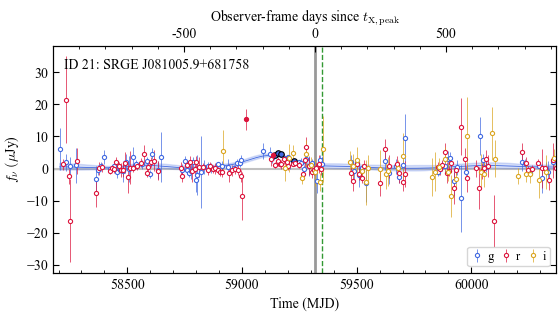}\\
    \includegraphics[width=0.33\textwidth]{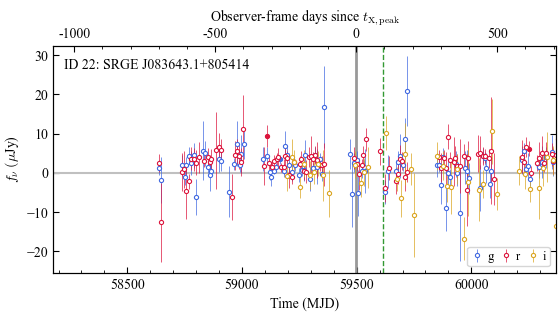}
    \includegraphics[width=0.33\textwidth]{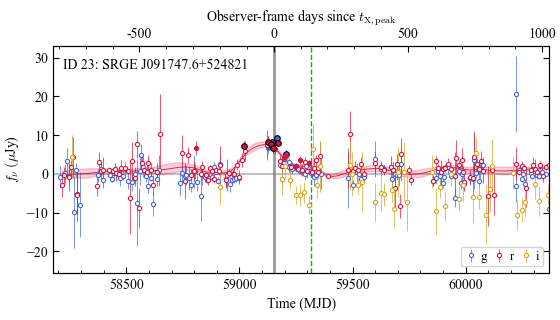}
    \includegraphics[width=0.33\textwidth]{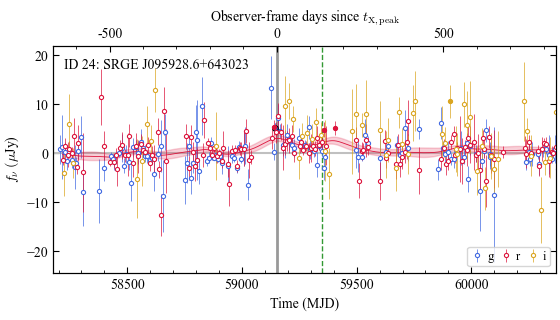}\\
    \includegraphics[width=0.33\textwidth]{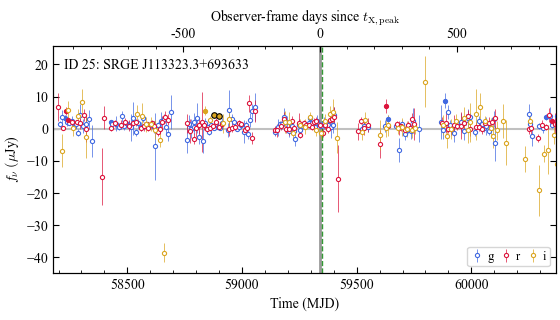}
    \includegraphics[width=0.33\textwidth]{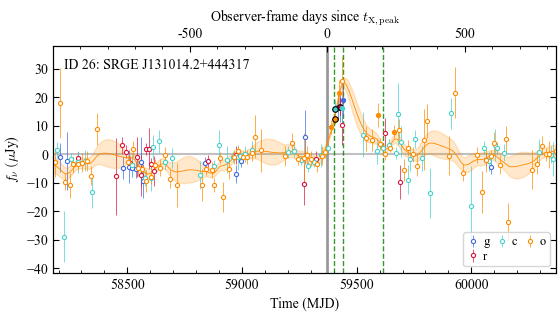}
    \includegraphics[width=0.33\textwidth]{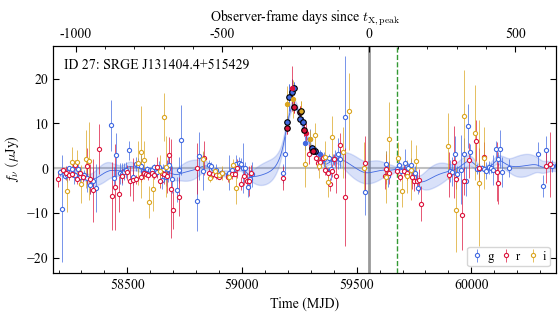}\\
    \includegraphics[width=0.33\textwidth]{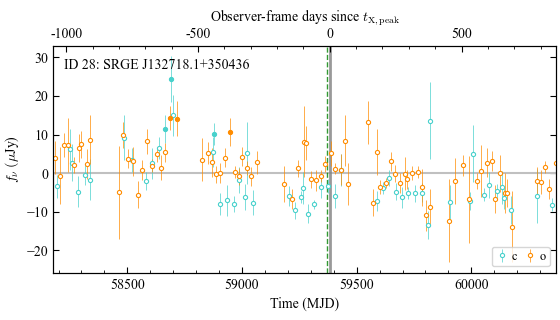}
    \includegraphics[width=0.33\textwidth]{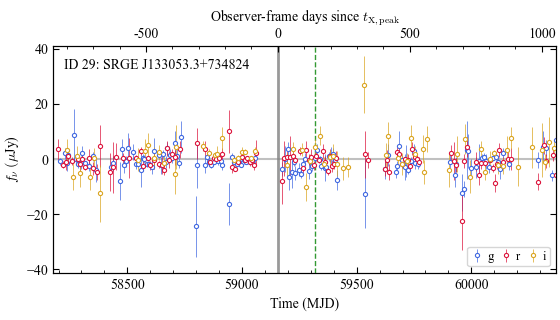}
    \includegraphics[width=0.33\textwidth]{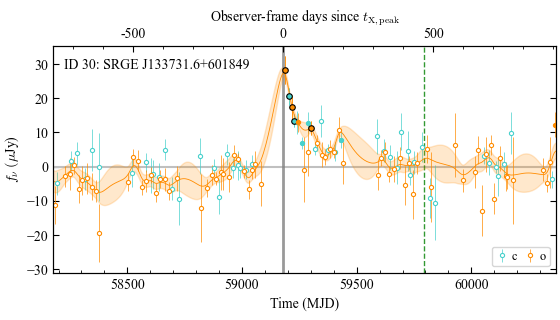}\\
    \includegraphics[width=0.33\textwidth]{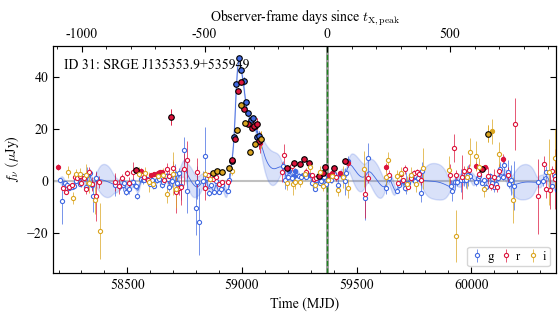}
    \includegraphics[width=0.33\textwidth]{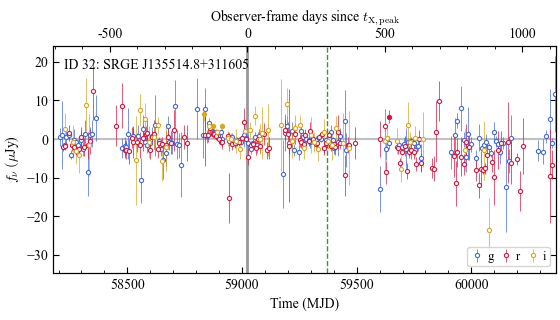}
    \includegraphics[width=0.33\textwidth]{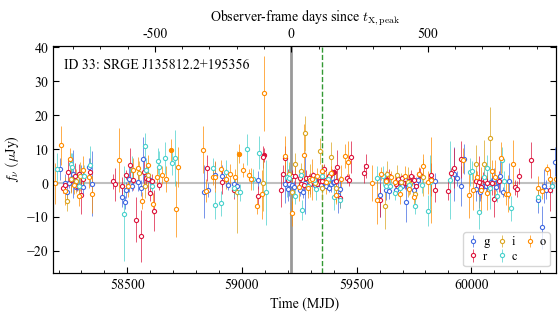}\\
    \includegraphics[width=0.33\textwidth]{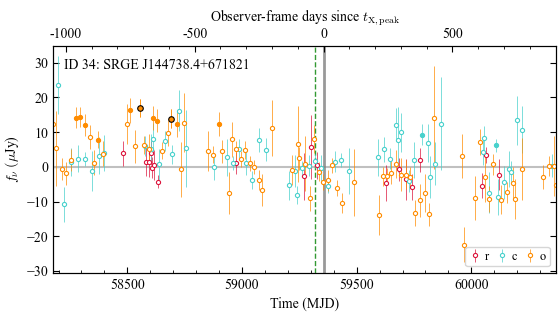}
    \includegraphics[width=0.33\textwidth]{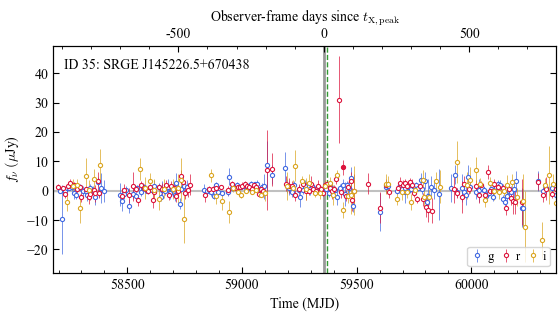}
    \includegraphics[width=0.33\textwidth]{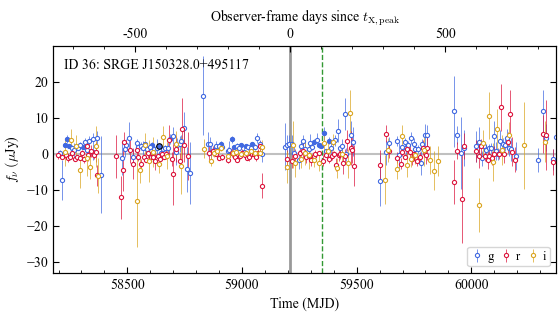}\\
    \includegraphics[width=0.33\textwidth]{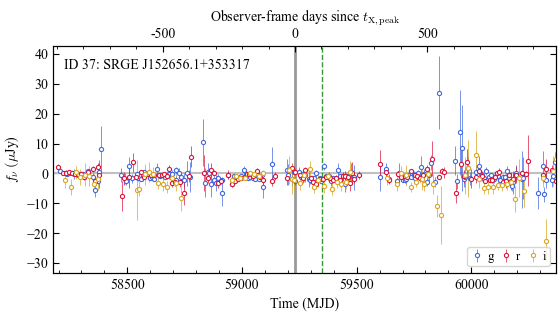}
    \includegraphics[width=0.33\textwidth]{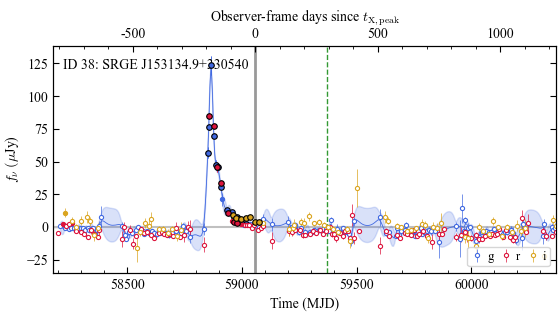}
    \includegraphics[width=0.33\textwidth]{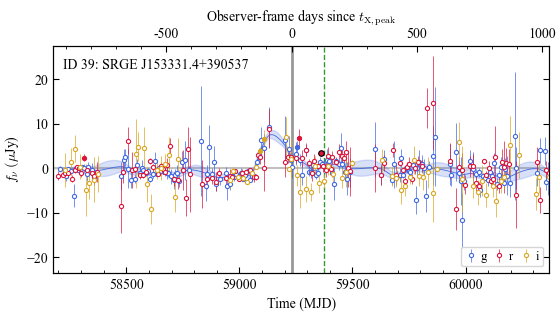}\\
    \caption{Continued figure of Fig.~\ref{fig:opt_diff_lc}. \label{fig:opt_diff_lc_2}}
\end{figure*}

\begin{figure*}
    \centering
    \includegraphics[width=0.33\textwidth]{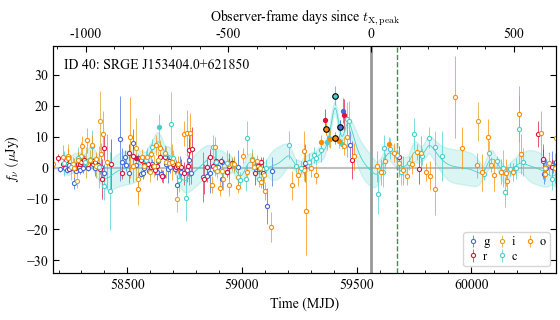}
    \includegraphics[width=0.33\textwidth]{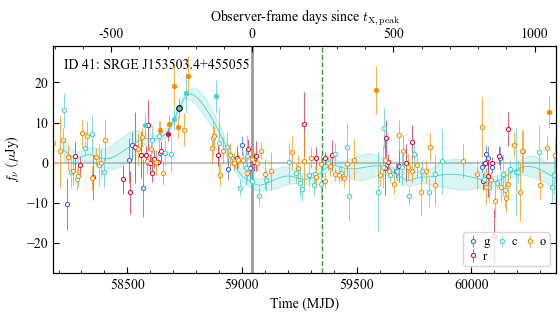}
    \includegraphics[width=0.33\textwidth]{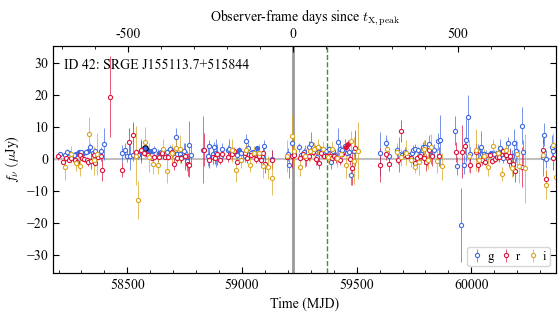}\\
    \includegraphics[width=0.33\textwidth]{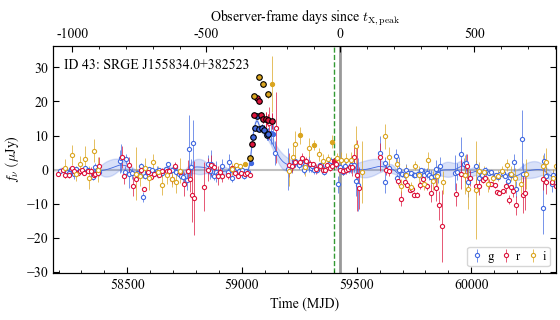}
    \includegraphics[width=0.33\textwidth]{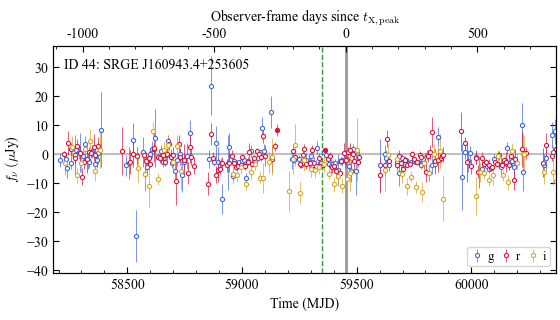}
    \includegraphics[width=0.33\textwidth]{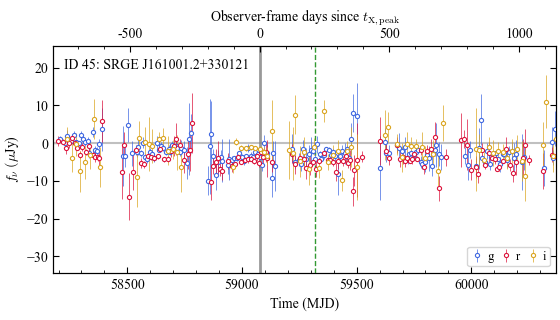}\\
    \includegraphics[width=0.33\textwidth]{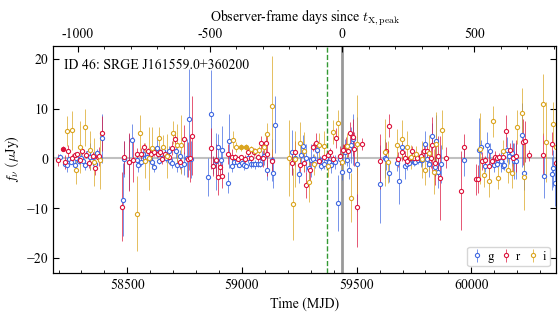}
    \includegraphics[width=0.33\textwidth]{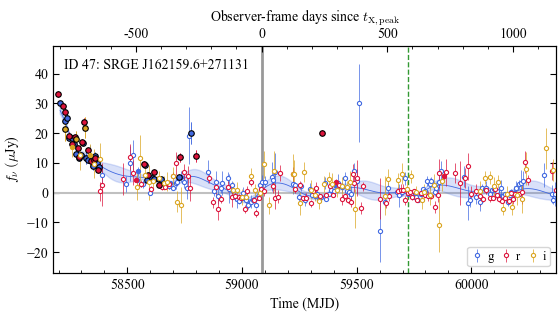}
    \includegraphics[width=0.33\textwidth]{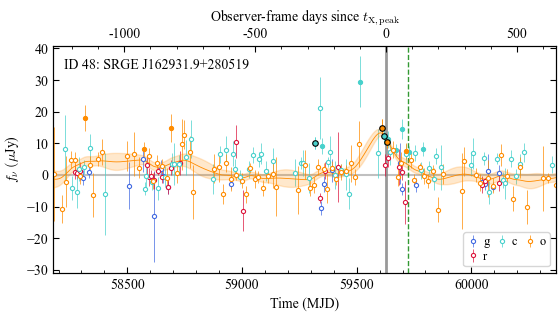}\\
    \includegraphics[width=0.33\textwidth]{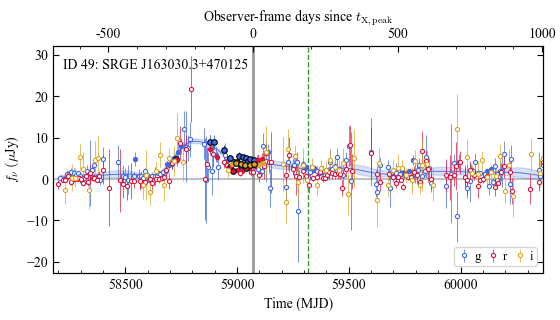}
    \includegraphics[width=0.33\textwidth]{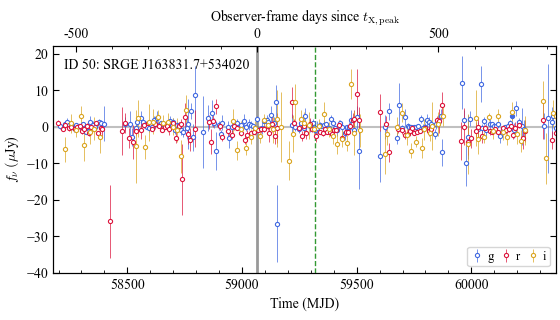}
    \includegraphics[width=0.33\textwidth]{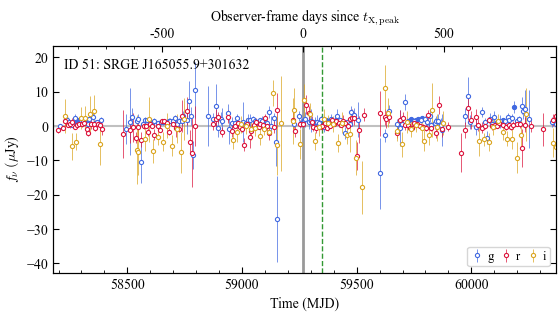}\\
    \includegraphics[width=0.33\textwidth]{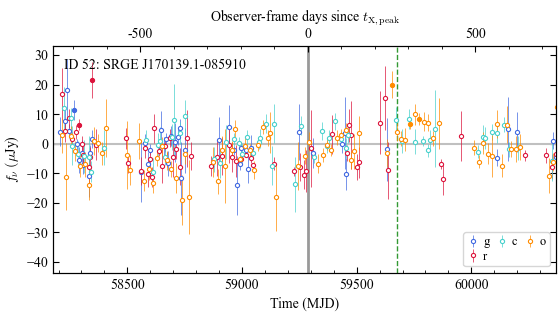}
    \includegraphics[width=0.33\textwidth]{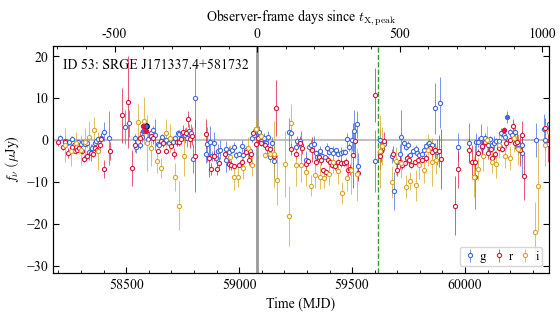}
    \includegraphics[width=0.33\textwidth]{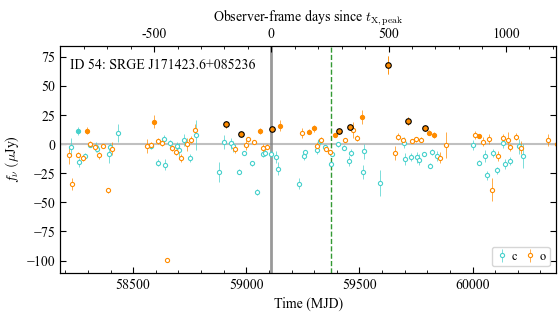}\\
    \includegraphics[width=0.33\textwidth]{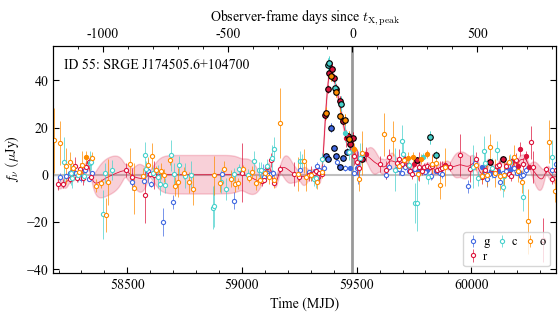}
    \includegraphics[width=0.33\textwidth]{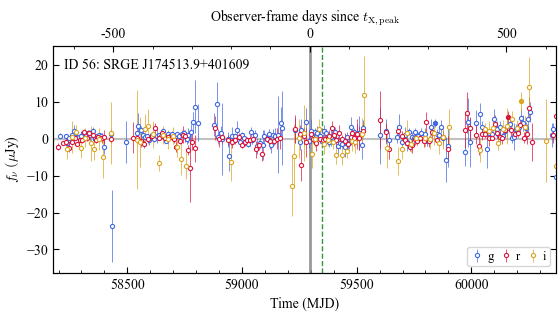}
    \includegraphics[width=0.33\textwidth]{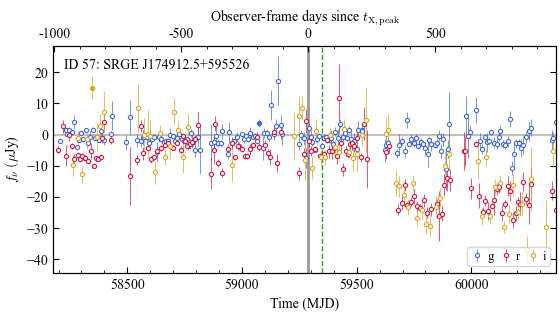}\\
    \includegraphics[width=0.33\textwidth]{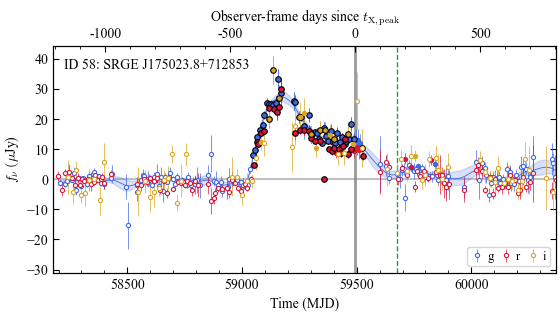}
    \includegraphics[width=0.33\textwidth]{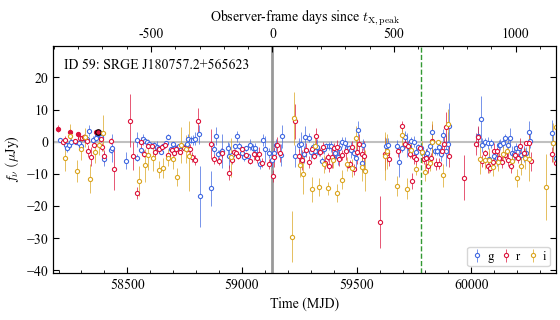}
    \includegraphics[width=0.33\textwidth]{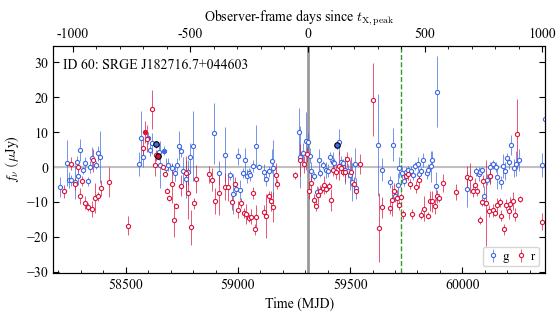}\\
    \caption{Continued figure of Fig.~\ref{fig:opt_diff_lc_2}. \label{fig:opt_diff_lc_3}}
\end{figure*}

\begin{figure*}
    \centering
    \includegraphics[width=0.33\textwidth]{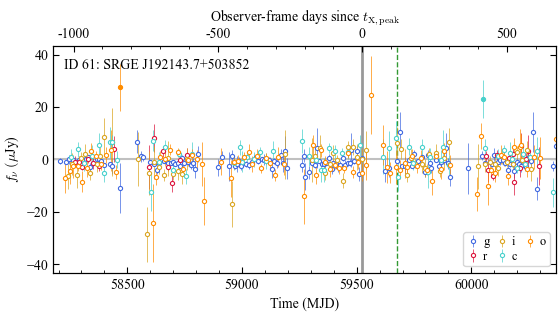}
    \includegraphics[width=0.33\textwidth]{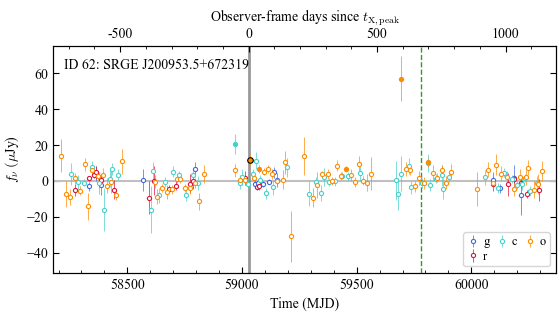}
    \includegraphics[width=0.33\textwidth]{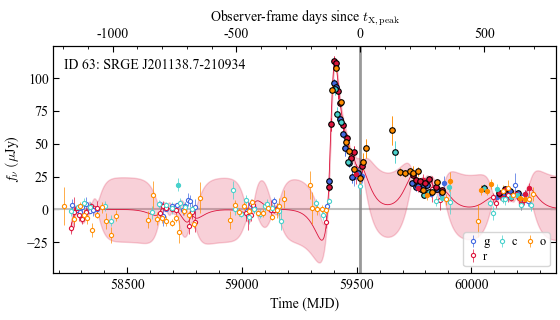}\\
    \includegraphics[width=0.33\textwidth]{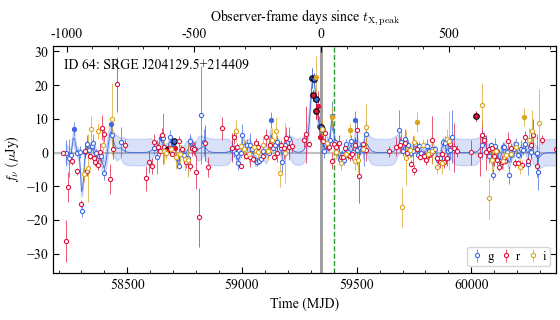}
    \includegraphics[width=0.33\textwidth]{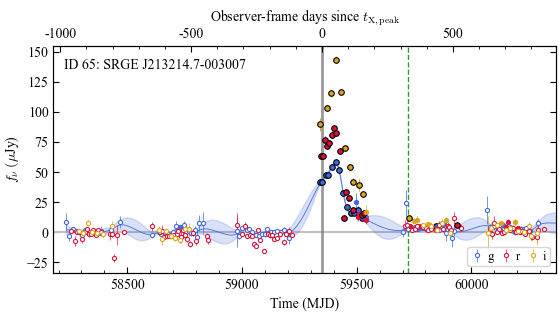}
    \includegraphics[width=0.33\textwidth]{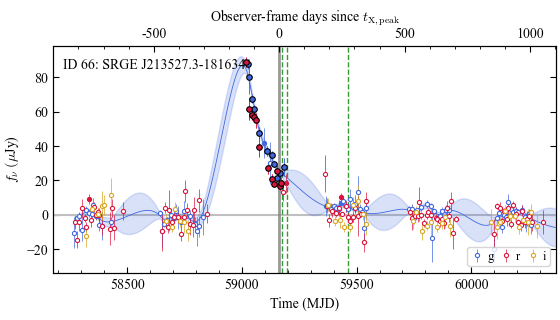}\\
    \includegraphics[width=0.33\textwidth]{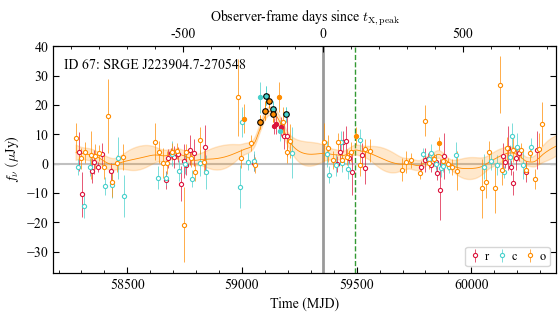}
    \includegraphics[width=0.33\textwidth]{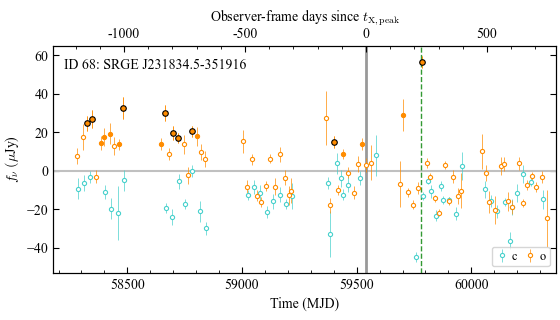}
    \includegraphics[width=0.33\textwidth]{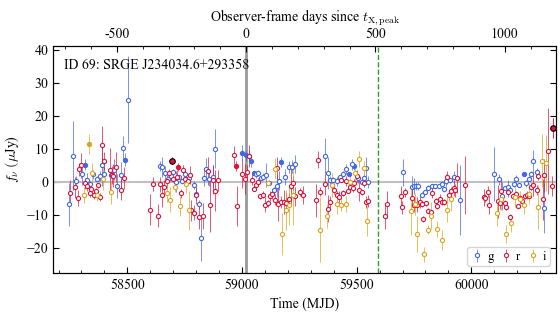}\\
    \includegraphics[width=0.33\textwidth]{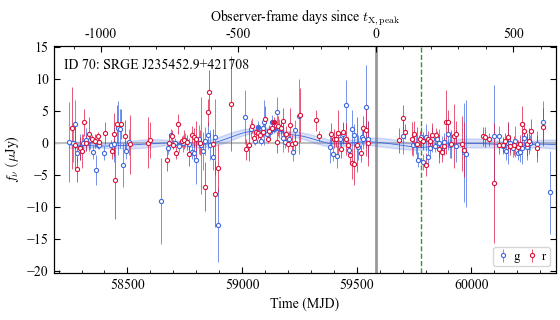}
    \caption{Continued figure of Fig.~\ref{fig:opt_diff_lc_3}. \label{fig:opt_diff_lc_4}}
\end{figure*}


\begin{figure*}[htbp]
    \raggedleft

    \includegraphics[width=0.24\textwidth]{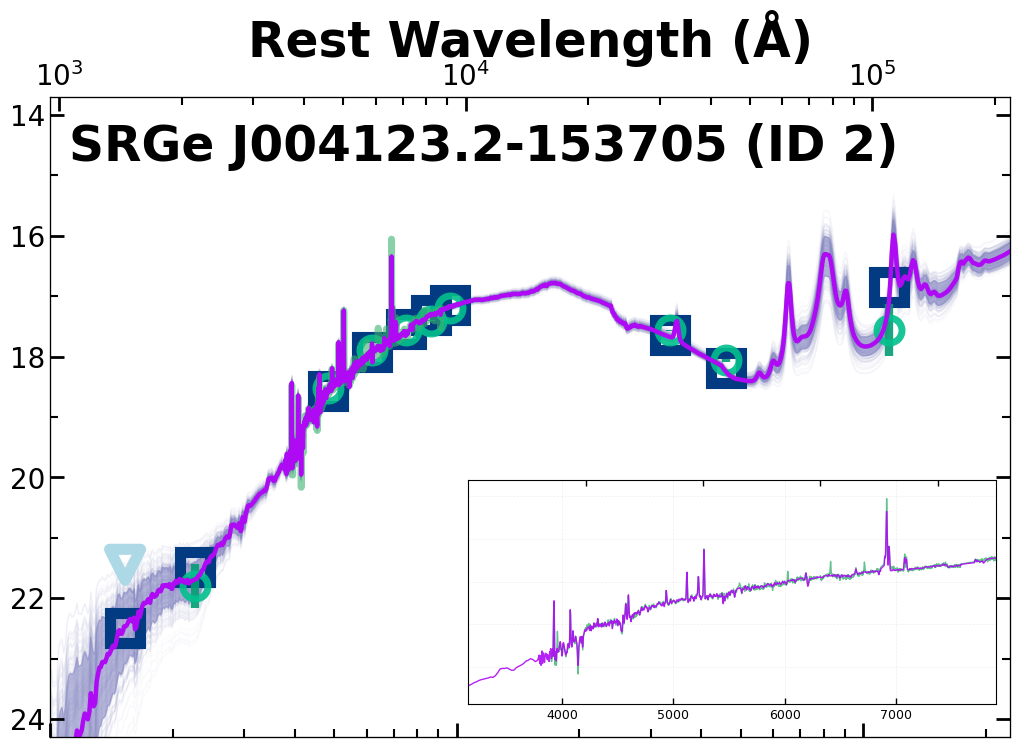}
    \includegraphics[width=0.24\textwidth]{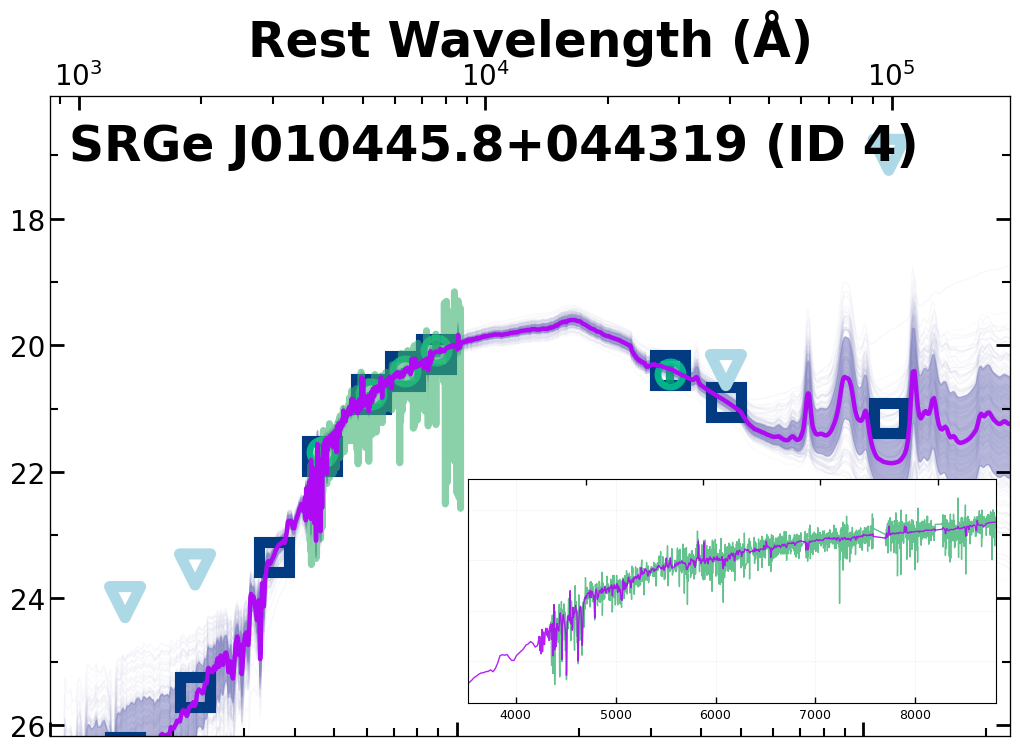}
    \includegraphics[width=0.24\textwidth]{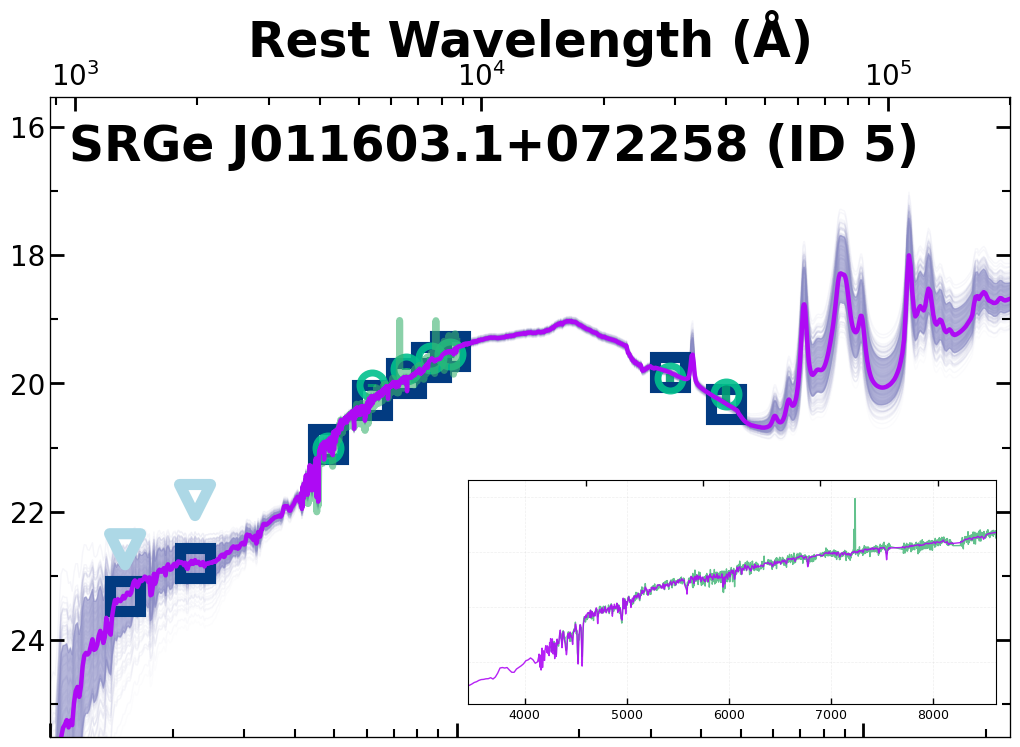}
    \includegraphics[width=0.24\textwidth]{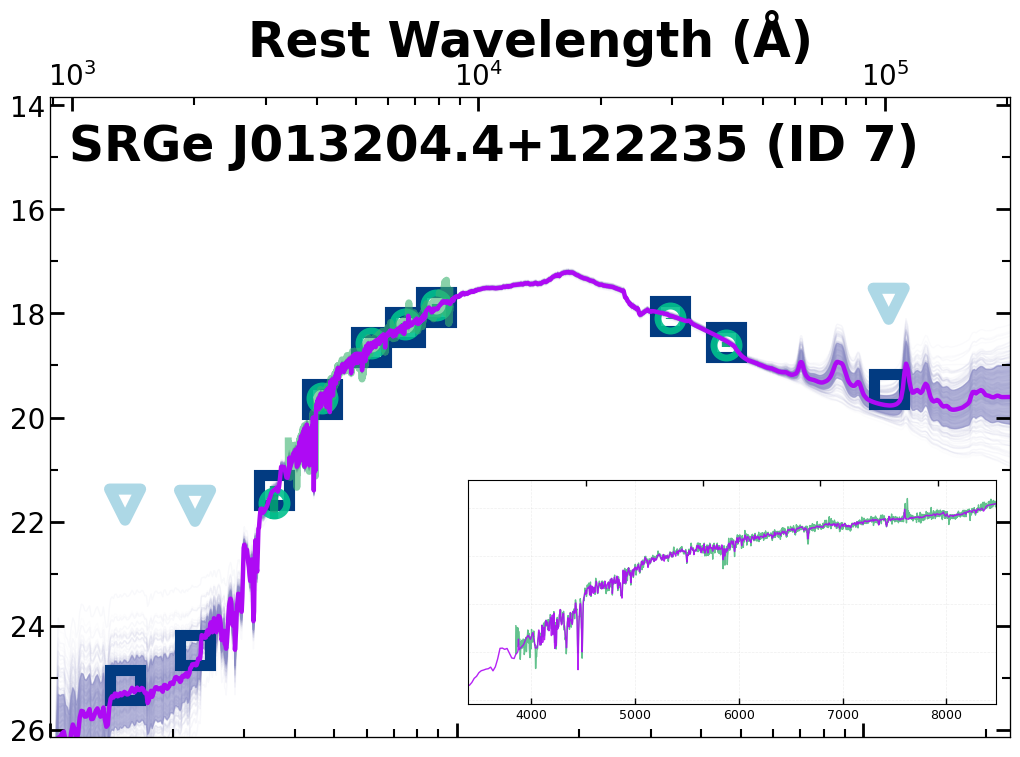}
    \vspace{0.2cm}
    
    \includegraphics[width=0.24\textwidth]{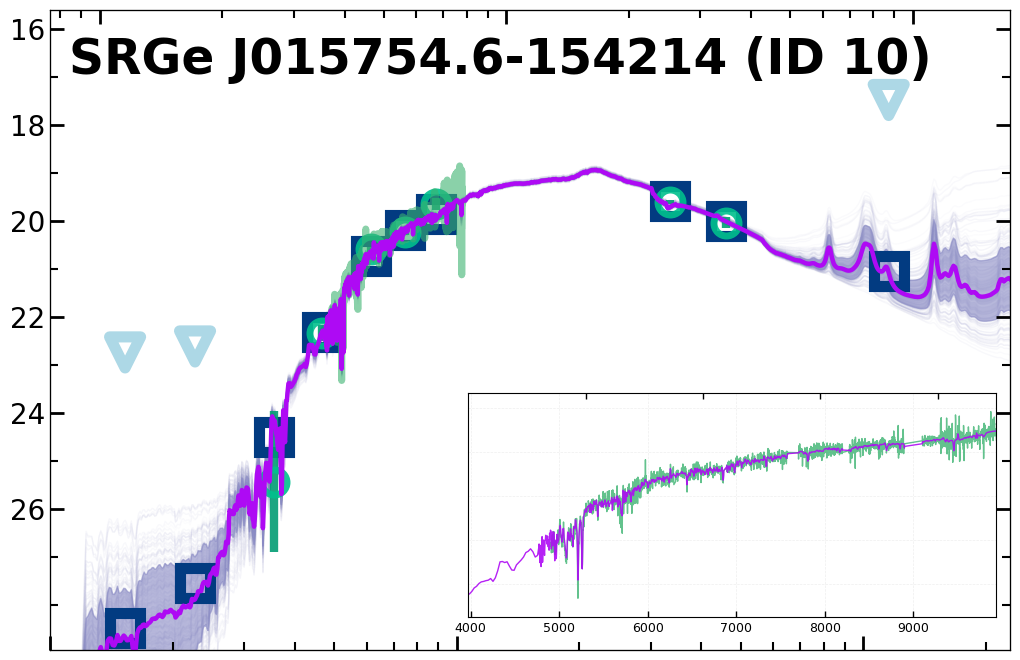}
    \includegraphics[width=0.24\textwidth]{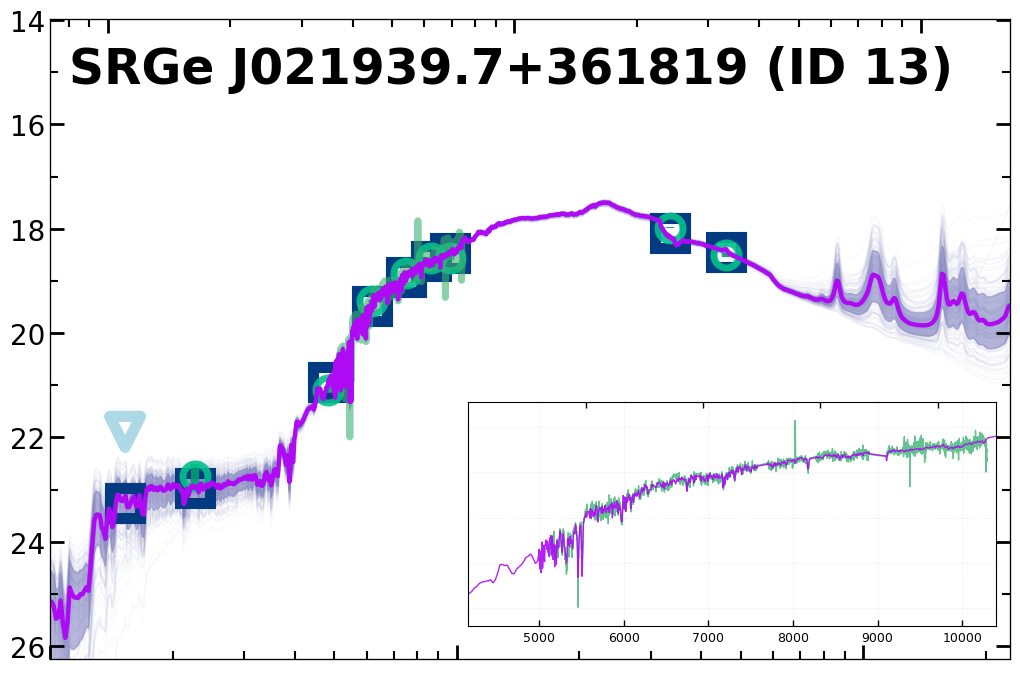}
    \includegraphics[width=0.24\textwidth]{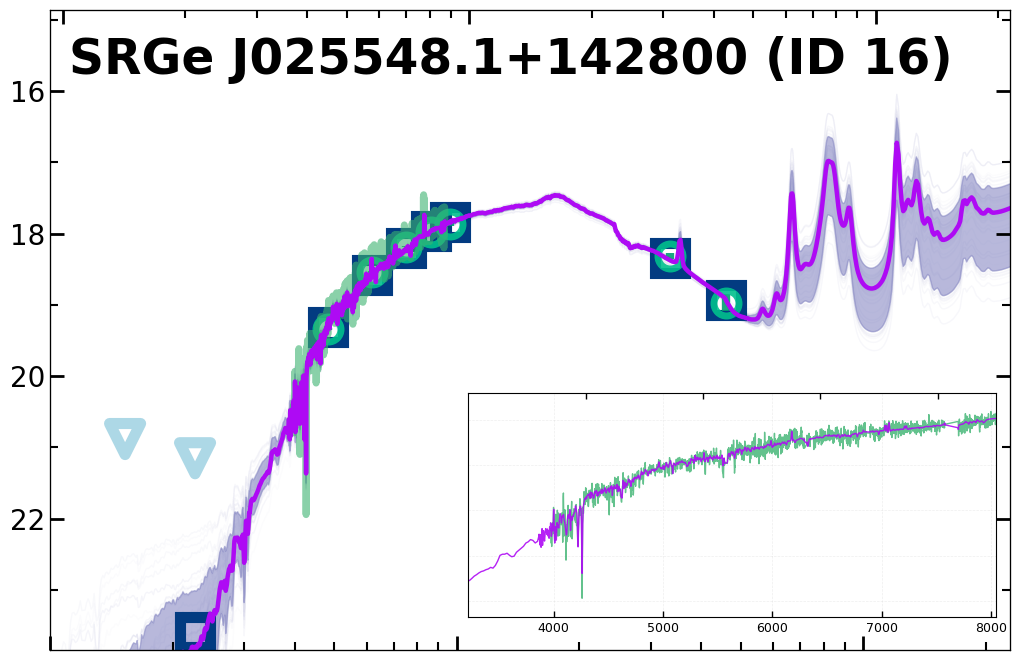}
    \includegraphics[width=0.24\textwidth]{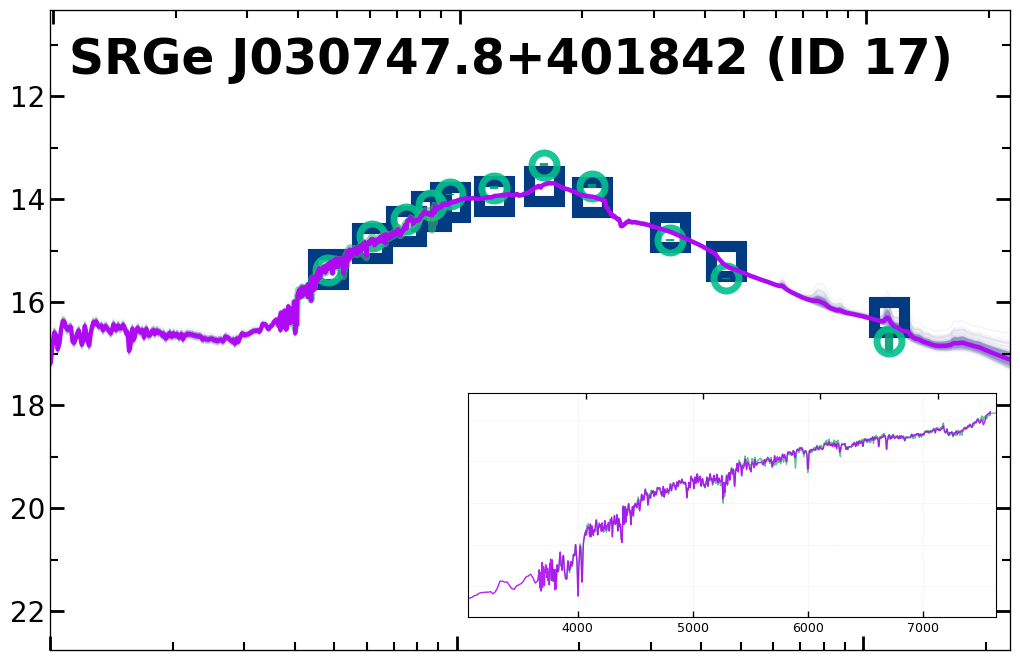}
    \vspace{0.2cm}
    
    \includegraphics[width=0.24\textwidth]{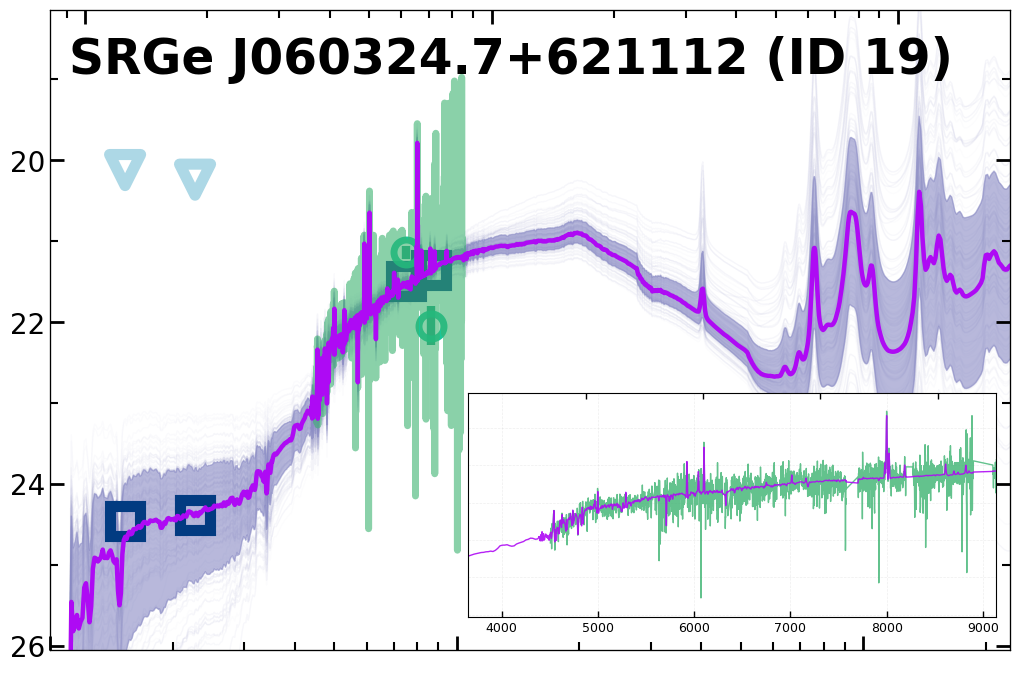}
    \includegraphics[width=0.24\textwidth]{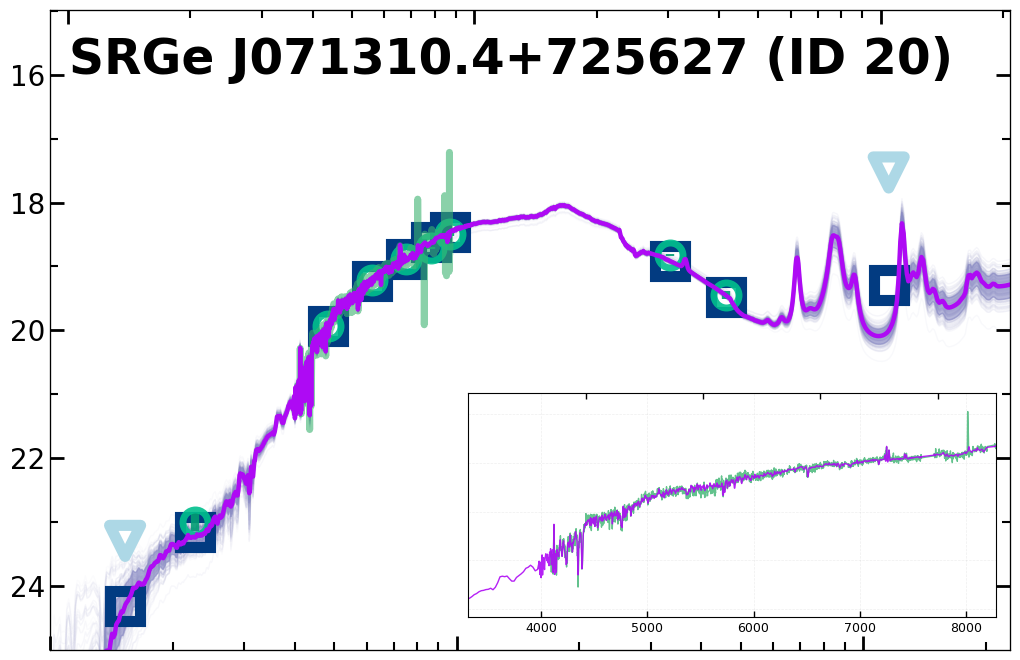}
    \includegraphics[width=0.24\textwidth]{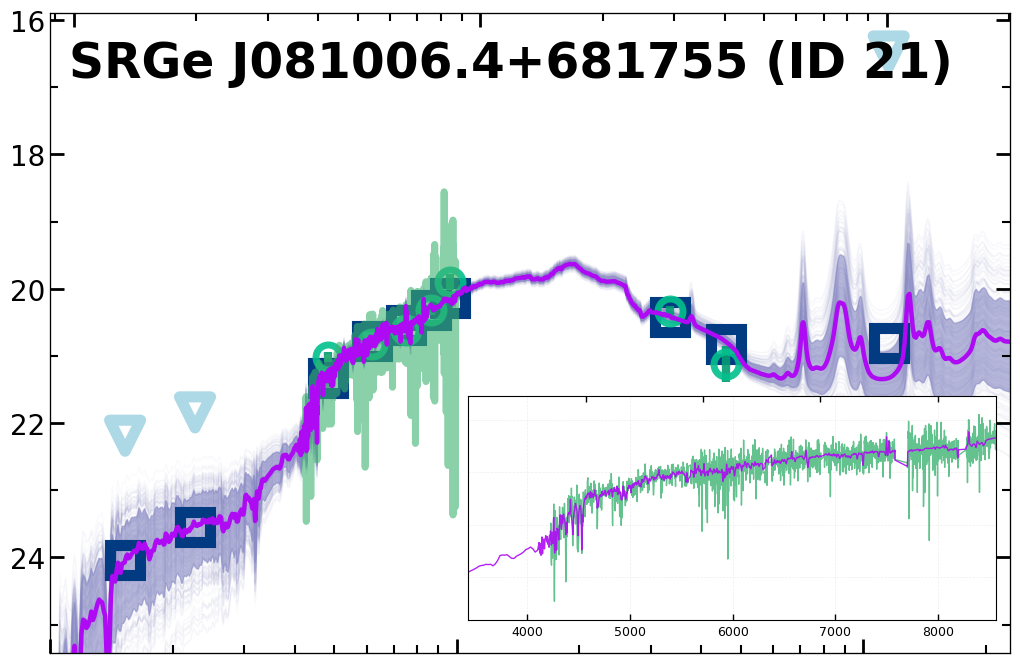}
    \includegraphics[width=0.24\textwidth]{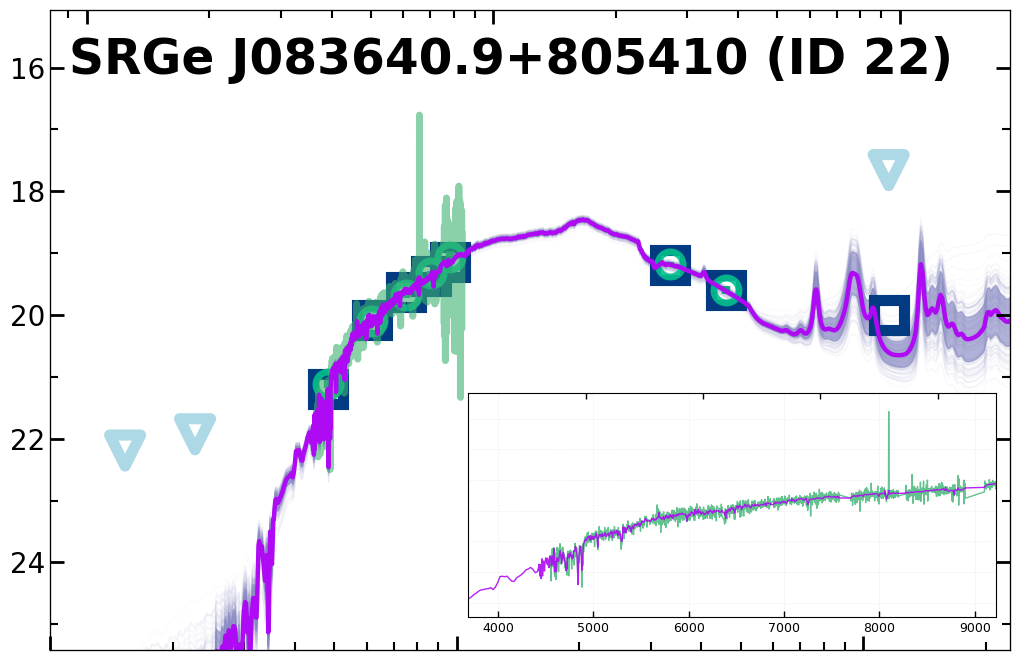}
    \vspace{0.2cm}
    
    \includegraphics[width=0.255\textwidth]{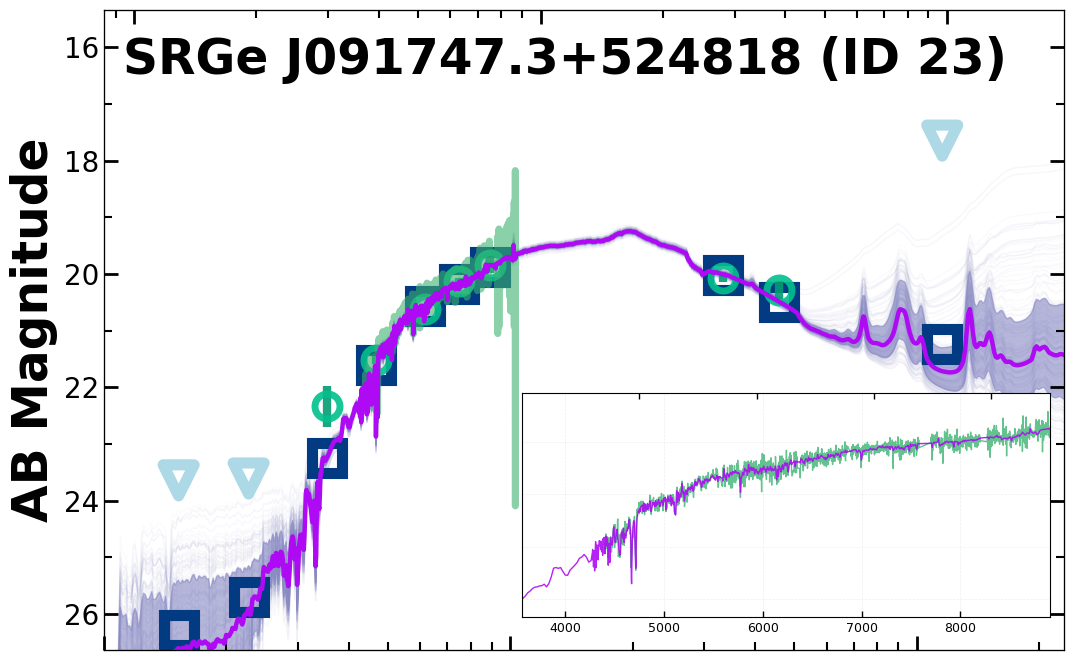}
    \includegraphics[width=0.24\textwidth]{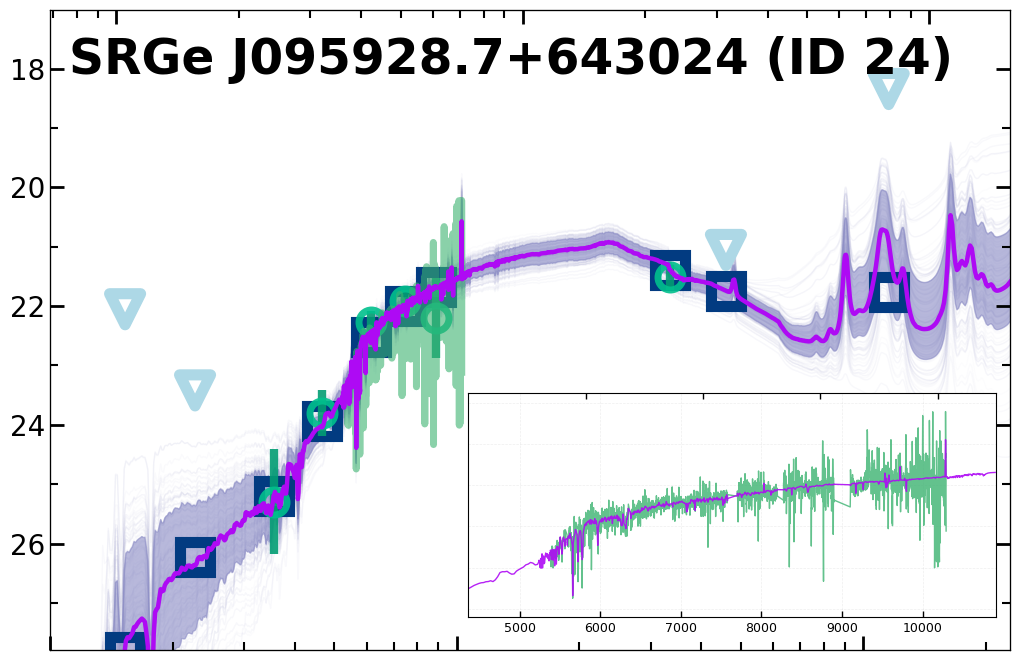}
    \includegraphics[width=0.24\textwidth]{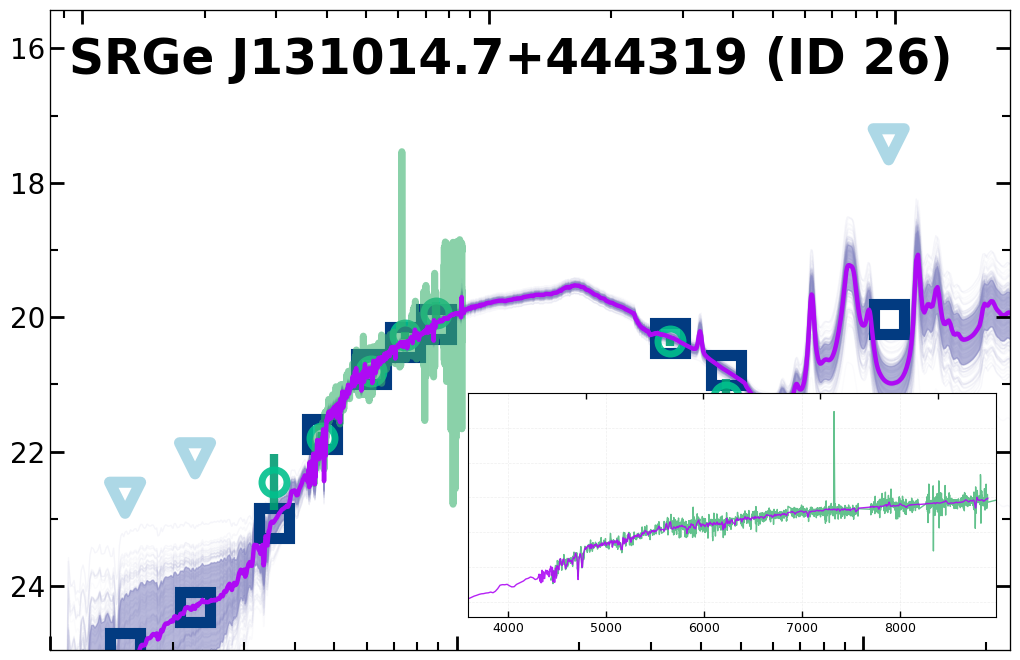}
    \includegraphics[width=0.24\textwidth]{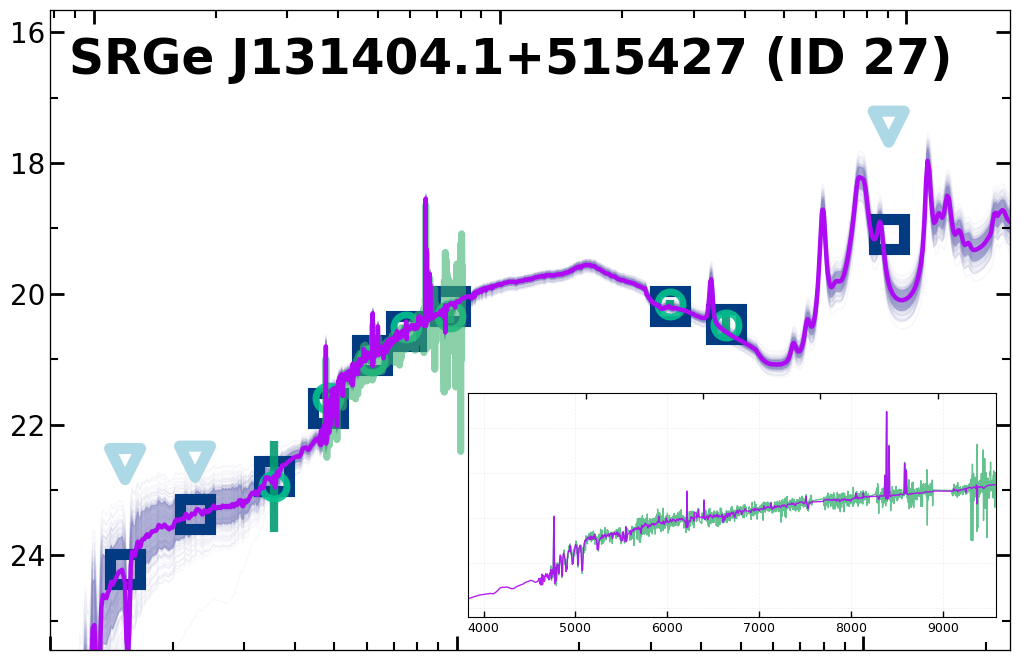}
    \vspace{0.2cm}
    
    \includegraphics[width=0.24\textwidth]{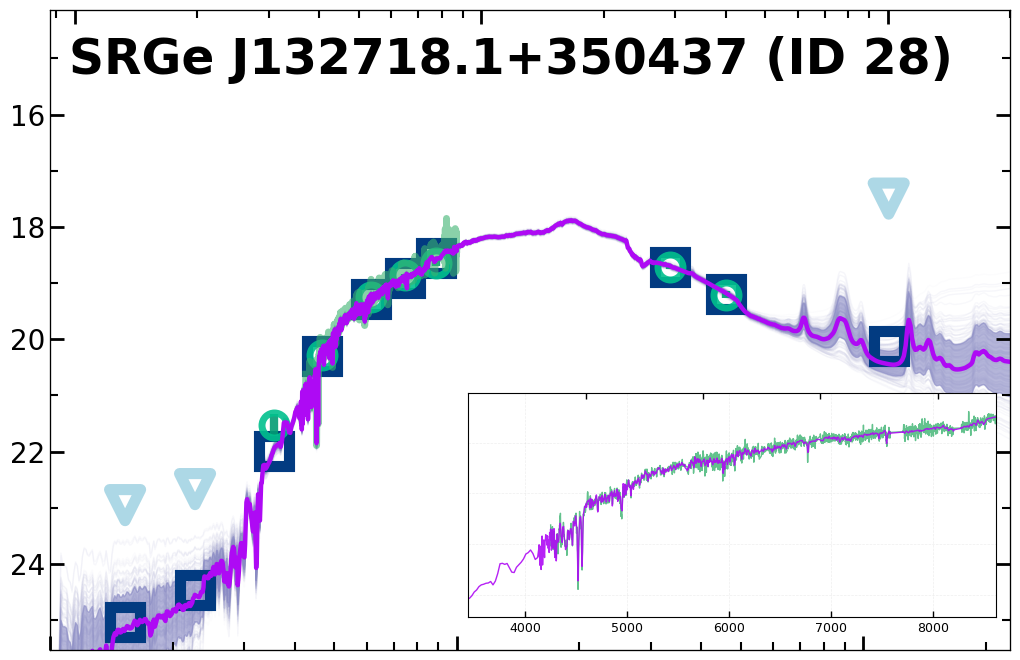}
    \includegraphics[width=0.24\textwidth]{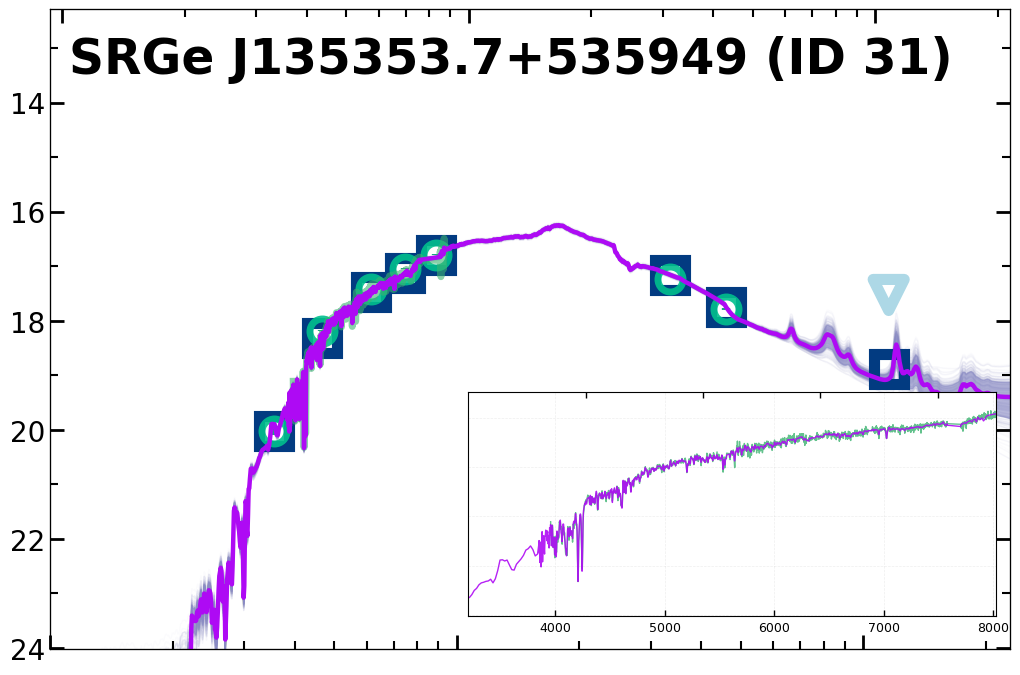}
    \includegraphics[width=0.24\textwidth]{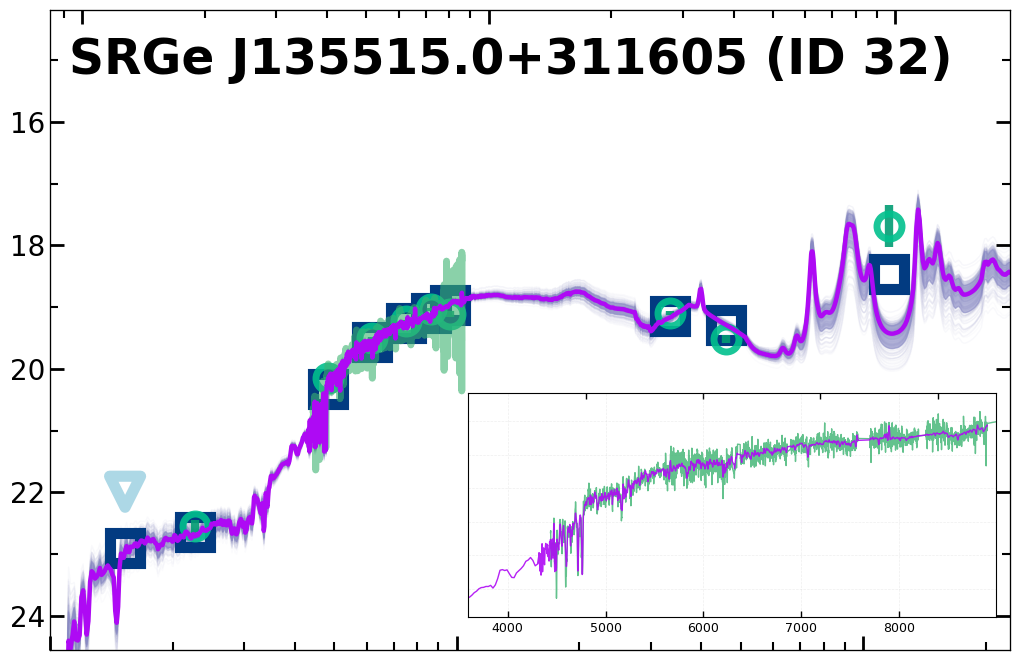}
    \includegraphics[width=0.24\textwidth]{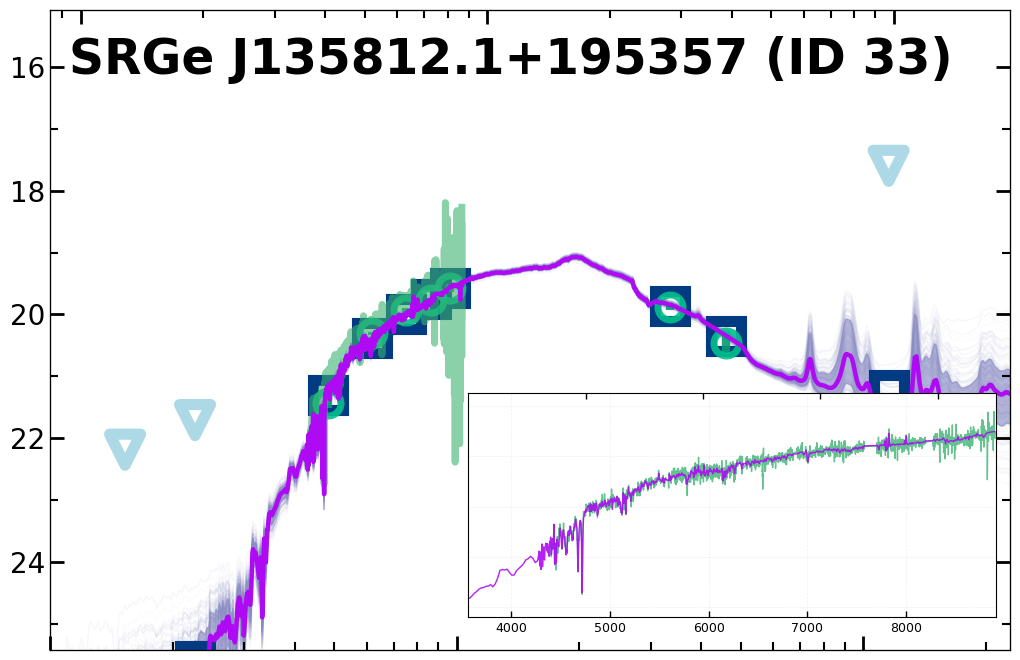}
    \vspace{0.2cm}
    
    \includegraphics[width=0.24\textwidth]{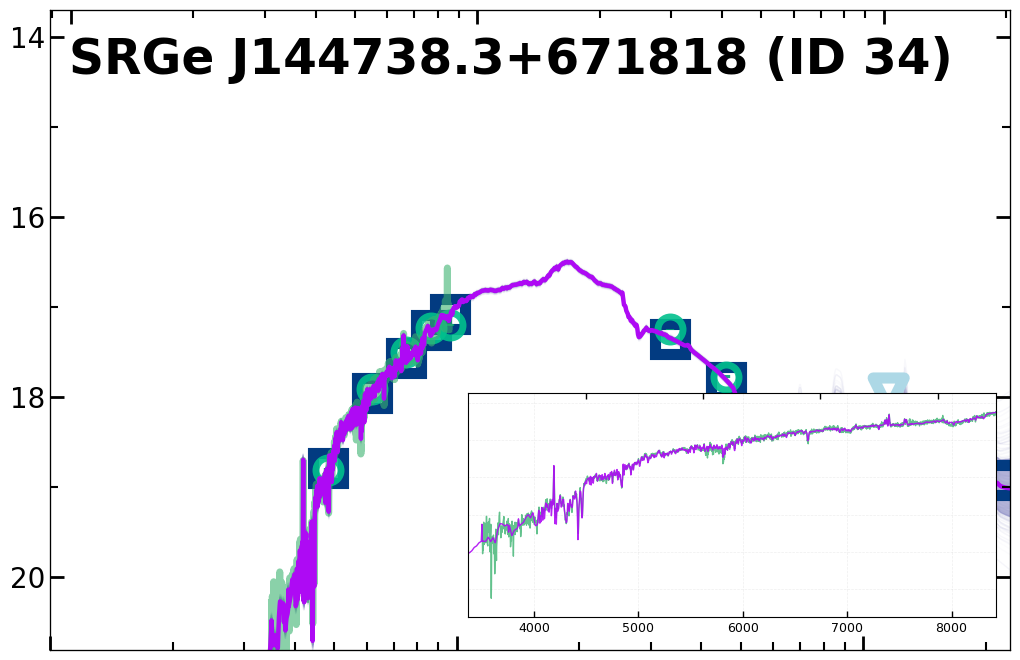}
    \includegraphics[width=0.24\textwidth]{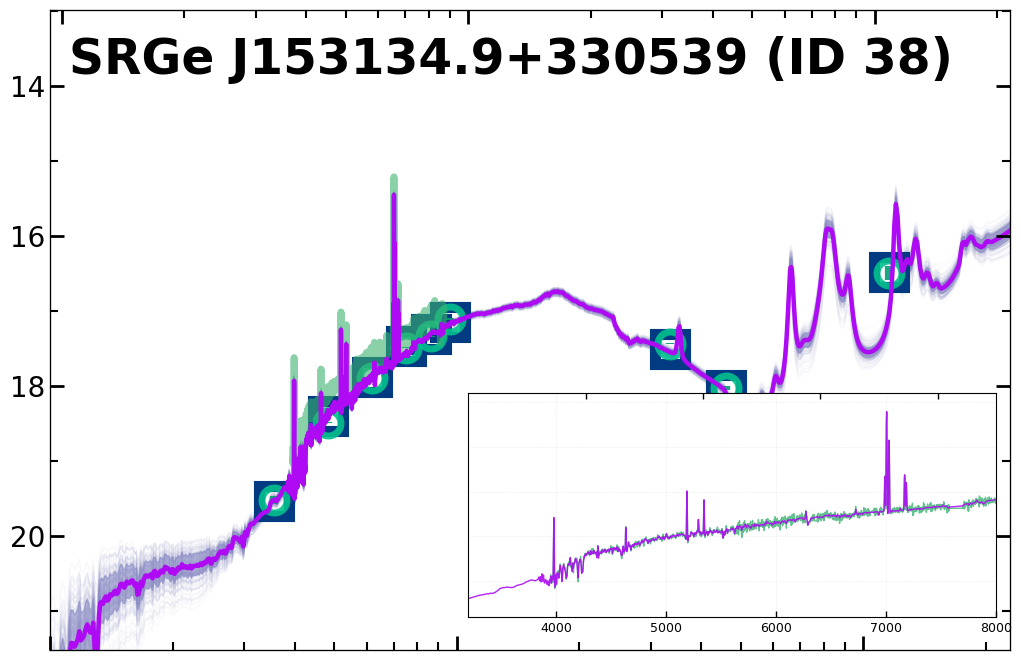}
    \includegraphics[width=0.24\textwidth]{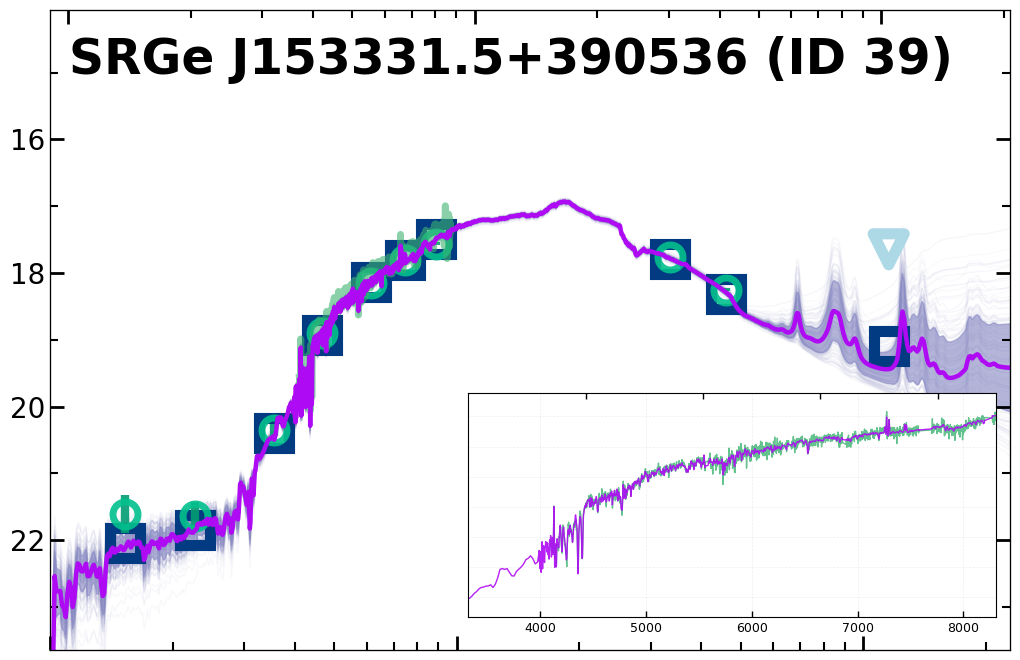}
    \includegraphics[width=0.24\textwidth]{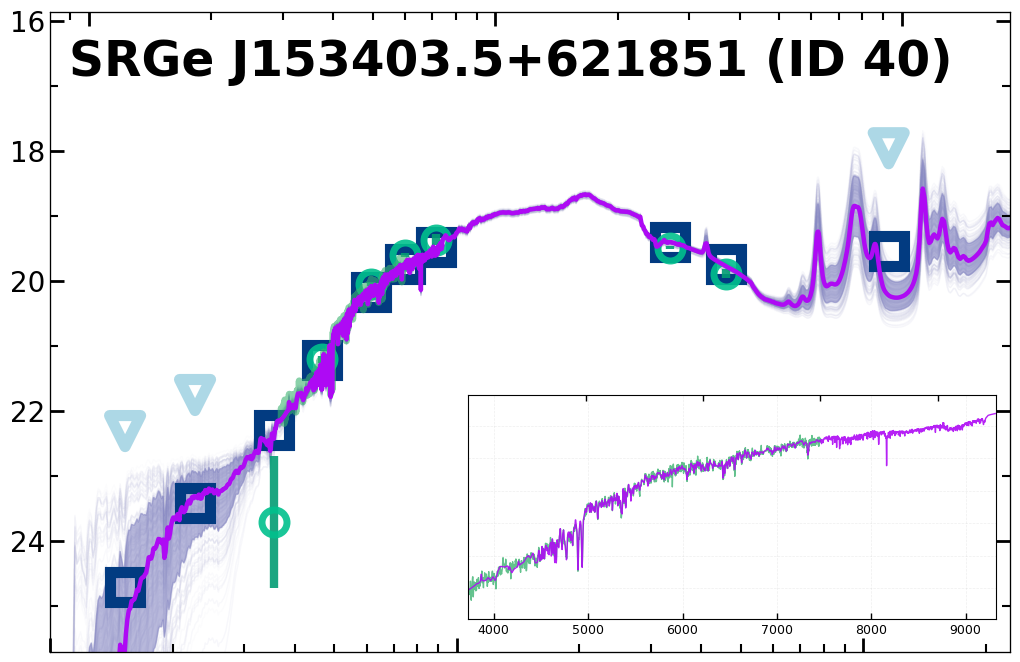}
    \vspace{0.2cm}
    
    \includegraphics[width=0.24\textwidth]{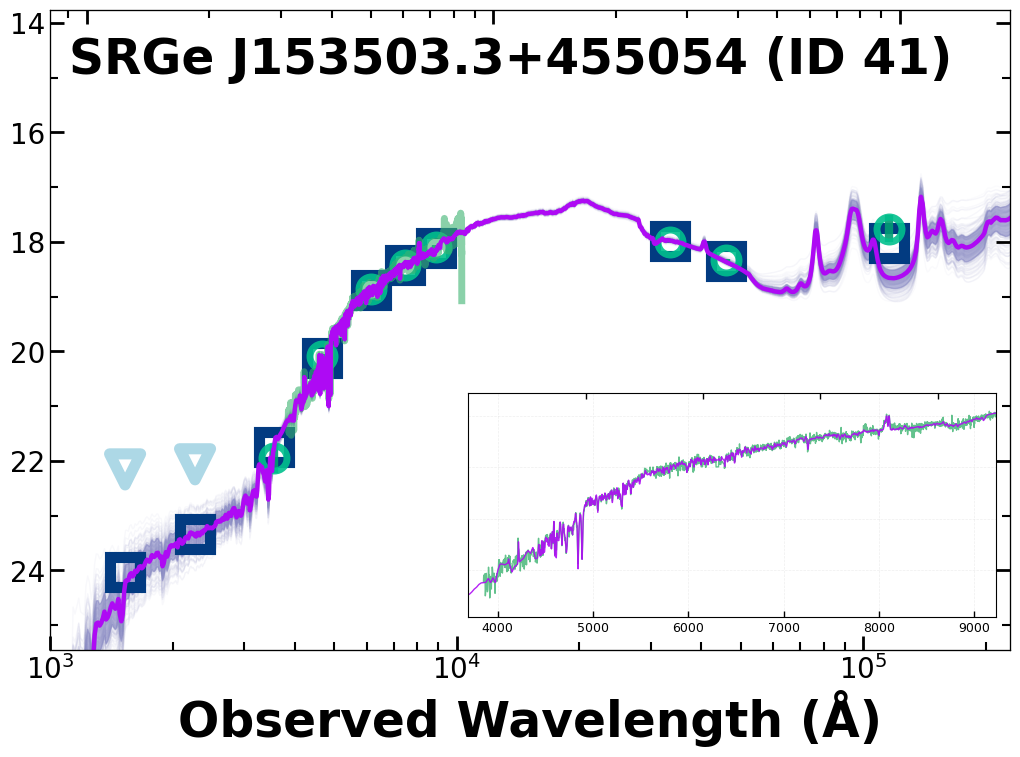}
    \includegraphics[width=0.24\textwidth]{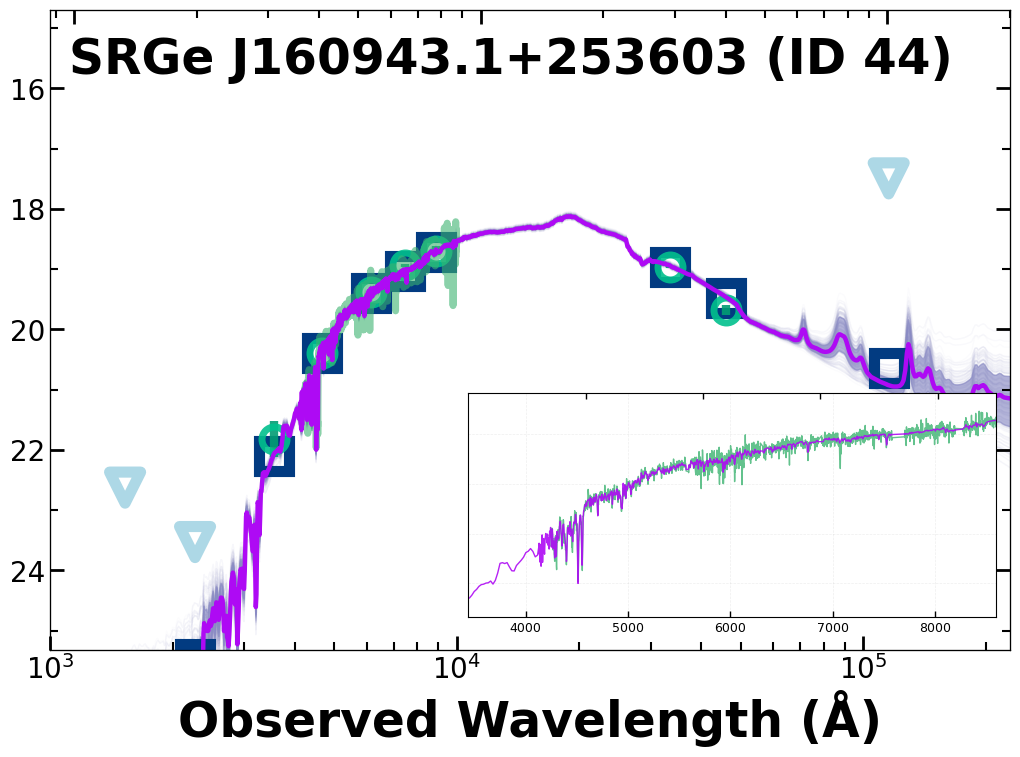}
    \includegraphics[width=0.24\textwidth]{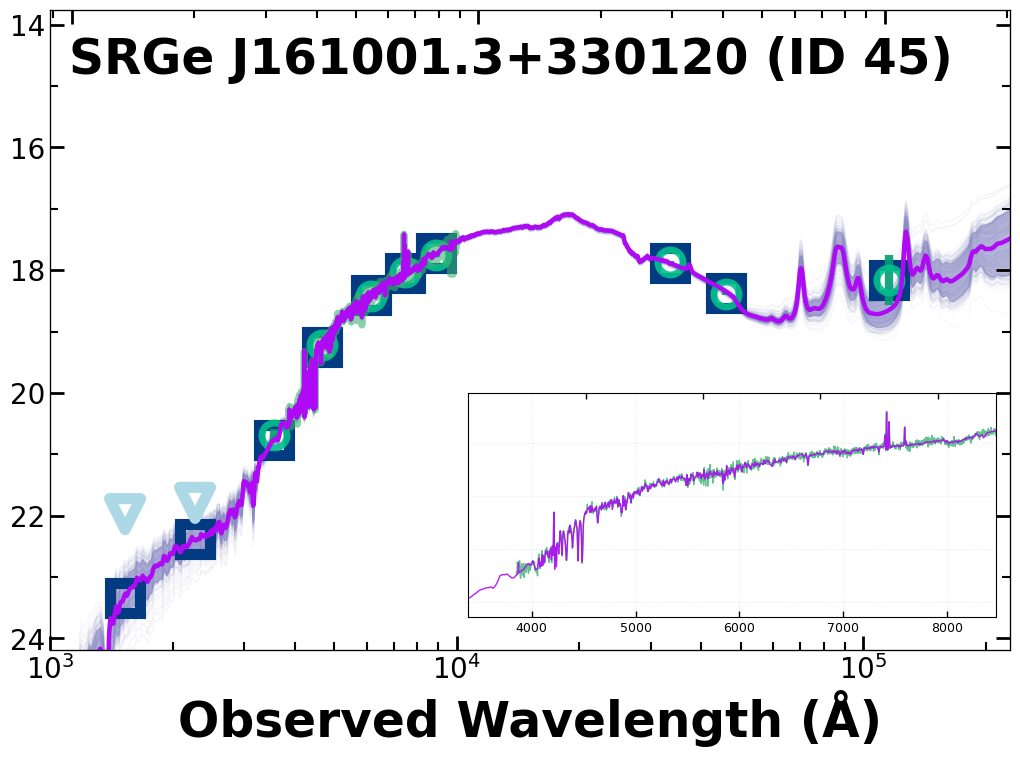}
    \includegraphics[width=0.24\textwidth]{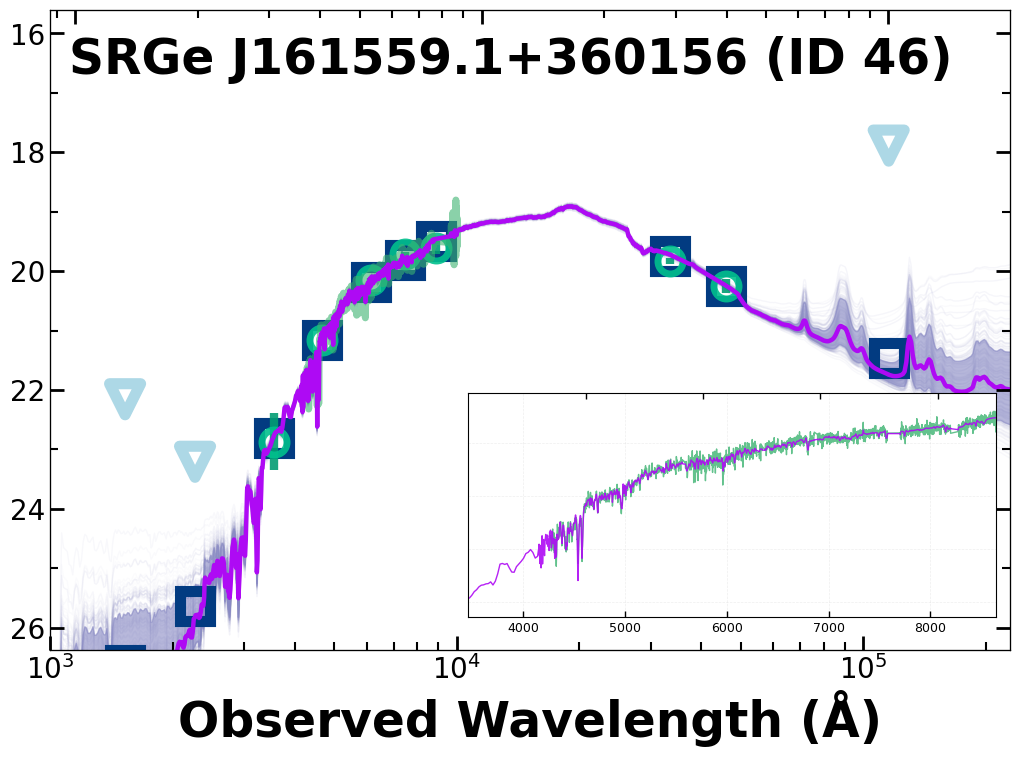}

    \caption{\texttt{Prospector} SED fitting results for our gold and silver sample, and we show the gold sample first. Observed photometry (green circles), upper limits (light blue triangles), and model spectra (purple lines with gray 1σ uncertainty bands) are shown. The inset zooms into the rest-frame 3000-7500\AA\ region (lower axis: observed frame; upper axis: rest frame). SRGe J163831.9+534018 (ID 50) uses photometry-only fitting due to spectroscopic convergence failure. \label{fig:sed_fits}}
\end{figure*}

\begin{figure*}
    \raggedleft

    \includegraphics[width=0.24\textwidth]{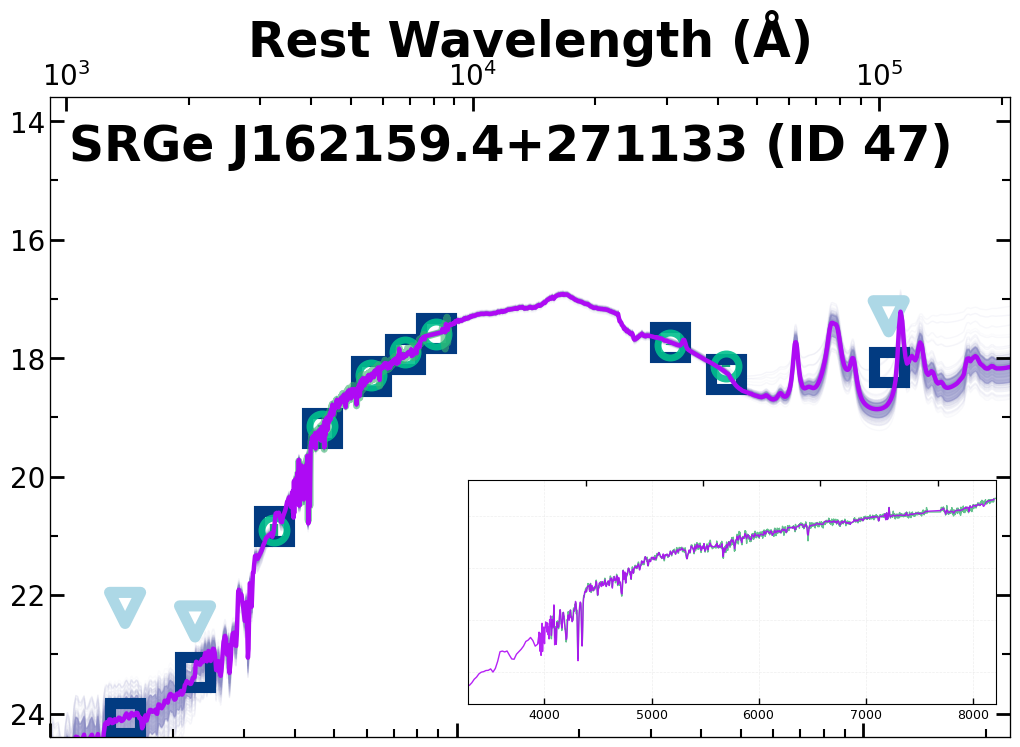}
    \includegraphics[width=0.24\textwidth]{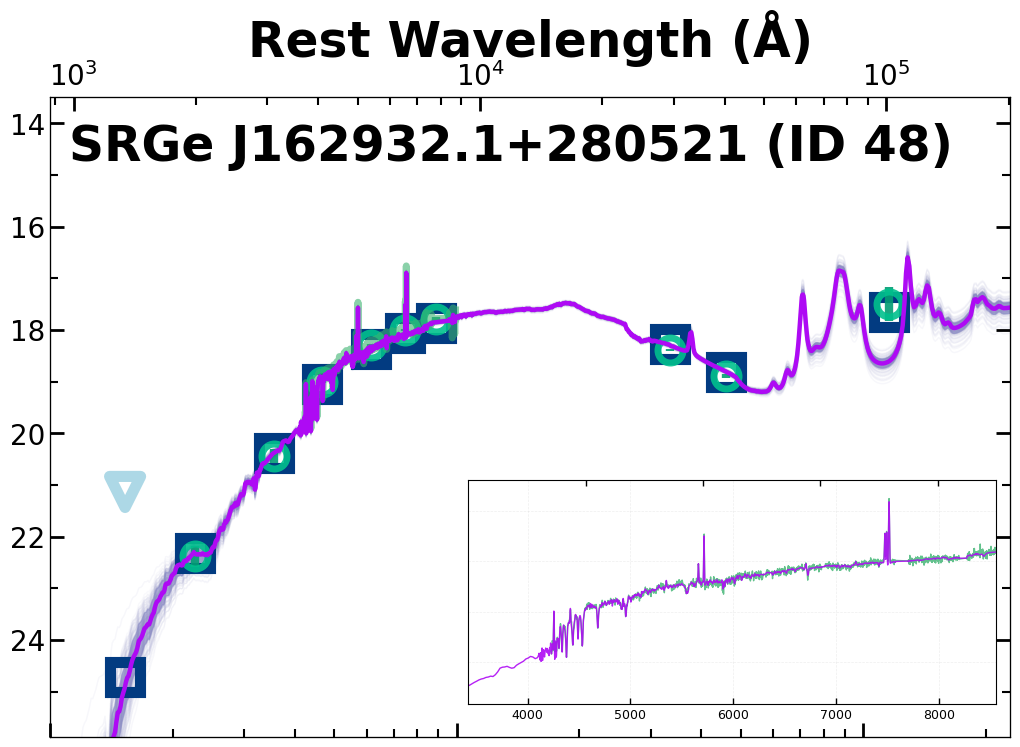}
    \includegraphics[width=0.24\textwidth]{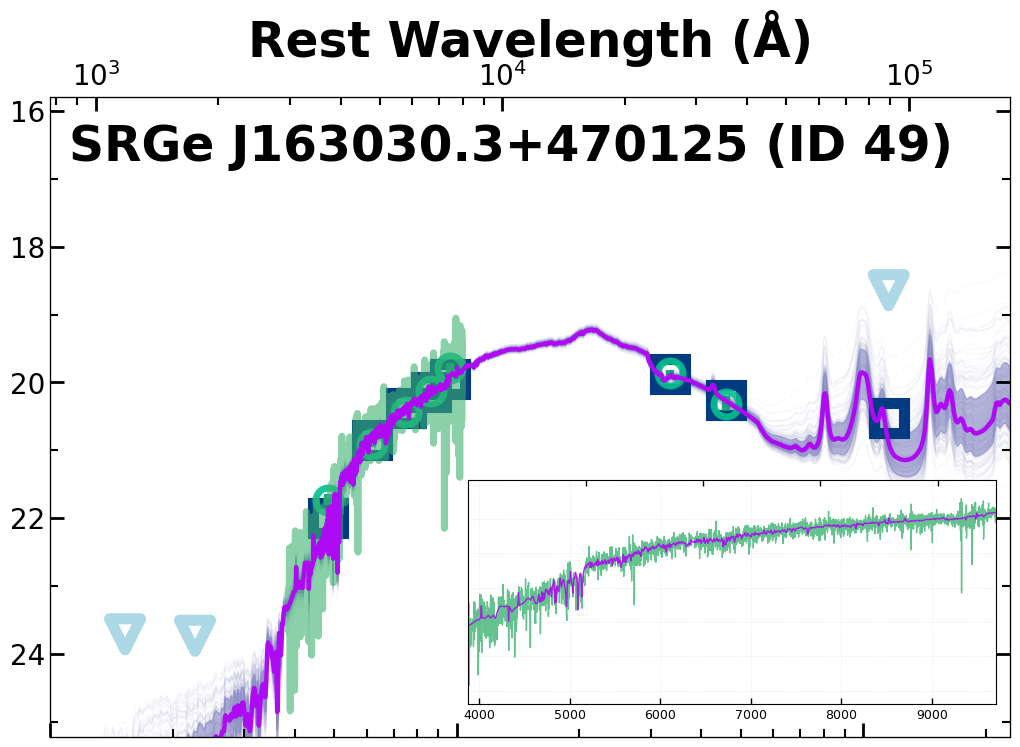}
    \includegraphics[width=0.24\textwidth]{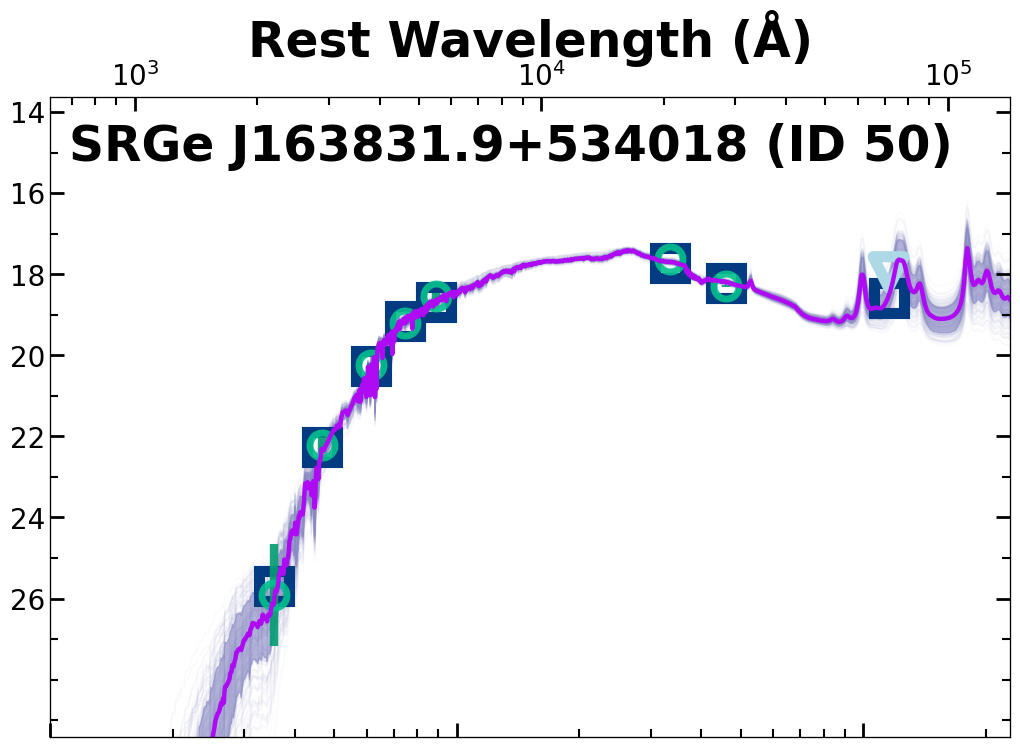}

    \includegraphics[width=0.24\textwidth]{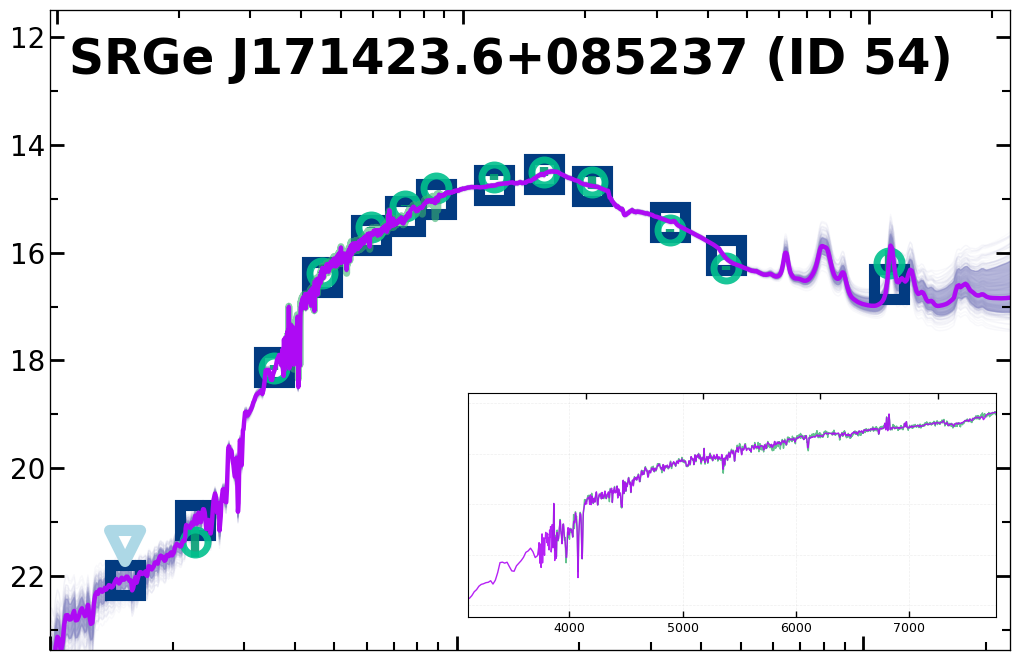}
    \includegraphics[width=0.24\textwidth]{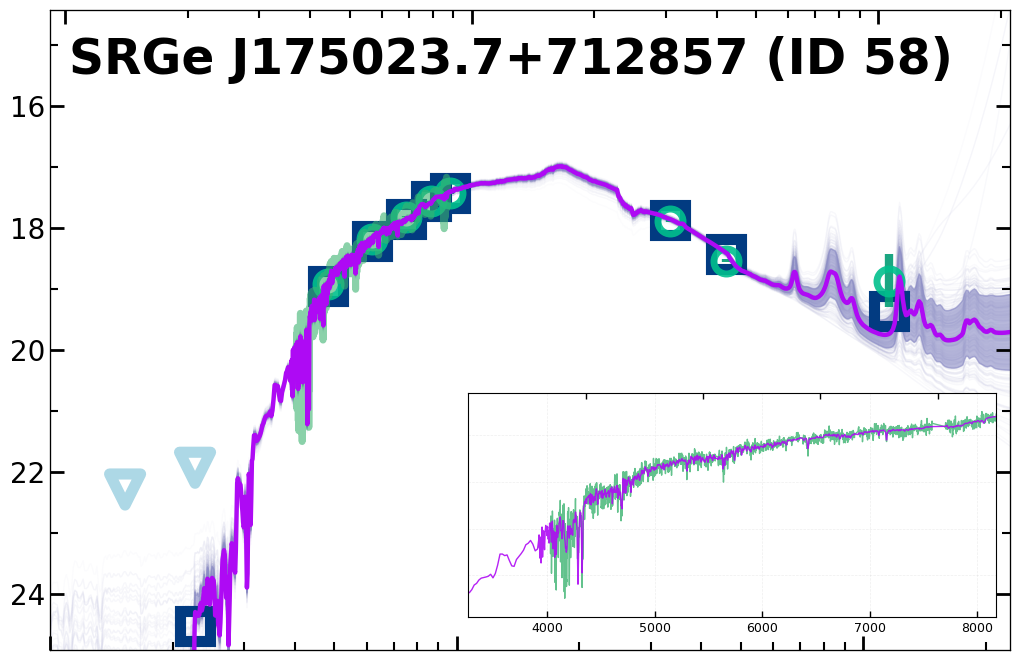}
    \includegraphics[width=0.24\textwidth]{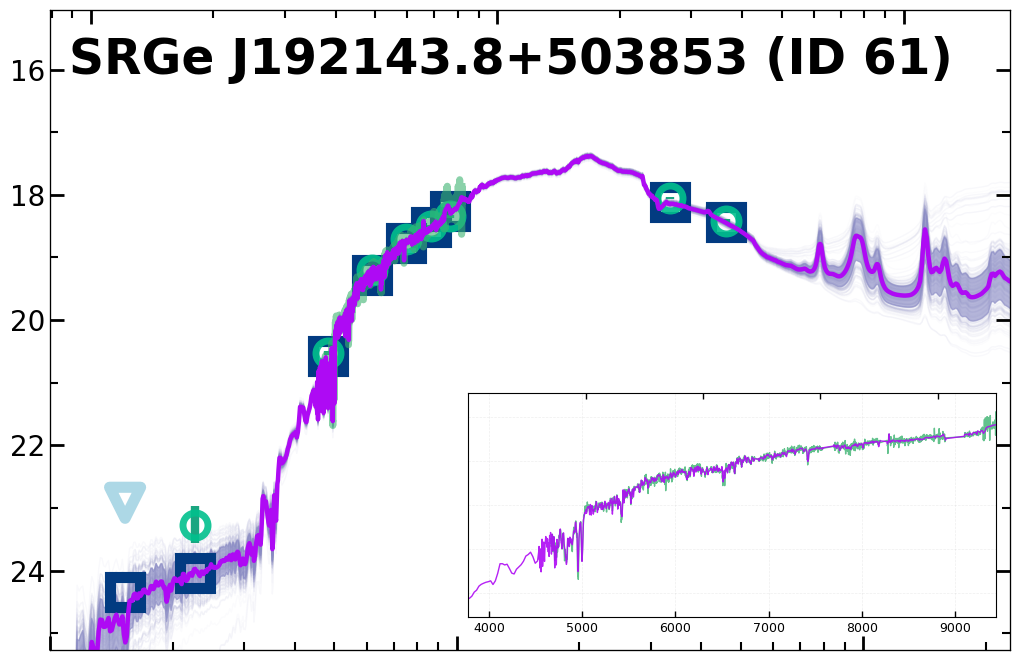}
    \includegraphics[width=0.24\textwidth]{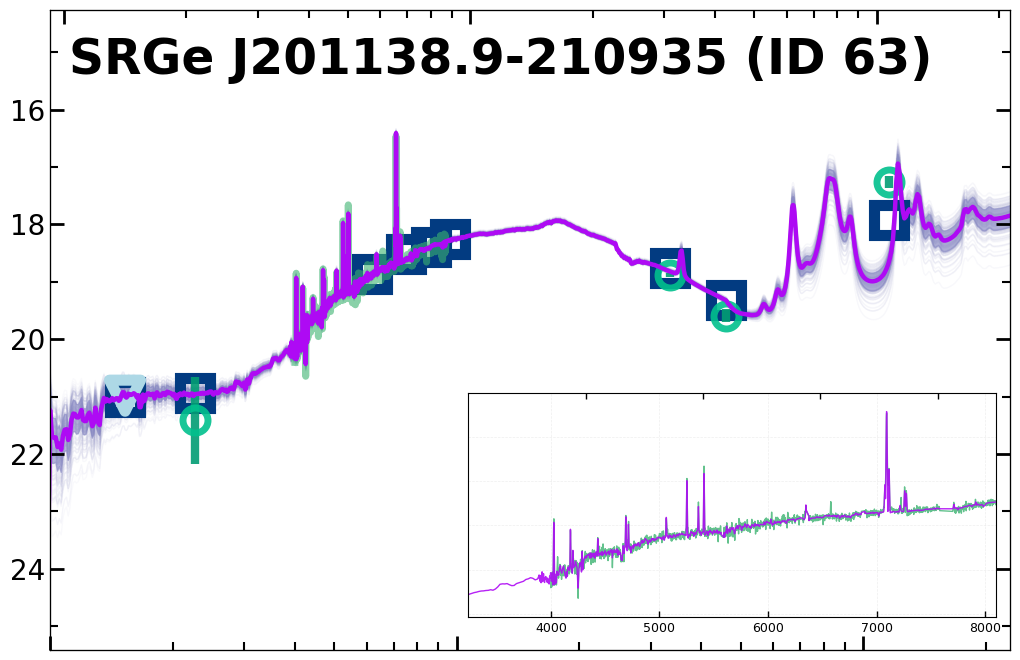}

    \includegraphics[width=0.255\textwidth]{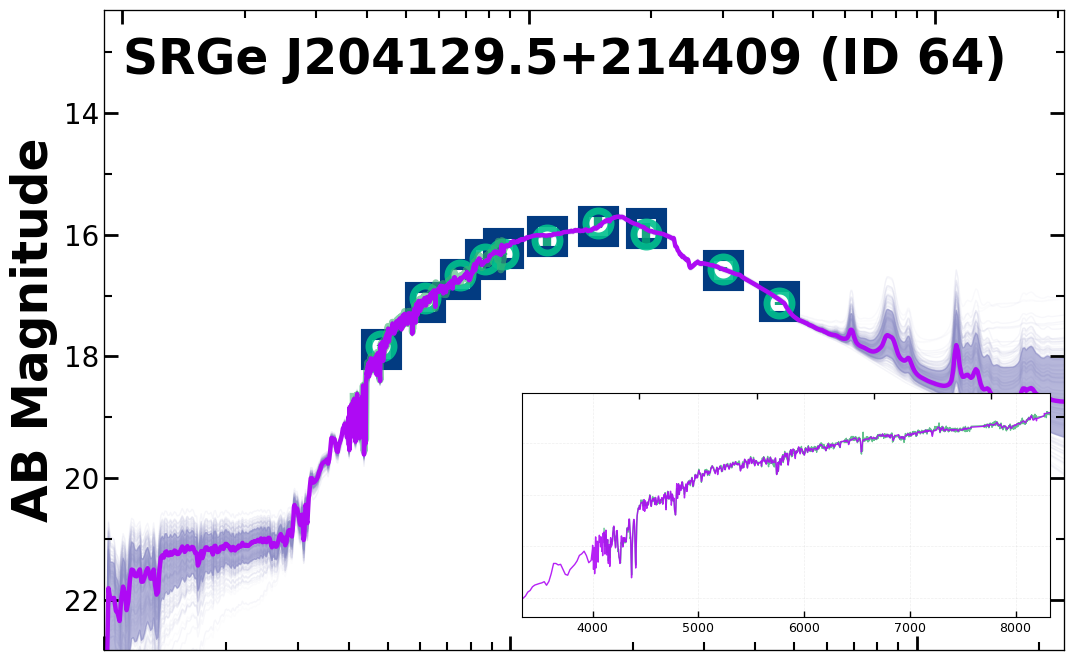}
    \includegraphics[width=0.24\textwidth]{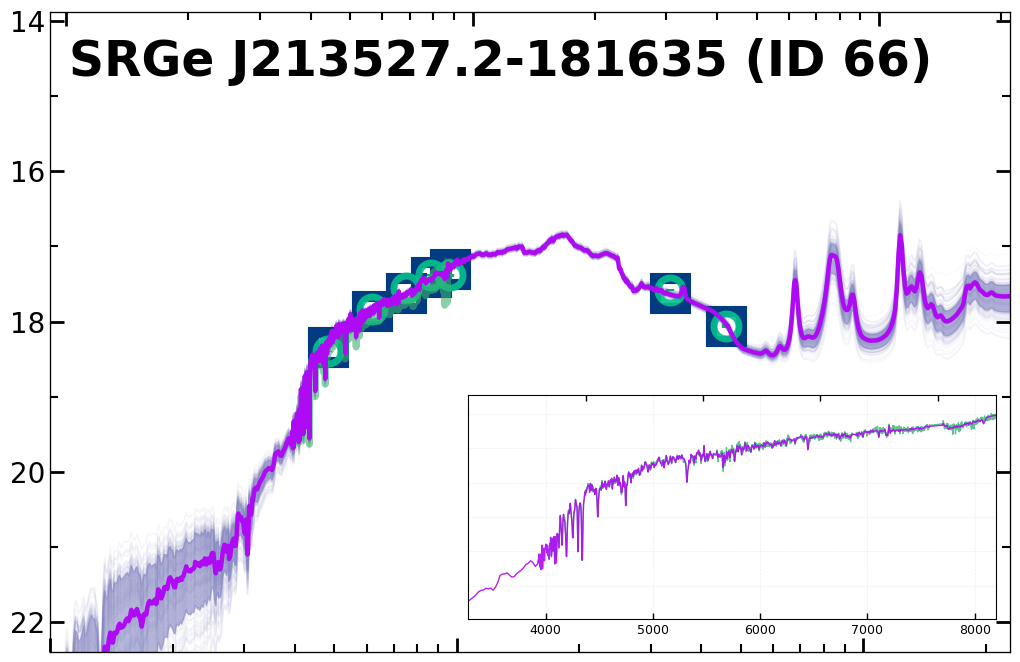}
    \includegraphics[width=0.24\textwidth]{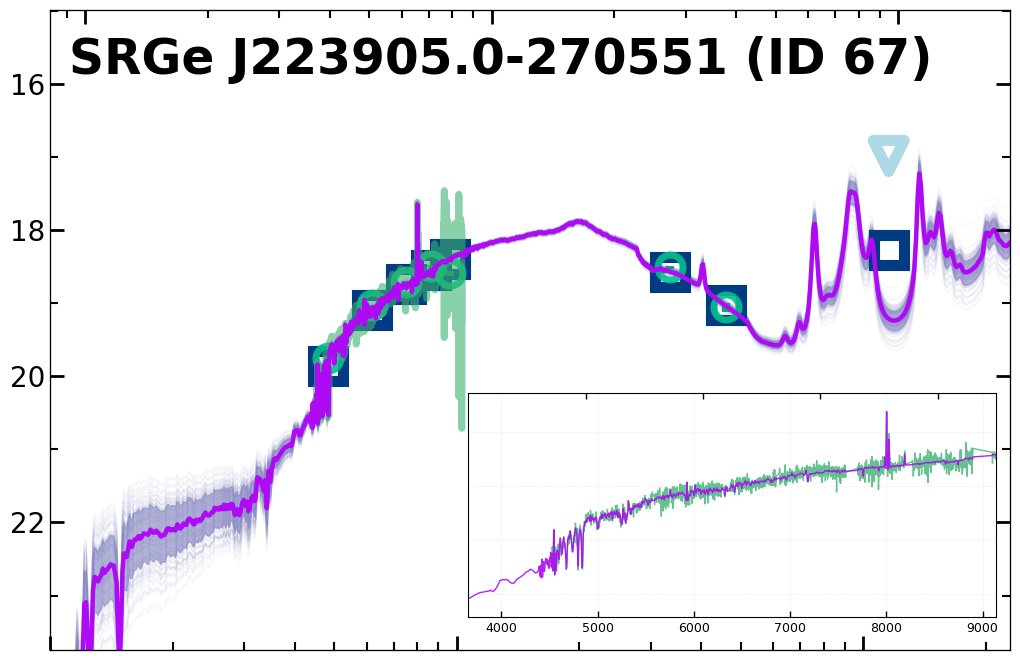}
    \includegraphics[width=0.24\textwidth]{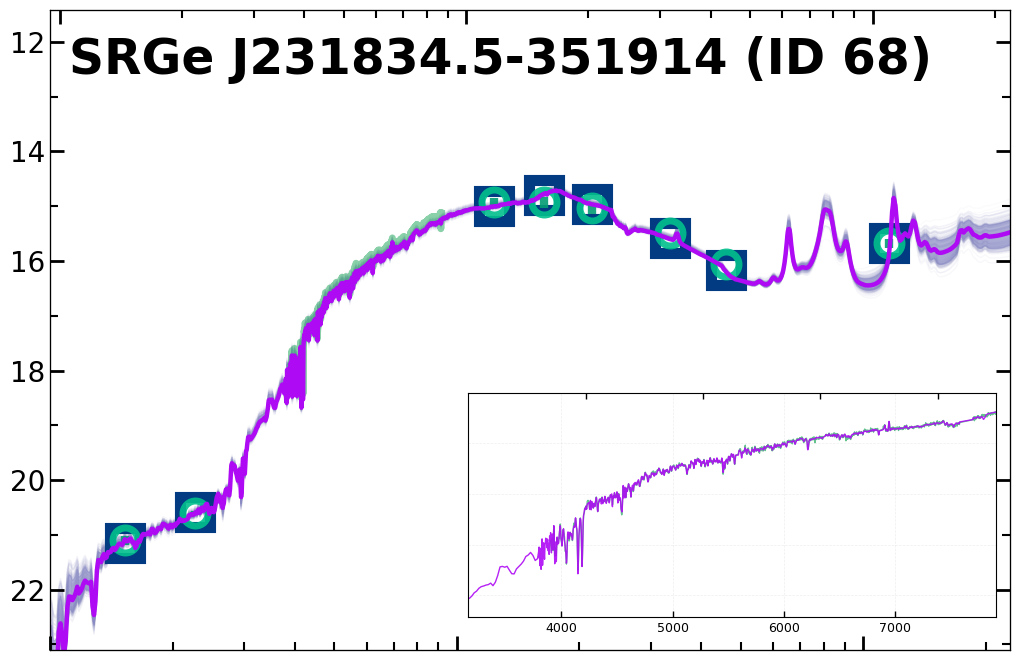}
    
    \includegraphics[width=0.24\textwidth]{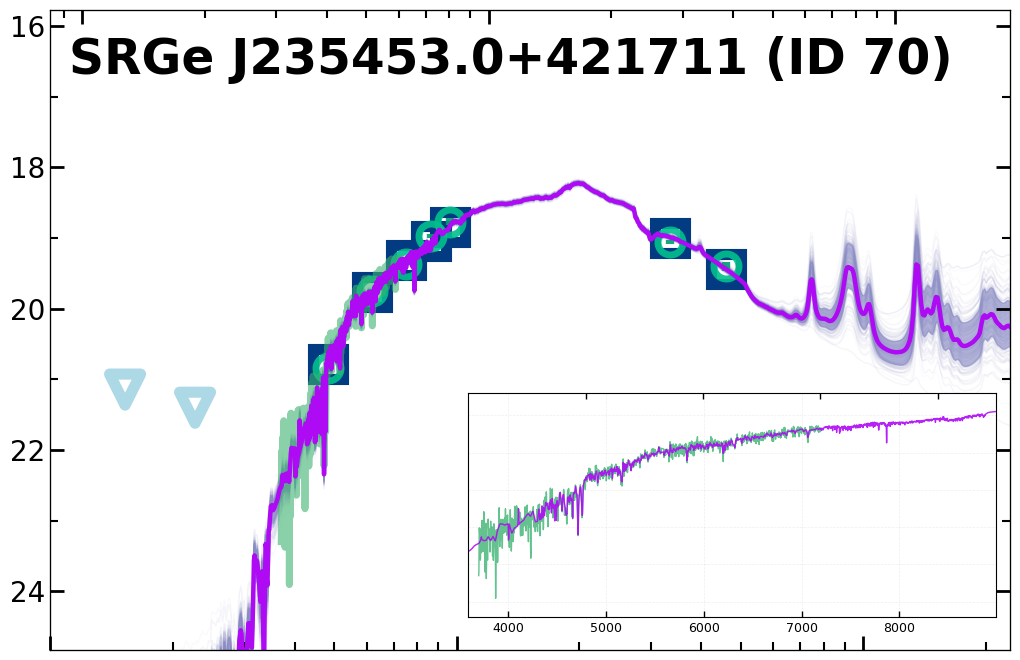}
    \includegraphics[width=0.24\textwidth]{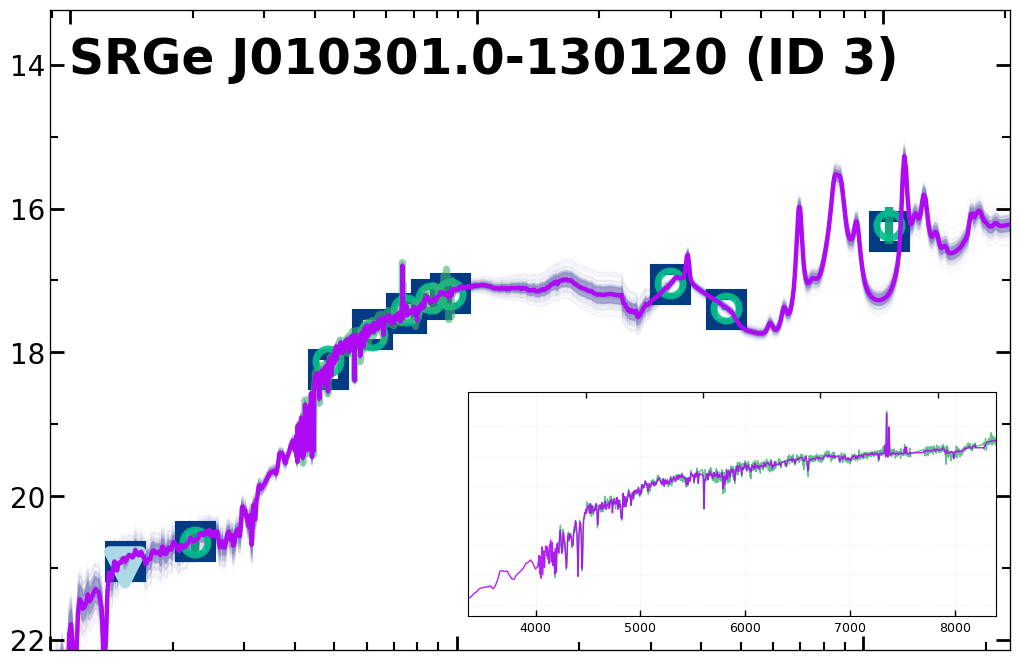}
    \includegraphics[width=0.24\textwidth]{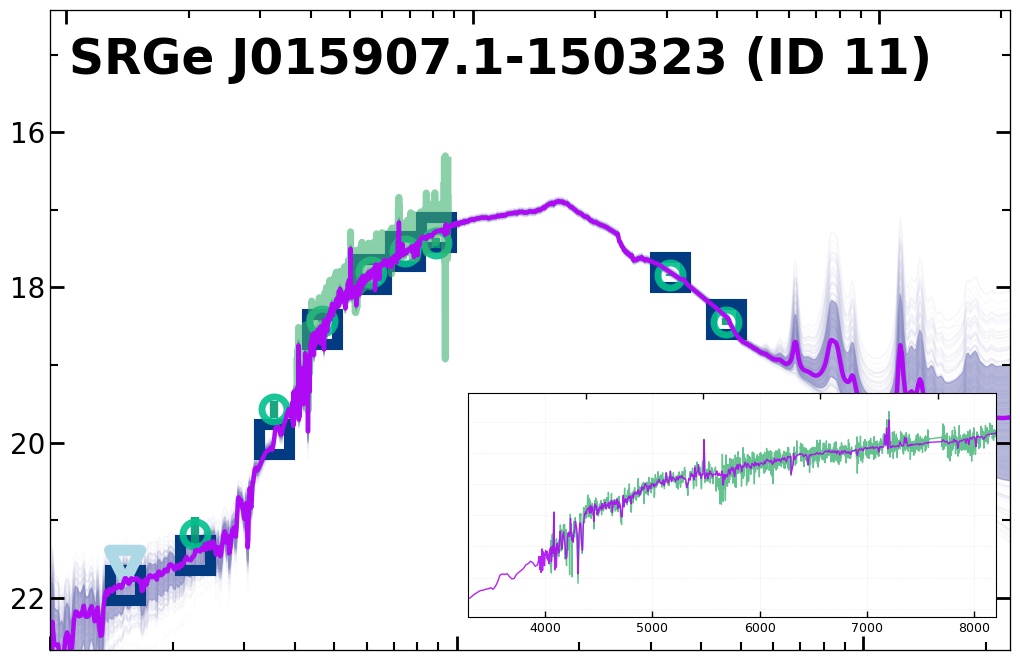}
    \includegraphics[width=0.24\textwidth]{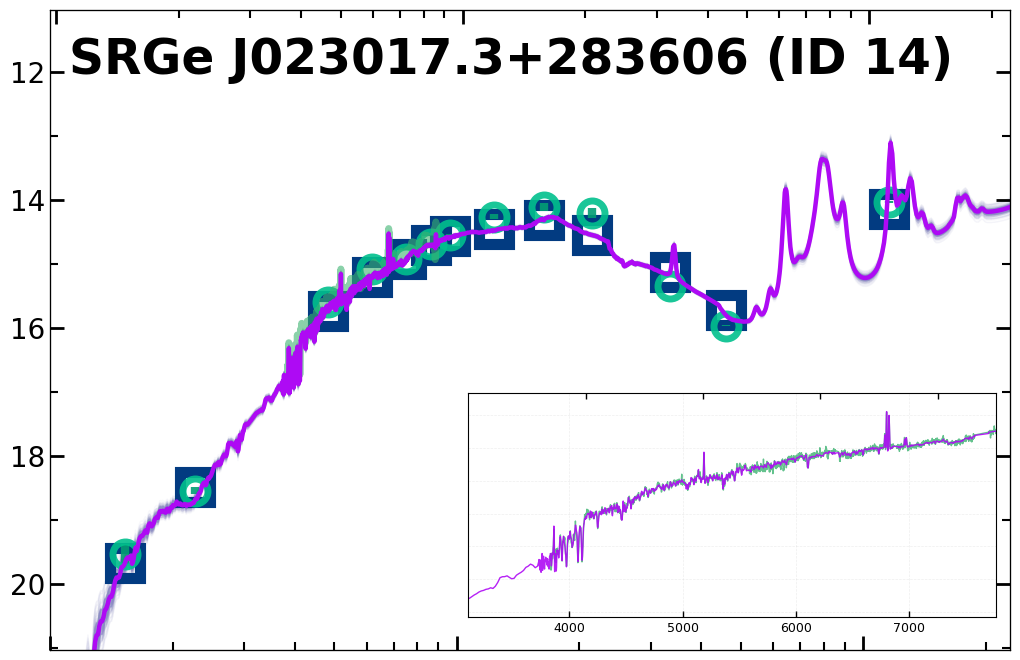}
    
    \includegraphics[width=0.24\textwidth]{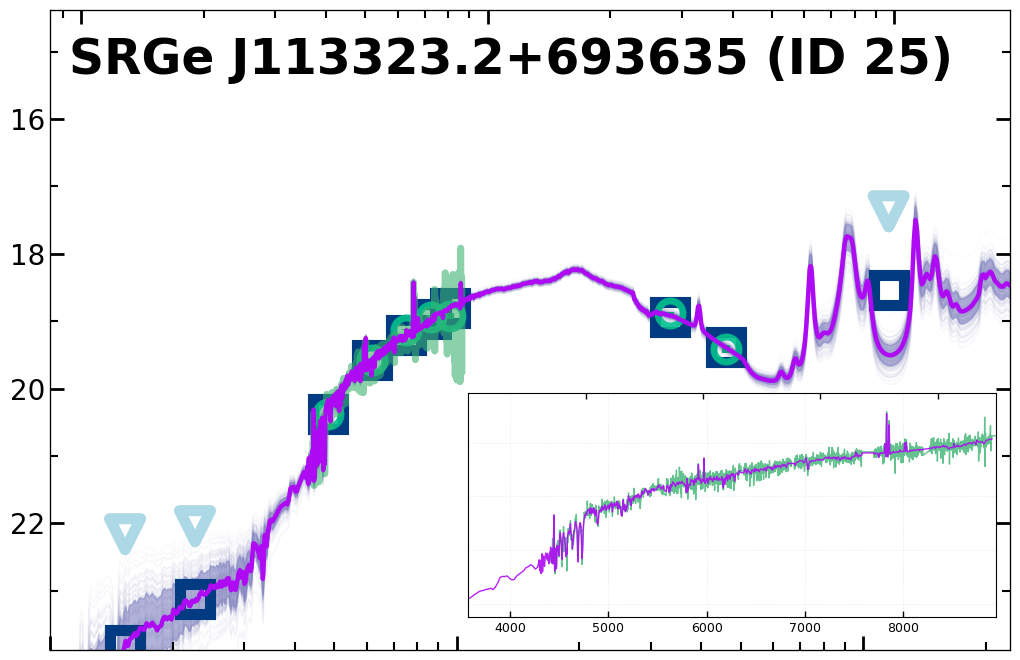}
    \includegraphics[width=0.24\textwidth]{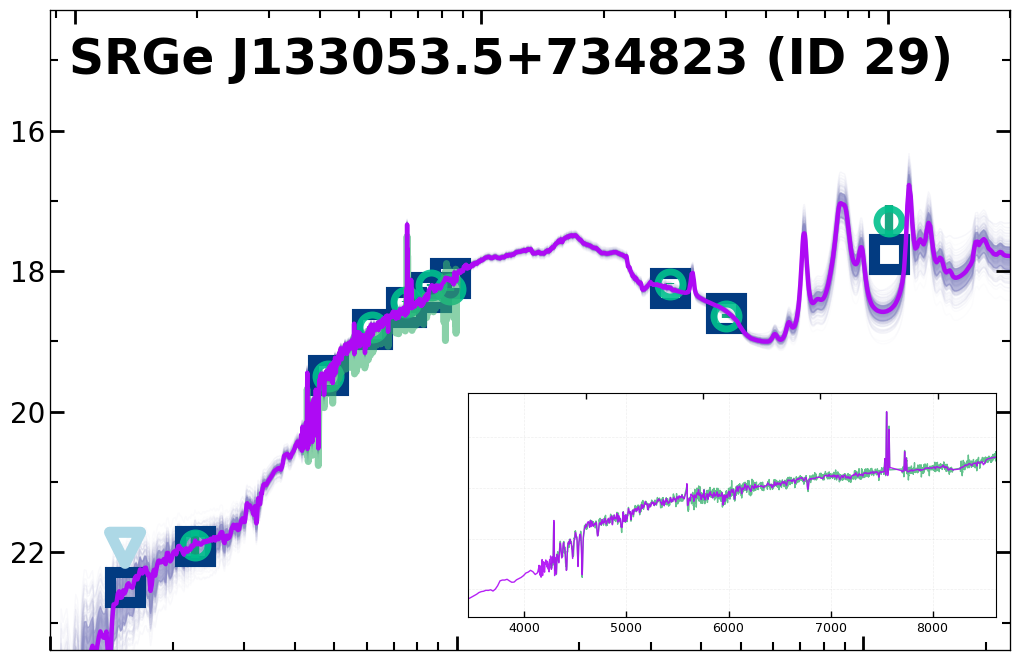}
    \includegraphics[width=0.24\textwidth]{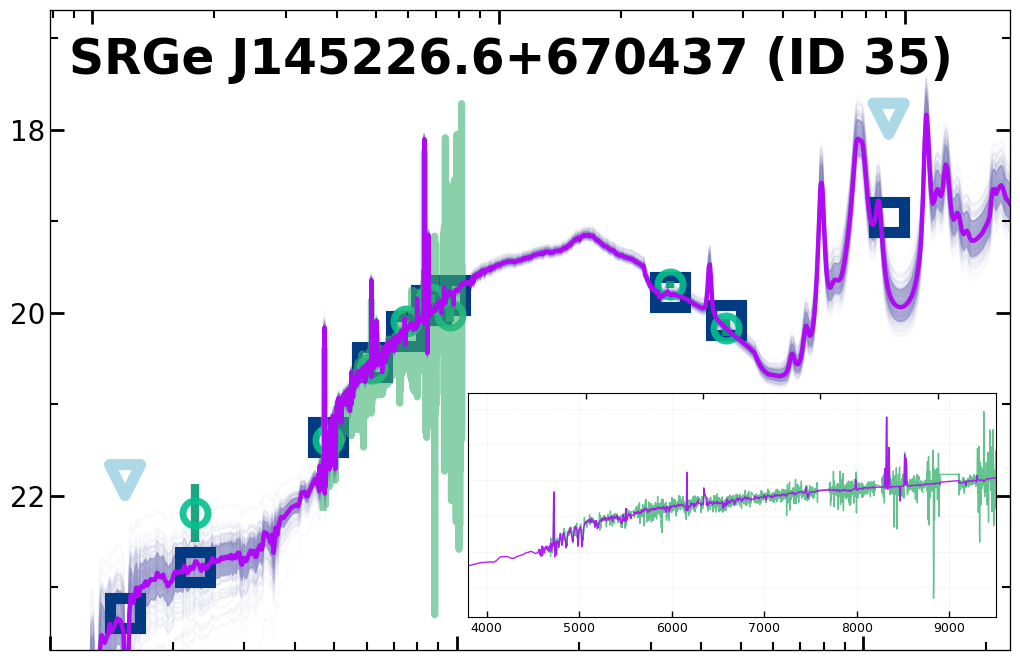}
    \includegraphics[width=0.24\textwidth]{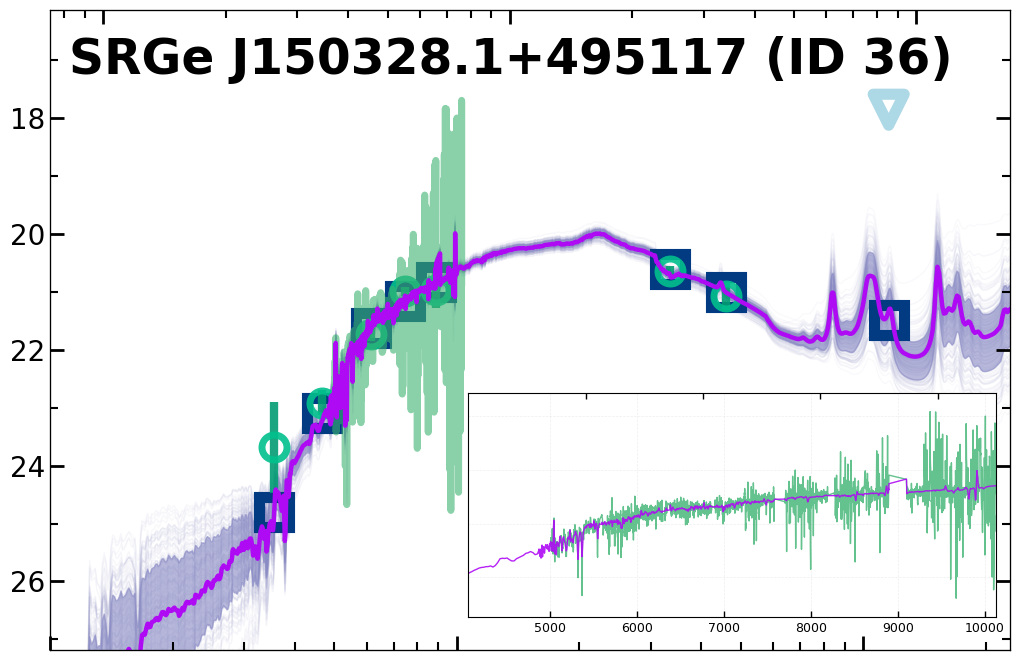}
    
    \includegraphics[width=0.24\textwidth]{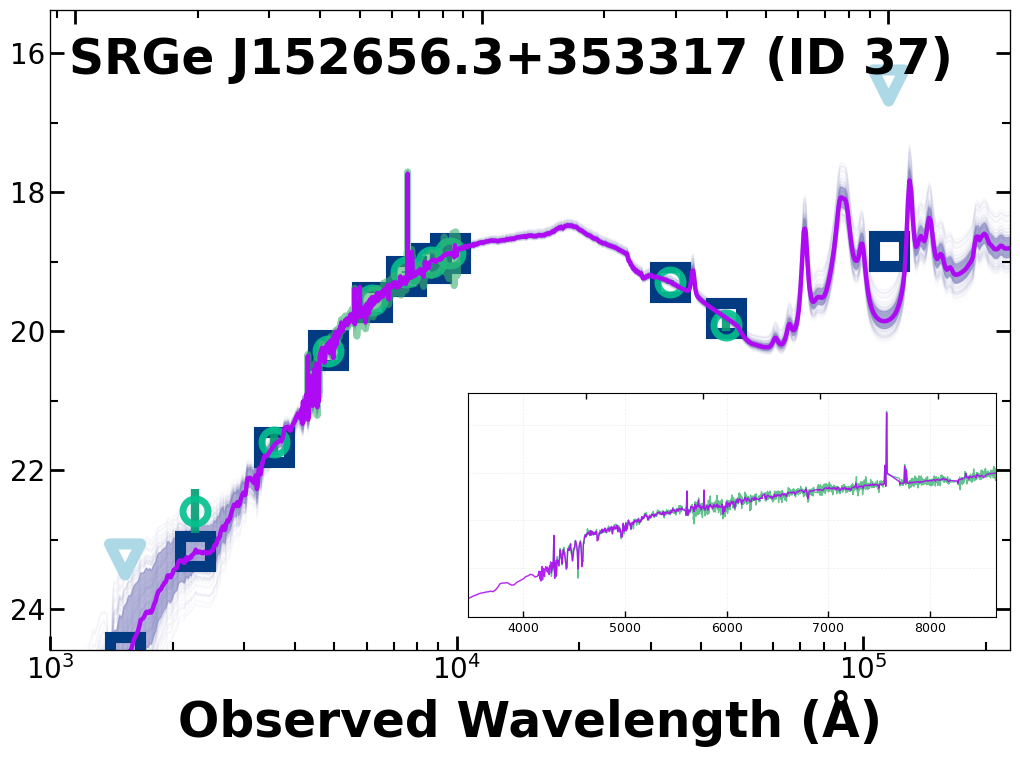}
    \includegraphics[width=0.24\textwidth]{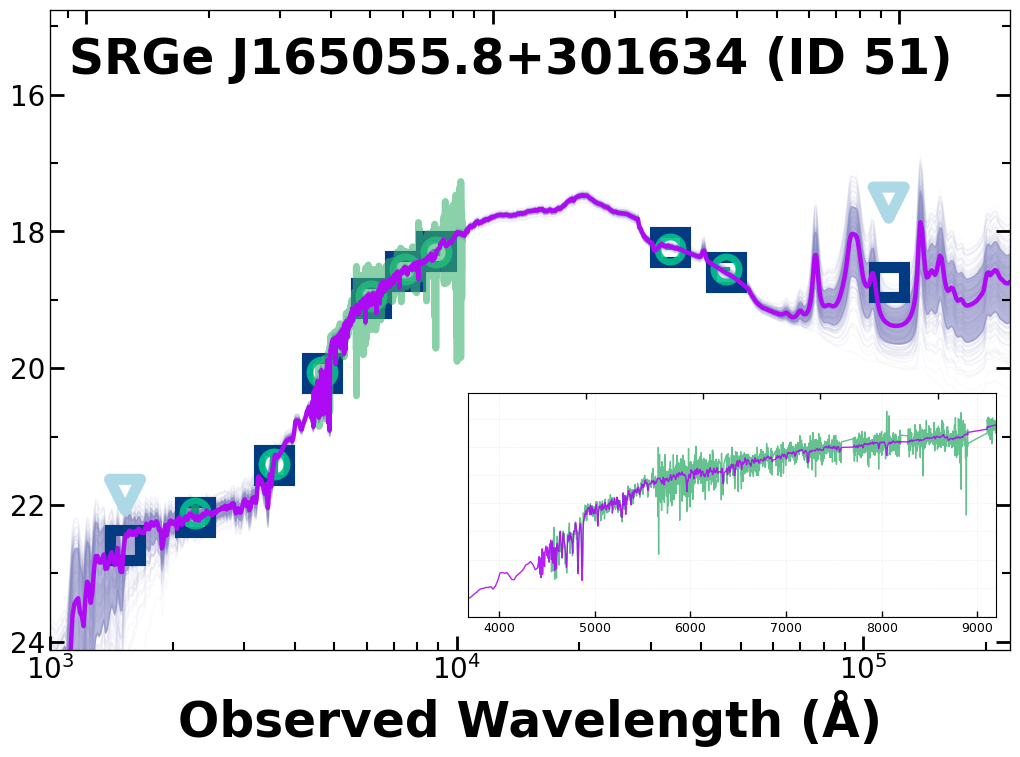}
    \includegraphics[width=0.24\textwidth]{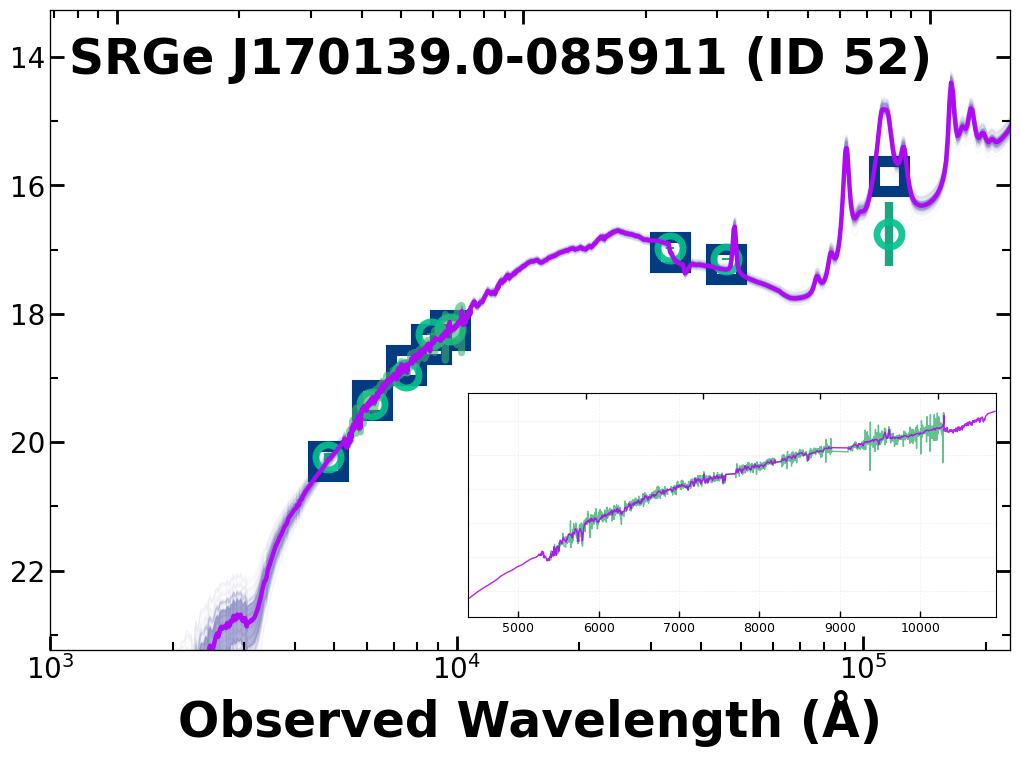}
    \includegraphics[width=0.24\textwidth]{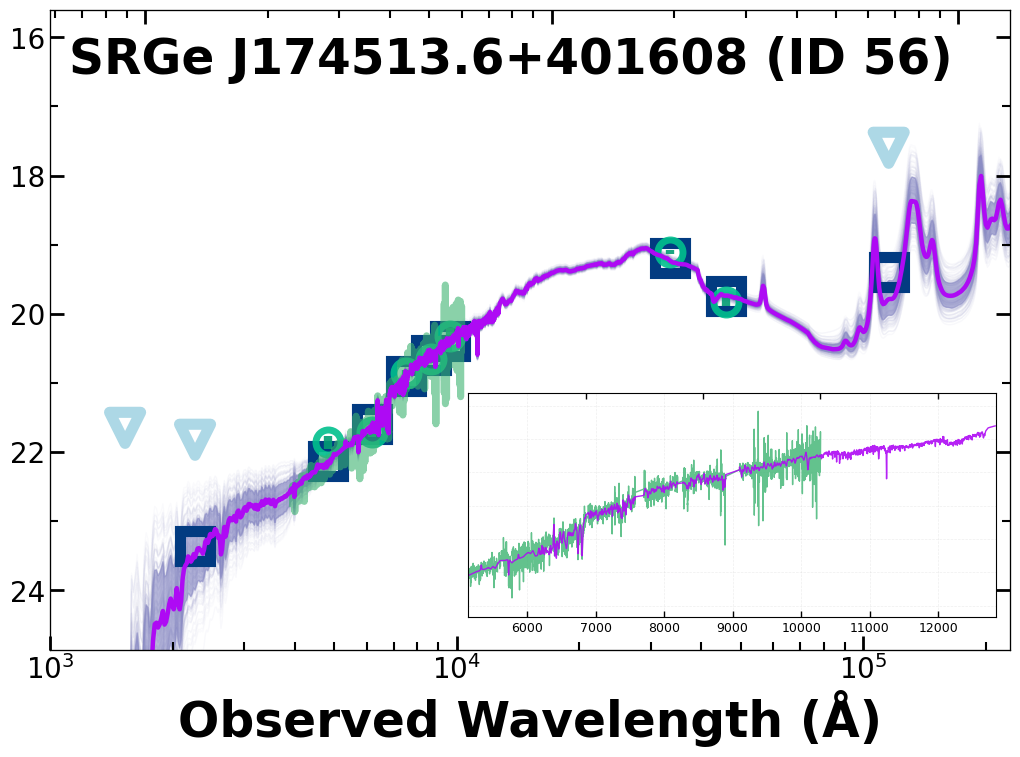}
    
\caption{Continued figure of Fig.~\ref{fig:sed_fits}. \label{fig:sed_fits_2}}
\end{figure*}

\section{Supplementary Tables} \label{sec:sup_tab}
Table~\ref{tab:he_h_flux} shows the properties of broad TDE emission lines detected in our Keck-I/LRIS spectra.

\begin{table*}[htbp]
\caption{Properties of emission lines in TDE spectra modeled with Gaussian profiles. \label{tab:he_h_flux}}
\small
\begin{center}
\begin{tabular}{clccccc}
  \toprule
  TDE class & Phase & Line & $\lambda_0$ & EW & FWHM & Luminosity \\
   & (days) &  & (\AA) & (\AA) & (km s$^{-1}$) & ($10^{38}$ erg s$^{-1}$) \\
  \midrule
  \multicolumn{7}{l}{\textbf{SRGe J011603.1+072258 (ID 5)}} \\
  TDE-H+He & 40 & He\,II $\lambda$4686 & 4685.7  & 72.68 $\pm$13.24 & 19381 $\pm$969  & 1344 $\pm$ 44.39 \\
   &  & H$\beta$ broad & 4861.3   & 6.33 $\pm$1.19  &  2435 $\pm$122  & 72.60 $\pm$ 18.42 \\
   &  & H$\alpha$ broad & 6562.8   & 11.95 $\pm$2.45  & 3758 $\pm$188 & 96.16 $\pm$ 17.11 \\
   & 65 & He\,II $\lambda$4686 & --- & --- & --- & < 24.89\\
   &  & H$\alpha$ broad & --- & --- & --- & < 25.32\\
   &  & H$\beta$ broad & --- & --- & --- & <19.51 \\
   
  \midrule   
  \multicolumn{7}{l}{\textbf{SRGe J131014.7+444319 (ID 26)}} \\
  TDE-H+He & 29 & He\,II $\lambda$4686 & 4685.7 & 40.69  $\pm$6.43  & 17263 $\pm$1726 & 1353$\pm$ 203\\
   &  & H$\beta$ broad & 4861.3   & 12.44$\pm$1.92  & 11616$\pm$1162 & 367.3 $\pm$ 55.1   \\
   &  & H$\alpha$ broad & 6562.8   & 25.13 $\pm$3.90  & 11235$\pm$1124   & 473.7 $\pm$ 51.72  \\
   &  & Bowen blend & 4566.1$\pm$36.5  & 9.12$\pm$1.33 & 5639$\pm$564 & 299.1$\pm$44.87  \\
   & 67 & He\,II $\lambda$4686 & 4685.7 & 33.91 $\pm$5.14  & 12779$\pm$1278  & 690.7 $\pm$ 103.6  \\
   &  & H$\beta$ broad & 4861.3   & 22.44$\pm$3.46   & 11616 $\pm$1026  & 408.3$\pm$61.25    \\
   &  & H$\alpha$ broad & 6562.8   & 48.07$\pm$7.61 & 11103  $\pm$1110  & 606.7$\pm$51.43   \\
   &  & Bowen blend & 4600.3$\pm$27.5  & 9.27$\pm$1.50 & 4220$\pm$422 & 184.5$\pm$27.67\\
   & 244 & He\,II $\lambda$4686 & --- & ---  & --- & < 31.30  \\
   &  & H$\alpha$ broad & ---  & ---  & --- & < 49.26    \\
   &  & H$\beta$ broad & ---  & ---  & --- & < 12.63    \\
   \midrule   
    \multicolumn{7}{l}{\textbf{SRGe J153331.5+390536 (ID 39)}} \\
  --- & 142 & He\,II $\lambda$4686 & 4685.7 &  25.76 $\pm$ 4.99 & 28474$\pm$1424   & 188.7$\pm$150.5 \\
   &  & H$\alpha$ broad & 6541.13$\pm$22.36 & 3.83$\pm$1.95 & 3758$\pm$189 & 42.30$\pm$21.15 \\
  \midrule   
  \multicolumn{7}{l}{\textbf{SRGe J153503.3+455054 (ID 41)}} \\
  TDE-H? & 303 & H$\alpha$ broad & 6562.8 & 9.95$\pm$1.52 & 2679 $\pm$134 & 912.5$\pm$136.2 \\
   &  & He\,II $\lambda$4686 & --- & ---  & --- & <  165.0 \\
  
  \midrule   
  
   \multicolumn{7}{l}{\textbf{SRGe J175023.7+712857 (ID 58)}} \\
  TDE-H+He & 182 & He\,II $\lambda$4686 & 4685.7  & 25.77$\pm$4.86  &  22597$\pm$1130  & 692.6$\pm$142.9 \\
   &  & H$\alpha$ blue & 6477.2$\pm$5.7  & 3.08 $\pm$2.34  & 1255 $\pm$815 & 85.28 $\pm$ 64.21 \\
   &  & H$\alpha$ broad & 6525.0$\pm$10.8   & 8.76$\pm$7.91    & 5365 $\pm$3752  & 241.2$\pm$216.5 \\
   &  & H$\alpha$ red$_1$ & 6594.4$\pm$7.2 & 3.48$\pm$2.36  & 1326$\pm$742 & 94.91$\pm$63.54 \\
   &  & H$\alpha$ red$_2$ & 6651.6$\pm$8.2 & 6.85$\pm$3.22  & 2278$\pm$783 & 184.9$\pm$85.09 \\   
   
   & 1333 & He\,II $\lambda$4686 & --- & ---  & --- & < 26.57  \\
   &  & H$\alpha$ & ---  & ---  & --- & < 31.56  \\
  \midrule   
  \multicolumn{7}{l}{\textbf{SRGe J213527.2-181635 (ID 66)}} \\
  TDE-H+He & 11 & He\,II $\lambda$4686 & 4690.0$\pm$2.67  & 39.45 $\pm$7.52   & 30132$\pm$3366 & 2420 $\pm$ 674   \\
   &  & H$\alpha$ broad & 6588.2$\pm$110.4  & 20.20   $\pm$4.29  & 12029 $\pm$4033 & 1018 $\pm$ 209 \\
   &  & H$\alpha$ rest & 6567.1$\pm$17.2   & 3.82 $\pm$2.35   & 1909 $\pm$1380  & 193$\pm$119 \\
   &  & H$\alpha$ blue & 6472.6$\pm$20.8 & 2.69$\pm$2.02  & 2238    $\pm$1289 & 127$\pm$105 \\
   & 33 & He\,II $\lambda$4686 & 4690.00$\pm$0.85  & 15.18 $\pm$6.32& 22598 $\pm$4185 & 604.2$\pm$251.5\\
   &  & H$\alpha$ broad & 6562.8$\pm$126.5 & 14.90$\pm$4.98    & 9265  $\pm$3765 & 450$\pm$151 \\
   &  & H$\alpha$ rest & 6567.1 $\pm$19.7   & 3.67  $\pm$ 1.89   & 1855 $\pm$209 & 111$\pm$57  \\
   &  & H$\alpha$ blue & 6472.6$\pm$18.9 & 2.84 $\pm$ 1.45 & 2047 $\pm$417 & 83.3$\pm$53.9 \\
   & 302 & He\,II $\lambda$4686 & ---  & ---  & --- & < 43.57 \\
   &  & H$\alpha$ broad & --- & ---  & --- & < 52.19 \\

  \midrule   
  \multicolumn{7}{l}{\textbf{SRGe J235453.0+421711 (ID 70)}} \\
  TDE-He & 197 & He\,II $\lambda$4686 & 4515.7$\pm$12.5 &  69.97 $\pm$ 13.48  & 24262 $\pm$ 7809 & 966.9$\pm$179.2 \\
   &  & H$\beta$ broad & 4861.3   & 5.73 $\pm$2.22   & 1595 $\pm$ 607 & 79.88$\pm$32.15  \\ 
   &  & H$\alpha$ broad & --- & ---  & --- & < 48.81  \\
  \bottomrule
  \end{tabular}
\end{center}
  \textbf{Notes.}
  For SRGe J235453.0+421711 (ID 70), the skewness of He\,II $\lambda$4686 line is $-0.36 \pm 0.29$.
  Given the non-detection of H$\alpha$, we classify this object as TDE-He. 
  While a Gaussian centered on H$\beta$ is favored by BIC ($\Delta {\rm BIC} = 11.42 > 10$), we note that this line detection is of very low significance ($2.6\sigma$). 
  Similarly, given the low significance of line EWs of SRGe J153331.5+390536 (ID 39), we do not assign a spectroscopic subtype for this event. 
\end{table*}

\end{appendix}

\end{document}